\documentclass[fleqn,10pt]{article}
\usepackage{latexsym, graphicx, epsfig, amsmath, amssymb,amsfonts}
\usepackage{natbib,amsthm,version}
\usepackage{amsbsy,bm,multirow,enumerate}
\usepackage[titletoc,page]{appendix}
\usepackage[mathscr]{eucal}
\usepackage{mathtools}
\usepackage{color}
\usepackage[utf8]{inputenc}
\usepackage[english]{babel}
\usepackage{amsthm}
\usepackage{enumerate,enumitem}
\usepackage[hidelinks]{hyperref}
\usepackage{url}
\usepackage{subfigure}
\usepackage[lined,boxed,linesnumbered,ruled]{algorithm2e}
\usepackage{float}
\usepackage{arydshln}
\usepackage{dsfont,bbm}

\DeclareMathOperator*{\argmin}{arg\,min}

\newcommand{\mbs}[1]{\mathbf{#1}}

\newtheorem{remark}{Remark}[section]

\theoremstyle{definition}

\title{
  On Computing the Hyperparameter of Extreme Learning Machines:
  Algorithm and Application to Computational PDEs,
  and Comparison with Classical and High-Order Finite Elements
} 
\author{
  Suchuan Dong\thanks{Author of correspondence. Email: sdong@purdue.edu},
  \ \ Jielin Yang
  \\
  Center for Computational and Applied Mathematics \\
  Department of Mathematics \\
  Purdue University, USA 
 } 

\date{(October 26, 2021)}

\begin{document}
\maketitle


\begin{abstract}

  We consider the use of extreme learning machines (ELM) for
  computational partial differential equations (PDE).
  In ELM the hidden-layer
  coefficients in the neural network are assigned to random values generated on
  $[-R_m,R_m]$ and fixed, where $R_m$ is a user-provided
  constant, and the output-layer coefficients are trained by
  a linear or nonlinear least squares computation.
  We present a method for computing the optimal or near-optimal
  value of $R_m$ based on the differential evolution algorithm.
  This method seeks the optimal $R_m$ by minimizing the
  residual norm of the linear or nonlinear algebraic system
  that results from the ELM representation of the PDE solution and that
  corresponds to the ELM least squares solution to the system.
  It amounts to a pre-processing procedure that determines a
  near-optimal $R_m$, which can be used in ELM for solving
  linear or nonlinear PDEs.
  The presented method enables us to illuminate the characteristics of the optimal
  $R_m$ for two types of ELM configurations: (i) Single-Rm-ELM,
  corresponding to the conventional ELM method in which a single $R_m$
  is used for generating the random coefficients in all the hidden layers,
  and (ii) Multi-Rm-ELM, corresponding to a modified ELM method
  in which multiple $R_m$ constants are involved with each used for
  generating the random coefficients of a different hidden layer.
  We adopt the optimal $R_m$ from this method and also incorporate other
  improvements into the ELM implementation.
  In particular, here
  we compute all the differential operators involving
  the output fields of the last hidden layer by a forward-mode
  auto-differentiation, as opposed to the reverse-mode auto-differentiation
  in a previous work. These improvements significantly reduce
  the network training time and enhance the ELM performance.
  We systematically compare the computational
  performance of
  the current improved ELM with that of the finite element method
  (FEM), both the classical second-order FEM and the high-order FEM
  with Lagrange elements of higher degrees, for solving a number of linear
  and nonlinear PDEs. It is shown that the current improved ELM far outperforms the
  classical FEM. Its computational performance is comparable to
  that of the high-order FEM for smaller problem sizes, and for larger
  problem sizes the ELM markedly outperforms the high-order FEM.

\end{abstract}


\vspace{0.05cm}
Keywords: {\em
  extreme learning machine,
  local extream learning machine,
  neural network,
  least squares,
  nonlinear least squares,
  differential evolution
}

\tableofcontents


\section{Introduction}
\label{sec:intro}

This work focuses on the use of extreme learning machines (ELM)
for computational partial differential equations (PDE).
ELM was originally developed in~\cite{HuangZS2006,HuangCS2006} for linear
classification/regression problems with single hidden-layer
feed-forward neural networks, and has since found wide
applications in a number of fields; see~the reviews
in~\cite{HuangHSY2015,Alabaetal2019} and the references therein.
Two strategies underlie ELM: (i) random but fixed (non-trainable)
hidden-layer coefficients,
and (ii) trainable linear output-layer coefficients
determined by a linear least squares method or by using
the pseudo-inverse of the coefficient matrix (for linear problems).
The ELM type idea has also been developed for nonlinear
problems; see e.g.~\cite{DongL2020,DongL2021} for
solving stationary and time-dependent nonlinear PDEs in which
the neural network is trained by a nonlinear least squares method.
Following~\cite{DongL2021}, we broadly refer to
the artificial neural network-based methods exploiting these strategies
as ELM type methods, including those employing neural networks with
multiple hidden layers and those for nonlinear
problems~\cite{TangDH2015,TisseraM2016,DongL2020,DongL2021,FabianiCRS2021}.


Our research is motivated by the following questions:
\begin{itemize}
\item Can artificial neural networks provide a competitive technique for scientific
  computing and in particular for computational PDEs?
\item Can we devise a neural network-based method for approximating PDEs
  that can  outcompete traditional numerical techniques in computational performance? 
\end{itemize}
These questions have been hanging in the air
ever since the early studies on neural networks
for differential equations in the 1990s
(see e.g.~\cite{LeeK1990,MeadeF1994,MeadeF1994b,YentisZ1996,LagarisLF1998}).
The remarkable success of deep learning~\cite{GoodfellowBC2016} in the last decade or so 
has stimulated a significant amount of efforts in the development
of deep neural network (DNN) based PDE solvers and scientific
machine learning~\cite{Karniadakisetal2021,SirignanoS2018,RaissiPK2019,EY2018,WanW2020}.
By exploiting the universal approximation property of neural
networks~\cite{HornikSW1989,HornikSW1990,Cotter1990,Li1996},
these solvers transform the PDE solution problem into an optimization
problem. The field function is represented by a neural network,
whose weight/bias coefficients are adjusted to minimize an appropriate
loss function. The differential equation, the boundary and initial
conditions are then encoded into the loss function by penalizing some
residual norms of these quantities in strong or weak
forms~\cite{LagarisLF1998,LagarisLP2000,SirignanoS2018,RaissiPK2019,EY2018,ZangBYZ2020,KrishnapriyanGZKM2021}.
The differential operators involved therein are usually computed
analytically with shallow neural networks in the early works
(see e.g.~\cite{LagarisLF1998}), and in modern implementations
they are typically computed by auto-differentiation~\cite{BaydinPRS2018}
available from the common machine learning libraries
such as Tensorflow (www.tensorflow.org)
and PyTorch (pytorch.org). The minimization of the loss function is
performed by an optimizer, which is usually based on some flavor of
gradient descent or back propagation type techniques~\cite{Werbos1974,Haykin1999}.
The optimization
process constitutes the predominant computations in the
neural network-based PDE solvers, and it is commonly known as
the training of the neural network. Upon convergence of the optimization process,
the PDE solution is given by the neural network, with
the adjustable weight/bias parameters set based on their
converged values.
A number of prominent works on DNN-based PDE solvers have appeared
in the literature, and we refer the reader to
e.g.~\cite{RaissiPK2019,SirignanoS2018,EY2018,RuddF2015,WinovichRL2019,HeX2019,KharazmiZK2019,ZangBYZ2020,WangL2020,Samaniegoetal2020,LiTWL2020,JagtapKK2020,TangWL2021,CaiCLL2020,WangYP2020,DongN2020}
(among others), and also the review article~\cite{Karniadakisetal2021}
and the references contained therein.

As discussed in~\cite{DongL2020},
while their computational performance is promising, the existing
DNN-based PDE solvers suffer from several drawbacks: limited
accuracy, general lack of convergence with a certain convergence rate,
and extremely high computational cost (very long time to train).
We refer the reader to~\cite{DongL2020} for more detailed discussions
of these aspects.
These drawbacks make such solvers numerically less than
attractive and computationally uncompetitive.
There is mounting evidence  that
these solvers, in their current state, seem to fall short
and cannot compete with traditional numerical methods for
commonly-encountered computational PDE
problems (especially in low dimensions)~\cite{DongL2020}.

The pursuit for higher accuracy and more competitive performance
with neural networks for computational PDEs has led us
in~\cite{DongL2020,DongL2021} to explore 
randomized neural networks (including ELM)~\cite{ScardapaneW2017,FreireRB2020}.
Since optimizing the entire set of weight/bias coefficients
in the neural network can be extremely hard and costly,
perhaps randomly assigning and fixing a subset of
the network's weights will make the resultant optimization task
of network training simpler, and ideally linear, without severely
sacrificing the achievable approximation capacity.
This is the basic strategy in randomizing the neural networks.
In ELM one assigns random values to and fixes the hidden-layer coefficients,
and only allows the output-layer (assumed to be linear)
coefficients to be trainable.
For linear problems, the resultant system becomes linear
with respect to the output-layer coefficients, which can then
be determined by a linear least squares
method~\cite{HuangZS2006,PanghalK2020,DwivediS2020,DongL2020,DongL2021,CalabroFS2021}.
Random-weight neural networks similarly possess a universal approximation property.
It has been shown in~\cite{HuangCS2006} that a single hidden-layer
feed-forward neural network having random but fixed (not trained) hidden nodes
can approximate any continuous function to any desired
degree of accuracy, provided that the number of hidden units
is sufficiently large.


In~\cite{DongL2020} we have developed a local version
of the ELM method (termed locELM) for solving linear
and nonlinear PDEs, which combines
the ideas of ELM, domain decomposition, and local neural networks.
%
We use a local feed-forward neural network to represent the field
solution on each sub-domain, and impose $C^k$ (with an appropriate $k$)
continuity conditions on the sub-domain boundaries.
The weight/bias coefficients in all the hidden layers of the local neural networks
are preset to random values generated on $[-R_m,R_m]$ ($R_m$ denoting
a user-provided constant) and fixed, and the weights of the linear
output layers of the local networks are trained by
a linear least squares method for linear PDEs or by a nonlinear least
squares method for nonlinear PDEs.
For time-dependent linear/nonlinear PDEs, we have introduced a block
time marching scheme together with locELM for long-time dynamic
simulations.
Note that locELM reduces to (global) ELM if a single subdomain is used
in the domain decomposition.
Most interesting is that locELM is highly accurate and computationally fast.
This method exhibits a clear sense of convergence with respect to
the degrees of freedom in the neural network. For smooth PDE solutions,
its numerical errors decrease exponentially or nearly exponentially
as the number of training parameters or the number of training data points
increases, reminiscent of the spectral convergence of traditional
high-order methods such as the spectral, spectral element or hp-finite element
(high-order finite element) type
techniques~\cite{KarniadakisS2005,SzaboB1991,ZhengD2011,YuKK2017,DongS2012,Dong2018,Dong2015clesobc,LinYD2019,YangD2020}.
When the number of degrees of freedom (number of training collocation points,
number of training parameters) becomes large, the errors of locELM can
reach a level close to the machine zero.

More importantly, it is shown in~\cite{DongL2020} that the computational
performance (accuracy, computational cost) of locELM is on par
with that of the classical FEM (2nd order, linear elements), and
locELM outperforms the classical FEM for larger problem sizes.
Here for locELM the computational cost refers to the time for training
the neural network and related computations, and for FEM it refers to
the FEM computation time (see~\cite{DongL2020} for details).
By ``outperform'' we mean that one method achieves a better accuracy
under the same computational budget/cost or incurs a lower computational
cost to achieve the same accuracy.
More specifically, there is a cross-over point in the relative
performance between locELM and the FEM with respect to the problem size.
The classical FEM typically outperforms locELM for smaller
problem sizes, and for larger problem sizes locELM outperforms the classical
FEM~\cite{DongL2020}.

Some comparisons between locELM and the high-order FEM (employing high-order
Lagrange elements) for the 2D Poisson equation have also
been conducted in~\cite{DongL2020}.
It is observed that the locELM method can outperform the
Lagrange elements of degree $2$, but can barely outperform
the Lagrange elements of degree $3$. The method as implemented
in~\cite{DongL2020} cannot outcompete high-order FEM with element degrees
$4$ or larger. Overall the method of~\cite{DongL2020} seems competitive
to some degree when compared with the high-order FEM, but it is
in general not as efficient as the latter as of
the writing of~\cite{DongL2020}.
This inefficiency in comparison with high-order FEM is the primary motivator
for the current work.

We would like to mentioned that in~\cite{DongL2020}
a systematic comparison between locELM and two state-of-the-art
DNN-based PDE solvers, the deep Galerkin method (DGM)~\cite{SirignanoS2018}
and the physics-informed neural network (PINN) method~\cite{RaissiPK2019},
has also been performed. It is shown that locELM outperforms
PINN and DGM by a considerable degree. The numerical errors and the computational
cost (network training time) of locELM are considerably smaller,
typically by orders of magnitude, than those of DGM and PINN.


In the current paper we present improvements to the ELM technique
(which also apply to
locELM) in two aspects. First, we present a method
for computing the optimal (or near-optimal) value of
the $R_m$ constant in ELM, i.e.~the maximum magnitude of
the random hidden-layer coefficients.
Note that in~\cite{DongL2020}
$R_m$ is estimated by using a preliminary simulation with manufactured
solutions for a given problem. The method presented here is based on
the differential evolution algorithm~\cite{StornP1997},
and seeks the optimal $R_m$ by minimizing the residual norm of
the linear/nonlinear algebraic system that results from the ELM representation
of the PDE solution and that corresponds to the ELM least squares solution
to the system. This method amounts to a pre-processing procedure
that determines a near-optimal value for $R_m$, which can be used
in ELM for solving linear or nonlinear PDEs.

The procedure for computing the optimal $R_m$
enables us to investigate two types of ELM methods
based on how the random hidden-layer coefficients are assigned:
Single-Rm-ELM and Multi-Rm-ELM.
The Single-Rm-ELM configuration corresponds to the conventional ELM method,
in which the weight/bias coefficients in all the hidden layers are
set to random values generated on $[-R_m,R_m]$, with a single $R_m$ for all
hidden layers. The Multi-Rm-ELM configuration corresponds to a modified ELM method,
in which the weight/bias coefficients for any given hidden layer
are set to random values generated on $[-R_m,R_m]$, with a different $R_m$ value
for a different hidden layer. Therefore, multiple $R_m$ constants are involved
in Multiple-Rm-ELM for assigning the random coefficients,
with each corresponding to a different hidden layer.
The characteristics of the optimal $R_m$ corresponding to
these two types of ELM configurations are studied and illuminated.
The Multi-Rm-ELM configuration leads to more accurate simulation
results than Single-Rm-ELM.

The second aspect of improvement is in the implementation of the ELM method.
A major change in the current work lies in that here  we compute
all the differential operators
involving the output fields of the last hidden layer of the neural network
by the forward-mode auto-differentiation, implemented by the
``ForwardAccumulator'' in the Tensorflow library.
In contrast, in the ELM implementation of~\cite{DongL2020} these
differential operators are computed by the default reverse-mode
auto-differentiation (``GradientTape'') in Tensorflow.
Because in ELM the number of nodes in the last hidden layer is typically
much larger than that of the input layer,
this change reduces the ELM network training time dramatically.

Building upon these improvements, in the current paper
we systematically compare the current
improved ELM with the classical FEM (2nd-order)
and the high-order FEM employing Lagrange elements of higher degrees
(with both h-type and p-tye refinements)
for solving a number of linear and nonlinear PDEs.
We show that the improved ELM far outperforms the classical FEM.
The ELM's computational performance is comparable to that of the high-order FEM
for smaller problem sizes, and for larger problem sizes the ELM
markedly outperforms the high-order FEM.
Overall, the current ELM method is computationally far more competitive
than the classical FEM, and is more competitive than or as competitive as 
the high-order FEM.

As in~\cite{DongL2020}, the ELM method here is implemented in Python using the
Tensorflow and Keras (keras.io) libraries.
The classical FEM and the high-order FEM are implemented in Python
using the FEniCS library, in which the linear and higher-order Lagrange
elements are available.

The rest of this paper is structured as follows.
In Section \ref{sec:method} we present the method for computing
the optimal $R_m$ constant(s) with the Single-Rm-ELM and Multi-Rm-ELM
configurations based on the differential evolution algorithm.
In Section \ref{sec:tests} we investigate the
characteristics of the optimal $R_m$
and study the effect of the network/simulation
parameters on the optimal $R_m$
for function approximation and for solving linear/nonlinear PDEs.
We compare systematically the computational performance
of the ELM and the classical/high-order FEMs for solving
the differential equations.
In Section \ref{sec:summary} we summarize the common characteristics
of the optimal $R_m$ in Single-Rm-ELM and Multi-Rm-ELM
and also the performance comparisons between
ELM and classical/high-order
FEM to conclude the presentations.


\section{Computing the Optimal $R_m$ Constant(s) in ELM}
\label{sec:method}

\subsection{The Maximum Magnitude of Random Coefficients ($R_m$)}

When the ELM method is used to
to solve linear/nonlinear PDEs, the hidden-layer
coefficients are set to uniform random values generated on
the interval $[-R_m,R_m]$, where $R_m$ is a user-defined constant
(see~\cite{DongL2020,DongL2021}).
The $R_m$ constant is a hyperparameter of the ELM method.
Its value can have a marked influence on the
accuracy of the ELM results~\cite{DongL2020}. The best accuracy is typically
associated with $R_m$ from a range of moderate values,
while very large or very small $R_m$ values
can result in simulation results with poor
or poorer accuracy (see~\cite{DongL2020} for details).
For a given problem, in~\cite{DongL2020}
the $R_m$ (or the optimal range of $R_m$) is estimated
by preliminary simulations using some manufactured solution.


One goal of this work is to devise a method to attain
the optimal or near-optimal value of the constant
$R_m$ for a given problem.
This method enables us to explore and study the characteristics
of  and the effects of the simulation/network parameters
on the optimal $R_m$. In particular, it enables us to look into 
a modified ELM method, which employs multiple $R_m$
constants, 
each for generating the random coefficients of a different hidden layer,
in a deeper neural network.
The use of multiple $R_m$ constants  in ELM
leads to more accurate
simulation results. It would be extremely difficult,
and often practically impossible, to determine
an optimal vector of $R_m$ values without the automatic procedure.

The method for computing the optimal $R_m$ presented here amounts to
a pre-processing procedure. It only needs to be performed
when a given problem (or some configuration)
is considered for the first time.
The $R_m$ in ELM can then be fixed to the value obtained by the method
in subsequent computations.
As will become clear in later sections, the optimal $R_m$ is insensitive
to the number of training collocation points and has a quite weak dependence
on the number of training parameters.
In addition, the $R_m$ values in a range around the optimal $R_m$
lead to comparable and essentially the same accuracy
as the optimal $R_m$.
Therefore, for a given problem one can basically employ a specific set of collocation
points and a specific number of training parameters to compute
the optimal $R_m$ using the method developed here.
Then in subsequent computations
one can fix the $R_m$ in ELM at the returned value
for a different set of collocation points or
different number of training parameters.
This is especially useful in long-time dynamic
simulations of time-dependent PDEs by the block time marching scheme~\cite{DongL2020}.
The pre-processing computation for the optimal $R_m$ only needs to be performed using
the spatial-temporal domain of the first time block.
Then the $R_m$ fixed at the returned value can be used for ELM computations
on all the time blocks.

\subsection{ELM Configuration with a Single $R_m$ Constant (Single-Rm-ELM)}
\label{ord_elm}

We now develop a procedure for computing the optimal $R_m$ constant
in ELM for solving partial differential equations.
Consider a domain $\Omega$ in $d$ ($d=1$, $2$ or $3$) dimensions,
and the following generic boundary value problem on $\Omega$,
\begin{subequations}\label{eq_1}
  \begin{align}
    & \mbs L(\mbs u) = \mbs f(\mbs x), \label{eq_1a} \\
    & \mbs B(\mbs u) = \mbs g(\mbs x), \quad \text{on}\ \partial\Omega, \label{eq_1b}
  \end{align}
\end{subequations}
where $\mbs L$ and $\mbs B$ are differential (or algebraic) operators
that may be linear or nonlinear, $\mbs u(\mbs x)$ is the field function to be solved for,
$\mbs f$ and $\mbs g$ are given functions (source
terms)  in the domain $\Omega$
or on the domain boundary $\partial\Omega$.
We assume that this problem is well-posed.

We solve this problem using a feed-forward neural network by
the ELM method (see~\cite{DongL2020}).
Let the vector $[M_0, M_1, \dots, M_L]$ denote the architecture
of the neural network, where $(L+1)$ is the number of
layers in the neural network with $L\geqslant 2$,
and $M_i\geqslant 1$ denotes the number of
nodes in layer $i$ ($0\leqslant i\leqslant L$).
The input layer (i.e.~layer $0$) to the neural network
represents the coordinate $\mbs x$, with $M_0=d$.
The output layer (i.e.~layer $L$) represents the field solution
$\mbs u(\mbs x)$, with $M_L$ being the dimension of $\mbs u$.
Those layers in between are the hidden layers.

The neural network logically represents a parameterized function
constructed through repeated function compositions
with a nonlinear activation function  and repeated affine
transforms~\cite{GoodfellowBC2016}.
More specifically, we choose a set of $Q$ discrete points (collocation points)
on the domain $\Omega$, among which $Q_b$ ($1\leqslant Q_b<Q$) points
reside on the boundary $\partial\Omega$, and these are the training data points.
We use
\begin{equation}
  Z = \{\mbs x_i\in\Omega, \ 1\leqslant i\leqslant Q \}, \quad\text{and}\
  Z_b = \{\mbs x_i\in Z\ \text{and}\ \mbs x_i\in\partial\Omega,\ 1\leqslant i\leqslant Q_b  \}
\end{equation}
to denote the set of all collocation points and and set of boundary
collocation points, respectively.
Let the matrix $\mbs X$ of dimension
$Q\times M_0$ denote the coordinates of the collocation points,
which are the input data to the network.
Let the matrix $\mbs U$ of dimension $Q\times M_L$ denote
the output data of the neural network, which represent $\mbs u(\mbs x)$
on the collocation points.
Let the matrix $\bm\Phi_l$ of dimension $Q\times M_l$
denote the output data of layer $l$ ($0\leqslant l\leqslant L$),
with $\bm\Phi_0=\mbs X$ and $\bm\Phi_L=\mbs U$.
Then the logic of the hidden layer $l$ ($1\leqslant l\leqslant L-1$)
is represented by
\begin{equation}\label{eq_2}
  \bm\Phi_l = \sigma\left(\bm\Phi_{l-1}\mbs W_l + \mbs b_l  \right),
  \quad 1\leqslant l\leqslant L-1,
\end{equation}
where $\sigma(\cdot)$ is the activation function,
$\mbs W_l$ is a constant matrix of dimension $M_{l-1}\times M_l$ representing
the weight coefficients of layer $l$, and $\mbs b_l$ is a
row vector of dimension $1\times M_l$ representing the biases of
this layer. Note that we have used the convention
(as in the Python language) here that
when computing the right hand side of \eqref{eq_2} the data in $\mbs b_l$
will be propagated along the first dimension to form a $Q\times M_l$ matrix.

With the ELM method, we
follow~\cite{DongL2020} to set the weight/bias coefficients
in all the hidden layers, $(\mbs W_l, \mbs b_l)$ for $1\leqslant l\leqslant L-1$,
to uniform random values generated on $[-R_m,R_m]$, and fix their
values (not trainable) once they are set.
The output layer is required to be linear, i.e.~without the
activation function, with zero bias.
The logic of the output layer is given by,
\begin{equation}\label{eq_4}
  \mbs U = \bm\Phi_{L-1}\mbs W_L
\end{equation}
where $\mbs W_L$ is a $M_{L-1}\times M_L$ constant matrix denoting the
weights of the output layer, which are the trainable parameters of
the ELM neural network.

On the continuum level, the relation \eqref{eq_4}
becomes the following in terms of the coordinate $\mbs x$,
\begin{equation}\label{eq_5}
  u_i(\mbs x) = \sum_{j=1}^M V_j(\mbs x) \beta_{ji},
  \quad 1\leqslant i\leqslant M_L,
\end{equation}
where $\mbs u = (u_1,u_2,\dots,u_{M_L})$,
$\mbs W_L=\left[\beta_{ij}\right]_{M_{L-1}\times M_L}$, and
$M=M_{L-1}$ denotes the number of nodes in the last hidden layer, which
can be large in the ELM neural network.
$V_j(\mbs x)$ ($1\leqslant j\leqslant M$) are the output
fields of the last hidden layer, whose data on the collocation points are given
by $\bm\Phi_{L-1}$.


Substituting the expression \eqref{eq_5} for $\mbs u(\mbs x)$
into the system \eqref{eq_1}, enforcing \eqref{eq_1a} on
all the collocation points in $Z$ and enforcing \eqref{eq_1b}
on all the boundary collocation points in $Z_b$, we have
\begin{subequations}\label{eq_6}
  \begin{align}
    & \mbs L\left(\sum_{j=1}^MV_j(\mbs x_p)\bm\beta_j  \right) = \mbs f(\mbs x_p),
    \quad \text{for all}\ \mbs x_p\in Z, \ 1\leqslant p\leqslant Q, \\
    & \mbs B\left(\sum_{j=1}^MV_j(\mbs x_q)\bm\beta_j   \right) = \mbs g(\mbs x_q),
    \quad \text{for all}\ \mbs x_q\in Z_b, \ 1\leqslant q\leqslant Q_b,
  \end{align}
\end{subequations}
where $\bm\beta_j = (\beta_{j1}, \beta_{j2},\dots,\beta_{jM_L})$.
This is a system of $(Q+Q_b)$ algebraic equations about the training parameters
$\mbs W_L=\left[\beta_{ij} \right]_{M_{L-1}\times M_L}$.
This system is linear with respect to $\beta_{ij}$ if the original system~\eqref{eq_1}
is linear with respect to $\mbs u$, and it is nonlinear if the original
system~\eqref{eq_1} is nonlinear.
Following \cite{DongL2020},
we solve this algebraic system \eqref{eq_6} for $\beta_{ij}$
by the linear least squares (LLSQ)
method if it is linear, and by the
nonlinear least squares method with perturbations (NLLSQ-perturb) if it is
nonlinear; see~\cite{DongL2020} for details.

What has been discussed so far is the main idea
of the ELM method from~\cite{DongL2020}
for solving the system~\eqref{eq_1}. Let us now consider how to
determine the optimal value for the $R_m$ constant.

We first modify the generation of the random
hidden-layer coefficients as follows. Let
$
N_h = \sum_{l=1}^{L-1}\left(M_{l-1}+1\right)M_l
$
denote the total number of hidden-layer coefficients (weights/biases)
in the neural network.
We first generate a set of $N_h$ uniform random values on the interval
$[-1,1]$, which will be denoted by the vector $\bm\xi$ of length $N_h$.
Once $\bm\xi$ is generated, it will be fixed throughout the computation.
Given a constant $R_m$, the vector $R_m\bm\xi$ contains a set of $N_h$
random values on $[-R_m,R_m]$, and we will set
the random hidden-layer coefficients,
$(\mbs W_l,\mbs b_l)$ for $1\leqslant l\leqslant L-1$,
by the random vector $R_m\bm\xi$.
With this modification the $R_m$ constant becomes a scaling coefficient,
and it is not confined to positive values. In
the following discussions $R_m$ can in principle
assume positive, zero, or negative values.

Let $\beta_{ij}^{LS}$ ($1\leqslant i\leqslant M_{L-1}$, $1\leqslant j\leqslant M_L$)
denote the lest squares solution to the system~\eqref{eq_6}
obtained by the linear or nonlinear least squares method.
Define the residual vector $\mbs r$, with length $(Q+Q_b)$,
of the system~\eqref{eq_6} at the least
squares solution,
\begin{equation}\label{eq_7}
  \mbs r(R_m) = \left[\begin{array}{c}
    \vdots \\
    \mbs L\left(\sum_{j=1}^MV_j\left(\mbs x_p,R_m\right)\bm\beta_j^{LS}  \right) - \mbs f(\mbs x_p) \\
    \vdots \\ \hdashline[2pt/2pt]
    \vdots \\
    \mbs B\left(\sum_{j=1}^MV_j\left(\mbs x_q,R_m\right)\bm\beta_j^{LS}   \right) - \mbs g(\mbs x_q) \\
    \vdots
    \end{array}\right]
  = \left[\begin{array}{c}
    \vdots \\
    \mbs L\left(\mbs u^{LS}(\mbs x_p, R_m)  \right) - \mbs f(\mbs x_p) \\
    \vdots \\ \hdashline[2pt/2pt]
    \vdots \\
    \mbs B\left(\mbs u^{LS}(\mbs x_q, R_m)   \right) - \mbs g(\mbs x_q) \\
    \vdots
    \end{array}\right]
\end{equation}
where $\bm\beta_j^{LS} = \left(\beta_{j1}^{LS}, \beta_{j2}^{LS},\dots,\beta_{jM_L}^{LS}\right)$,
$\mbs u^{LS}=\sum_{j=1}^M V_j\bm\beta_j^{LS}$ is the least squares solution,
$\mbs x_p\in Z$ and $\mbs x_q\in Z_b$.
Note that here the dependence of the residual $\mbs r (R_m)$,
the output fields of the last hidden layer $V_j(\mbs x,R_m)$,
and the least squares solution $\mbs u^{LS}(\mbs x,R_m)$ on the scaling
coefficient $R_m$ 
has been made explicit.
Let
\begin{equation}\label{eq_8}
  \mathcal{N}(R_m) = \left\|\mbs r(R_m)  \right\|
\end{equation}
denote the Euclidean norm of the residual vector $\mbs r(R_m)$.

We seek the optimal value $R_{m0}$ for $R_m$
such that the norm $\mathcal{N}(R_m)$ is minimized, i.e.
\begin{equation}\label{eq_9}
  R_{m0} = \argmin_{R_m} \mathcal{N}(R_m).
\end{equation}
This is an optimization problem of a scalar
function with a single variable. Note that the derivative
$\mathcal{N}'(R_m)$ may be approximated by finite difference.
But in general $\mathcal{N}'(R_m)$
cannot be computed directly based on the system~\eqref{eq_6},
if this system is nonlinear. This is because of the nonlinear
least squares solution $\beta_{ij}^{LS}$ involved therein.
If this system is linear 
$\mathcal{N}'(R_m)$ can be computed from~\eqref{eq_6} in principle.

A number of methods can be used for solving
the problem~\eqref{eq_9}.
In the current work we adopt the differential evolution (DE)
algorithm~\cite{StornP1997} for computing $R_{m0}$.
Differential evolution is
a population-based
evolutionary (genetic) algorithm for continuous functions.
The implementation of differential evolution is
available in several scientific libraries,
e.g.~the Tensorflow-Probability library (www.tensorflow.org/probability)
and the scipy library.
In the current paper we employ the scipy implementation
for the differential evolution algorithm. 
This algorithm requires, as an input argument,
a routine for evaluating the cost function $\mathcal{N}(R_m)$ for
any given $R_m$. The procedure for computing $\mathcal{N}(R_m)$
is summarized in Algorithm~\ref{alg_1}.

\begin{algorithm}[tb]
  \DontPrintSemicolon
  \SetKwInOut{Input}{input}\SetKwInOut{Output}{output}

  \Input{$R_m$; input data $\mbs X$ to neural network;
    fixed random vector $\bm \xi$, of length $N_h$ containing
    uniform random values on $[-1,1]$.
  }
  \Output{$\mathcal{N}(R_m)$.}
  \BlankLine\BlankLine
  update the hidden-layer coefficients, $(\mbs W_l,\mbs b_l)$ for $1\leqslant l\leqslant L-1$,
  by $R_m\bm\xi$\;
  compute $\bm\Phi_{L-1}$ by evaluating the neural network (first $(L-1)$ layers) on
  the input data $\mbs X$\;
  solve system \eqref{eq_6} for $\beta_{ij}^{LS}$ by the linear or nonlinear least squares method
  from~\cite{DongL2020}\;
  update the output-layer coefficients $\mbs W_L$
  by $\beta_{ij}^{LS}$ ($1\leqslant i\leqslant M_{L-1}$,
  $1\leqslant j\leqslant M_L$)\;
  compute $\mbs u^{LS}$, $\mbs L(\mbs u^{LS})$ and $\mbs B(\mbs u^{LS})$ by evaluating
  the neural network on $\mbs X$ and by auto-differentiation\;
  compute the residual vector $\mbs r(R_m)$ by equation \eqref{eq_7}\;
  compute $\mathcal{N}(R_m)$ by equation \eqref{eq_8}
  
  \caption{Computing the cost function $\mathcal{N}(R_m)$ in Single-Rm-ELM}
  \label{alg_1}
\end{algorithm}

\begin{remark}\label{rem_1}
  If the system \eqref{eq_6} is linear, then in line $2$ of the Algorithm~\ref{alg_1}
  one also needs to the compute the derivatives of $\bm\Phi_{L-1}$ involved in
  the $\mbs L(\mbs u)$ and $\mbs B(\mbs u)$ operators. This can be done by
  auto-differentiation.
  In this case, in line $5$ of Algorithm~\ref{alg_1}, one can compute
  $\mbs u^{LS}$ by equation \eqref{eq_4}, and compute $\mbs L(\mbs u^{LS})$
  and $\mbs B(\mbs u^{LS})$ by multiplying $\mbs W_L$ to the appropriate
  derivatives already computed in line $2$.

  When solving a linear system~\eqref{eq_6}, it is important to avoid implementations
  of linear least squares solvers employing normal equations (e.g.~the Tensorflow's
  ``lstsq'' routine with the default ``fast'' option), which can
  lead to severe ill-conditioning and significantly lower
  accuracy,   even if the original system is only
  moderately ill-conditioned~\cite{GillMW2021}.
  In the current paper (and
  in~\cite{DongL2020}), we employ the linear least squares
  routine from the LAPACK, available through the wrapper function
  in the scipy package (``scipy.linalg.lstsq'').

\end{remark}

\begin{remark}\label{rem_2}
  The computation for $R_{m0}$ amounts to a pre-processing procedure, which can be
  performed when a given problem setting or neural network setup is considered
  for the first time.
  In subsequent computations the $R_m$ in ELM can be fixed to
  the attained $R_{m0}$ (or a value nearby).
  Numerical experiments indicate that the optimal $R_m$
  in Single-Rm-ELM is not sensitive
  to the number of collocation points and only weakly depends on the number of
  training parameters, and that the $R_m$ values in a range
  around the optimum $R_{m0}$ lead to essentially the same accuracy as $R_{m0}$.
  Therefore, in general one can use a specific set of degrees of freedom
  (number of collocation points and training parameters) in the pre-processing
  run to compute the optimal $R_m$. Then the $R_m$ fixed at the
  obtained value can 
  be used in subsequent computations with other sets
  degrees of freedom or simulation parameters.
  Oftentimes one would like to perform a series of
  simulations with a simulation parameter
  varied in a range of values, e.g.~varying the number of training parameters
  between $10$ and $500$.
  In this case,
  one can compute the $R_{m0}$ in the pre-processing run with a
  representative value for this parameter. One can typically use
  a larger value (e.g.~the largest or close to the largest value)
  for this simulation parameter. The resultant $R_{m0}$ can be used
  for $R_m$ to perform the planed series of simulations.
  
  When computing the $R_{m0}$ during pre-processing
  we require that the number of collocation
  points be sufficiently large such that in the system~\eqref{eq_6}
  the number of equations is larger than the number of unknowns.
  This is to try to
  avoid the regime where, if the system is linear,
  its coefficient matrix may become rank deficient.
    
\end{remark}

\begin{remark}\label{rem_rand}
  The methods for computing the optimal $R_m$ and for solving PDEs using ELM
  all involve random number generators.
  Given a specific problem to be solved,
  in the current work we require that  the seed for the random number
  generator be fixed when computing the $R_{m0}$ during pre-processing
  and when using the ELM with a fixed $R_m$ to solve the
  PDE subsequently. In other words,
  the random number generator should be initialized
  by the same fixed seed in all these computations.
  The specific seed value is unimportant, as long as
  the random numbers are all generated by that seed. 
  This is to ensure that the random coefficients of the neural network
  in the pre-processing run for computing the $R_{m0}$, and in the subsequent ELM
  runs using the attained $R_{m0}$ value, are generated by the same seed.
  A fixed and known seed value also makes all the ELM simulation
  results (their numerical values)
  deterministic and exactly reproducible.
  We follow this convention in all the numerical experiments
  reported in Section \ref{sec:tests}.

\end{remark}

\begin{remark}
  In addition to differential evolution, we have also considered the
  simplicial homology global optimization (SHGO) algorithm~\cite{EndresSF2018}
  (implementation also available from scipy)
  for computing the $R_{m0}$. 
  The results from SHGO and from differential evolution are comparable.
  We only consider the results from the differential evolution algorithm
  in Section \ref{sec:tests}.
  When one uses the differential evolution implementation from scipy
  a pair of  values, $(R_{\min},R_{\max})$, needs to be provided to serve as
  the lower/upper bounds for the range of $R_m$ values.
  
\end{remark}

\begin{remark}\label{rem_rm}
  One should note that the computed $R_{m0}$ is
  but a reference value in practice.
  The $R_m$ values in a neighborhood of $R_{m0}$ typically lead to
  simulation results with comparable or essentially the same accuracy.
  Therefore, in ELM simulations
  one can usually employ a ``nicer'' $R_m$ value that is close to $R_{m0}$,
  instead of $R_{m0}$ itself. For example, with an $R_{m0}=1.24931$
  obtained from the method, one can typically employ
  $R_m=1.25$ in subsequent ELM computations to attain results with
  the same or similar accuracy.

\end{remark}

\begin{remark}\label{rem_6a}
  For a nonlinear system~\eqref{eq_6}, 
  when computing $R_{m0}$ one can turn off 
  the initial-guess perturbations and the associated sub-iterations
  in the NLLSQ-perturb method (see~\cite{DongL2020} for details).
  In other words, in the $R_{m0}$ computation
  we can solve the nonlinear algebraic system~\eqref{eq_6}
  for $\beta_{ij}$
  using the nonlinear least squares method  without perturbations.
  This is because only the relative merits
  of different $R_m$ values are important when computing the optimal $R_m$.
  Once the $R_{m0}$ is obtained, one can turn on the
  perturbations in the subsequent ELM computations with the NLLSQ-perturb method.

\end{remark}

\begin{remark}\label{rem_7}
  If the PDE is time-dependent and the temporal domain $[0,T]$ is small,
  we can treat the time variable $t$ in the same way as
  the spatial coordinate $\mbs x$,
  e.g.~by treating $t$ as the $(d+1)$-th independent variable,
  and generate collocation points in the spatial-temporal domain
  $\Omega \times [0,T]$. Therefore, the foregoing discussions equally apply
  to solving
  the initial-boundary value problems and for computing the  optimal $R_m$
  with time-dependent PDEs.
  If the temporal domain is large (large $T$),
  we employ the block time marching (BTM) scheme (see~\cite{DongL2020} for details)
  together with the ELM method for solving the problem. The temporal domain
  is divided into a number of windows (time blocks) and the problem is
  computed block by block~\cite{DongL2020}.
  
  When using ELM together with block time marching for time-dependent PDEs, 
  in the pre-processing run for computing the optimal $R_m$
  one only needs to use the first time block.
  In other words, 
  when computing $R_{m0}$ we can use a smaller spatial-temporal domain
  in the computation, which consists of only the first time
  block (not the entire temporal domain).
  The resultant $R_{m0}$
  can then be used subsequently in the ELM/BTM simulations for all the time
  blocks in the entire spatial-temporal domain.
  
\end{remark}

\begin{remark}\label{rem_8a}
  When the domain $\Omega$ is partitioned into multiple sub-domains and
  the system~\eqref{eq_1} is solved by the localELM method (see~\cite{DongL2020}
  for details),
  the procedure for computing the optimal $R_m$ remains essentially the same.
  The modification lies in that in the system~\eqref{eq_6}
  one needs to additionally include the $C^k$ continuity
  conditions (see~\cite{DongL2020}) on the collocation points of
  the sub-domain boundaries.
  The residual vector in~\eqref{eq_7} needs to be modified accordingly to
  include these additional equations.
  
\end{remark}

\subsection{ELM Configuration with Multiple $R_m$ Constants (Multi-Rm-ELM)}
\label{sec:multi}

We next consider a modified ELM method that involves multiple $R_m$ constants
for setting the random hidden-layer coefficients,
and present a procedure for computing the
optimal values of these constants. 
This modified ELM has the advantage over
the conventional ELM from Section~\ref{ord_elm} that it 
leads to generally more accurate simulation results.
The notation below follows
that of Section~\ref{ord_elm}.

We modify the ELM configuration for solving the system~\eqref{eq_1} as follows.
Instead of setting the coefficients for all the hidden layers
to random values from $[-R_m,R_m]$ with a single $R_m$,
we set the weight/bias coefficients
for each different hidden layer to random values
generated on an interval with a different $R_m$.
Specifically, we set the weight/bias coefficients in hidden layer $l$,
$(\mbs W_l,\mbs b_l)$ for $1\leqslant l\leqslant L-1$,
to uniform random values generated on $[-R_m^{(l)},R_m^{(l)}]$,
where $R_m^{(l)}$ ($1\leqslant l\leqslant L-1$) are user-prescribed constants
(hyperparameters). The random hidden-layer coefficients are again
fixed once they are assigned.
This modified ELM provides increased freedom for generating
the random hidden-layer coefficients.

Let us consider how to determine the
optimal or near-optimal values for these $R_m^{(l)}$ constants.
We first generate, for each hidden layer $l$ ($1\leqslant l\leqslant L-1$),
a set of $(M_{l-1}+1)M_l$ uniform random values on the interval $[-1,1]$,
which will be denoted by the vector $\bm\xi_l$ of length $(M_{l-1}+1)M_l$.
Once the random vectors $\bm\xi_l$ ($1\leqslant l\leqslant L-1$)
are generated, they will be
fixed throughout the computation.
Given the constants $R_m^{(l)}$ ($1\leqslant l\leqslant L-1$),
the random hidden-layer coefficients $(\mbs W_l,\mbs b_l)$
will be set by the random vector $R_m^{(l)}\bm\xi_l$
for $1\leqslant l\leqslant L-1$.
Let $\mbs R_m=(R_m^{(1)},R_m^{(2)},\dots,R_m^{(L-1)})$.
Then $\mbs R_m$ represents the set of scaling parameters for
the random hidden-layer coefficients.

We use a procedure analogous to that of Section~\ref{ord_elm}
for computing the optimal $\mbs R_m$.
We solve system~\eqref{eq_6} by the linear or nonlinear least squares method,
and let $\beta_{ij}^{LS}$ ($1\leqslant i\leqslant M_{L-1}$, $1\leqslant j\leqslant M_L$)
denote its least squares solution.
Let $\mbs u^{LS}(\mbs x)=\sum_{j=1}^M V_j(\mbs x)\bm\beta_j^{LS}$
denote the least squares solution to the system~\eqref{eq_1}.
The residual vector of system~\eqref{eq_6} at the least squares
solution is given by
\begin{equation}\label{eq_10}
  \mbs r(\mbs R_m) = \left[\begin{array}{c}
    \vdots \\
    \mbs L\left(\sum_{j=1}^MV_j\left(\mbs x_p,\mbs R_m\right)\bm\beta_j^{LS}  \right) - \mbs f(\mbs x_p) \\
    \vdots \\ \hdashline[2pt/2pt]
    \vdots \\
    \mbs B\left(\sum_{j=1}^MV_j\left(\mbs x_q,\mbs R_m\right)\bm\beta_j^{LS}   \right) - \mbs g(\mbs x_q) \\
    \vdots
    \end{array}\right]
  = \left[\begin{array}{c}
    \vdots \\
    \mbs L\left(\mbs u^{LS}(\mbs x_p, \mbs R_m)  \right) - \mbs f(\mbs x_p) \\
    \vdots \\ \hdashline[2pt/2pt]
    \vdots \\
    \mbs B\left(\mbs u^{LS}(\mbs x_q, \mbs R_m)   \right) - \mbs g(\mbs x_q) \\
    \vdots
    \end{array}\right]
\end{equation}
where the dependence on $\mbs R_m$ has been explicitly specified.
Note that now $\mbs r$ depends on the multitude of $R_m^{(l)}$
($1\leqslant l\leqslant L-1$) constants.
Consider the Euclidean norm of the residual vector
\begin{equation}\label{eq_11}
  \mathcal{N}(\mbs R_m) = \left\|\mbs r(\mbs R_m)  \right\|.
\end{equation}
We seek the optimal value $\mbs R_{m0}$ for $\mbs R_m$ such that
$\mathcal{N}(\mbs R_m)$ is minimized, namely,
\begin{equation}\label{eq_12}
  \mbs R_{m0} = \argmin_{\mbs R_m} \mathcal{N}(\mbs R_m).
\end{equation}
We solve the optimization problem~\eqref{eq_12} again by
the differential evolution algorithm~\cite{StornP1997}
and employ its scipy implementation.
This algorithm requires the valuation routine for the cost
function $\mathcal{N}(\mbs R_m)$ for any given $\mbs R_m$.
$\mathcal{N}(\mbs R_m)$ can be evaluated by a modification
of the Algorithm~\ref{alg_1}, and the modified version
is summarized in Algorithm~\ref{alg_2}.

\begin{algorithm}[tb]
  \DontPrintSemicolon
  \SetKwInOut{Input}{input}\SetKwInOut{Output}{output}

  \Input{$\mbs R_m=(R_m^{(1)},\dots,R_m^{(L-1)})$; input data $\mbs X$;
    fixed random vectors $\bm \xi_l$, of length $(M_{l-1}+1)M_l$ and containing
    uniform random values on $[-1,1]$,
    for $1\leqslant l\leqslant N-1$.}
  \Output{$\mathcal{N}(\mbs R_m)$.}
  \BlankLine\BlankLine
  update the hidden-layer coefficients $(\mbs W_l,\mbs b_l)$ 
  by $R_m^{(l)}\bm\xi_l$, for $1\leqslant l\leqslant L-1$\;
  compute $\bm\Phi_{L-1}$ by evaluating the neural network (first $(L-1)$ layers) on
  the input data $\mbs X$\;
  solve system \eqref{eq_6} for $\beta_{ij}^{LS}$ by the linear or nonlinear least squares method
  from~\cite{DongL2020}\;
  update the output-layer coefficients $\mbs W_L$
  by $\beta_{ij}^{LS}$ ($1\leqslant i\leqslant M_{L-1}$,
  $1\leqslant j\leqslant M_L$)\;
  compute $\mbs u^{LS}$, $\mbs L(\mbs u^{LS})$ and $\mbs B(\mbs u^{LS})$ by evaluating
  the neural network on $\mbs X$ and by auto-differentiation\;
  compute the residual vector $\mbs r(\mbs R_m)$ by equation \eqref{eq_10}\;
  compute $\mathcal{N}(\mbs R_m)$ by equation \eqref{eq_11}
  
  \caption{Computing the cost function $\mathcal{N}(\mbs R_m)$ in Multi-Rm-ELM}
  \label{alg_2}
\end{algorithm}


The modified ELM method involves the multiple
components of $\mbs R_m$. An automatic procedure is essential for computing
the optimal or near-optimal $\mbs R_m$ in this case.
It would be extremely difficult,
  and practically impossible if the neural network becomes deeper,
  to determine a near-optimal $\mbs R_m$
  manually such as in~\cite{DongL2020}.
  

\begin{remark}\label{rem_3}
  For neural networks with a single hidden layer, the modified ELM
  method would contain a single $R_m$ and is therefore identical
  to the conventional ELM for this case.
  If the neural network consists of two or more hidden layers,
  the modified ELM and the conventional ELM will generally be different.
  It is observed that the modified ELM with multiple $R_m$ constants
  generally leads to a better accuracy than the conventional ELM,
  under the same network architecture and the same number of collocation points.
  The cost for computing the optimal $\mbs R_m$ 
  in the modified ELM
  is generally larger than that for computing the optimal $R_m$
  in the conventional ELM.
  Note that
  a list of lower-/upper-bound pairs, each for a component of $\mbs R_m$,
  needs to be provided when using the scipy
  routine of the differential evolution algorithm.

\end{remark}

\begin{remark}
  Several parameters are important and can influence
  the accuracy when the differential evolution routine in scipy
  (scipy.optimize.differential\_evolution) is invoked.
  These include the population size, the bounds for $R_m$ (or $\mbs R_m$),
  and the relative tolerance for convergence.
  The size of the population must be at least $4$, as required
  by the differential evolution algorithm~\cite{StornP1997}.
  We observe that a population size in the range of $6$ to $10$
  will typically suffice. While a large population size 
  can in principle produce more accurate results, in reality this
  can make the algorithm harder to converge and take significantly more
  iterations, with little improvement in the obtained result.
  As mentioned previously, a pair of bounds $[R_{\min},R_{\max}]$
  for $R_m$ in Single-Rm-ELM
  (or pairs of bounds for $\mbs R_m$ in Multi-Rm-ELM, each for a component
  of $\mbs R_m$) needs to be provided to the routine.
  Since $R_m$ is in a range of moderate values,
  we typically employ
  a range $[0.01, 5]$ (or $[0.01, 3]$) for $R_m$ and also for
  the $\mbs R_m$ components in the numerical tests.
  A larger range can ensure that the true optimum $R_{m0}$ will not be
  missed, but on the other hand may not produce a very accurate
  $R_{m0}$  under a given maximum number of iterations.
  An appropriate narrower range
  is conducive to attaining a more accurate $R_{m0}$.
  One can therefore also start out with a larger range to get a
  rough estimate for $R_{m0}$, and then narrow down the range
  based on the rough estimate to obtain a more accurate $R_{m0}$.
  The relative tolerance refers to the tolerance on
  the ratio between the standard deviation and the mean of
  the objective function values within the population.
  A small enough tolerance ensures that all members
  of the population will reach the minimum upon convergence.
  We observe that a relative tolerance around $0.1$ would
  typically lead to very good simulation results. Even
  smaller tolerance values can substantially increase the number
  of iterations, with little improvement in the obtained results.
  In the current paper we employ
  a relative tolerance $0.1$ for all the numerical tests in Section \ref{sec:tests}.
  In addition to the above parameters,
  the scipy routine can also optionally polish
  the obtained result in the end using a local minimizer. 
  We observe that the local polishing typically has little or
  no improvement on the result. In the current paper
  no polishing is performed
  on the result from the differential evolution algorithm.
  We typically employ a maximum of $50$ generations in
  the differential evolution algorithm for the numerical tests.

\end{remark}

\begin{remark}\label{rem_12}
  We would like to emphasize that
  there is one major difference between the current paper and
  our previous work~\cite{DongL2020}
  in terms of the ELM implementation, for computing
  the $V_j(\mbs x)$ (last hidden-layer output fields)
   and the differential operators involving $V_j$ (see equation
  \eqref{eq_6}).
  In \cite{DongL2020} these differential operators are computed by the default
  reverse-mode auto-differentiation (``GradientTape'') in Tensorflow.
  In the current
  work we have employed the forward-mode auto-differentiation
  (implemented by the ``ForwardAccumulator'' in Tensorflow)
  for computing the differential operators involving $V_j(\mbs x)$.
  This modification has sped up the computations and significantly
  reduced the ELM network training time in the current paper,
  when compared with that of~\cite{DongL2020}.
  This is because in ELM the number of nodes in the last hidden layer is typically
  much larger than that of the input layer, which is particularly suitable for
  forward-mode auto-differentiations.

  In the current paper
  we have compared extensively the current implementation of the ELM method
  with the finite element method (FEM),
  including both the classical second-order FEM and
  the high-order FEM with Lagrange elements~\cite{Courant1943},
  in terms of their accuracy and computational cost (FEM computation time, ELM network
  training time). We observe that, for time-dependent PDEs, the ELM method combined
  with block time marching consistently and far outperforms the
  FEM (both 2nd-order and high-order FEM).
  For stationary PDEs, ELM outperforms FEM (both 2nd-order and high-order FEM)
  for essentially all problem sizes, except for a range of very small problem sizes.
  By ``outperform'' we mean that one method achieves a better accuracy under the same
  computational budget/cost or incurs a lower computational cost to achieve
  the same accuracy.
  These observations can be contrasted with those of~\cite{DongL2020},
  where the comparisons between ELM (locELM) and FEM are also performed.
  In~\cite{DongL2020} it is observed that: (i) ELM (with the implementation
  therein) outperforms the classical 2nd-order FEM for larger problem sizes;
  (ii) ELM is competitive to some degree compared with the high-order FEM, but
  is not as efficient as the latter.
  With the improvements in the algorithm and implementation
  in the current work,
  the ELM method far outperforms the classical second-order FEM. Furthermore,
  ELM can markedly outperform the high-order FEM. It is
  more efficient than or as efficient as the high-order FEM.
  The comparisons between ELM and the classical and high-order FEMs will be
  detailed in Section \ref{sec:tests}.

\end{remark}

\section{Numerical Examples}
\label{sec:tests}

In this section we use several numerical examples,
with linear/nonlinear and stationary/dynamic PDEs
in two dimensions (2D), or in one spatial dimension (1D) plus time
for dynamic problems,
to demonstrate the effectiveness of the presented
method for computing the optimal $R_m$. We also compare the
current improved  ELM method
with the classical second-order and high-order finite element methods (FEM)
with regard to their accuracy and computational cost.

\subsection{General Notes on the Implementations}
\label{sec:note}

We first provide some implementation notes on the ELM and FEM.
They apply to all the numerical tests in the following subsections.


The ELM method is implemented in Python, employing
the Tensorflow (www.tensorflow.org) and the Keras (keras.io)
libraries. In particular, the differential operators involving the output fields
of the last hidden layer are computed using the forward-mode
auto-differentiation employing the ``ForwardAccumulator'' in
Tensorflow, as stated before. We use the Gaussian activation function,
$\sigma(x) = e^{-x^2}$,
in all the hidden nodes
for all the test problems in Section \ref{sec:tests}.

In the pre-processing run for computing the optimum $R_{m0}$ (or $\mbs R_{m0}$),
we have monitored and
recorded the wall time for $R_{m0}$ (or $\mbs R_{m0}$)
computation using the ``timeit'' module in Python.
The $R_{m0}$/$\mbs R_{m0}$ computation time includes all the time spent in the
iterations with the differential evolution algorithm and the update of
the random hidden-layer coefficients with the final $R_{m0}$ (or $\mbs R_{m0}$)
value upon convergence.
Within every differential evolution iteration, the primary computations involve
the evaluation of $\mathcal{N}(R_m)$
or $\mathcal{N}(\mbs R_m)$
using the Algorithm~\ref{alg_1} or Algorithm~\ref{alg_2}
for a given $R_m$ or $\mbs R_m$.

When ELM is used to solve a PDE with a given $R_m$ (or $\mbs R_m$),
the computational cost refers to
the training time of the ELM neural network.
The ELM network training time
includes the computation time for the output fields of
the last hidden layer ($V_j(\mbs x)$) and the associated differential operators
involving these field functions, the computation time for the coefficient matrix
and the right hand side for the linear least squares problem,
the computation time for the residual of the nonlinear algebraic system and
the associated Jacobian matrix for the nonlinear least squares problem,
the solution time for the linear/nonlinear least squares problem,
and the update time of the output-layer coefficients by the linear/nonlinear
least squares solution.
Note that,
following the protocol in~\cite{DongL2020}, this time does not include,
after the network is trained, the evaluation time of the neural network
on a set of given data points for the output of the solution data.

\phantomsection\label{graph}
In the current paper, the computations for $V_j(\mbs x)$ (output fields of
the last hidden layer) and the associated differential operators
are implemented as ``Tensorflow Functions'' (tf.function) executed as
a computational graph~\cite{GoodfellowBC2016}.
When these functions are invoked for the first time,
the Tensorflow library builds the computational graph by ``autograph and tracing''
the operations contained in these functions, and performs graph optimizations.
When they are invoked subsequently, the computations are 
performed directly in the graph mode,
which generally speeds up the computations significantly.
The autograph/tracing operations during the first invocation of
the Tensorflow Functions can slow down the computations notably.
This means that the network training time 
when the ELM training routine is invoked for the first
time can be markedly larger than that when the training routine
is invoked subsequently.
We will illustrate this difference with some specific test problems in
the following subsections.
In the comparisons between ELM and FEM,
the ELM network training time refers to
the time obtained with the computations in the graph mode
(no autograph/tracing).

\phantomsection\label{fem}
The finite element method is implemented also in Python,
by using the FEniCS library (fenicsproject.org).
The FEM implementations for different problems
follow those in~\cite{DongL2020},
which we refer the reader to for more detailed discussions.

When FEM is used to solve a given PDE, the computational cost
refers to the FEM computation time. The computation time
includes the symbolic specifications of the mesh, the finite element space,
the trial/test function spaces, the variational problem,
the forming and solution of the linear system~\cite{DongL2020}.
All these operations are handled by FEniCS, and
are opaque to the user. 
Note that the FEM computation time does not include
the output of the solution data after the problem is solved.

As pointed out in~\cite{DongL2020}, when the FEM code is run
for the first time, the FEniCS library compiles the key operations
in the Python code into a C++ code using Just-in-Time (JIT) compilers,
which is in turn compiled
by the C++ compiler and then cached. In subsequent runs of the FEM code,
the cached operations are used directly in the FEM computation,
making it significantly faster.
Therefore, the FEM computation time when the code is run for the first time
(with JIT compilation)
is considerably larger than that when the code is run subsequently
(no JIT compilation).
In the comparisons between ELM and FEM for different test problems
in the following subsections,
the FEM computation time
refers to the time collected by ``timeit'' in subsequent runs of the FEM code
(other than the first run, no JIT compilation).
All the timing data in this paper
are collected on a MAC computer (Intel Core i5 CPU 3.2GHz, 24GB memory)
in the authors' institution.


Here are some further comments on the random number generators.
In the current ELM implementation,
the random vector $\bm \xi$ in Algorithm~\ref{alg_1},
the random vectors $\bm\xi_l$ ($1\leqslant l\leqslant L-1$)
in Algorithm \ref{alg_2}, and the random perturbations in
the NLLSQ-perturb method~\cite{DongL2020} for solving nonlinear PDEs,
are all generated by the random number generator from the Tensorflow library.
The random numbers involved in the differential evolution routine
of the scipy library are generated by the random number generator
from the numpy package in Python.
In order to make the simulation results reported here exactly reproducible,
we employ the same seed value for the random number generators
in both Tensorflow and numpy for all the numerical tests in Section \ref{sec:tests}.
In addition, the seed value is fixed for all the numerical experiments
within a subsection (see also Remark~\ref{rem_rand}).
Specifically, the seed value is $1$ for the numerical tests in
Section \ref{sec:func}, $10$ for those in Section \ref{sec:poisson},
$25$ for those in Section \ref{sec:nonl_helm},
and $100$ for those in Section \ref{sec:burger}, respectively.
This seed value is passed to the scipy differential evolution routine
when invoking that algorithm.


We would like to mention another implementation detail in ELM for all
the numerical tests in the following subsections.
When implementing the neural network model in Keras, between
the input layer (representing $(x,y)$ in 2D or $(x,t)$ in 1D plus time)
and the first hidden layer, we have added an affine mapping to
normalize the $(x,y)$ or $(x,t)$ data. 
Suppose $(x,y/t)\in [a_1,b_1]\times [a_2,b_2]$, then this affine mapping
normalizes the $(x,y)$ or $(x,t)$ data from $[a_1,b_1]\times [a_2,b_2]$
to the domain $[-1,1]\times [-1,1]$.
This mapping is implemented as a ``lambda'' layer in Keras.
Because of this affine mapping, all the data into the first hidden layer
are normalized.
Note that this lambda layer is not counted toward the number of
hidden layers in the neural network.

\subsection{Function Approximation}
\label{sec:func}

\begin{figure}
  \centerline{
    \includegraphics[width=2in]{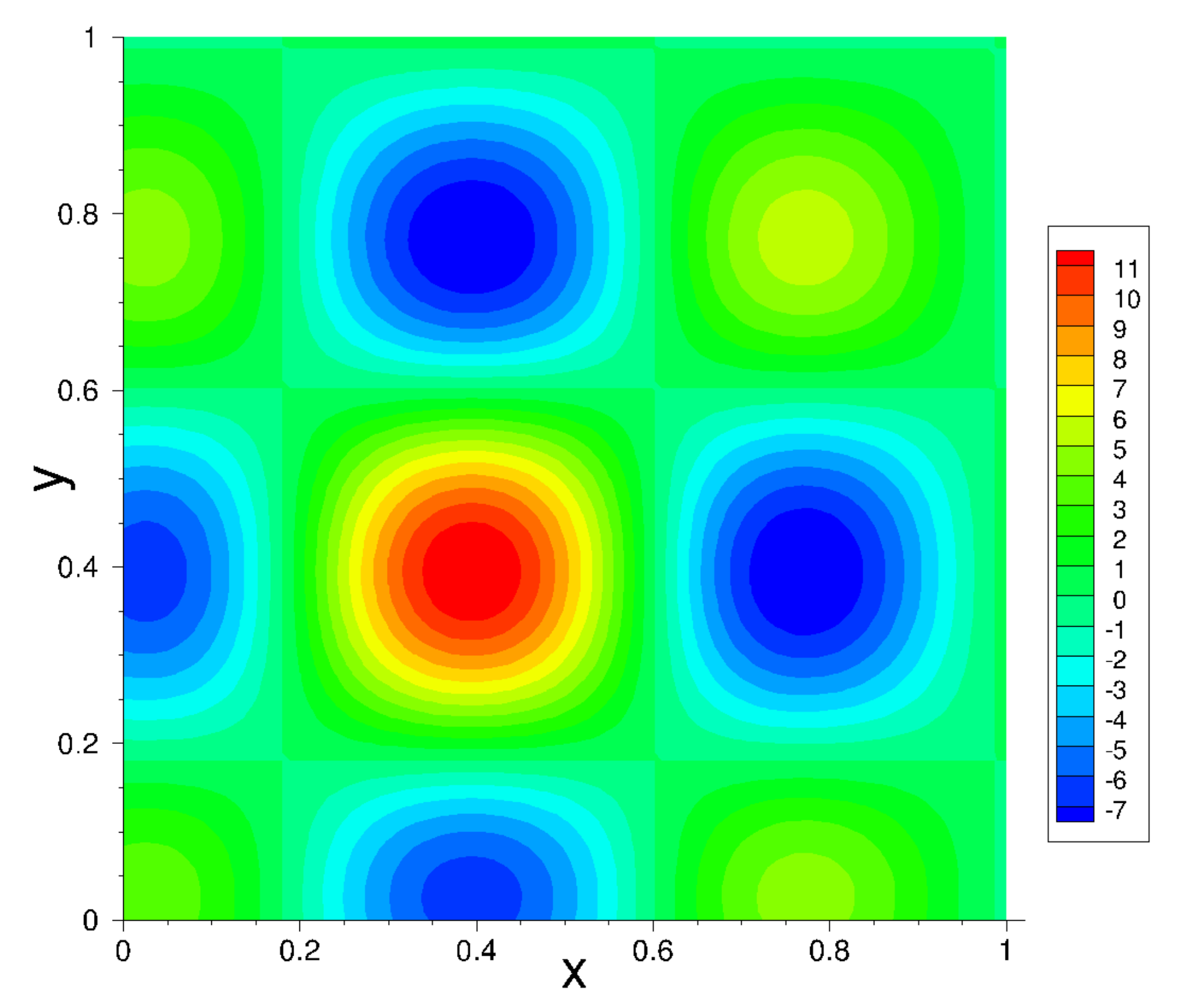}
  }
  \caption{Function approximation: distribution of the function to be approximated.}
  \label{fg_1}
\end{figure}

In the first example we consider the approximation of a 2D function
using ELM and illustrate the effects of the network/simulation parameters on
the optimal $R_m$.
The numerical results in this subsection are obtained using a seed value $1$
in the random number generators in both Tensorflow and Numpy.

We consider the unit square domain
$\Omega=\{(x,y)\ |\ 0\leqslant x,y\leqslant 1 \}$,
and the following function on $\Omega$,
\begin{equation}
  f(x,y) = \left[\frac32\cos\left(\frac32\pi x+\frac{9\pi}{20} \right)
    + 2\cos\left(3\pi x-\frac{\pi}{5} \right)
    \right] \left[\frac32\cos\left(\frac32\pi y+\frac{9\pi}{20} \right)
    + 2\cos\left(3\pi y-\frac{\pi}{5} \right)
    \right].
\end{equation}
Figure \ref{fg_1} illustrates the distribution of this function.
Given $f(x,y)$ on a set of data points, we would like to approximate $f$
by ELM.
The function approximation problem is equivalent to the problem~\eqref{eq_1},
with $\mbs L(u)=u$ and without the boundary condition~\eqref{eq_1b}.
In this case the PDE is reduced to an algebraic equation,
\begin{equation}\label{eq_14}
  u = f(x,y),
\end{equation}
where $u(x,y)$ is the approximant given in the form of an extreme learning machine.


We use a feedforward neural network to represent $u(x,y)$, with
a total of $(L+1)$ layers. The input layer (layer $0$)
contains two nodes, representing $(x,y)$. The output layer (layer $L$)
is linear and contains one node,
representing $u(x,y)$. The network consists of one or more hidden layers,
with the Gaussian activation function for all the hidden nodes.
As stated in Section \ref{sec:method}, the network architecture
is characterized by the vector $[M_0,M_1,\dots,M_L]$, where $M_i$
denotes the number of nodes in layer $i$ ($0\leqslant i\leqslant L$).
The specific architecture of the neural networks are given below.
We use $M=M_{L-1}$ to denote the width of the last hidden layer (or number of
training parameters). $M$ is either fixed or varied systematically below.


We employ a set of uniform grid points, $Q=Q_1\times Q_1$, as the collocation
points on $\Omega$, where $Q_1$ denotes the number of uniform collocation
points along both $x$ and $y$ directions.
Therefore, there are $Q_1$ uniform collocation points on each boundary
of $\Omega$.
The input data to the neural network consist of the coordinates
of all the collocation points.
$Q_1$ is either fixed or varied systematically in the numerical tests.
We assume that the function values $f(x,y)$ are given on the collocation
points.


With the above settings, we employ the Single-Rm-ELM and 
Multi-Rm-ELM configurations from Section \ref{sec:method}
to solve this problem. The difference of these two configurations lies in
the setting of the random hidden layer coefficients in the neural network,
as detailed in Section \ref{sec:method}.

\begin{figure}
  \centerline{
    \includegraphics[width=2in]{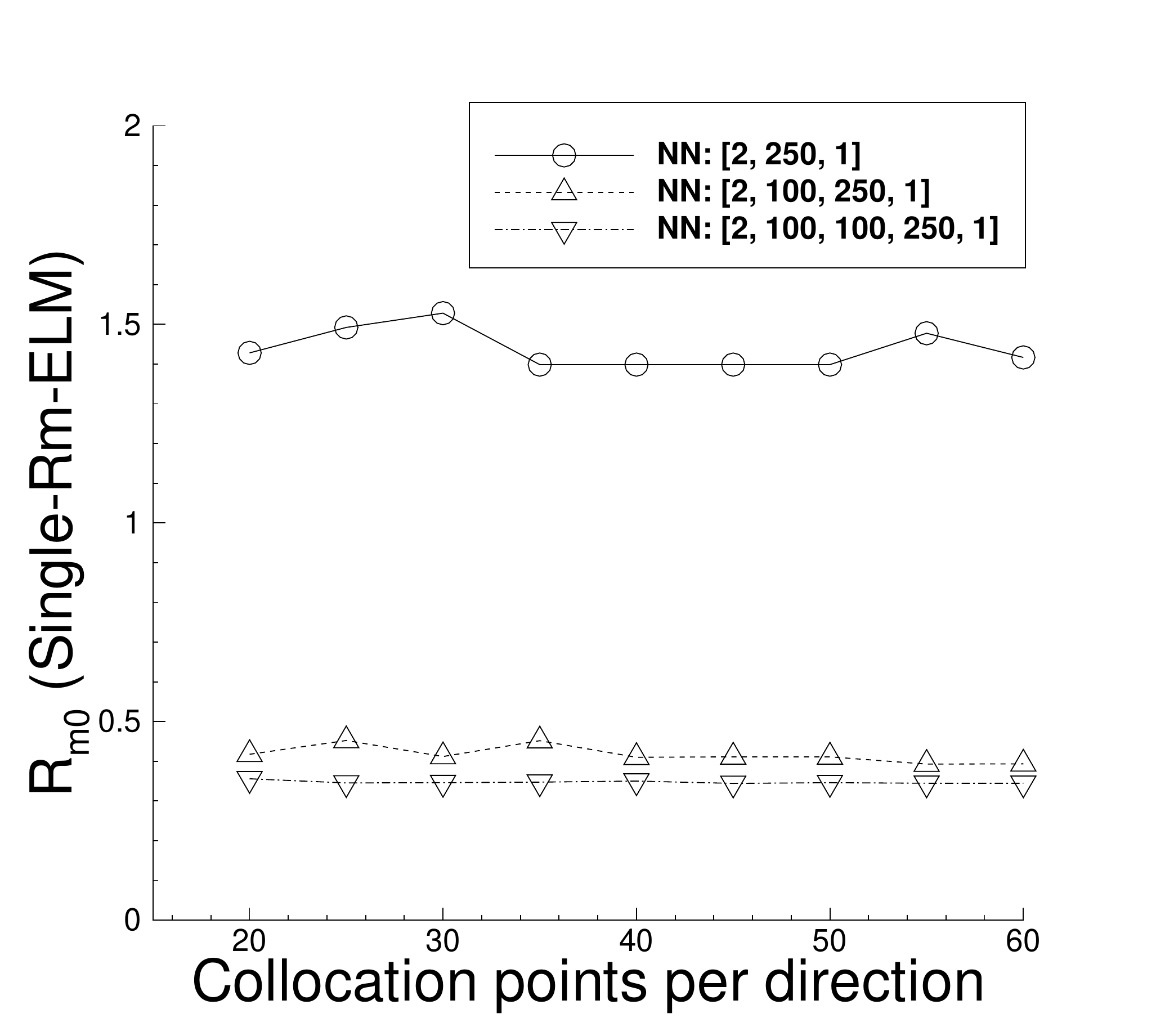}(a)
    \includegraphics[width=2in]{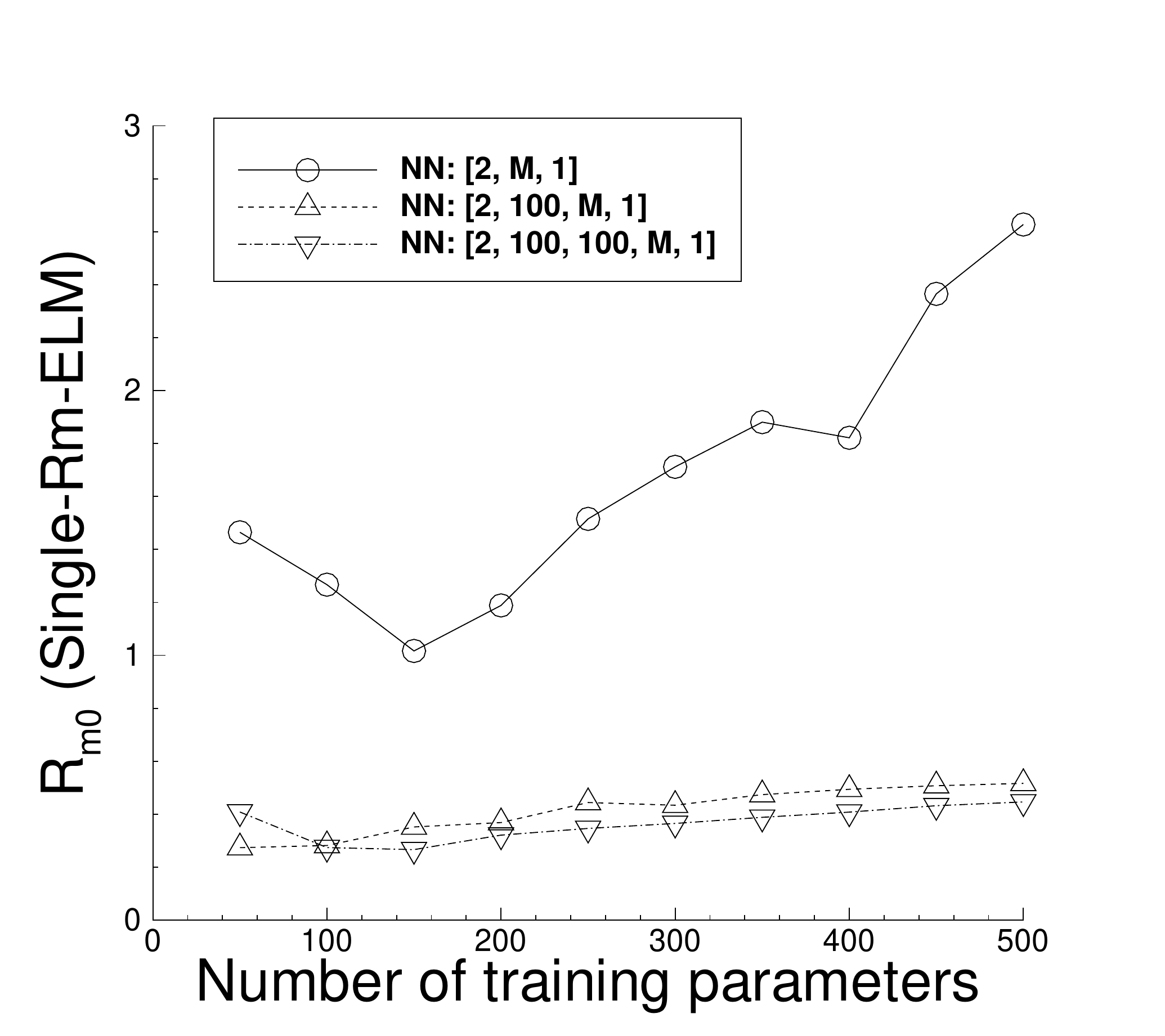}(b)
    \includegraphics[width=2in]{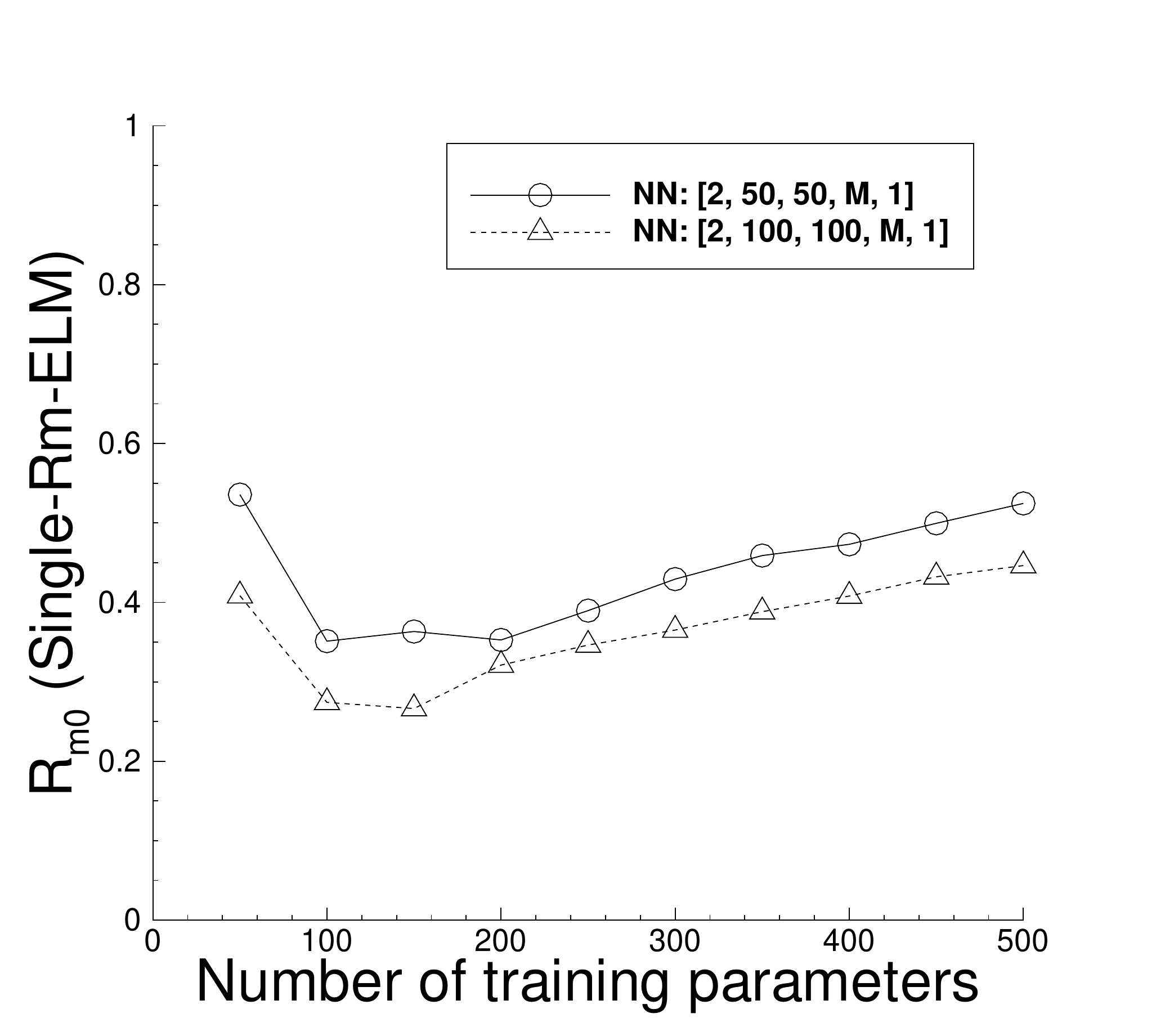}(c)
  }
  \caption{Function approximation (Single-Rm-ELM):
    The optimum $R_{m0}$ versus (a) the number of collocation points
    per direction and (b) the number of training parameters,
    with neural networks of different depth. 
    (c) $R_{m0}$ versus the number of training parameters with
    neural networks having the same depth but different width.
    $Q=31\times 31$ in (b,c), varied in (a).
    $M=250$ in (a), varied in (b,c).
  }
  \label{fg_2}
\end{figure}

Let us first look into the Single-Rm-ELM configuration, and we employ the method from
Section \ref{ord_elm} to compute the optimal $R_m$ based on the differential evolution
algorithm. We have considered several neural network architectures with
different depths and widths, and Figure \ref{fg_2} illustrates the characteristics
the optimum $R_{m0}$ obtained from the method.
%
%

Figure \ref{fg_2}(a) depicts the $R_{m0}$ as a function of the number of
uniform collocation points in each direction ($Q_1$) for three
neural networks as specified in the legend
with one to three hidden layers. The number of training parameters
is fixed at $M=250$, and the widths of the preceding
hidden layers are fixed at $100$ in them.
In this group of tests, for each neural network,
we vary the number of collocation points systematically between
$Q=20\times 20$ and $Q=60\times 60$, and
for each set of collocation points we compute
the optimimum $R_{m0}$.
We have employed a population size $6$, the bounds $[0.01, 3]$
for $R_m$,
and a relative tolerance $0.1$
in the scipy differential evolution routine.
We can make two observations from Figure~\ref{fg_2}(a).
First, $R_{m0}$ is largely insensitive to the number of
collocation points with Single-Rm-ELM. In other words,
$R_{m0}$ remains essentially the same as the number of
training data points varies.
Second, $R_{m0}$ tends to decrease with increasing number of layers in
the neural network. There is a big drop in $R_{m0}$
from a single hidden layer to two hidden layers. 
Beyond two hidden layers, 
as the network depth further increases,
the decrease in $R_{m0}$ is slight and
sometimes almost negligible.
These seem to be the common characteristics of
$R_{m0}$ with Single-Rm-ELM,
which will appear repeatedly in other test problems.

Figure \ref{fg_2}(b) shows the effect of the number of training parameters ($M$)
on $R_{m0}$, with three network architectures
as given in the legend.
In this group of tests, we employ a fixed set of $Q=31\times 31$ uniform
collocation points, and vary the number of training parameters
systematically between $M=50$ and $M=500$.
The widths of the preceding hidden layers are fixed at $100$ in the tests.
Corresponding to each $M$, 
we compute the optimum $R_{m0}$ 
using the differential evolution algorithm. 
The data in Figure \ref{fg_2}(b) correspond to a population size of $8$,
the $R_m$ bounds $[0.01, 3]$, and a relative tolerance $0.1$
in the scipy differential evolution routine.
We can make the following observations.
First, $R_{m0}$ decreases with increasing depth in the neural network.
$R_{m0}$ drops markedly from one hidden layer to two hidden layers
in the network. For neural networks with
two or more hidden layers, as the depth further increases, $R_{m0}$ decreases
only slightly. This observation is consistent with that from
Figure \ref{fg_2}(a).
Second, $R_{m0}$ has a dependence on
the number of training parameters.
For neural networks with a single hidden layer,
this dependence on $M$ is stronger.
Figure \ref{fg_2}(b) indicates that in this case
$R_{m0}$ generally increases with increasing $M$, except
in a range of smaller $M$ values
where $R_{m0}$ decreases with increasing $M$.
For neural networks with two or more hidden layers,
the dependence of $R_{m0}$ on $M$ is weak. 
$R_{m0}$ appears to increase only slightly as $M$ increases.

Figure \ref{fg_2}(c) illustrates the effect of the widths
of the preceding hidden layers on $R_{m0}$.
It depicts the $R_{m0}$ as a function of the number of training
parameters ($M$) for two neural networks, which contain three hidden layers
but have different widths ($50$ versus $100$) in the
preceding hidden layers (i.e.~other than
the last hidden layer).
These data are obtained again with a population size of $8$,
the $R_m$ bounds $[0.01, 3]$, and a relative tolerance $0.1$
in the differential evolution algorithm.
We observe that $R_{m0}$ generally decreases, albeit slightly,
as the width of the preceding hidden layers increases.
The characteristics observed from Figures \ref{fg_2}(b,c)
also seem  common to Single-Rm-ELM,
and they will appear repeatedly in other test problems.


\begin{figure}
  \centerline{
    \includegraphics[width=2in]{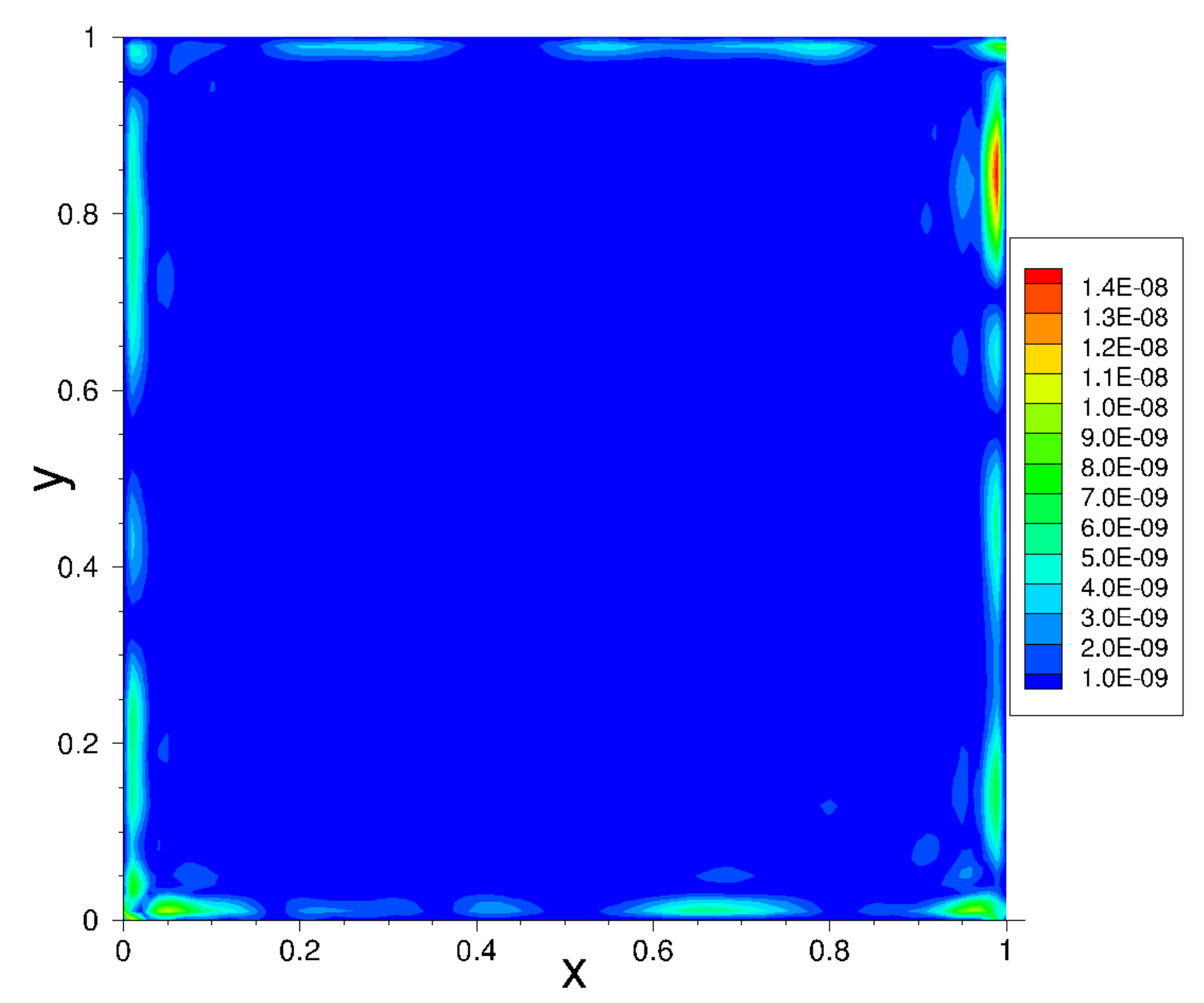}(a)
    \includegraphics[width=2in]{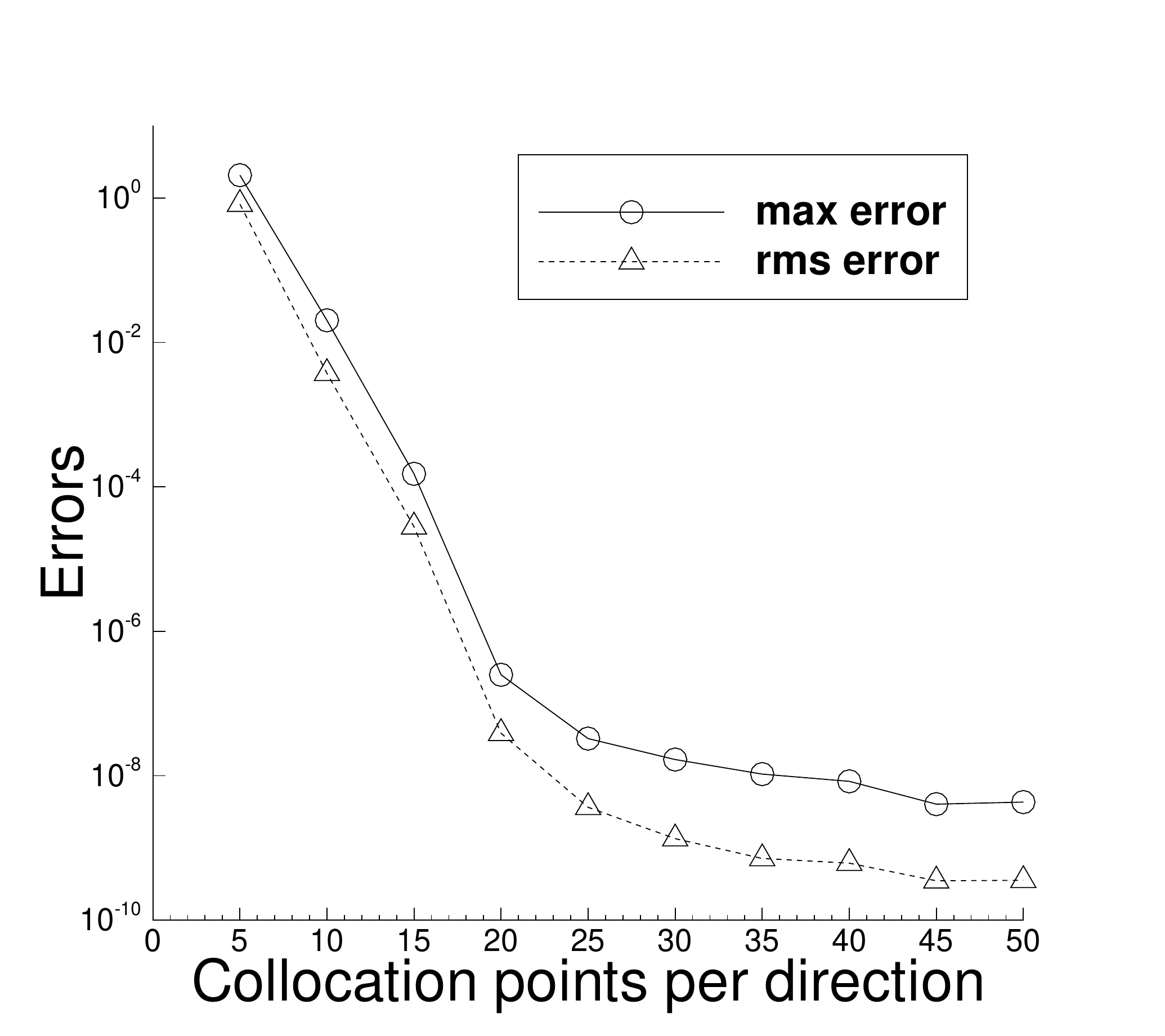}(b)
  }
  \centerline{
    \includegraphics[width=2in]{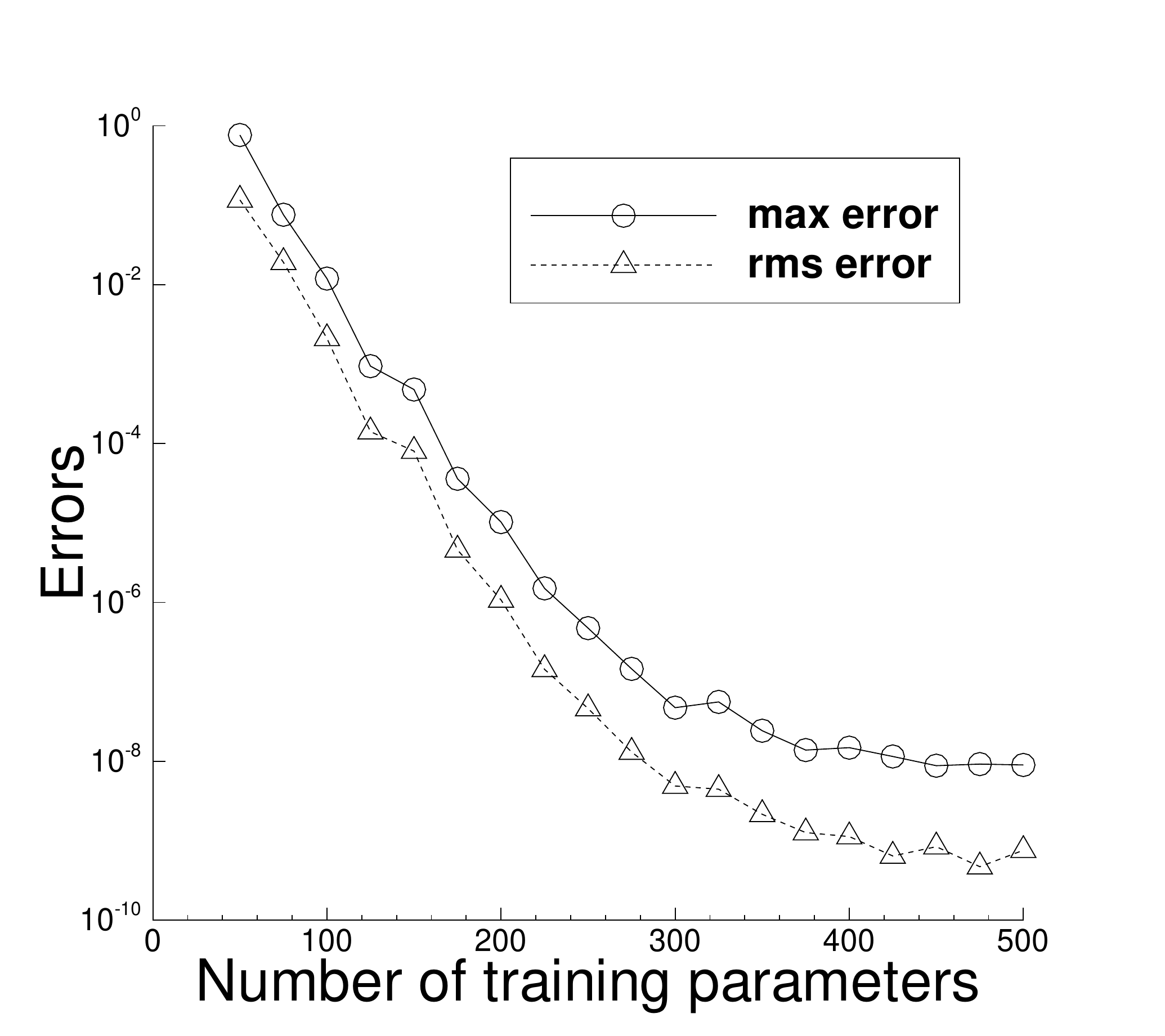}(c)
    \includegraphics[width=2in]{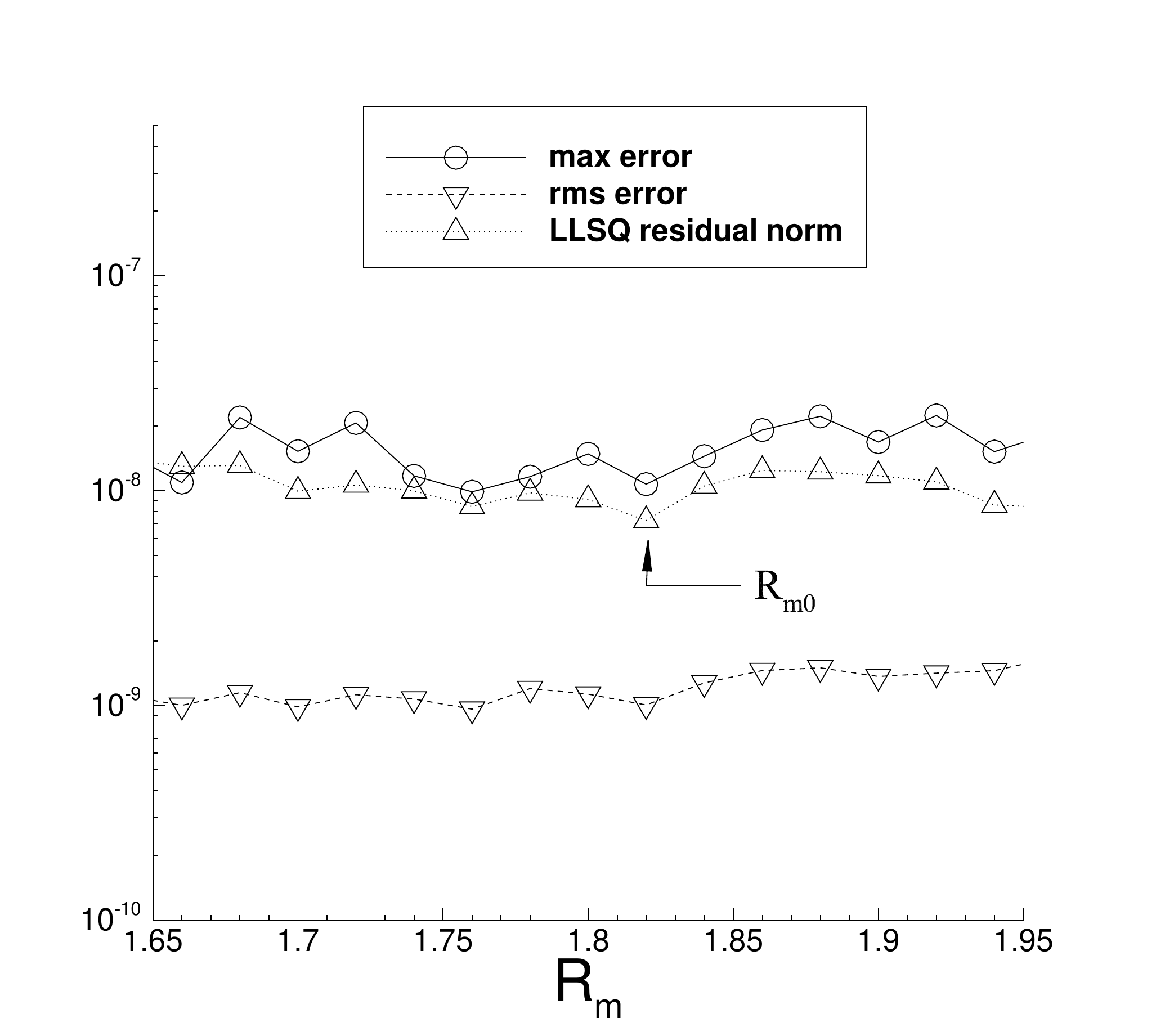}(d)
  }
  \caption{Function approximation (Single-Rm-ELM):
    (a) absolute-error distribution.
    The maximum and root-mean-squares (rms) errors in the domain
    versus (b) the number of collocation points per direction
    and (c) the number of training parameters ($M$).
    (d) The maximum/rms errors and the LLSQ residual norm
    versus $R_m$ in a neighborhood of $R_{m0}$.
    Network architecture: [2, $M$, 1].
    $Q=31\times 31$ in (a,c,d), varied in (b).
    $M=400$ in (a,b,d), varied in (c).
    $R_m=1.8$ in (a,b,c), varied in (d).
    $R_{m0}=1.822$ in (d).
  }
  \label{fg_3}
\end{figure}

Figure \ref{fg_3} illustrates the accuracy of the ELM approximant with $R_m$ near
the optimum $R_{m0}$ in Single-Rm-ELM.
In this group of tests, we employ a neural network with an architecture 
$[2, M, 1]$, where the number of training parameters $M$ is either fixed at
$M=400$ or varied between $M=50$ and $M=500$.
A set of $Q=Q_1\times Q_1$ uniform collocation points is employed,
where $Q_1$ is either fixed at $Q_1=31$ or varied between $Q_1=5$ and $Q_1=50$.
$R_m$ is either fixed at $R_m=1.8$, or varied in a neighborhood of
$R_{m0}=1.822$, which is attained
from the differential evolution algorithm
with $M=400$ and $Q=31\times 31$.
Figure \ref{fg_3}(a) shows the distribution in the $x$-$y$ plane
of the absolute error of the ELM approximant obtained with
$Q=31\times 31$, $M=400$ and $R_m=1.8$. It indicates that ELM can approximate
the function accurately, with the maximum error on
the order $10^{-8}$.
Figure \ref{fg_3}(b) shows the maximum and root-mean-squares (rms)
errors in the domain of the ELM approximant as a function of $Q_1$,
which is varied systematically here.
It is observed that the errors decrease exponentially with increasing $Q_1$
before they saturate gradually beyond about $Q_1\approx 25$.
Figure \ref{fg_3}(c) shows the maximum/rms errors of the ELM approximant
as a function
of $M$, which is varied systematically.
We observe an exponential decrease in the errors
with increasing $M$ before a gradual saturation
for $M$ beyond around $M\approx 300$.
Figure \ref{fg_3}(d) shows the maximum/rms errors of the ELM approximant,
as well as the residual norm of the linear least squares (LLSQ) problem
(i.e.~$\mathcal{N}(R_m)$ in equation \eqref{eq_8}),
as a function of $R_m$ in a neighborhood of $R_{m0}=1.822$.
The data indicate that the ELM accuracy is generally not sensitive to
$R_m$ within a neighborhood of $R_{m0}$. The ELM errors are largely
comparable with the $R_m$ values around $R_{m0}$.
These results attest to the point in Remark~\ref{rem_rm} that
one can generally employ an $R_m$ value around $R_{m0}$ in the
computations without seriously sacrificing the accuracy.

\begin{figure}
  \centerline{
    \includegraphics[width=2in]{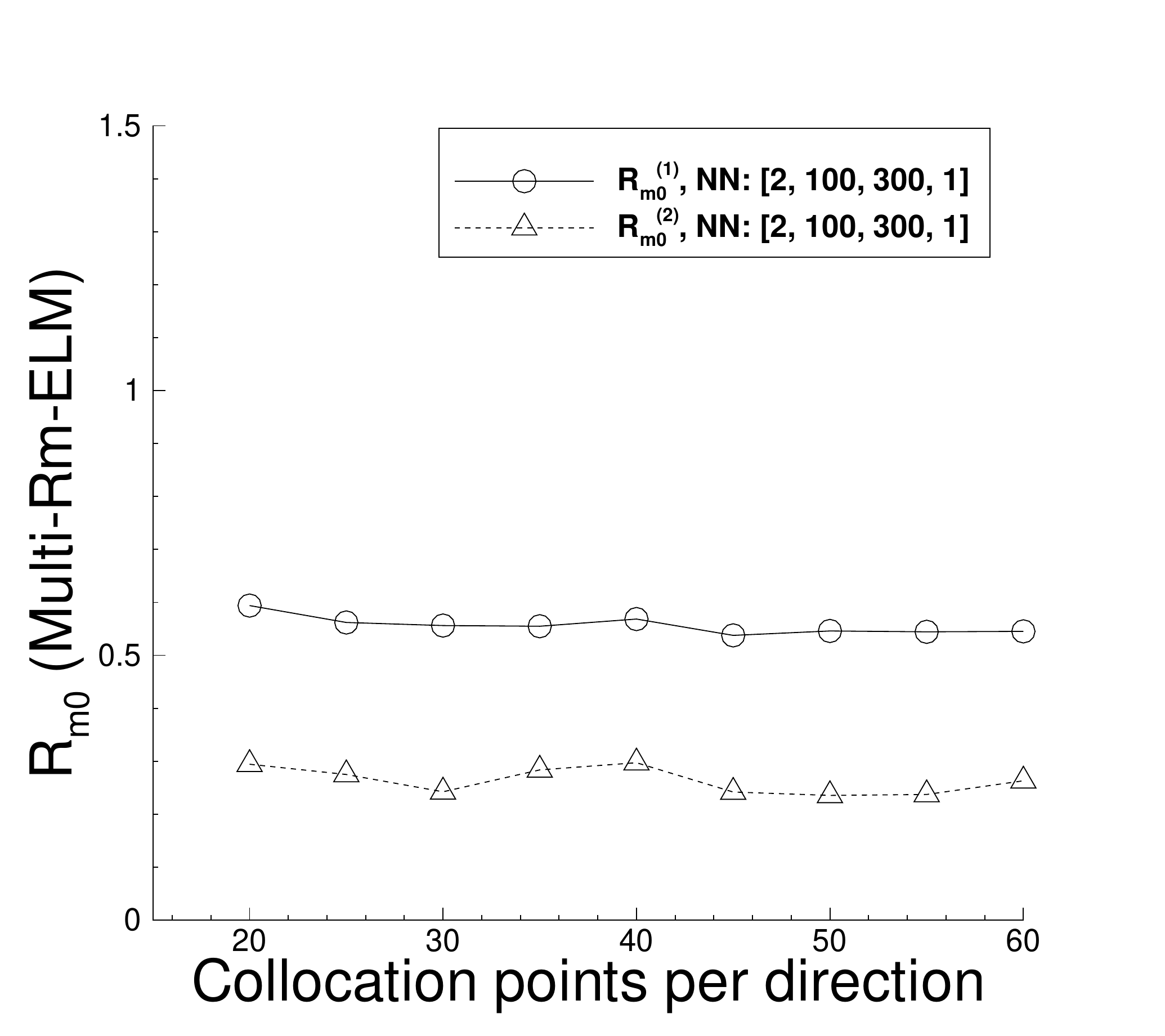}(a)
    \includegraphics[width=2in]{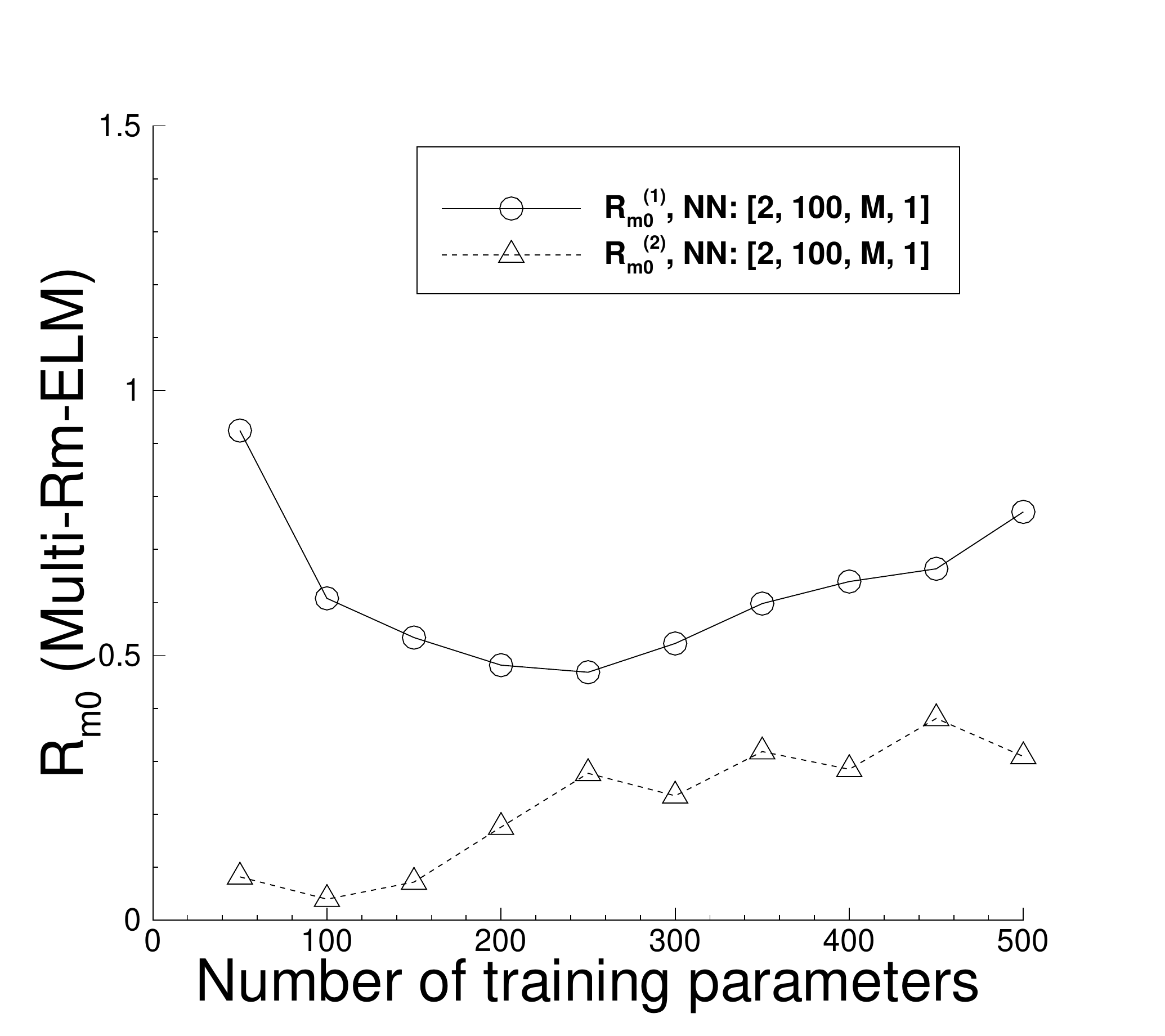}(b)
  }
  \centerline{
    \includegraphics[width=2in]{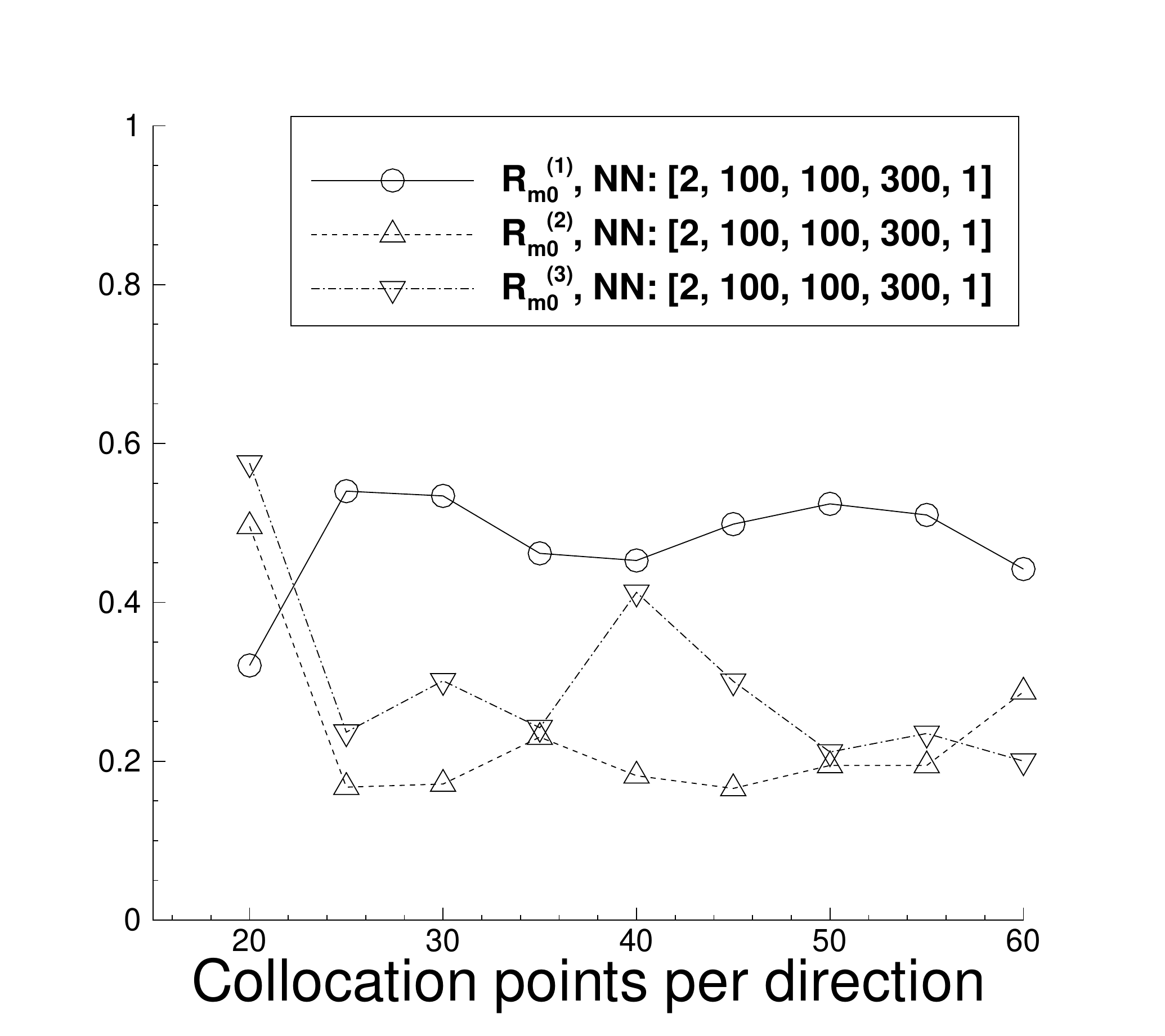}(c)
    \includegraphics[width=2in]{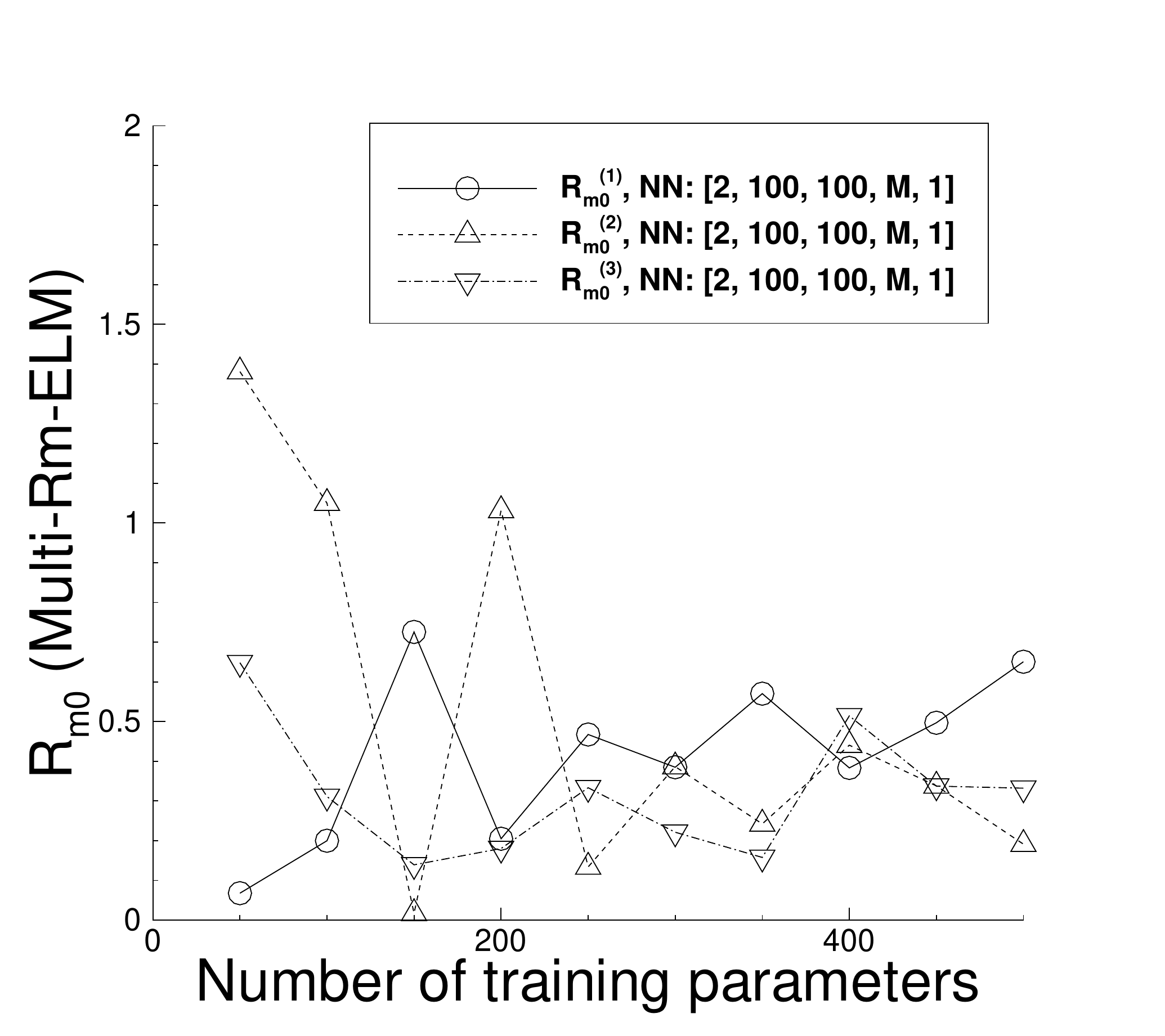}(d)
  }
  \caption{Function approximation (Multi-Rm-ELM):
    The $\mbs R_{m0}$ components versus the number of collocation points (a,c) and
    the number of training parameters (b,d), with neural
    networks having two (a,b)
    or three (c,d) hidden layers. The network architectures
    are given in the legends.
    $Q=31\times 31$ in (b,d), varied in (a,c).
    $M=300$ in (a,c), varied in (b,d).
  }
  \label{fg_4}
\end{figure}

Let us next consider the Multi-Rm-ELM configuration for the function
approximation problem~\eqref{eq_14}.
We employ the method from Section \ref{sec:multi} to compute
the optimum $\mbs R_{m0}$
for neural networks with two or three hidden layers.
The results are summarized in Figure \ref{fg_4}.
In this group of tests, the architecture of the neural networks
is characterized by $[2, 100, M, 1]$ or $[2, 100, 100, M, 1]$,
where  $M$ is either fixed
at $M=300$ or varied between $M=50$ and $M=500$.
As stated previously, the Gaussian activation function
has been employed in all the hidden nodes in this work.
We employ a set of $Q=Q_1\times Q_1$ uniform collocation points in the domain,
with $Q_1$  either fixed at $Q_1=31$ or varied 
between $Q_1=20$ and $Q_1=60$.

Figures \ref{fg_4}(a,b) illustrate
the optimum $\mbs R_{m0}=(R_{m0}^{(1)},R_{m0}^{(2)})$ for
the neural networks with two hidden layers,
which are obtained with a population size of $10$, the bounds
$[0.01, 3]$ for all components of $\mbs R_m$, and
a relative tolerance $0.1$ in the differential
evolution algorithm.
Figure \ref{fg_4}(a) depicts the $\mbs R_{m0}$ components
as a function of $Q_1$, with a fixed  $M=300$ in the neural network.
We observe that the values for $R_{m0}^{(1)}$ and
$R_{m0}^{(2)}$ are quite different, indicating that the random coefficients
for the first and second hidden layers could be
generated on two quite different intervals.
It is also observed that they are essentially independent of
the number of collocation points.
Figure \ref{fg_4}(b) shows the $\mbs R_{m0}$ components
as a function of the number of training parameters ($M$),
with a fixed $Q_1=31$ in these tests.
Both components of $\mbs R_{m0}$ appear to generally increase with
increasing $M$, except that $R_{m0}^{(1)}$ is observed
to decrease for a range of smaller $M$ values.

Figures \ref{fg_4}(c,d) show the corresponding optimum
$\mbs R_{m0}=(R_{m0}^{(1)},R_{m0}^{(2)}, R_{m0}^{(3)})$
for neural networks containing three hidden layers. These
are obtained with a population size of $12$, the bounds $[0.01, 3]$
for all components of $\mbs R_{m}$, and a relative tolerance $0.1$
in the differential evolution algorithm.
Figure \ref{fg_4}(c) depicts the $\mbs R_{m0}$ components as a function
of $Q_1$, with a fixed  $M=300$ in the neural network.
Figure \ref{fg_4}(d) depicts the $\mbs R_{m0}$ components as a function
of the number of training parameters $M$ in the neural network, with a fixed 
set of $Q=31\times 31$ uniform collocation points.
Overall the relations of the $\mbs R_{m0}$ components 
with respect to the collocation points and the training parameters
appear quite irregular.
The relation of $\mbs R_{m0}$ versus the number of collocation points seems somewhat
less irregular, and some $\mbs R_{m0}$ components appear to stay close
to a constant for a range of $Q_1$ values.
This is in stark contrast to those of Figures~\ref{fg_4}(a,b)
with two hidden layers in the neural network.

\begin{figure}
  \centerline{
    \includegraphics[width=2in]{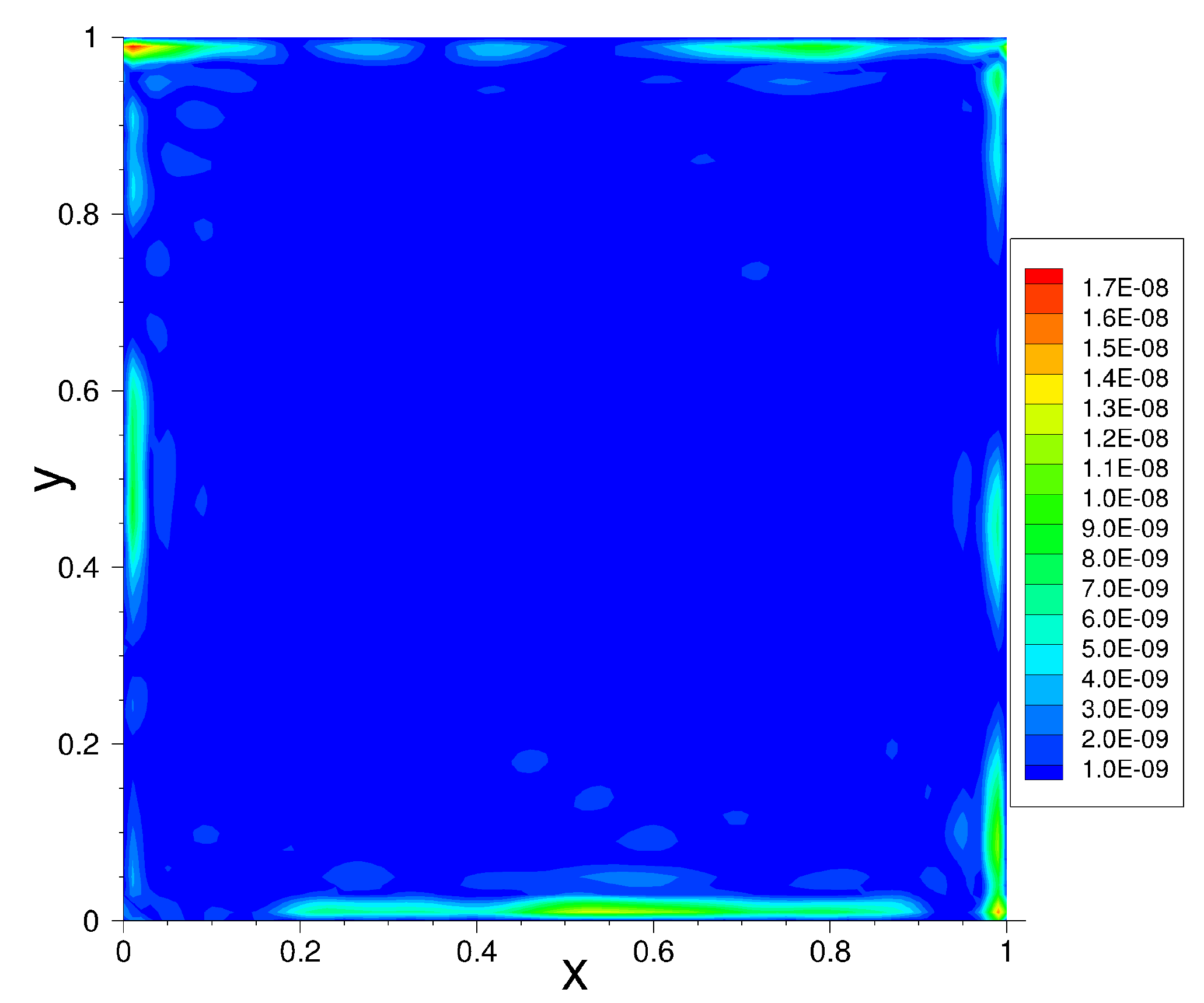}(a)
    \includegraphics[width=2in]{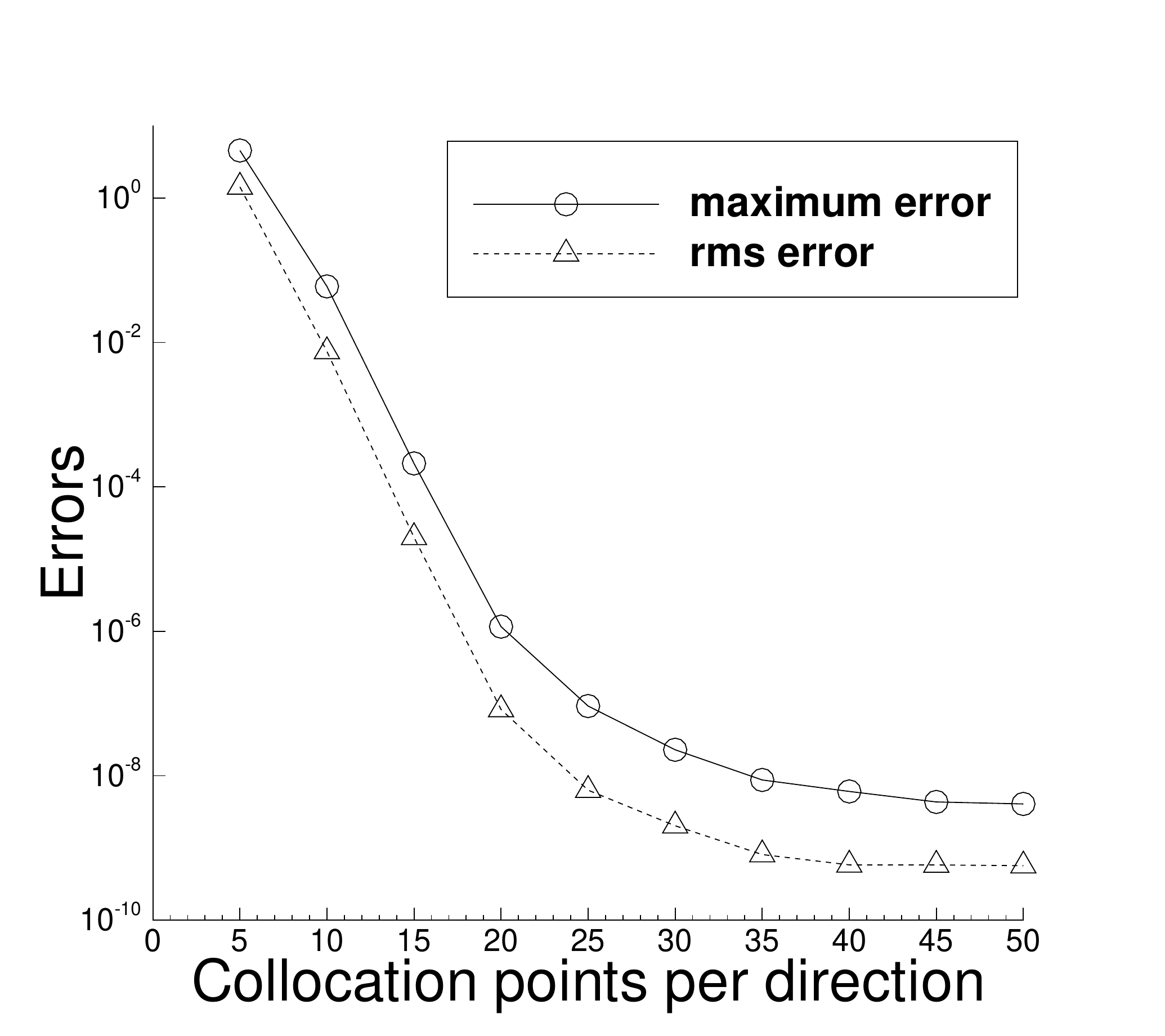}(b)
    \includegraphics[width=2in]{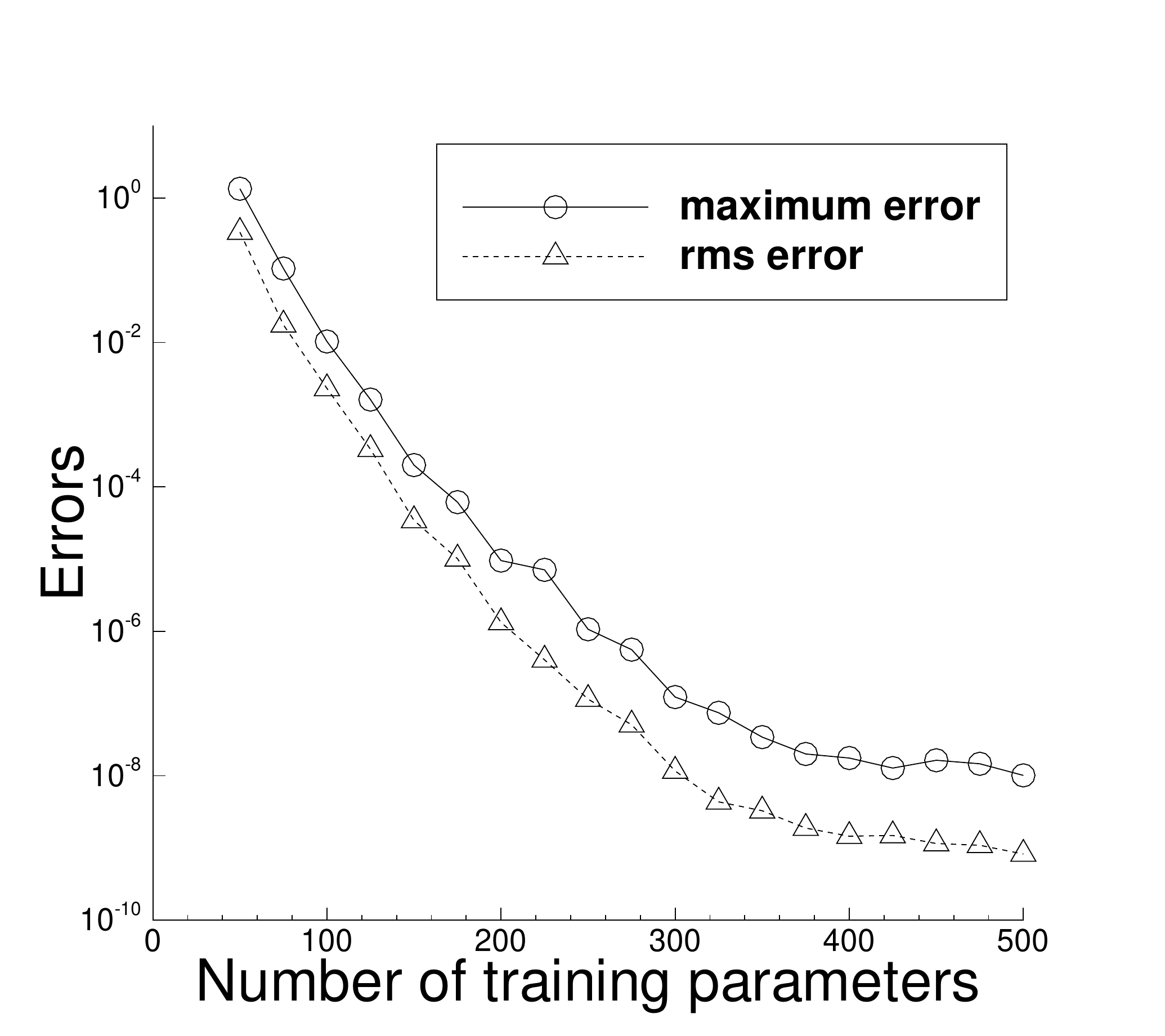}(c)
  }
  \caption{Function approximation (Multi-Rm-ELM):
    (a) absolute-error distribution of the ELM solution.
    The maximum/rms errors in the domain versus (b) the
    number of collocation points per direction,
    and (c) the number of training parameters $M$.
    Network architecture: [2, 100, $M$, 1].
    $Q=31\times 31$ in (a,c), varied in (b).
    $M=400$ in (a,b), varied in (c).
    $\mbs R_m=(0.6, 0.3)$ in (a,b,c).
  }
  \label{fg_5}
\end{figure}

Figure \ref{fg_5} illustrates the accuracy of the ELM approximant
obtained with Multi-Rm-ELM using an $\mbs R_m$
close to the optimum $\mbs R_{m0}$.
In this group of tests, we employ a neural network
with two hidden layers, with its architecture given
by $[2, 100, M, 1]$, where $M$ is either fixed at $M=400$ or varied systematically.
We employ a set of $Q=Q_1\times Q_1$ uniform collocation points,
where $Q_1$ is either fixed at $Q_1=31$ or varied systematically.
We employ a fixed $\mbs R_m=(0.6,0.3)$ in Multi-Rm-ELM,
which is close to the optimum $\mbs R_{m0}=(0.64,0.28)$
obtained from the differential evolution algorithm
corresponding to $M=400$ and $Q=31\times 31$.
Figure \ref{fg_5}(a) shows the absolute error distribution of
the Multi-Rm-ELM approximant obtained with $M=400$ and $Q=31\times 31$.
The approximation is observed to be highly accurate,
with a maximum error on the order $10^{-8}$ in the domain.
Figures \ref{fg_5}(b) and (c) show the maximum and rms errors
 of the Multi-Rm-ELM approximants as
a function of $Q_1$ (with a fixed $M=400$) and
as a function of $M$ (with a fixed $Q_1=31$), respectively.
The results demonstrate the exponential decrease (before saturation) in the errors
with respect to the collocation points and
the training parameters.

\begin{figure}
  \centerline{
    \includegraphics[width=2in]{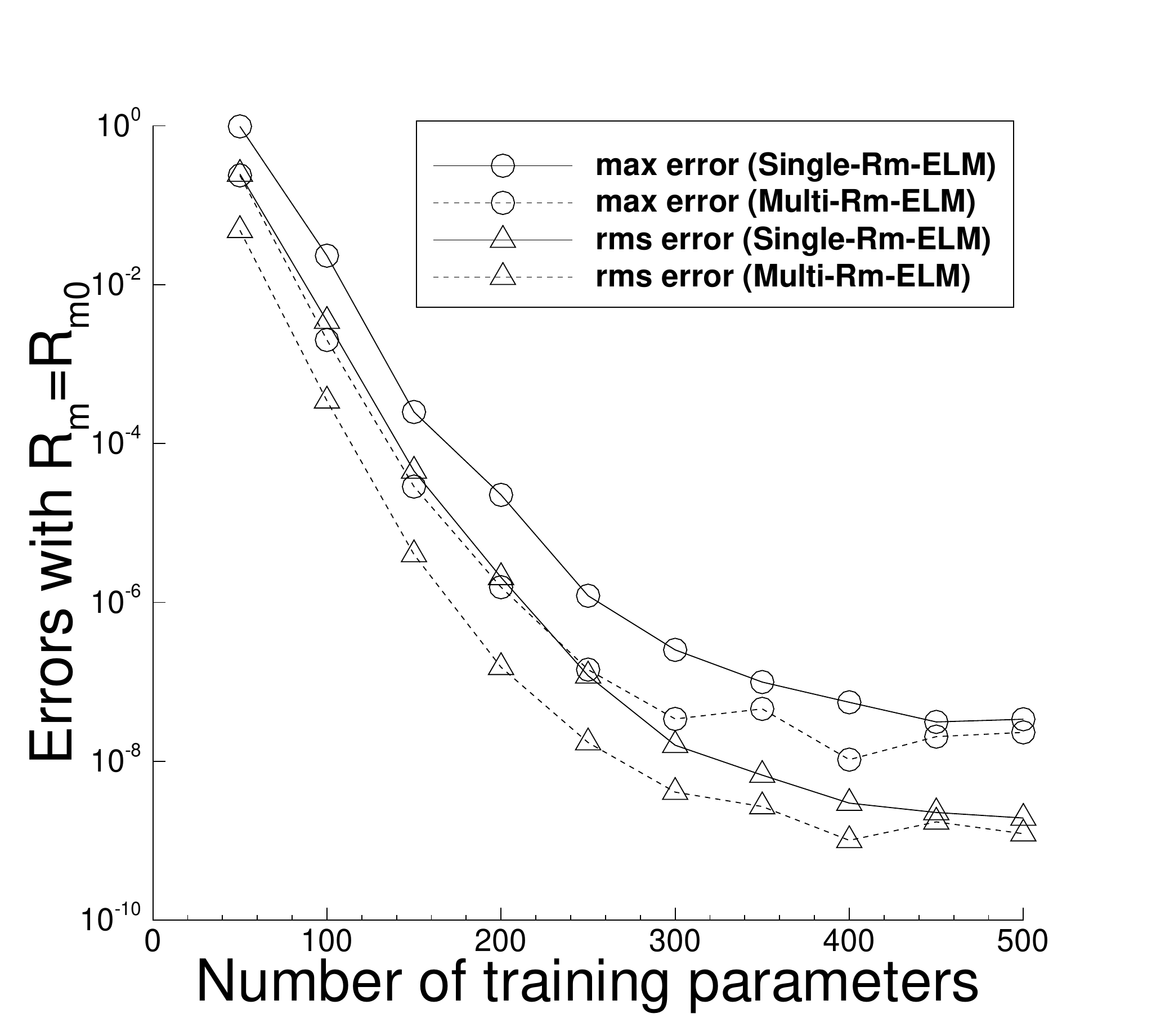}(a)
    \includegraphics[width=2in]{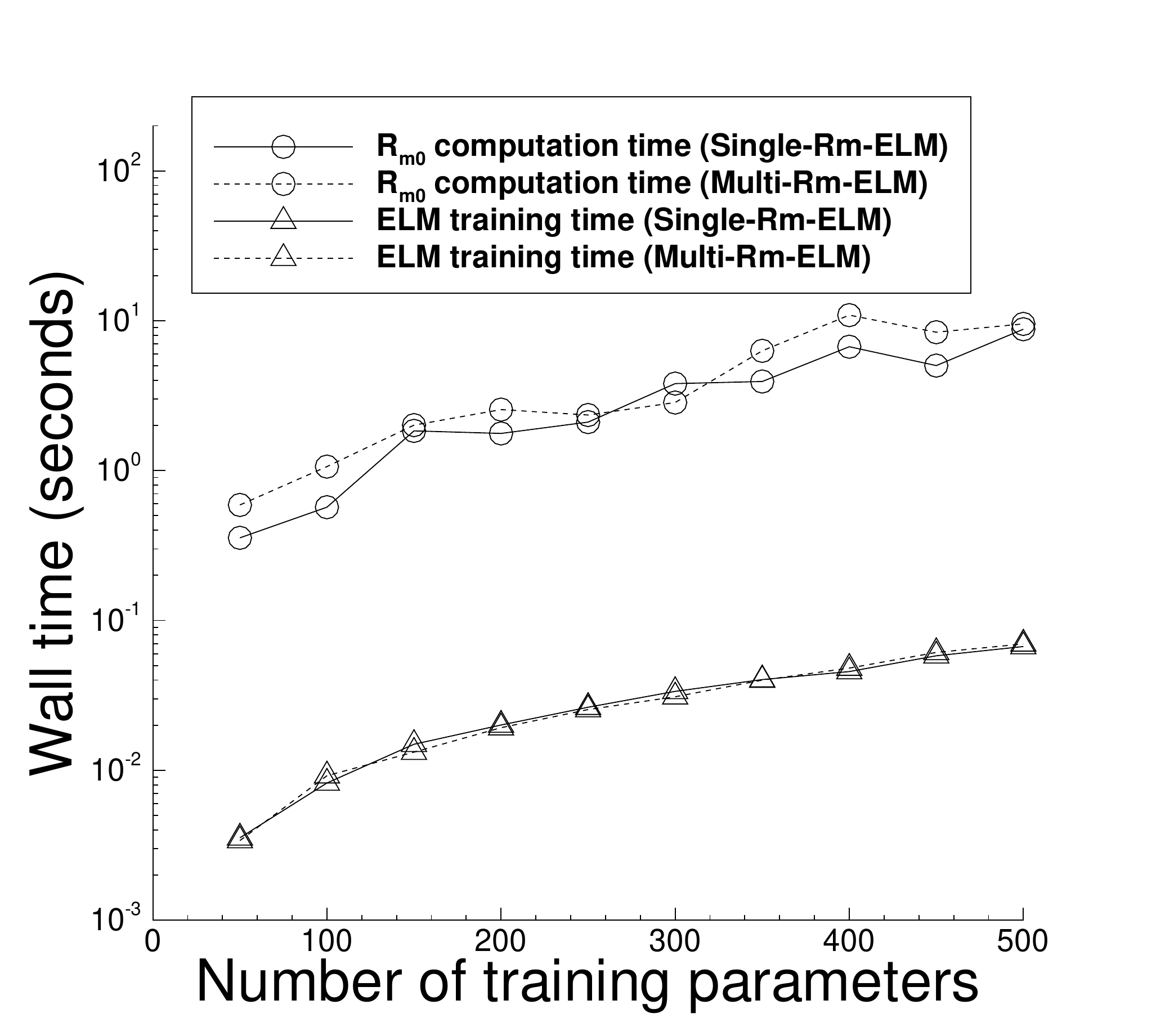}(b)
  }
  \caption{Function approximation: (a) The maximum/rms errors in the domain
    obtained with $R_m=R_{m0}$ in Single-Rm-ELM and with $\mbs R_m=\mbs R_{m0}$
    in Multi-Rm-ELM, versus the number of training parameters ($M$)
    in the neural network. (b) The  $R_{m0}$ (or $\mbs R_{m0}$) computation time
    and the ELM network training time  in Single-Rm-ELM and in
    Multi-Rm-ELM, versus the number of training parameters.
    Network architecture: [2, 100, $M$, 1].
    $Q=31\times 31$ in (a,b).
  }
  \label{fg_6}
\end{figure}

Figure \ref{fg_6} is a comparison of Single-Rm-ELM and
Multi-Rm-ELM, with regard to their accuracy
and cost for computing the optimal $R_m$/$\mbs R_m$
with the differential evolution algorithm.
In this group of tests we employ a set of $Q=31\times 31$
uniform collocation points, and a neural network with an
architecture  $[2, 100, M, 1]$, where $M$ is
varied between $50$ and $500$.

Figure \ref{fg_6}(a) shows the maximum/rms errors of
the Single-Rm-ELM (or Multi-Rm-ELM) configuration corresponding to $R_m=R_{m0}$
(resp.~$\mbs R_m=\mbs R_{m0}$), as a function of the number of
training parameters $M$ in the neural network.
The Multi-Rm-ELM errors are observed to be
consistently lower, sometimes by over an order of magnitude,
than the Single-Rm-ELM errors.
This shows that, by setting the hidden-layer coefficients
to random values with different maximum magnitudes for different
hidden layers, as in the Multi-Rm-ELM configuration,
one can achieve a better accuracy with the ELM method.

Figure \ref{fg_6}(b) shows a comparison of the
$R_{m0}$/$\mbs R_{m0}$ computation time by the differential evolution
algorithm, as well as the ELM training time of the neural network,
with the Single-Rm-ELM/Multi-Rm-ELM configurations.
For both Single-Rm-ELM and Multi-Rm-ELM, $R_{m0}$ and $\mbs R_{m0}$
are computed by using
a population size of $10$, $R_m$/$\mbs R_m$ bounds of $[0.01,3]$,
and a relative tolerance $0.1$ in the differential evolution algorithm.
Note that when computing the $R_{m0}$ (or $\mbs R_{m0}$) the differential
evolution algorithm would invoke the Algorithms~\ref{alg_1} or~\ref{alg_2}
(the ELM training routine) whenever the residual norm for
some $R_m$ (or $\mbs R_m$) needs to be evaluated.
The ELM training routine would be typically called
dozens of times by the differential evolution algorithm.
As shown by Figure \ref{fg_6}(b), the $R_{m0}$ (or $\mbs R_{m0}$)
computation time  for this function approximation problem is typically on
the order of $1$ to $10$ seconds.
In contrast, the ELM network training time for a given $R_m$
is typically on the order of $0.01$ to $0.1$ seconds
for this function approximation problem.
From Figure \ref{fg_6}(b) we can also observe  that
computing the $\mbs R_{m0}$ 
in Multi-Rm-ELM is generally more expensive than computing the $R_{m0}$
in Single-Rm-ELM.
Figure \ref{fg_6} indicates that there is a trade-off between
the accuracy and the cost for computing the optimal $R_m$ (or $\mbs R_m$).
While the Multi-Rm-ELM
is more accurate than the Single-Rm-ELM,
the cost for computing the optimal $\mbs R_m$ is also
generally larger.

\subsection{Poisson Equation}
\label{sec:poisson}

\begin{figure}
  \centerline{
    \includegraphics[width=2in]{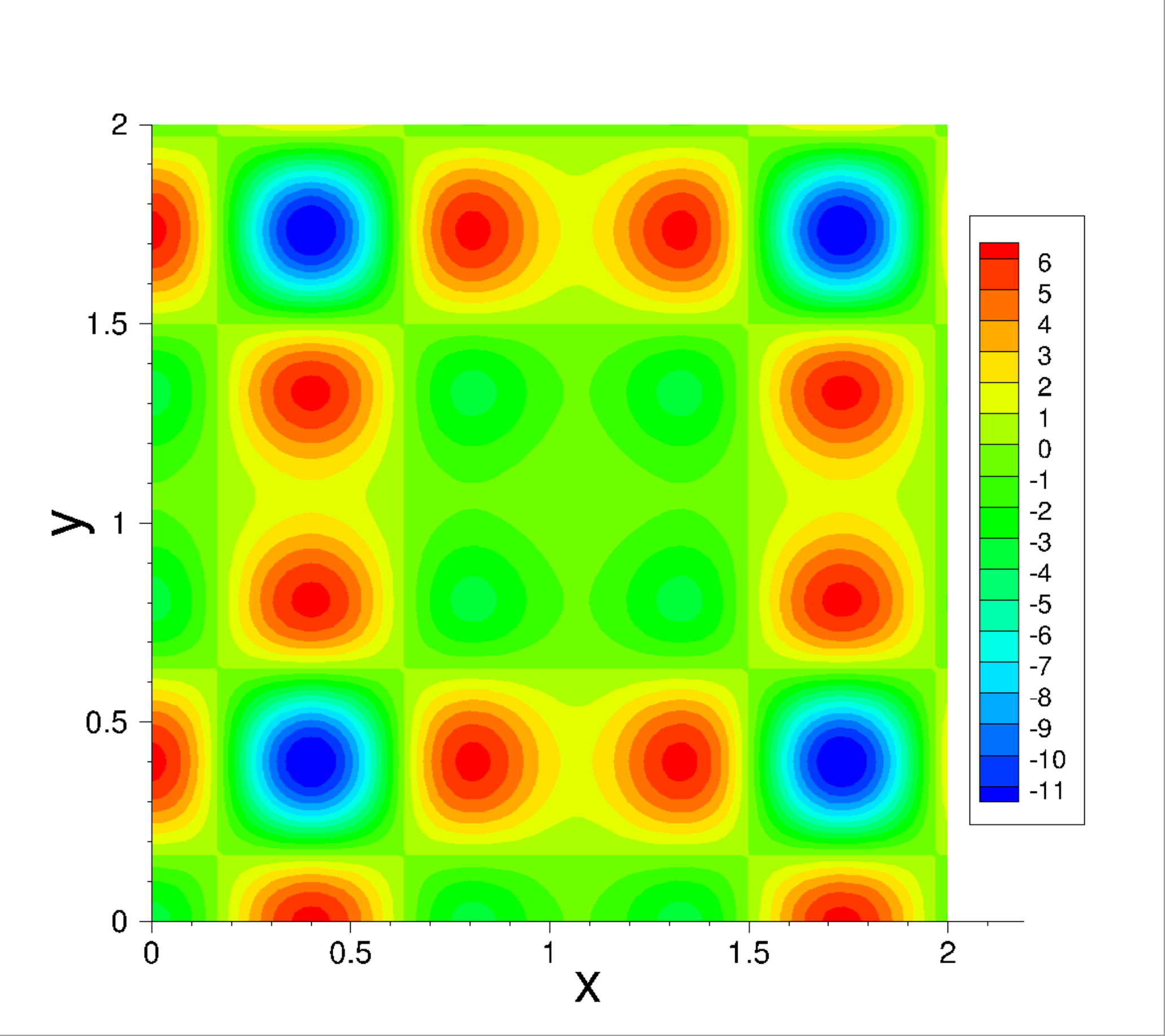}
  }
  \caption{Poisson equation: Distribution of the exact solution.}
  \label{fg_7}
\end{figure}

In this subsection we use the canonical 2D Poisson equation
to test the method for computing the optimal $R_m$ (or $\mbs R_m$)
and study the effect of the simulation parameters
on the optimum $R_{m0}$ (or $\mbs R_{m0}$).
We compare the current ELM method with the classical 
and high-order FEMs in terms of their computational performance.
A seed value $10$ is used
in the random number generators of Tensorflow and numpy
for all the numerical tests in this subsection.

Consider the 2D rectangular domain
$\Omega=\{ (x,y)\ |\ 0\leqslant x,y\leqslant 2 \}$
and the following boundary value problem with the Poisson equation
on $\Omega$ and Dirichlet boundary conditions on $\partial\Omega$,
\begin{subequations}\label{eq_15}
  \begin{align}
    & \frac{\partial^2 u}{\partial x^2}
    + \frac{\partial^2u}{\partial y^2} = f(x,y), \label{eq_15a} \\
    & u(x,0)=g_1(x),\quad u(x,2) = g_2(x),\quad
    u(0,y) = h_1(y), \quad u(2,y) = h_2(y). \label{eq_15b}
  \end{align}
\end{subequations}
Here $u(x,y)$ is the field to be solved for,
$f(x,y)$ is a prescribed source term, and $g_1$, $g_2$, $h_1$ and $h_2$
are the prescribed boundary distributions.
We choose $f(x,y)$ such that the following field satisfies \eqref{eq_15a},
\begin{align}
  &
  u(x,y) = -\left[ 2\cos\left(\frac32\pi x+\frac{2\pi}{5} \right)
    +\frac32\cos\left(3\pi x - \frac{\pi}{5} \right)
    \right] \left[ 2\cos\left(\frac32\pi y+\frac{2\pi}{5} \right)
    +\frac32\cos\left(3\pi y - \frac{\pi}{5} \right)
    \right]. \label{eq_16}
\end{align}
We set $g_1$, $g_2$, $h_1$ and $h_2$ by evaluating the expression
\eqref{eq_16} on the corresponding domain boundaries.
Under these settings the expression~\eqref{eq_16}
solves the boundary value problem~\eqref{eq_15}.
Figure \ref{fg_7} shows the distribution of the
analytic solution~\eqref{eq_16} in the $x$-$y$ plane.


We solve this problem using ELM. The neural network
has an input layer of two nodes (representing $x$ and $y$),
a linear output layer of one node (representing $u$),
and one or more hidden layers in between with the Gaussian
activation function. The specific architectures of the
neural networks will be provided below, again with
$M$ denoting the number of training parameters
(i.e.~the number of nodes of the last hidden layer).
We employ a set of $Q=Q_1\times Q_1$ uniform grid points on $\Omega$
as the collocation points,  where $Q_1$ denotes the number of
points in both $x$ and $y$ directions.
So there are $Q_1$ uniform collocation points on each domain boundary.
$Q_1$ and $M$ are varied systematically
in the numerical tests.
We employ the Single-Rm-ELM and Multi-Rm-ELM configurations
from Section \ref{sec:method} for setting the
random hidden-layer coefficients based on a single $R_m$
or a vector $\mbs R_m$, respectively.


\begin{figure}
  \centerline{
    \includegraphics[width=2in]{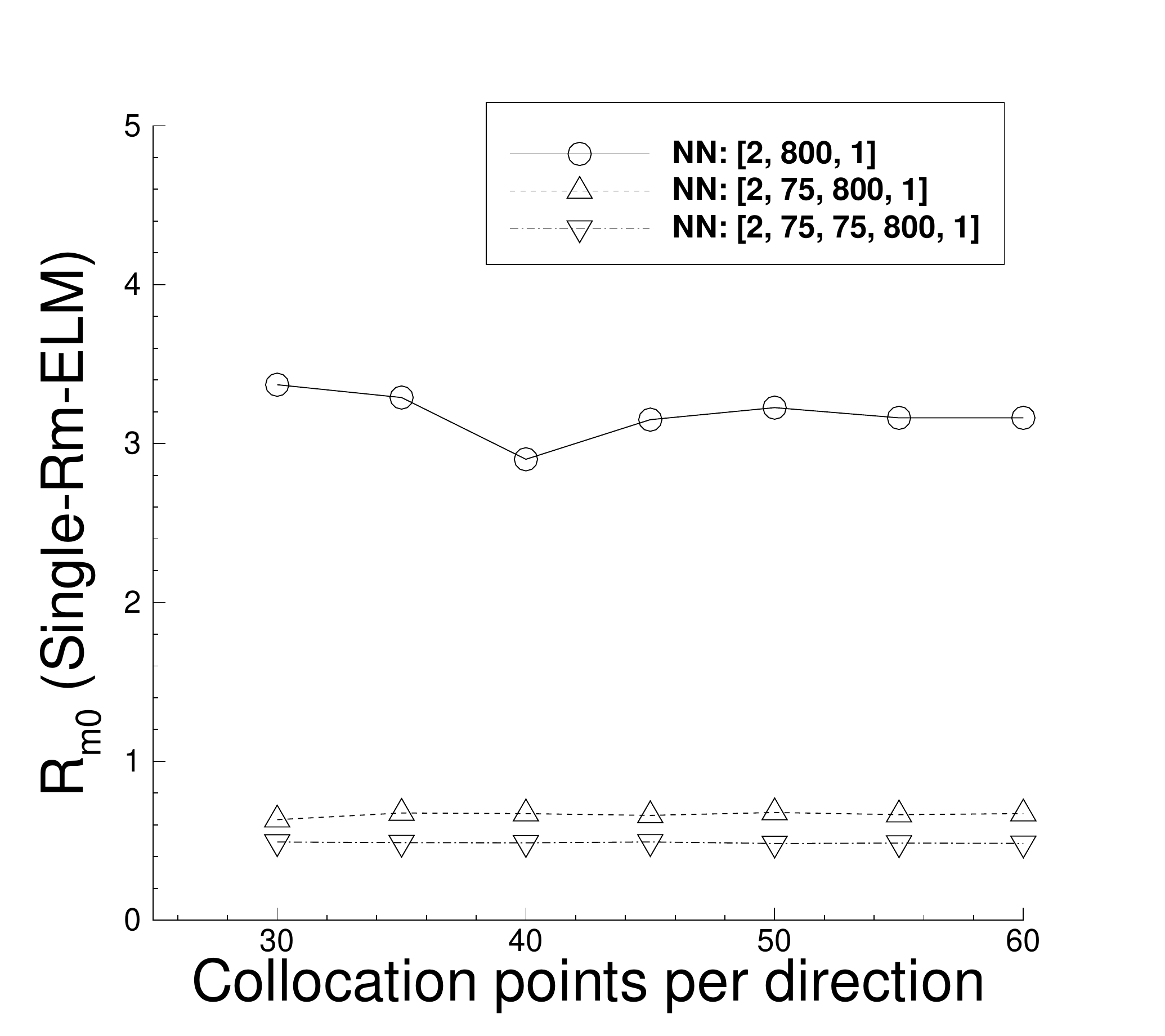}(a)
    \includegraphics[width=2in]{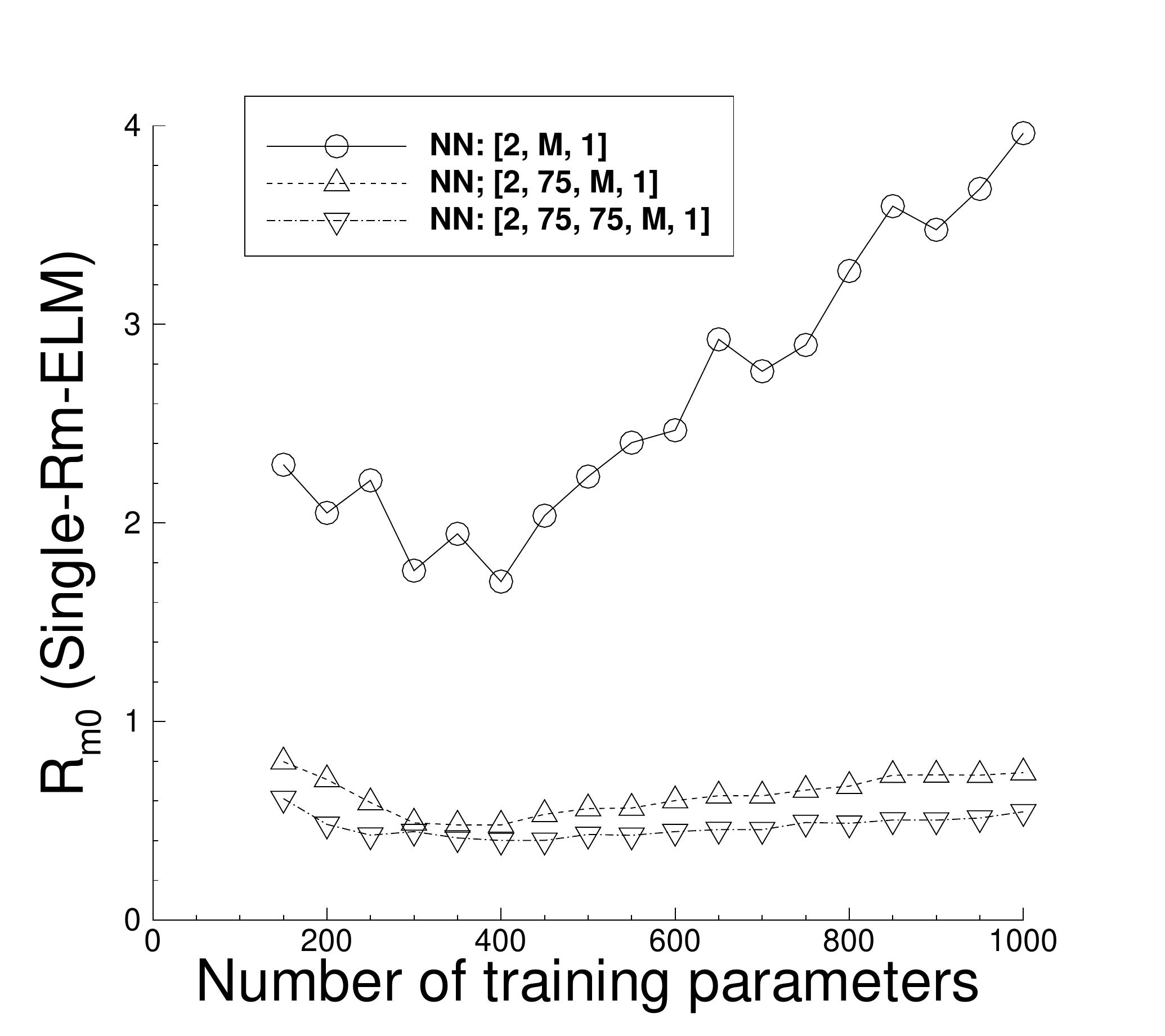}(b)
    \includegraphics[width=2in]{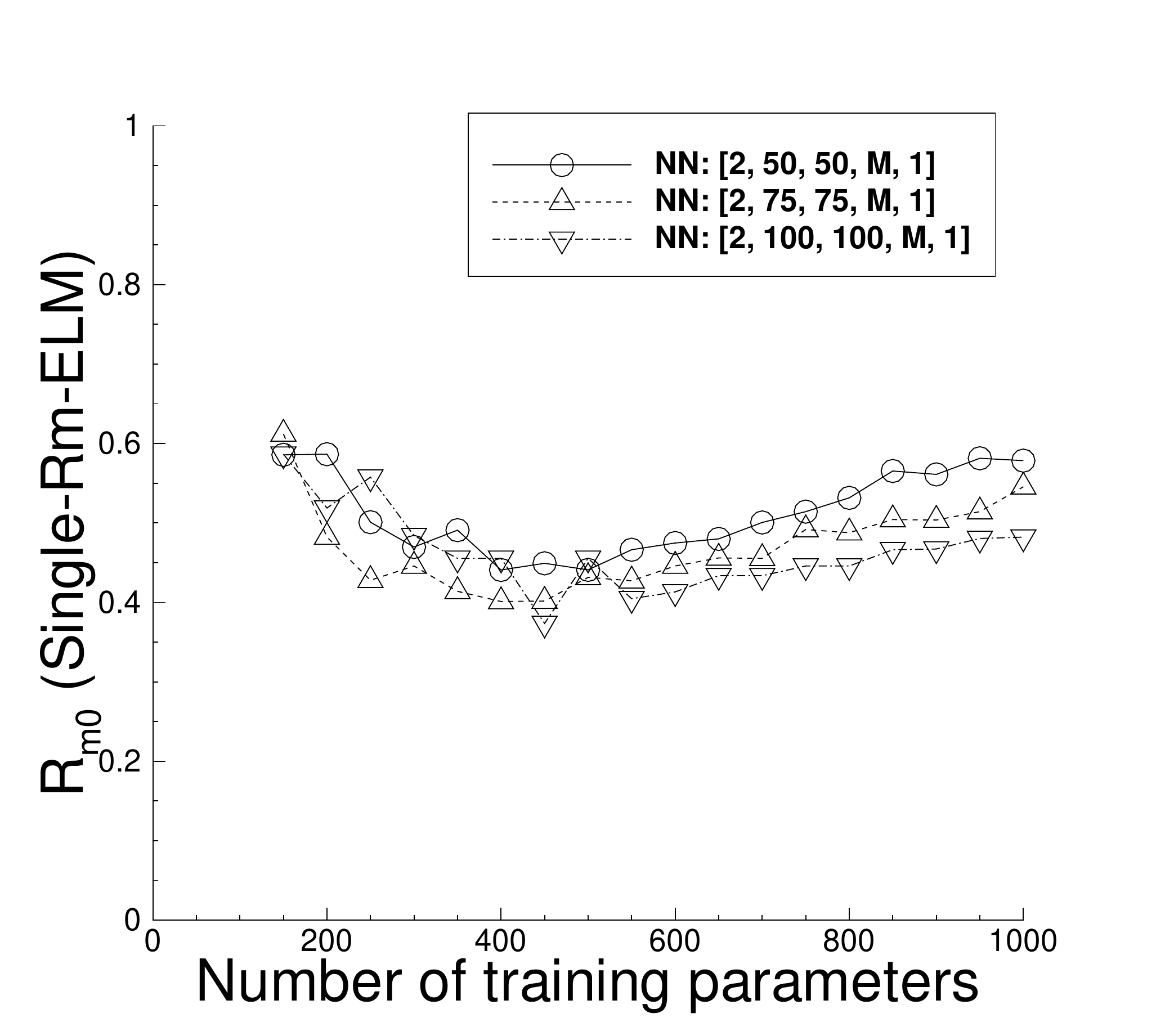}(c)
  }
  \caption{Poisson equation (Single-Rm-ELM):
    $R_{m0}$ versus (a) the number of collocation points
    per direction and (b) the number of training parameters,
    with neural networks of different depth.
    (c) $R_{m0}$ versus the number of training parameters
    for neural networks with the same depth but  different width. 
    $Q=35\times 35$ in (b,c), varied in (a).
    $M=800$ in (a), varied in (b,c).
    Network architectures are listed in the legends.
  }
  \label{fg_8}
\end{figure}

Figure \ref{fg_8} illustrates the characteristics of
the optimum $R_{m0}$ obtained
with the differential evolution algorithm for
the Poisson equation with the Single-Rm-ELM configuration.
Figure \ref{fg_8}(a) depicts $R_{m0}$ as a function
of $Q_1$ for several neural networks with different depth.
The network architectures are given in the legend, and 
$Q_1$ is varied systematically between $30$ and $60$.
Figure \ref{fg_8}(b) shows the $R_{m0}$
as a function of the number of training parameters $M$
for several neural networks with different depth.
Figure \ref{fg_8}(c) shows the computed $R_{m0}$
as a function of $M$ for several neural networks
that contain three hidden layers with the same $M$
but different width for the preceding
hidden layers.
In Figures \ref{fg_8}(b,c) a fixed set of $Q=35\times 35$
uniform collocation points is employed.
In the differential evolution algorithm we have employed
a population size of $6$, the $R_m$ bounds $[0.1,5]$ and
a relative tolerance of $0.1$ for these numerical tests.

We have the following observations from Figure \ref{fg_8}.
First, $R_{m0}$ is  essentially independent of the number of
collocation points in the simulation.
Second, $R_{m0}$ has a stronger dependence on the number of
training parameters $M$ for neural networks containing
a single hidden layer, and its dependence on $M$ is quite weak 
for neural networks with two or more hidden layers.
$R_{m0}$ generally increases with increasing $M$, except in
a range with smaller $M$ values where $R_{m0}$ is observed
to decrease as $M$ increases.
With two or more hidden layers in the neural network,
$R_{m0}$ can be approximated by essentially a constant for a wide
range of $M$ values.
Third, $R_{m0}$ generally decreases with increasing depth of
the neural network.
It drops significantly from a single hidden layer to two
hidden layers,  and then decreases only
slightly as the depth further increases.
Fourth, $R_{m0}$ has only a weak dependence on the
width of the preceding hidden layers (other than the last one),
and tends to decrease slightly with increasing width of the preceding
hidden layers.
These observations are consistent with those from
the function approximation problem in Section \ref{sec:func}.

\begin{figure}
  \centerline{
    \includegraphics[width=2in]{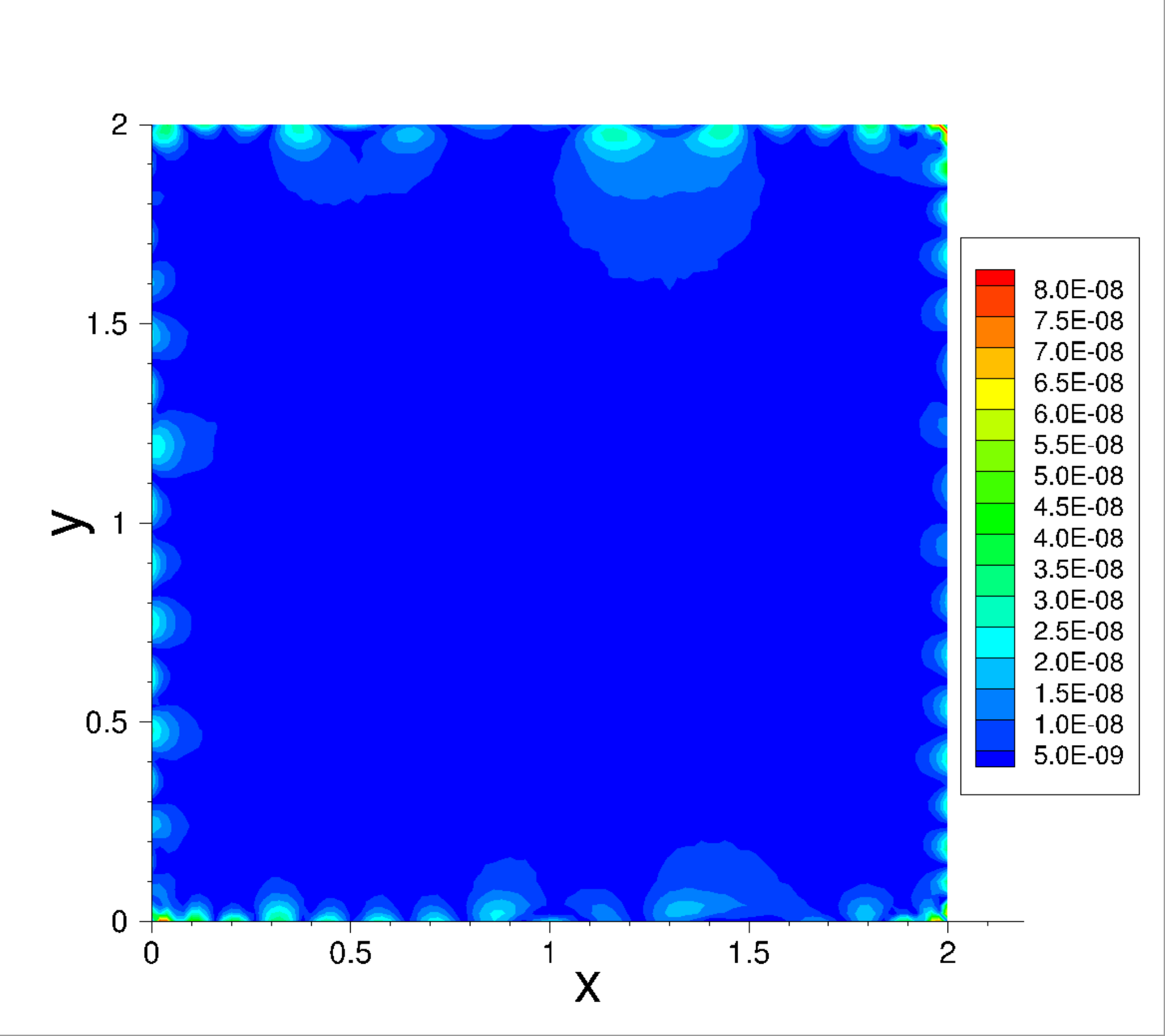}(a)
    \includegraphics[width=2in]{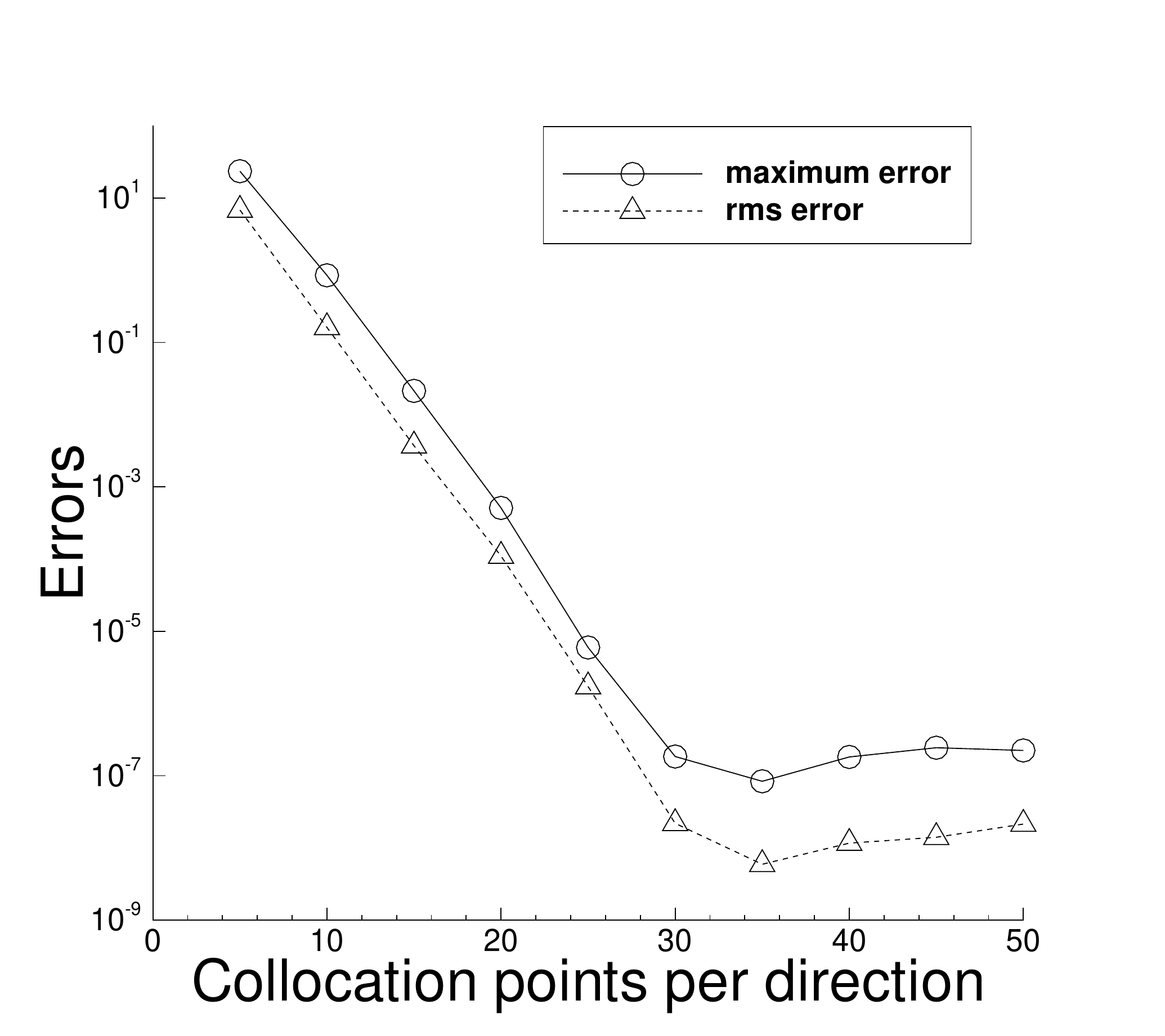}(b)
    \includegraphics[width=2in]{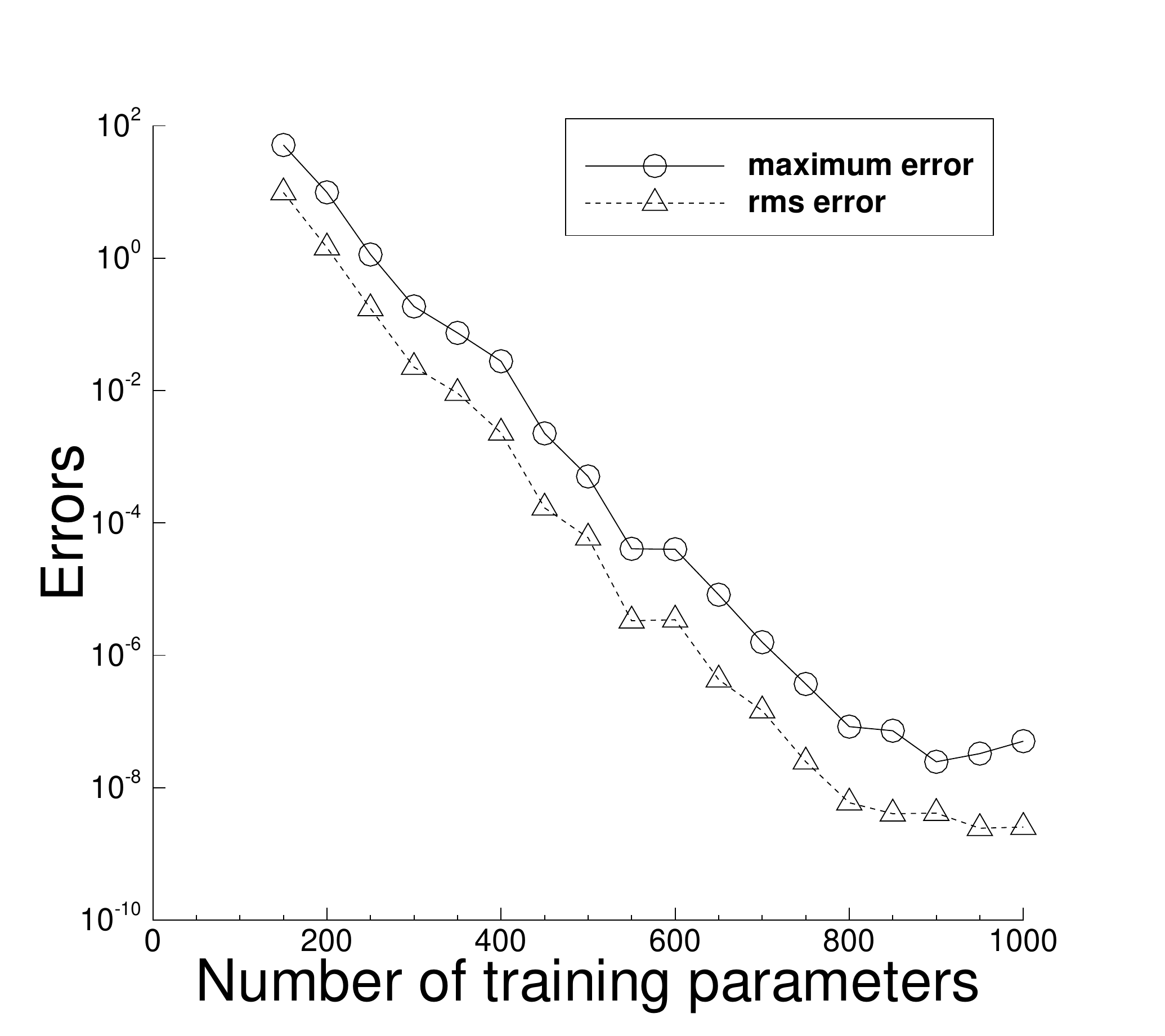}(c)
  }
  \caption{Poisson equation (Single-Rm-ELM):
    (a) Absolute error distribution of the ELM solution.
    The maximum/rms errors in the domain versus (b) the number of collocation
    points per direction and (c) the number of training parameters $M$.
    Network architecture: $[2,M,1]$.
    $Q=35\times 35$ in (a,c), varied in (b).
    $M=800$ in (a,b), varied in (c).
    $R_m=3.36$ in (a,b,c).
  }
  \label{fg_9}
\end{figure}

Figure \ref{fg_9} illustrates the solution accuracy  obtained
with the Single-Rm-ELM configuration.
In this group of tests we employ a neural network with
an architecture $[2, M, 1]$ and the Gaussian activation function,
where the number of training parameters
is either fixed at $M=800$ or varied systematically.
The set of uniform collocation points is either fixed at
$Q=35\times 35$ or varied between $Q=5\times 5$ and $Q=50\times 50$.
We employ a fixed $R_m=3.36$, which is close to the optimum $R_{m0}$
from the differential evolution algorithm, for generating the random
hidden-layer coefficients in Single-Rm-ELM.
Figure \ref{fg_9}(a) shows the absolute error distribution
of the ELM solution in the $x$-$y$ plane, which corresponds to
a fixed $M=800$ and $Q=35\times 35$.
It indicates that ELM produces an accurate solution,
with the maximum error on the order $10^{-8}$.
Figures \ref{fg_9}(b) and (c) depict the maximum and rms errors in
the domain as a function of the number of
collocation points and the number of training parameters,
respectively. One can clearly observe that the errors decrease
exponentially (before saturation) with increasing number of
collocation points and training parameters.

\begin{figure}
  \centerline{
    \includegraphics[width=2in]{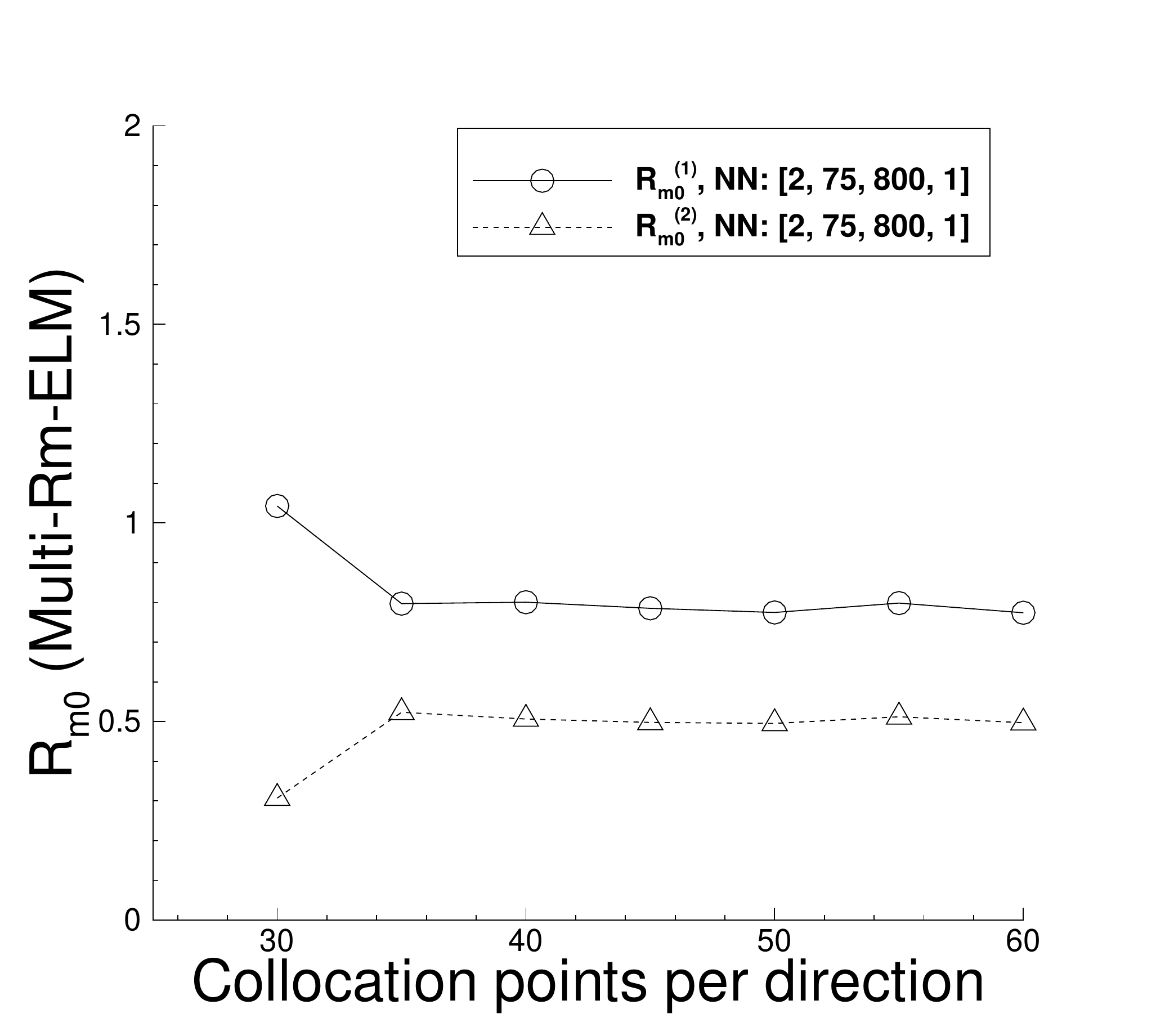}(a)
    \includegraphics[width=2in]{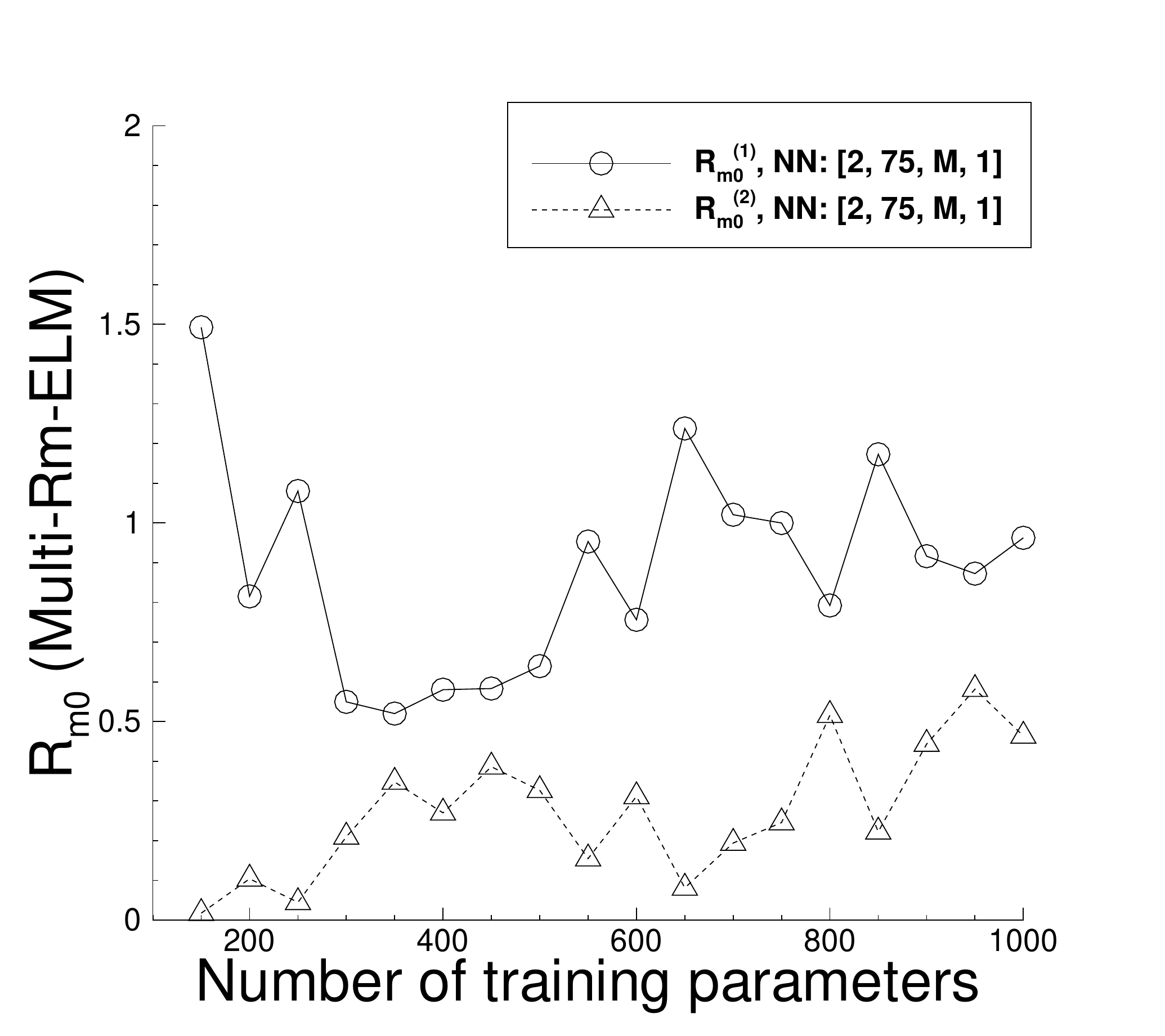}(b)
  }
  \centerline{
    \includegraphics[width=2in]{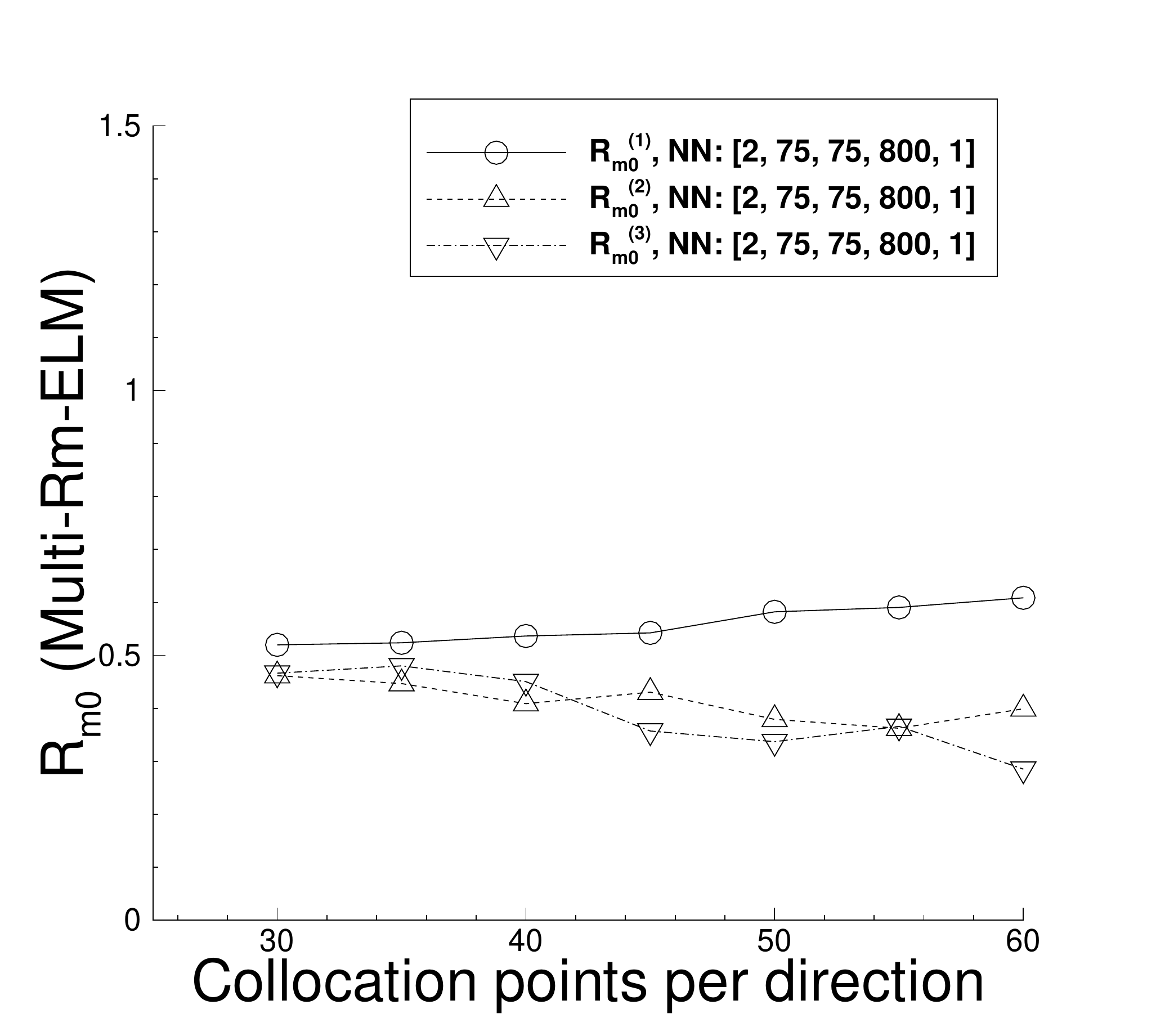}(c)
    \includegraphics[width=2in]{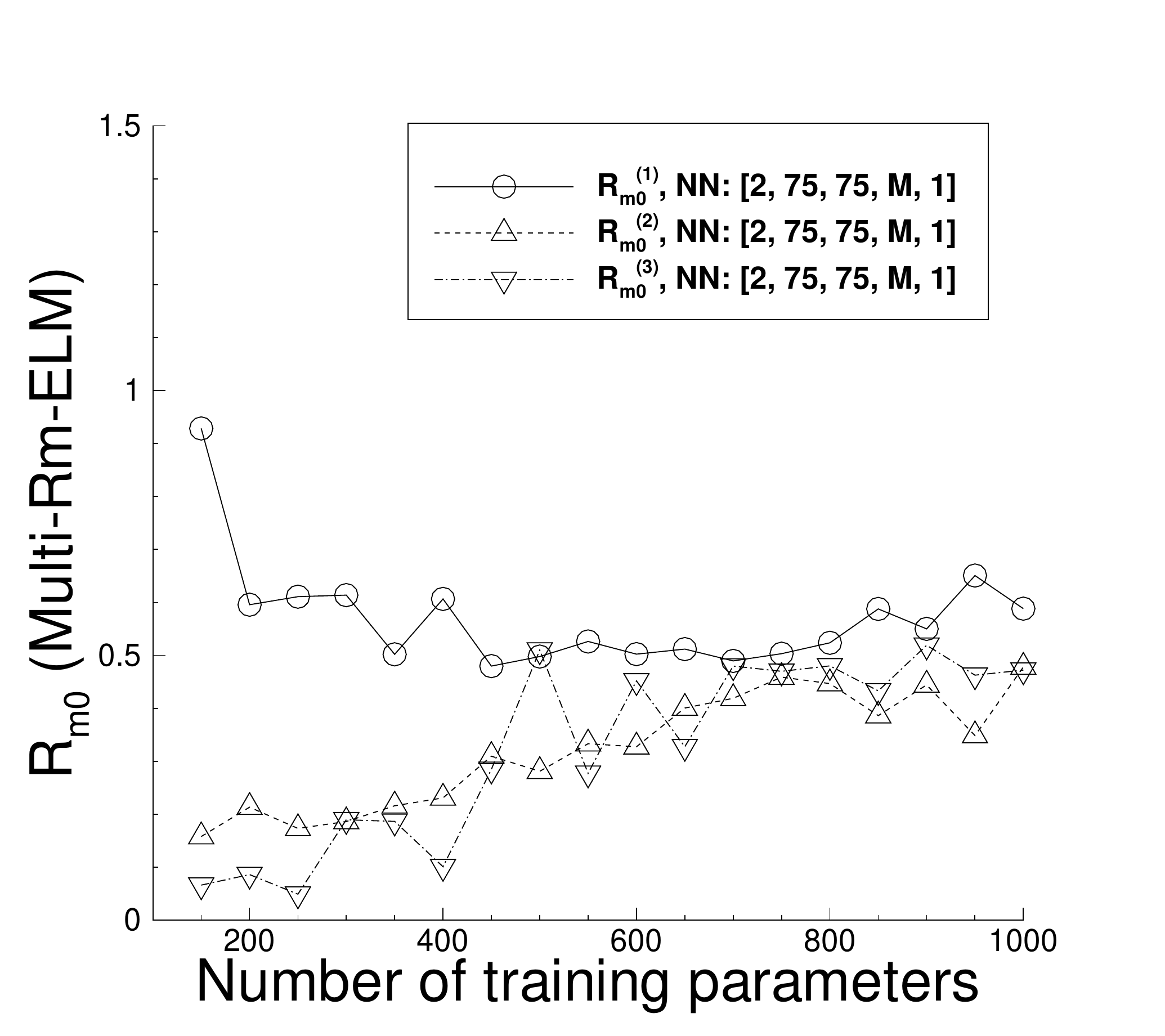}(d)
  }
  \caption{Poisson equation (Multi-Rm-ELM):
    The optimum $\mbs R_{m0}$ versus the number of collocation points
    per direction (a,c) and the number of training parameters (b,d),
    with neural networks
    having two (a,b) and three (c,d) hidden layers. The network architectures
    are given in the legends.
    $Q=35\times 35$ in (b,d), varied in (a,c). $M=800$ in (a,c),
    varied in (b,d).
  }
  \label{fg_10}
\end{figure}

Figure \ref{fg_10} illustrates the characteristics of the optimum
$\mbs R_{m0}$ for the Multi-Rm-ELM configuration obtained with
the differential evolution algorithm.
Here we have considered two neural networks with
two and three hidden layers, whose architectures are characterized by
$[2, 75, M, 1]$ and $[2, 75, 75, M, 1]$, respectively,
where $M$ is either fixed at $M=800$ or varied systematically.
A set of uniform collocation points is employed, either fixed at
$Q=35\times 35$ or varied systematically between $Q=30\times 30$
and $Q=60\times 60$.
Figures \ref{fg_10}(a) and (b) show the components of
$\mbs R_{m0}=(R_{m0}^{(1)},R_{m0}^{(2)})$
versus the number of collocation points and the number of training parameters
for the neural network with two hidden layers, respectively.
These are obtained with a population size of $8$, the bounds $[0.01, 3]$
for all $\mbs R_m$ components, and a relative tolerance of $0.1$
in the differential evolution algorithm.
Figures \ref{fg_10}(c) and (d) show the components of
$\mbs R_{m0}=(R_{m0}^{(1)},R_{m0}^{(2)},R_{m0}^{(3)})$ versus
the number of collocation points and the number of training parameters
for the neural network with three hidden layers, respectively.
They are obtained using a population size of $9$, the bounds $[0.01, 3]$
for all the $\mbs R_m$ components, and a relative tolerance of $0.1$
in the differential evolution algorithm.
One can see that the $\mbs R_{m0}$ components
exhibit a fairly weak dependence (Figure \ref{fg_10}(c)) or
essentially no dependence (Figure \ref{fg_10}(a)) on
the number of the collocation points in the domain.
The relation between $\mbs R_{m0}$
and the number of training parameters, on the other hand,  appears 
quite irregular.
The $\mbs R_{m0}$ components tend to increase as the
number of training parameters $M$
increases, except for some component, which appears to decrease
in a range of smaller $M$ values.

\begin{figure}
  \centerline{
    \includegraphics[width=2in]{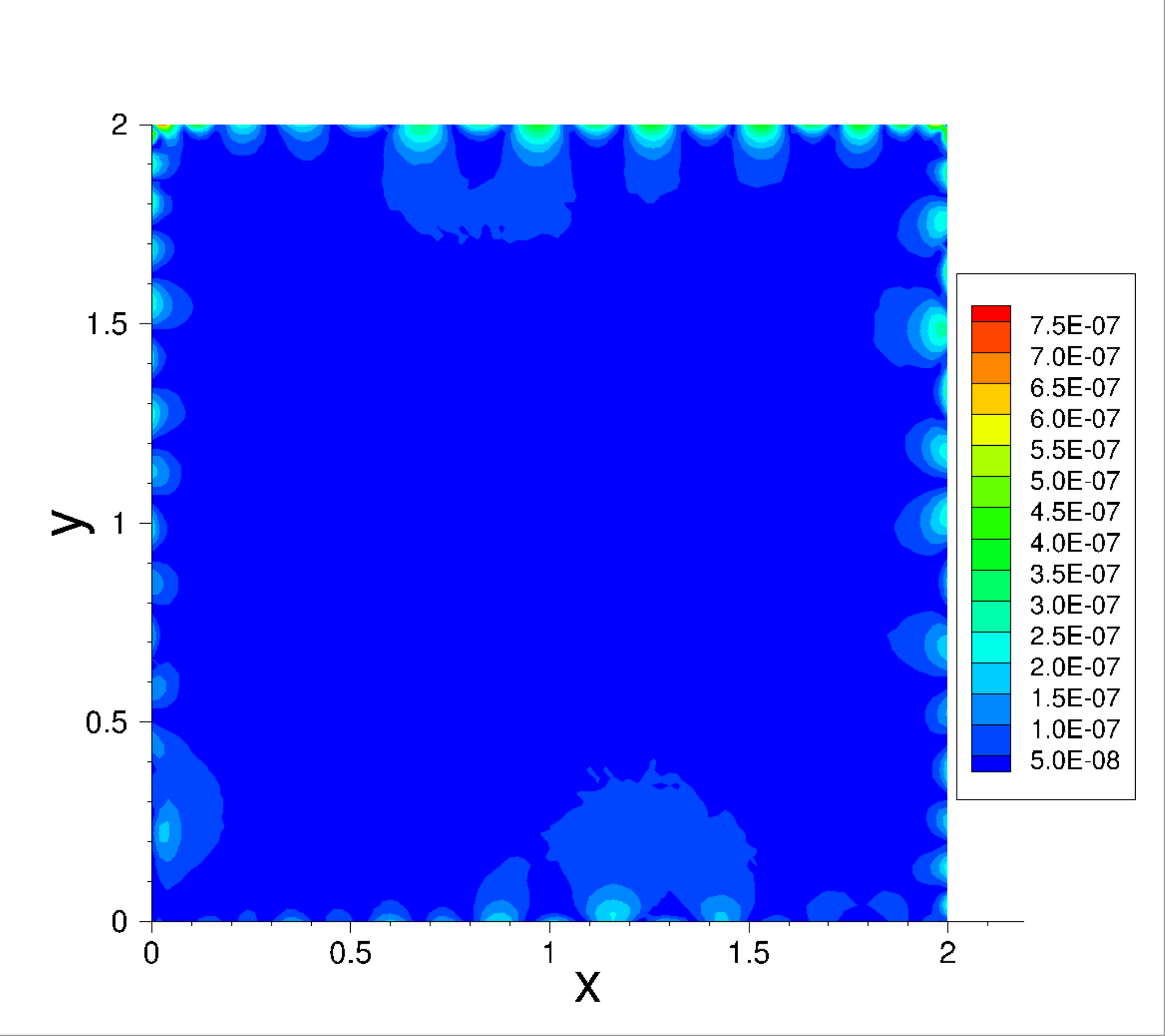}(a)
    \includegraphics[width=2in]{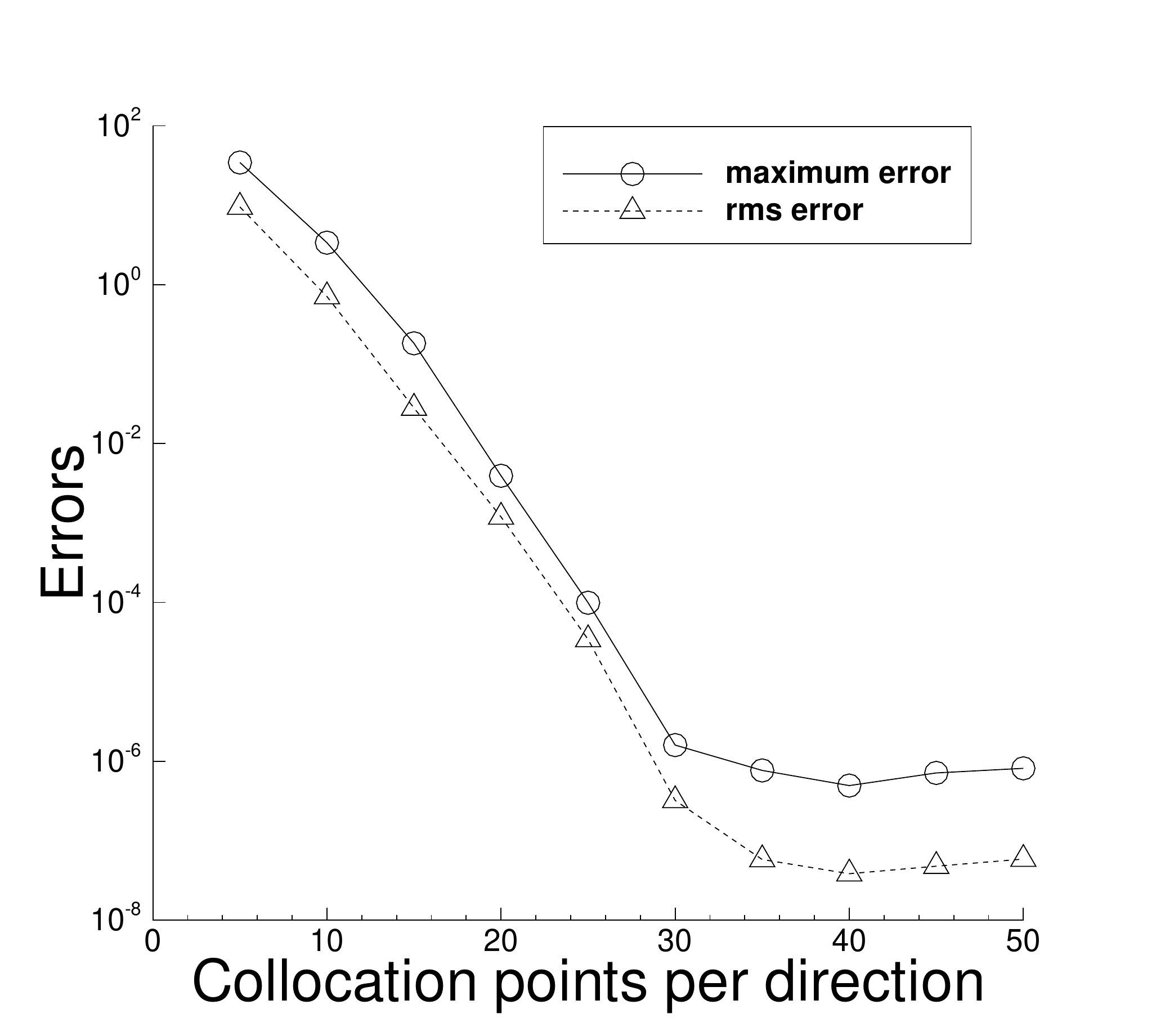}(b)
    \includegraphics[width=2in]{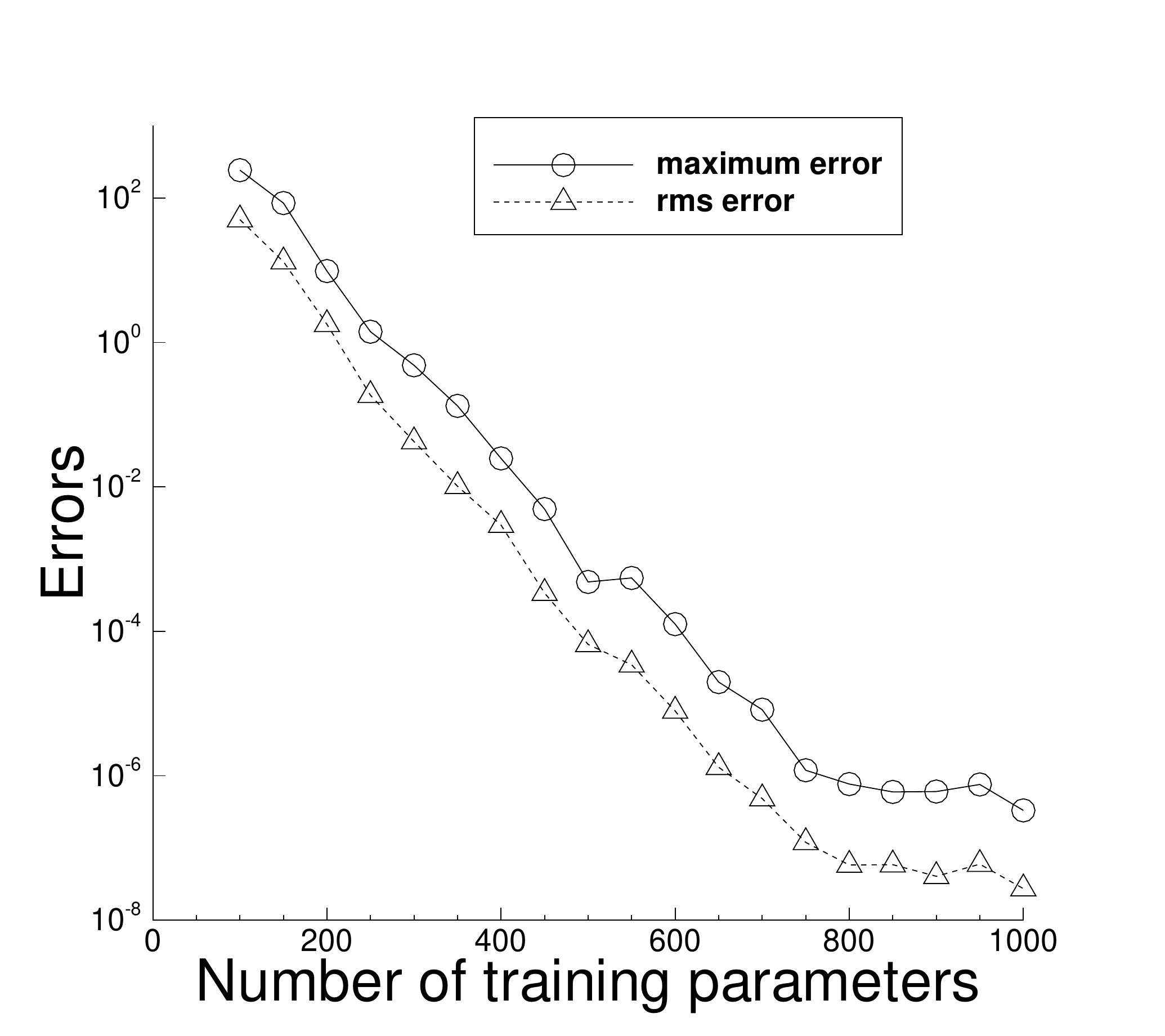}(c)
  }
  \caption{Poisson equation (Multi-Rm-ELM):
    (a) Absolute error distribution of the Multi-Rm-ELM solution.
    The maximum/rms errors in the domain versus
    (b) the number of collocation
    points per direction, and (c) the number of training parameters.
    Network architecture: $[2, 75, M, 1]$.
    $Q=35\times 35$ in (a,c), varied in (b).
    $M=800$ in (a,b), varied in (c).
    $\mbs R_m=(0.8, 0.5)$ in (a,b,c).
  }
  \label{fg_11}
\end{figure}

Figure \ref{fg_11} illustrates the solution accuracy obtained with
the Multi-Rm-ELM configuration.
In this group of tests we employ a neural network with two hidden layers,
with an architecture $[2, 75, M, 1]$,
where $M$ is either fixed at $M=800$ or
varied systematically.
The set of uniform collocation points is either fixed at
$Q=35\times 35$ or varied between $Q=5\times 5$ and $Q=50\times 50$.
We employ a fixed $\mbs R_m=(0.8,0.5)$ here, close
to the $\mbs R_{m0}$ obtained 
corresponding to $M=800$ and $Q=35\times 35$.
Figure \ref{fg_11}(a) shows the distribution of the absolute
error of the Multi-Rm-ELM solution corresponding to $M=800$
and $Q=35\times 35$, suggesting a quite high accuracy,
with the maximum error on the order $10^{-7}$.
Figures \ref{fg_11}(b) and (c) depict the maximum/rms errors in the
domain as a function of the number of collocation points and the training parameters,
respectively.
The exponential convergence of the errors (before saturation)
with respect to the collocation points
and the training parameters is evident.

\begin{figure}
  \centerline{
    \includegraphics[width=2in]{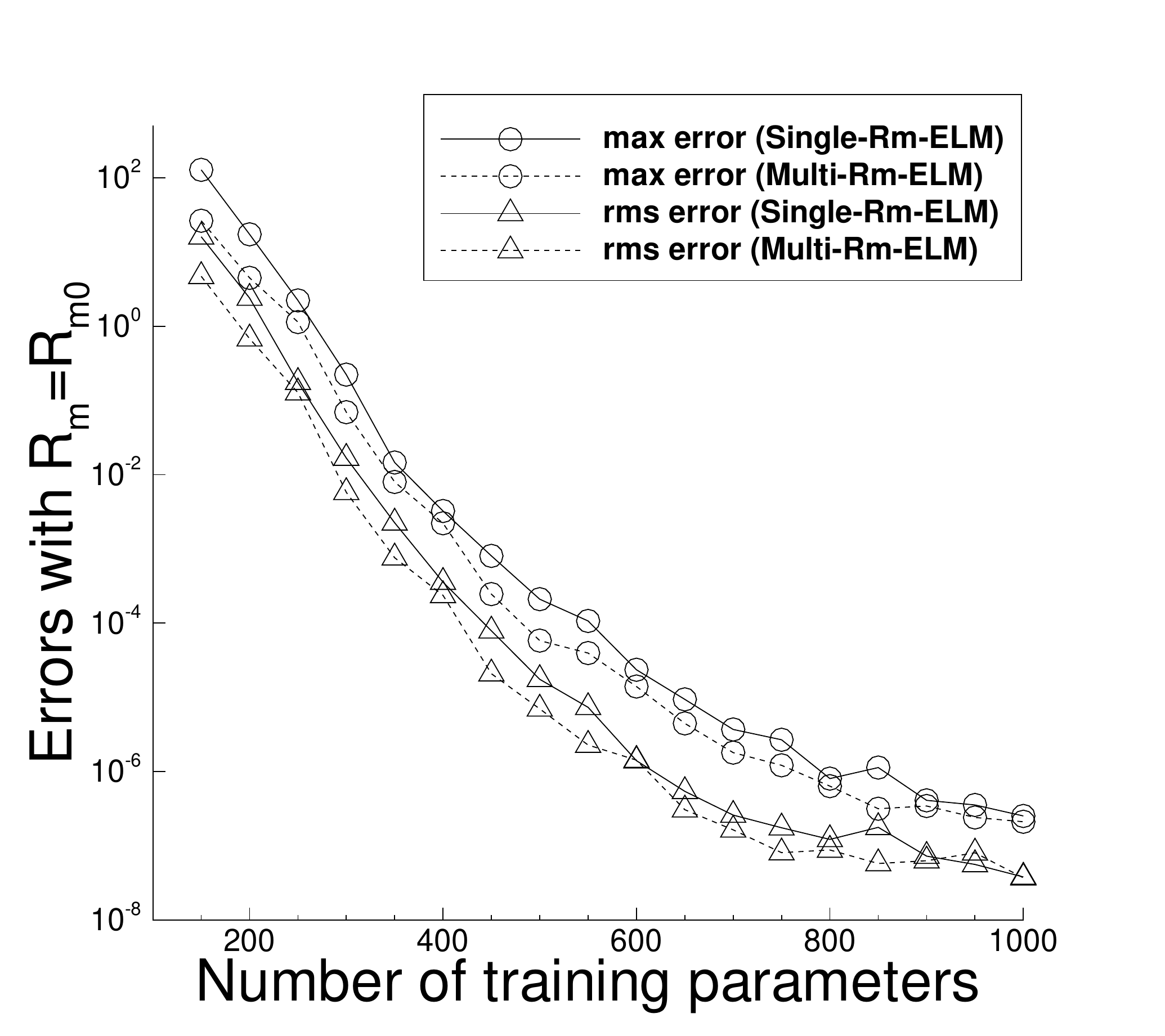}(a)
    \includegraphics[width=2in]{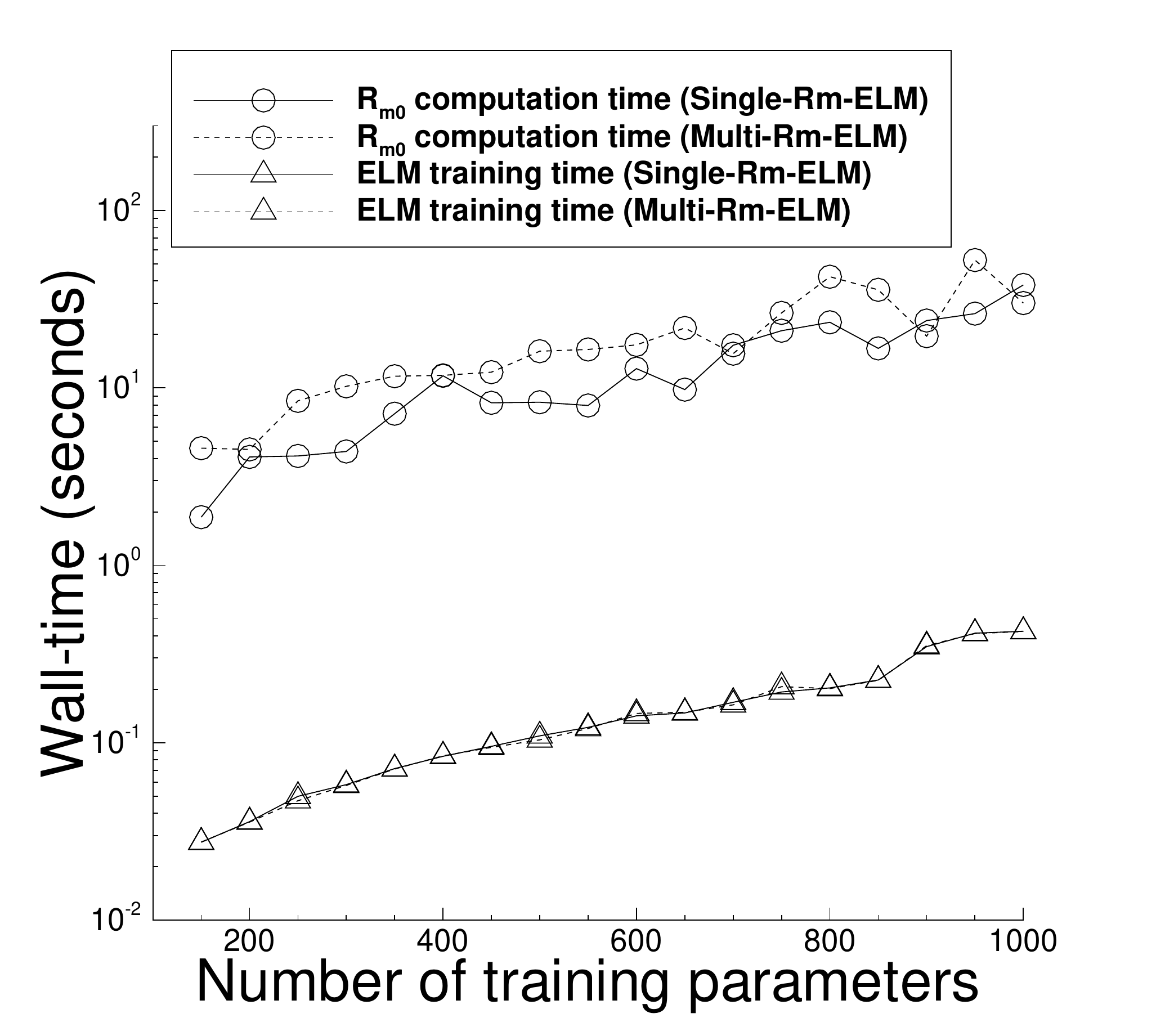}(b)
  }
  \centerline{
    \includegraphics[width=2in]{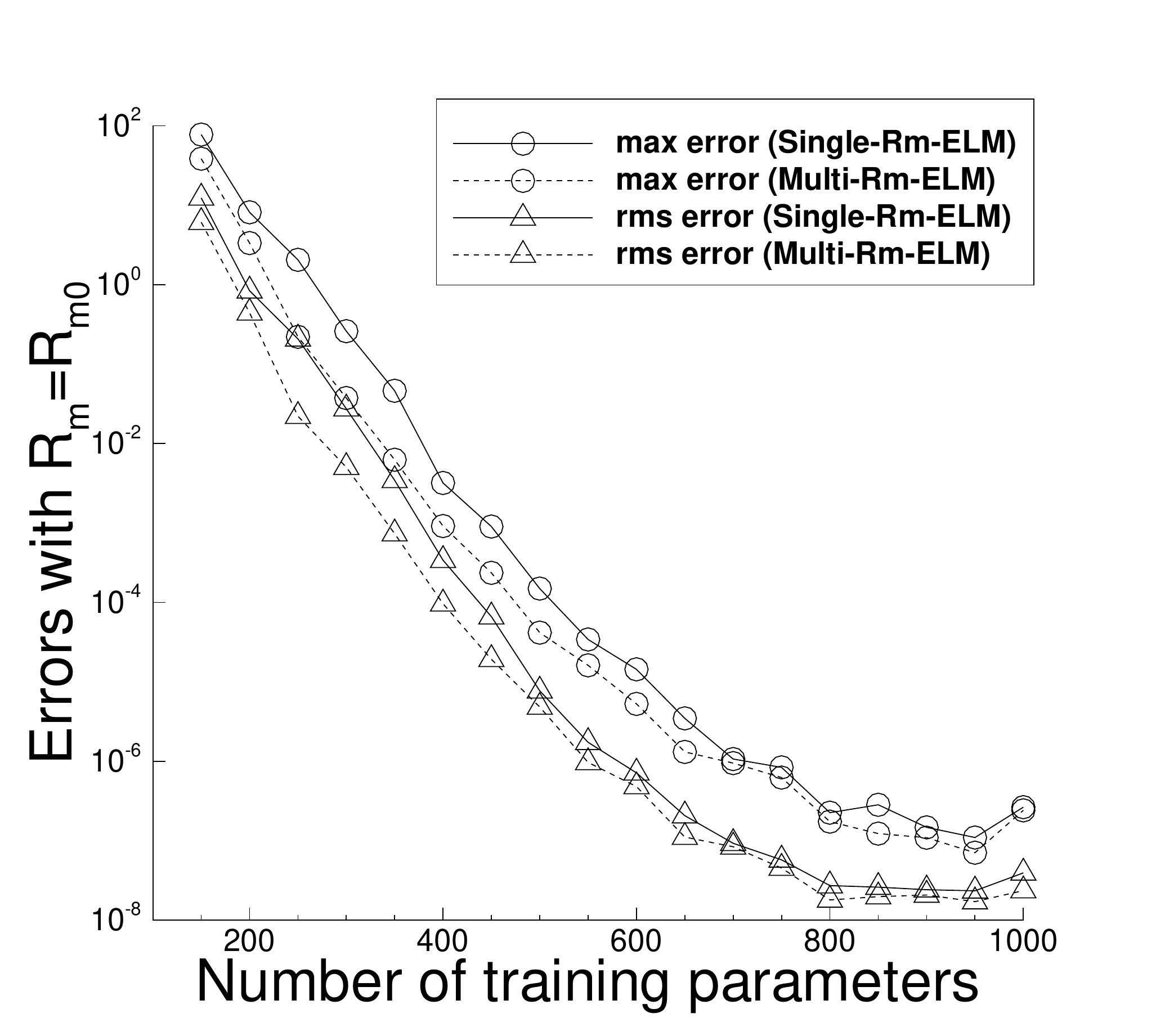}(c)
    \includegraphics[width=2in]{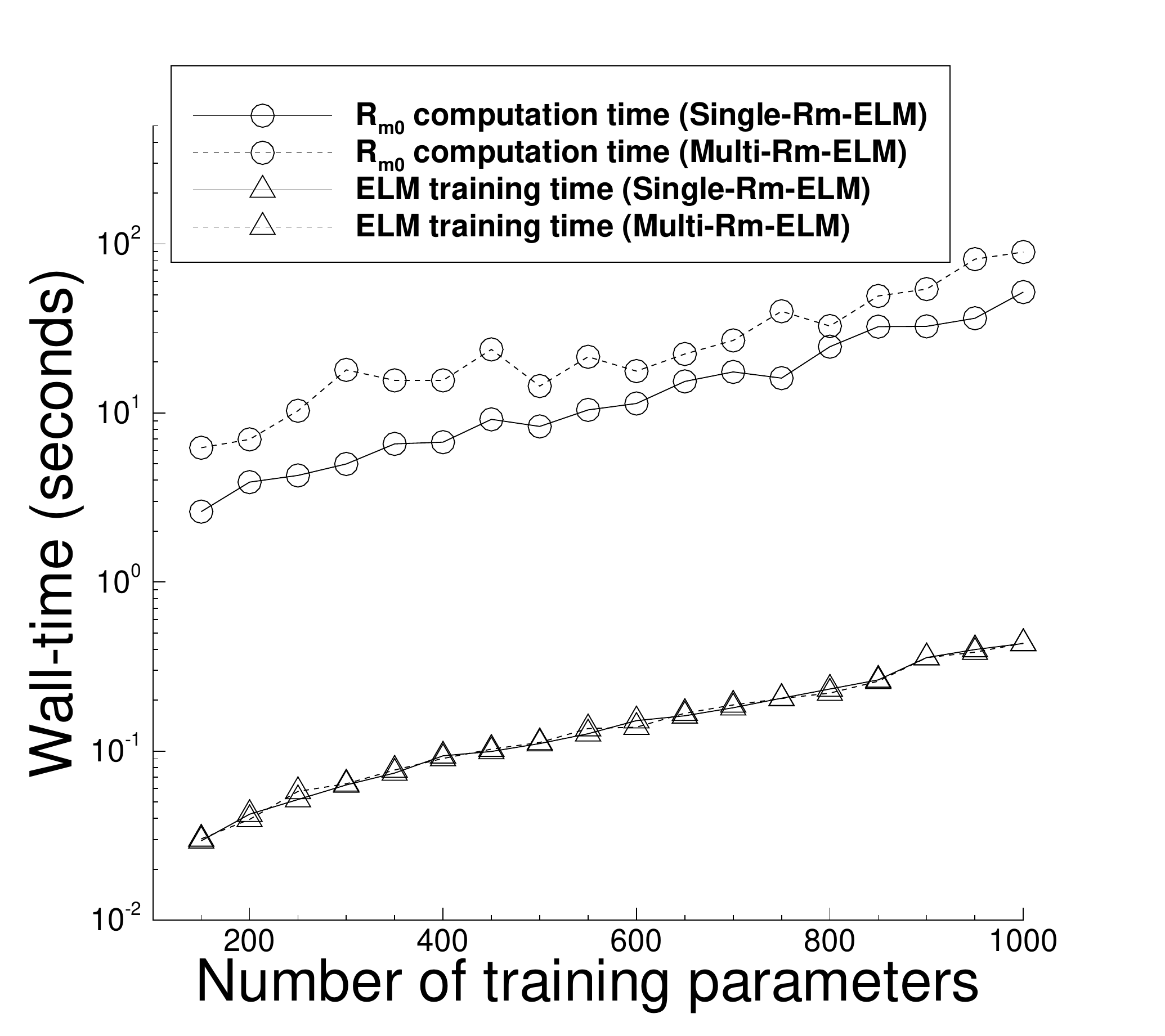}(d)
  }
  \caption{Poisson equation:
    (a,c) The maximum/rms errors in the domain corresponding to
    $R_m=R_{m0}$ in Single-Rm-ELM and $\mbs R_m=\mbs R_{m0}$ in Multi-Rm-ELM,
    versus the number of the training
    parameters ($M$).
    (b,d) The $R_{m0}$ (or $\mbs R_{m0}$) computation time and the
    ELM network training
    time in Single-Rm-ELM and Multi-Rm-ELM,
    versus the number of training parameters ($M$).
    Network architecture: $[2,75,M,1]$ in (a,b),
    $[2,75,75,M,1]$ in (c,d).
    $Q=35\times 35$ in (a,b,c,d).
  }
  \label{fg_12}
\end{figure}

Figure \ref{fg_12} is a comparison between the Single-Rm-ELM and
the Multi-Rm-ELM methods in terms of their accuracy and
$R_{m0}$/$\mbs R_{m0}$ computation cost.
Here we consider two neural networks
with architectures $[2, 75, M, 1]$ and $[2, 75, 75, M, 1]$,
respectively, where $M$ is
varied systematically.
We employ a set of $Q=35\times 35$ uniform collocation points in the domain.
We look into the numerical errors corresponding to
$R_m=R_{m0}$ in Single-Rm-ELM and $\mbs R_m=\mbs R_{m0}$ in Multi-Rm-ELM,
and the time spent on computing $R_{m0}$ and $\mbs R_{m0}$ with
the differential evolution algorithm,
as well as the networking training time with ELM
for solving the Poisson equation with the obtained $R_{m0}$ or $\mbs R_{m0}$.
Figures \ref{fg_12}(a) and (c) depict the maximum/rms errors in the domain as
a function of the number of training parameters $M$
for these two neural networks, respectively,
obtained with $R_m=R_{m0}$ in Single-Rm-ELM
and $\mbs R_m=\mbs R_{m0}$ in Multi-Rm-ELM.
Figures \ref{fg_12}(b) and (d) depict the corresponding $R_{m0}$ and $\mbs R_{m0}$
computation time with the differential evolution algorithm,
as well as the ELM network training time, versus $M$.
The $R_{m0}$ and $\mbs R_{m0}$ computations in Figures \ref{fg_12}(a) and (b),
for the neural network with two hidden layers,
correspond to a population size of $10$, the $R_m$ and $\mbs R_m$
bounds $[0.01, 3]$, and a relative tolerance $0.1$ in the differential evolution
algorithm with both Single-Rm-ELM and Multi-Rm-ELM.
The $R_{m0}$ and $\mbs R_{m0}$ computations in Figures \ref{fg_12}(c) and (d),
for three hidden layers in the neural network,
correspond to a population size of $9$ and the same bounds and relative
tolerance as in (a,b) for both Single-Rm-ELM and Multi-Rm-ELM.

We can make the following observations from Figure \ref{fg_12}.
First, the Multi-Rm-ELM method consistently leads to smaller
numerical errors than Single-Rm-ELM.
By setting the weight/bias coefficients in different hidden layers
to random values with different maximum magnitudes as given by $\mbs R_m$,
Multi-Rm-ELM can
produce more accurate results than Single-Rm-ELM,
which sets the weight/bias coefficients in all hidden layers
to random values with the same maximum magnitude $R_m$.
Second, the cost for computing $\mbs R_{m0}$ in Multi-Rm-ELM 
is generally higher than that for computing $R_{m0}$ in Single-Rm-ELM.
Third, for a given $R_m$ in Single-Rm-ELM and given $\mbs R_m$
in Multi-Rm-ELM, the ELM network training time for solving the Poisson equation
is essentially the same.
Fourth, the $R_{m0}$/$\mbs R_{m0}$ computation cost with differential evolution
is markedly higher than the ELM network training cost
for solving the PDE with a given $R_m$ or $\mbs R_m$.

\begin{figure}
  \centerline{
    \includegraphics[width=2in]{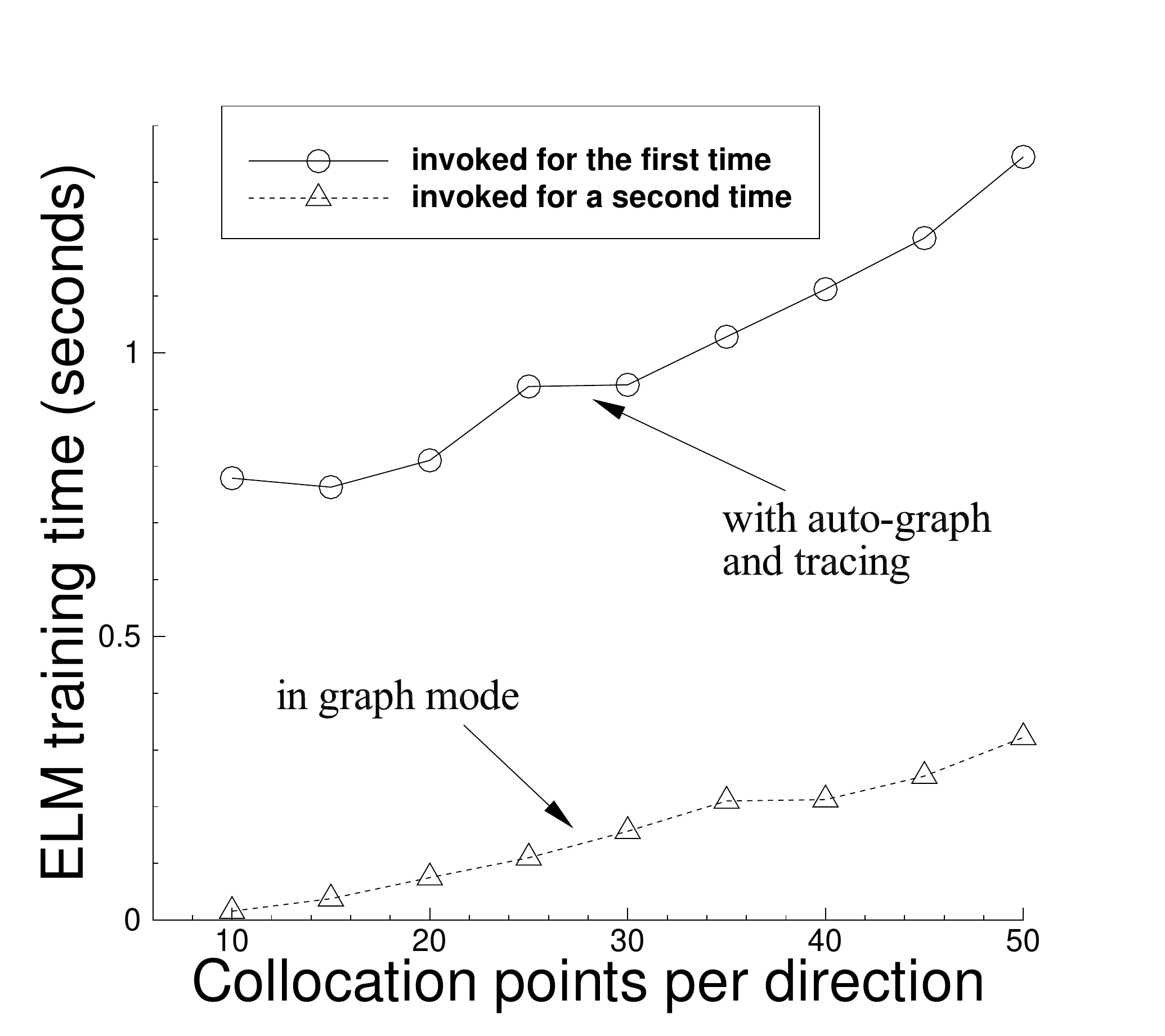}(a)
    \includegraphics[width=2in]{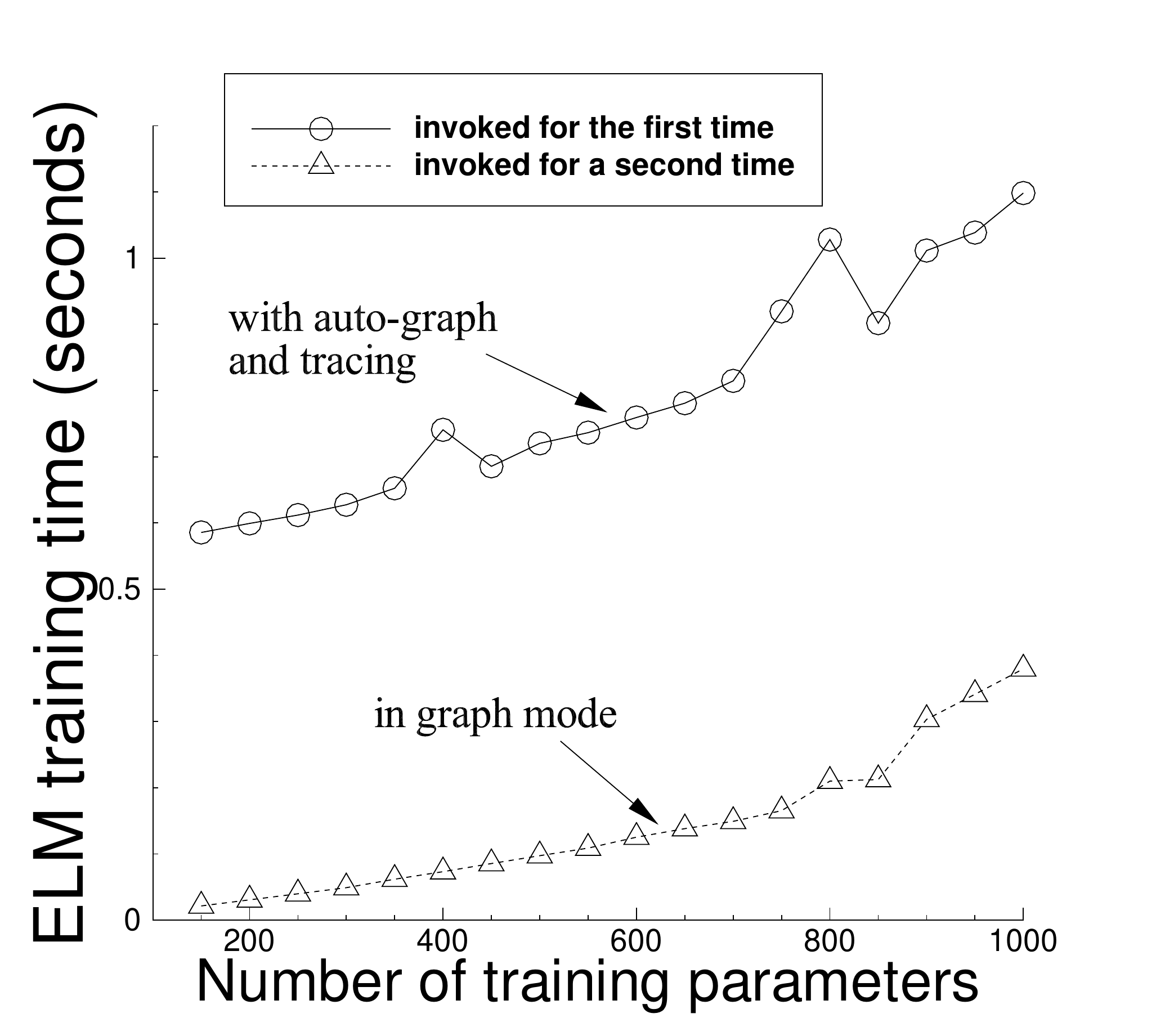}(b)
  }
  \caption{Poisson equation (Single-Rm-ELM):
    ELM network training time versus (a) the number of collocation points
    per direction and (b) the number of training parameters,
    obtained with the training routine invoked for the first time
    or subsequently. In the first invocation
    auto-graph/tracing occurs to build the
    computational graphs, which are used in the graph mode in subsequent
    invocations. The settings and parameters here
    correspond to those of Figures \ref{fg_9}(b,c).
  }
  \label{fg_13}
\end{figure}

As discussed in Section \ref{sec:note},
the computations for the output fields of the last hidden layer
and the associated differential operators are implemented
as ``Tensorflow Functions'' in this paper,
which are executed as a computational graph.
When these functions are invoked for the first time,
autograph/tracing occurs in Tensorflow
to build the computational graph,
which can slow down the computations.
Subsequent invocations of these functions are executed in the graph mode,
which is much faster. 
Figure \ref{fg_13} illustrates this effect
for solving the Poisson equation.
Figure \ref{fg_13}(a) depicts the ELM network training time
with the Single-Rm-ELM
configuration, with the training routine invoked for the first time
and invoked subsequently, as a function of the number of collocation
points in each direction.
Figure \ref{fg_13}(b) depicts the corresponding
ELM network training time
as a function of the number of training parameters in the neural network.
The settings and the simulation parameters here
correspond to those of Figures \ref{fg_9}(b) and (c), respectively.
One can observe that the ELM training time is reduced dramatically when
these computations are performed in the graph mode (without autograph/tracing).


\begin{figure}
  \centerline{
    \includegraphics[width=2in]{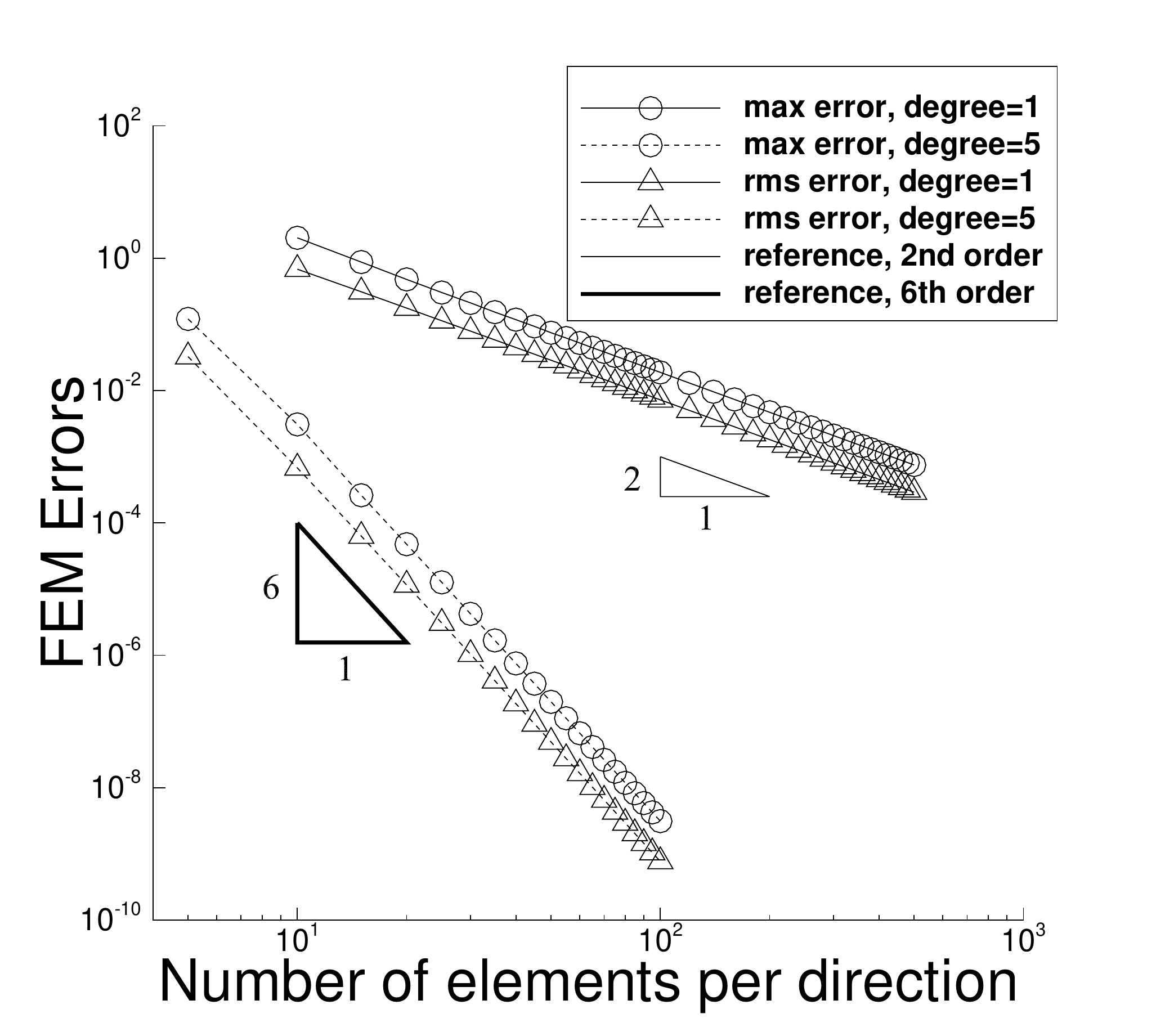}(a)
    \includegraphics[width=2in]{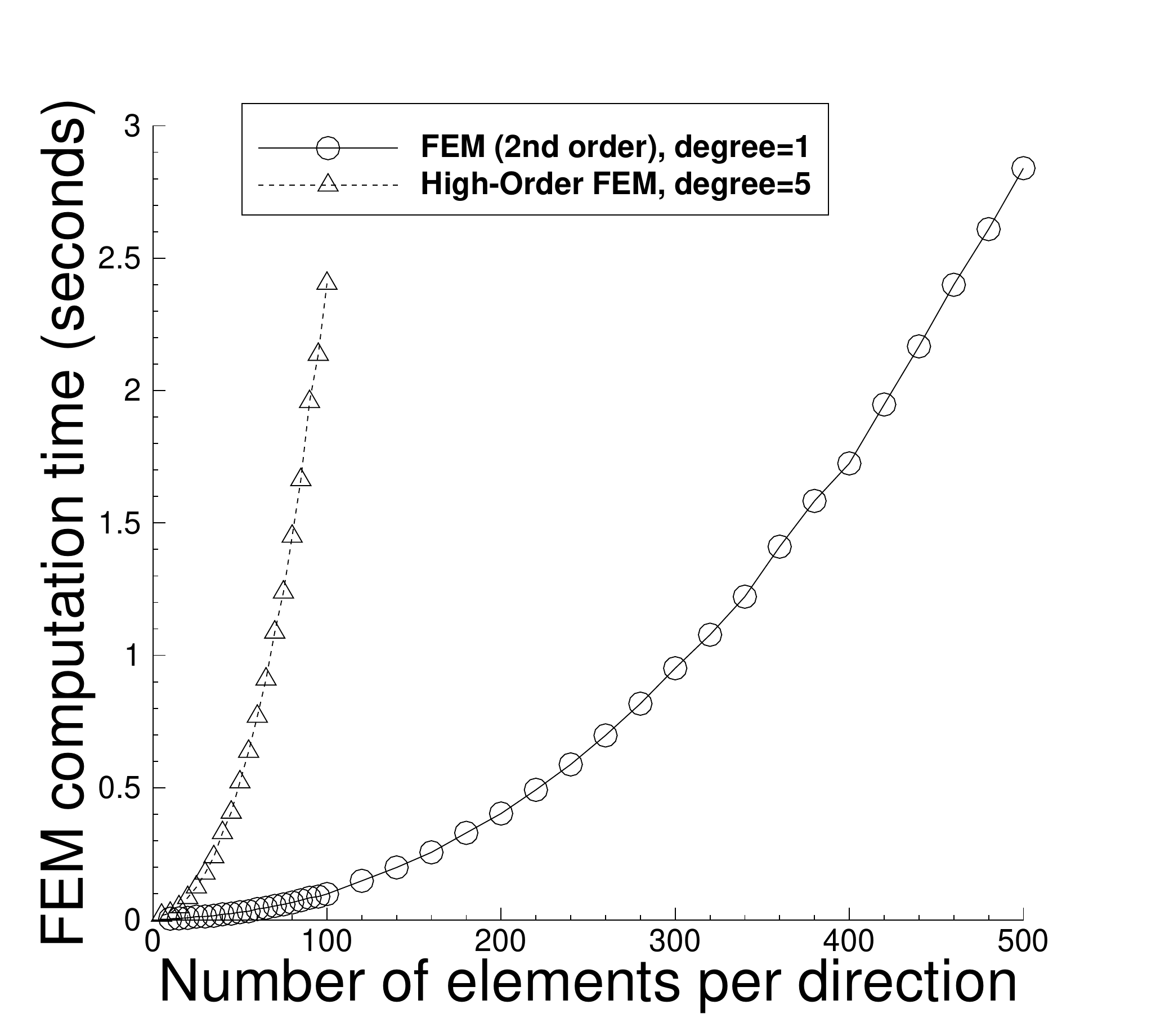}(b)
  }
  \centerline{
    \includegraphics[width=2in]{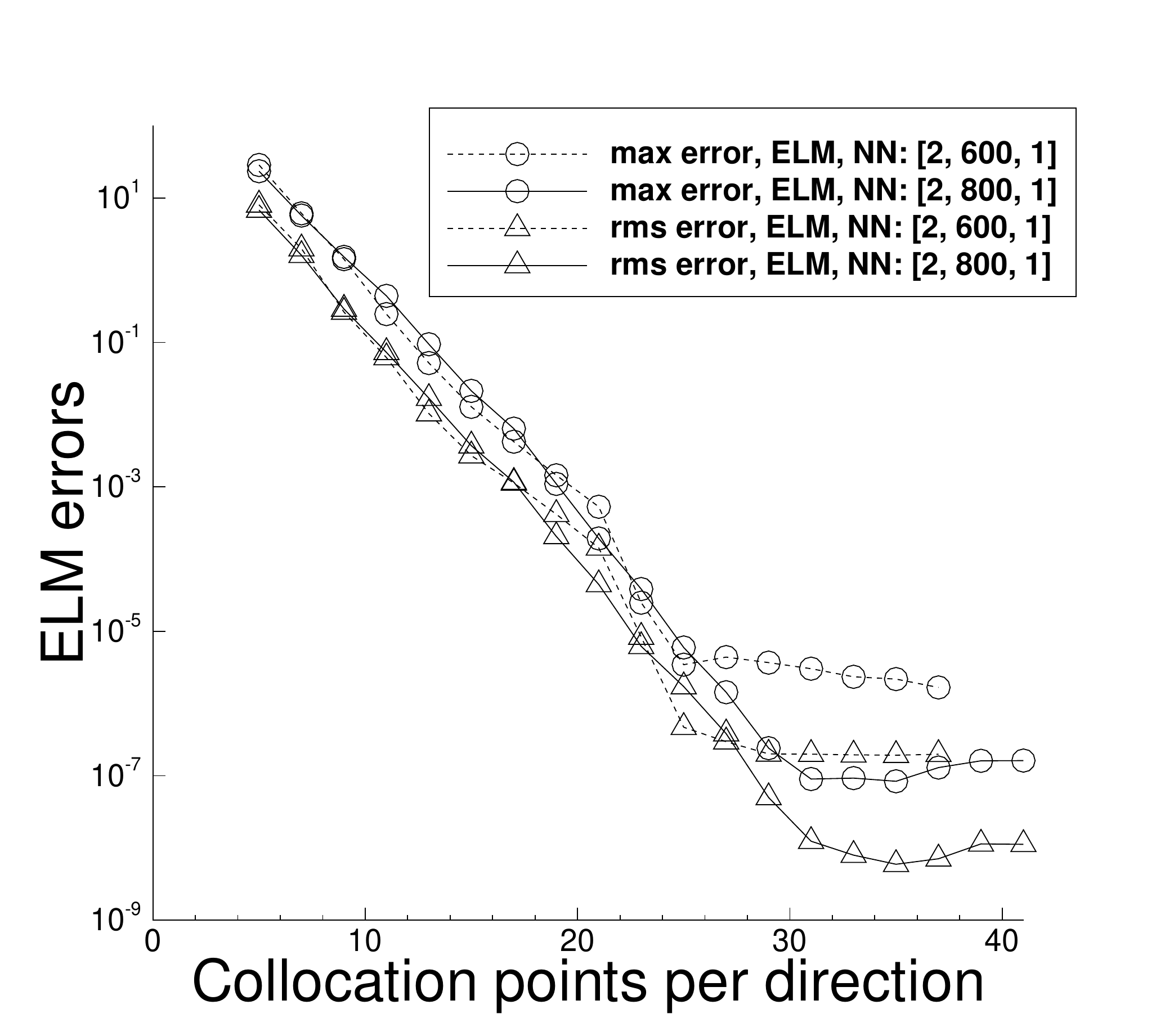}(c)
    \includegraphics[width=2in]{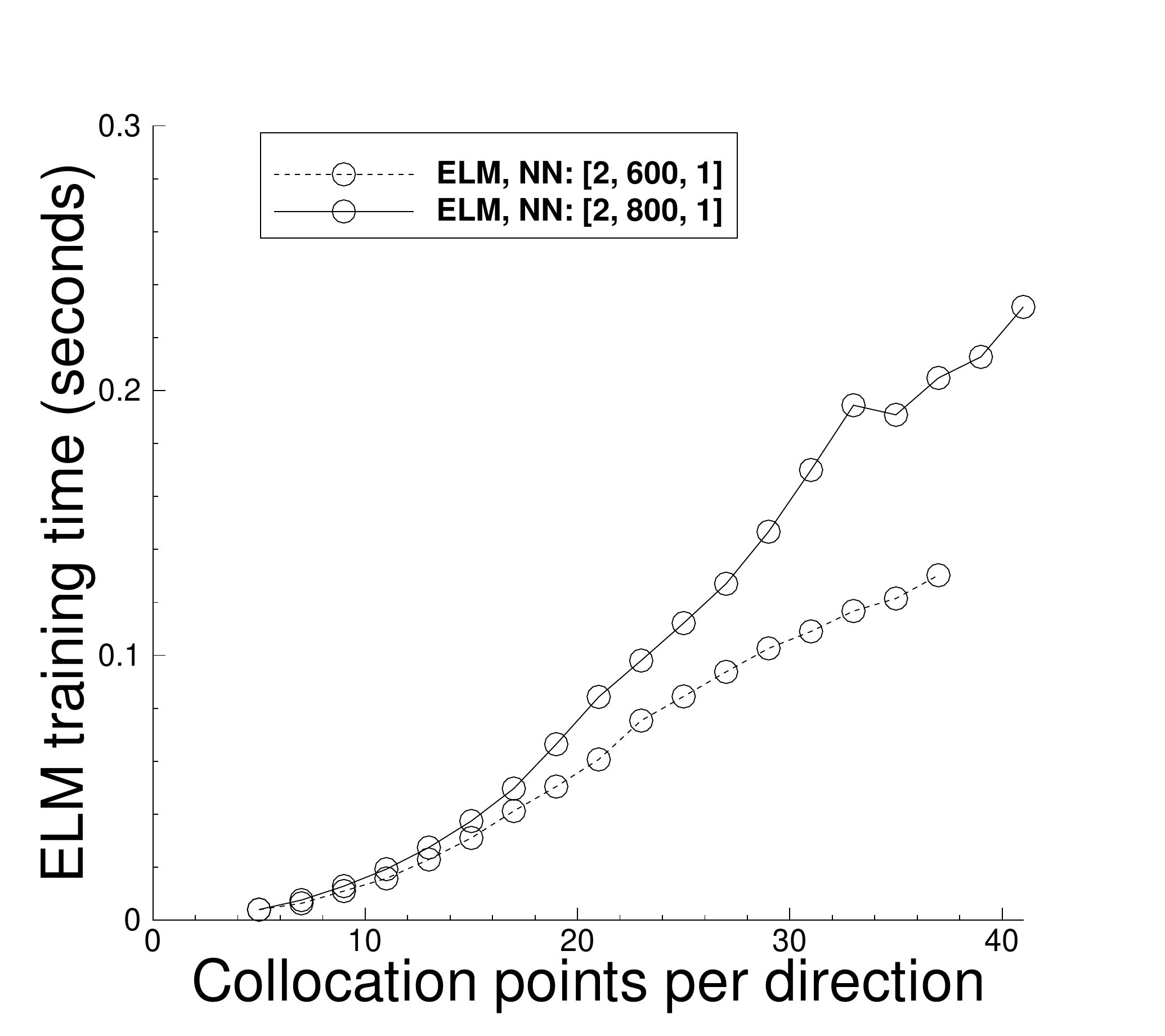}(d)
  }
  \caption{Poisson equation:
    The numerical errors (a) and the computation time (b) of
    the classical FEM (2nd-order, linear elements or degree=1) and
    the high-order FEM (Lagrange elements, degree $5$),
    versus the number of elements in each direction.
    The numerical errors (c) and the network training time (d) of ELM
    (Single-Rm-ELM) versus the number of collocation points per direction.
    ELM network architectures are given in the legends of (c,d).
    $R_m=2.76$ for $M=600$ and $R_m=3.36$ for $M=800$ in (c,d).
    The ELM network training time is the time obtained
    in the graph mode (no autograph/tracing).
  }
  \label{fg_16}
\end{figure}

We next compare the computational performance,
accuracy and computational cost,
between the current implementation of ELM  and
the finite element method (classical second-order FEM, and
high-order FEM) for solving the Poisson equation.
For ELM we use the Single-Rm-ELM configuration in the following comparisons.
For FEM, as stated in Section \ref{sec:note},
it is implemented using the FEniCS library as in~\cite{DongL2020}.
The classical FEM employs Lagrange elements of degree one (linear elements),
and high-order FEM employs Lagrange elements~\cite{Courant1943}
with degrees larger than one from the FEniCS library.
When solving the boundary value problem~\eqref{eq_15} using FEM,
we partition the domain $\Omega$ into an $N_1\times N_1$ rectangular mesh,
where $N_1$ is the number of rectangles in each direction.
Each rectangle is further partitioned into two triangular elements
along the diagonal. So a total of $2N_1^2$ triangular elements
are involved in the FEM computation.
For convenience we will loosely refer to $N_1$ as the number of elements
in each direction.
In the FEM tests we vary the number of elements per direction
$N_1$ and the degree of the Lagrange elements systematically.

Another implementation detail with FEM concerns the evaluation
of the source term and the Dirichlet boundary data
in equations~\eqref{eq_15a}--\eqref{eq_15b}.
These terms each is implemented as a FEniCS ``Expression'', in which
the degree parameter is specified as the element degree plus one
when solving the Poisson equation.
We observe that if the degree parameter in these FEniCS Expressions
is specified to be
 equal to the element degree or less, one cannot
seem to quite achieve the expected
convergence rate as the number of elements increases,
especially when the mesh size
is not very large.

Figure \ref{fg_16} provides an overview of the numerical errors of FEM and ELM,
as well as their computational cost (FEM computation time, ELM network training time).
In these tests the number of the elements in the FEM mesh and the number of collocation
points in ELM are varied systematically.

Figure \ref{fg_16}(a) shows the maximum/rms errors in the domain as
a function of the number of elements in each direction ($N_1$)
with the classical FEM (degree=1) and the high-order FEM with Lagrange
elements of degree=5.
One can clearly observe a second-order convergence rate and
a sixth-order convergence rate with these two types of elements.
Figure \ref{fg_16}(b) shows the corresponding FEM computation time
versus the number of elements per direction with these two types of elements.
The FEM computation time grows quite rapidly with increasing number of
elements. The cost of the high-order FEM grows much faster
than that of the classical FEM.

Figure \ref{fg_16}(c) shows the ELM maximum/rms errors in the domain
as a function of the number of uniform collocation points in each direction,
obtained using two neural networks
with the architectures $[2, M, 1]$ with $M=600$ and $M=800$,
respectively. We have employed $R_m=2.76$ for $M=600$ and $R_m=3.36$ for
$M=800$, close to their optimal $R_{m0}$, for generating the random hidden layer
coefficients.
A set of $Q=Q_1\times Q_1$ uniform collocation
points is employed and $Q_1$ is varied systematically.
On can clearly observe an exponential decrease in the ELM errors before saturation.
As $Q_1$ becomes sufficiently large, the ELM errors saturate at a higher level with
$M=600$ than with $M=800$. 
Figure \ref{fg_16}(d) shows the corresponding ELM network training time
versus the number of collocation points per direction with these
two neural networks.
Here the ELM training time refers to the time obtained with the graph mode
(no autograph/tracing).
They appear to grow quasi-linearly with increasing number of collocation
points.

\begin{figure}
  \centerline{
    \includegraphics[width=2in]{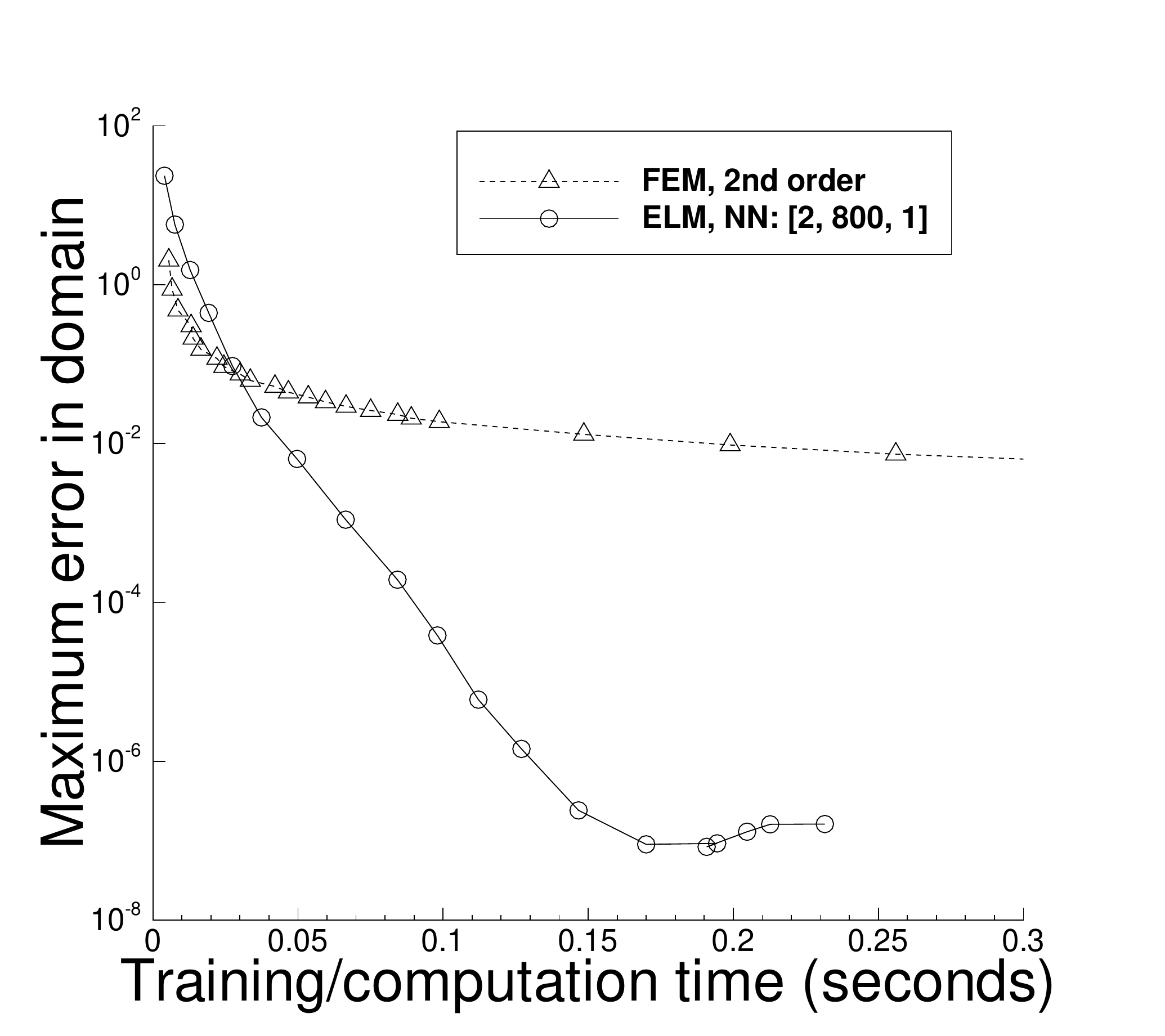}(a)
    \includegraphics[width=2in]{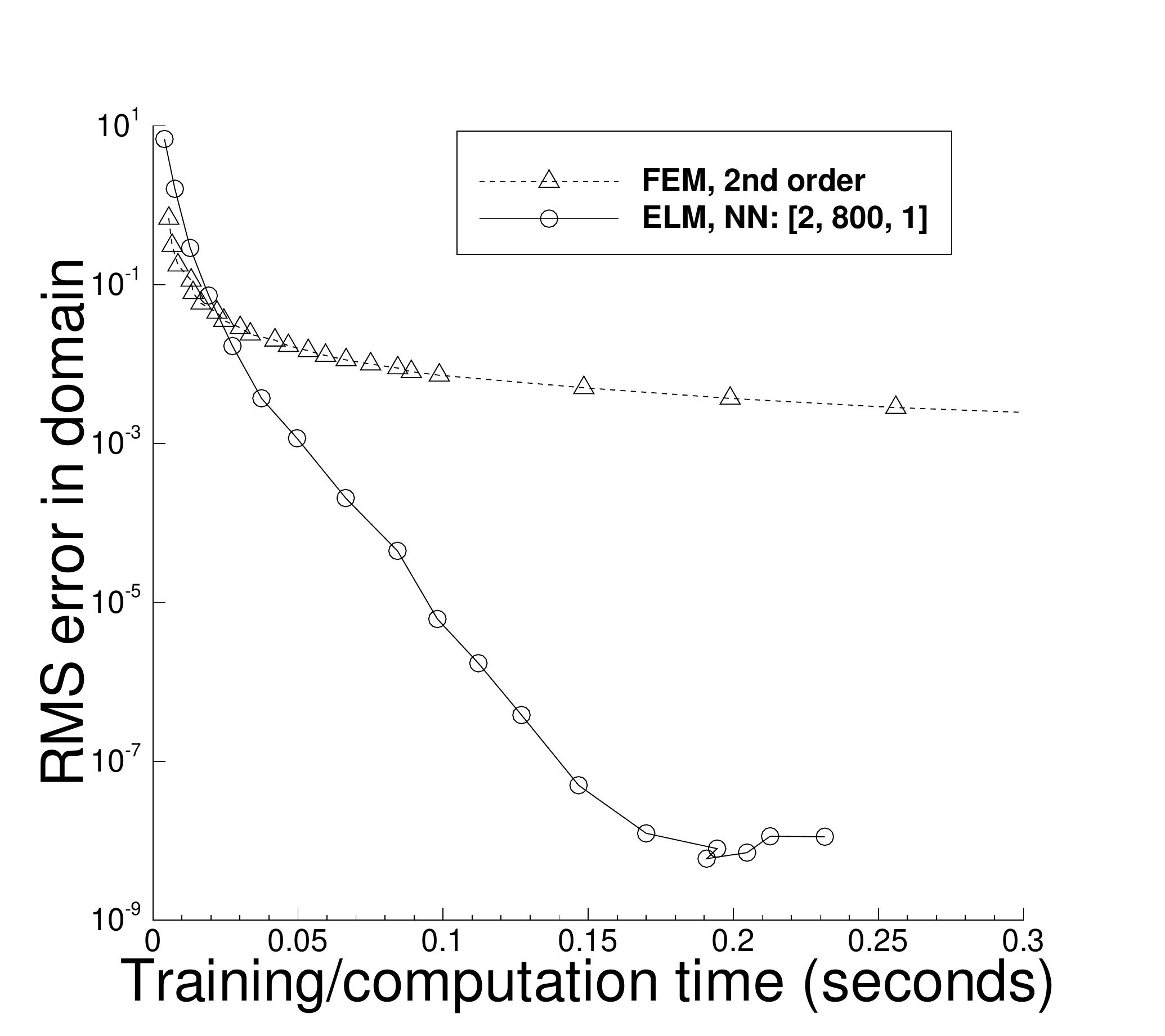}(b)
  }
  \caption{Poisson equation (comparison between ELM and classical FEM):
    The (a) maximum error and (b) rms error in the domain versus the computational
    cost (ELM training time, FEM computation time) for ELM and
    the classical FEM.
    The FEM data correspond to those of Figures \ref{fg_16}(a,b) with degree=1.
    The ELM data correspond to those of Figures \ref{fg_16}(c,d) with $M=800$.
  }
  \label{fg_15}
\end{figure}

Figure \ref{fg_15} compares the computational performance
of the ELM and the classical FEM.
The two plots show the maximum and rms errors in the domain
of the ELM and FEM versus their computational cost
(FEM computation time, ELM network training time).
The FEM data here correspond to those contained in
Figures~\ref{fg_16}(a,b) with degree=1, and the ELM data
here correspond to those of Figures~\ref{fg_16}(c,d)
with $M=800$.
We observe that the ELM far outperforms the classical FEM in essentially
all cases, except for a narrow range with very small problem sizes
(FEM mesh size below around $50\times 50$, ELM collocation points below
around $13\times 13$; error level above around $5\times 10^{-2}$; wall time
below around $0.03$ seconds).
With the same computational cost/budget, the ELM achieves a considerably
better accuracy (typically by orders of magnitude) than the classical FEM,
and to achieve the same accuracy the ELM incurs a much lower computational cost than
the classical FEM. 
Even in the narrow range of small problem sizes,
where the classical FEM is a little better,
the FEM performance and the ELM performance are quite close.

\begin{figure}
  \centerline{
    \includegraphics[width=2in]{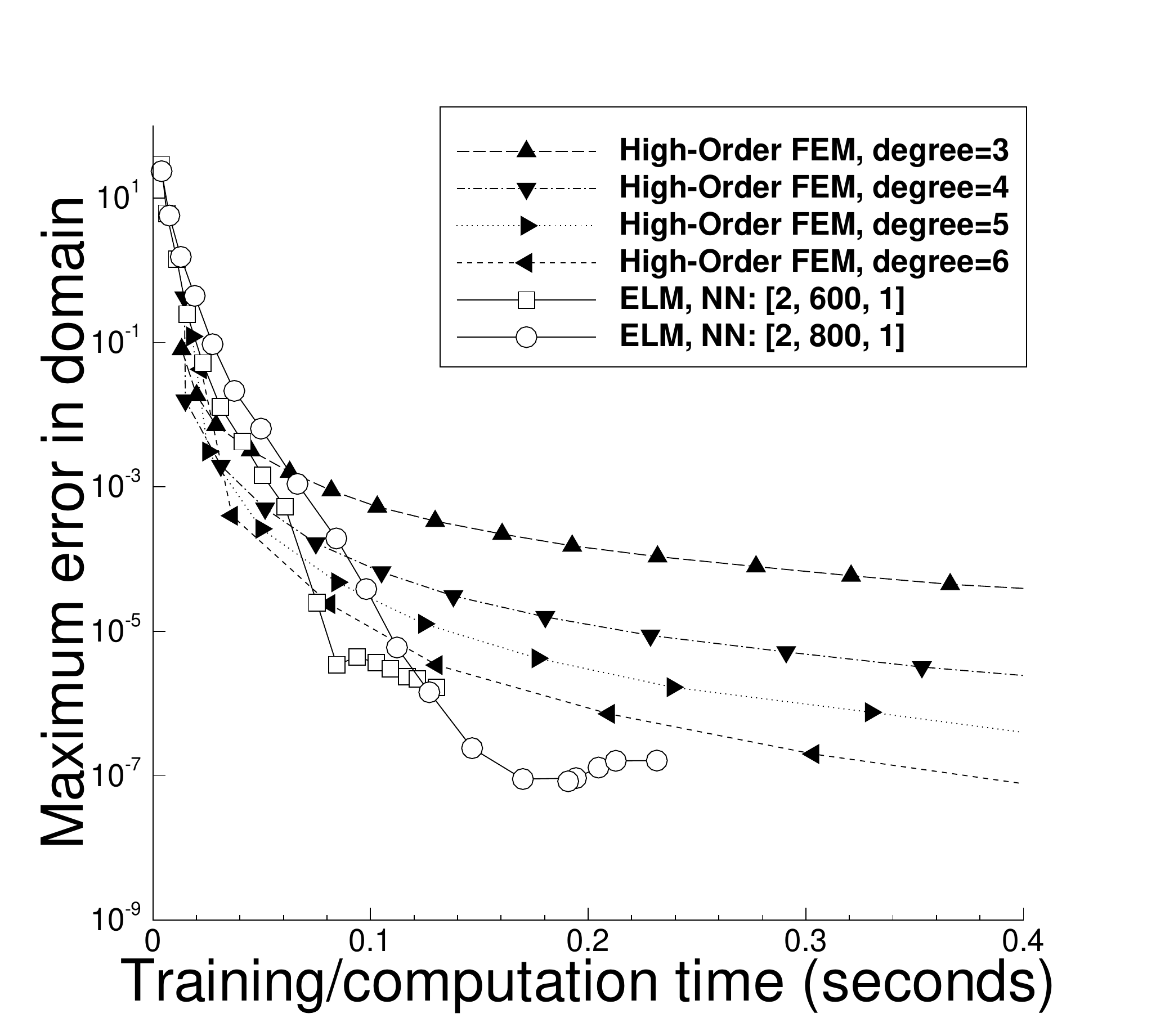}(a)
    \includegraphics[width=2in]{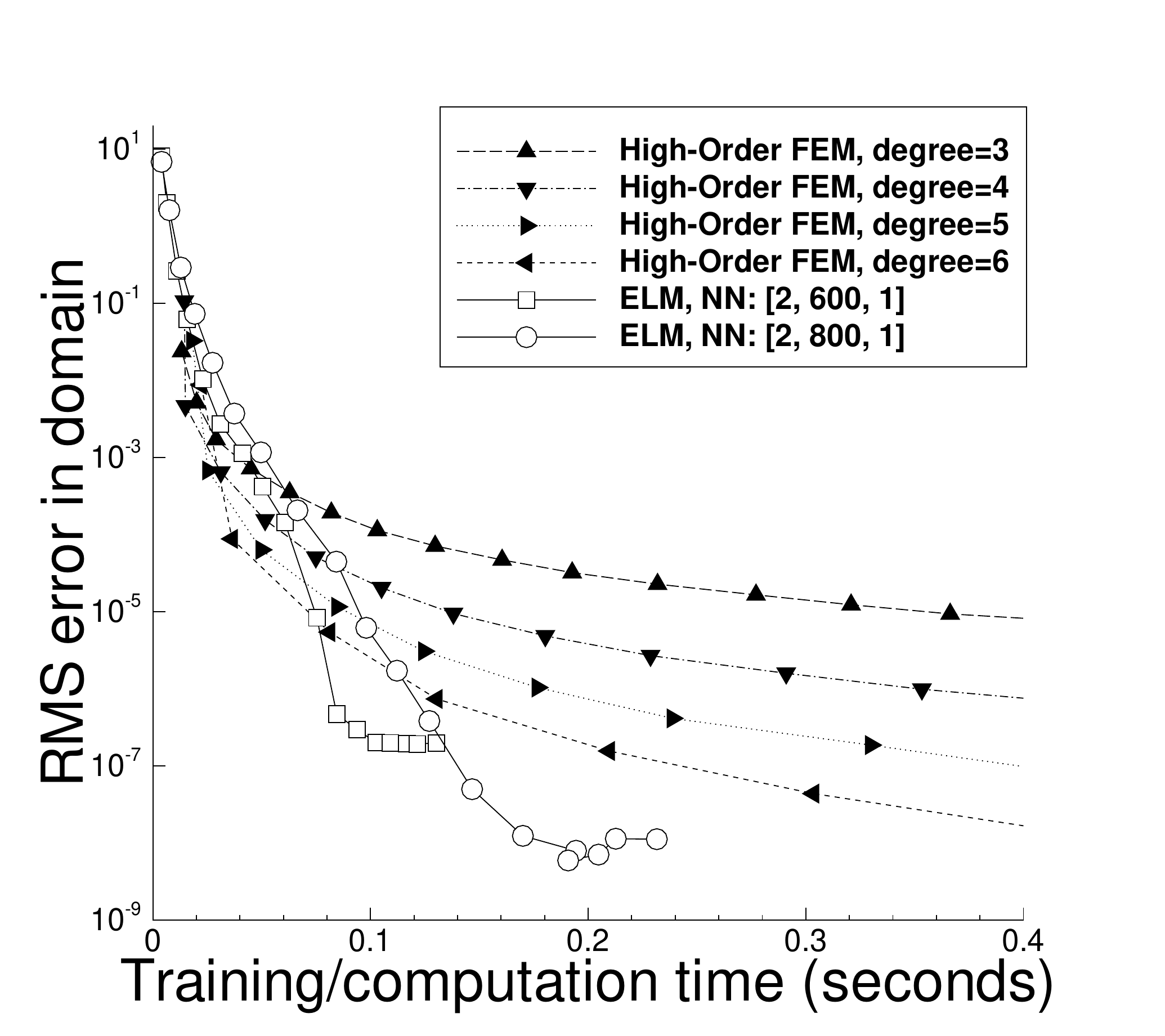}(b)
  }
  \centerline{
    \includegraphics[width=2in]{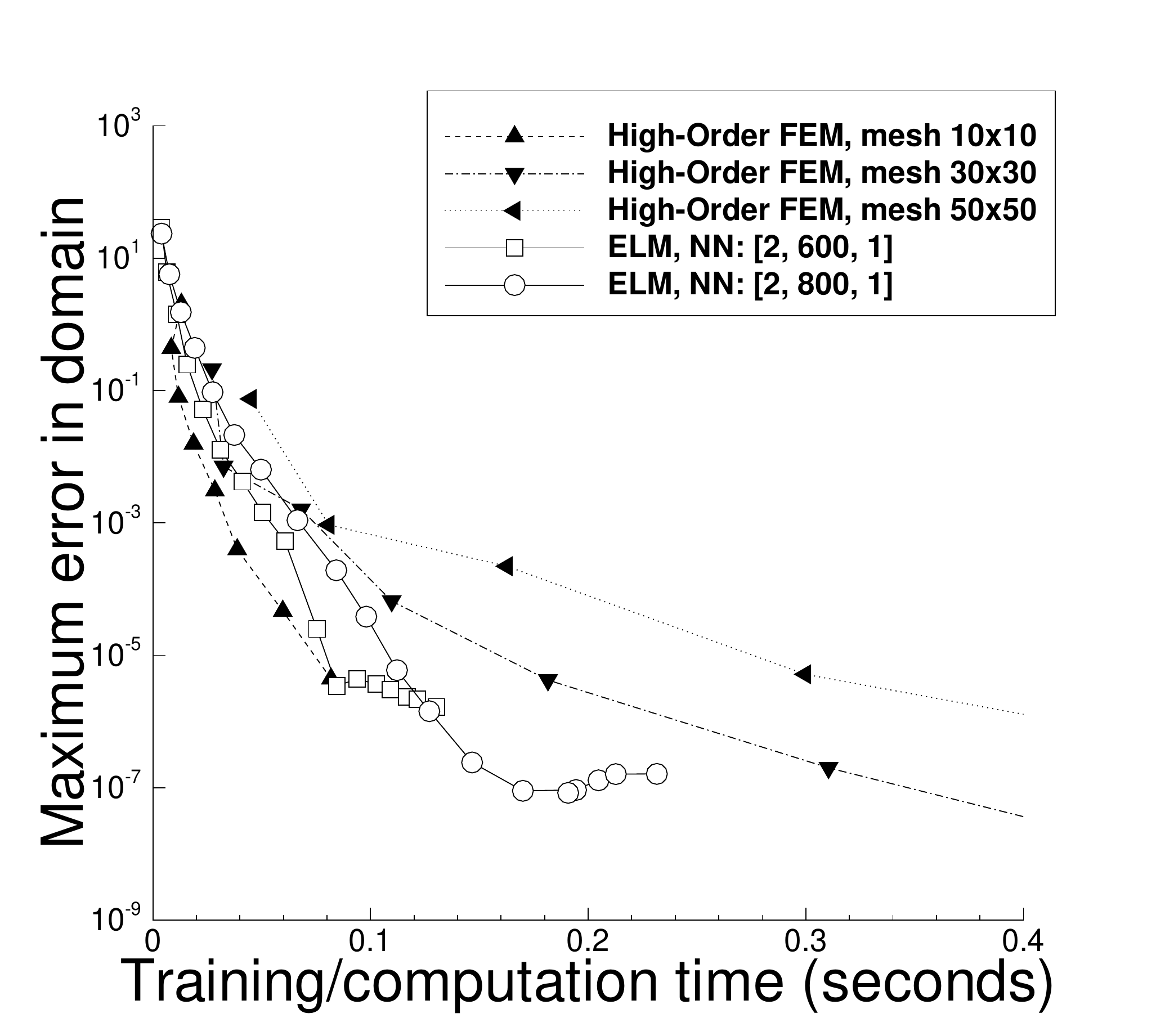}(c)
    \includegraphics[width=2in]{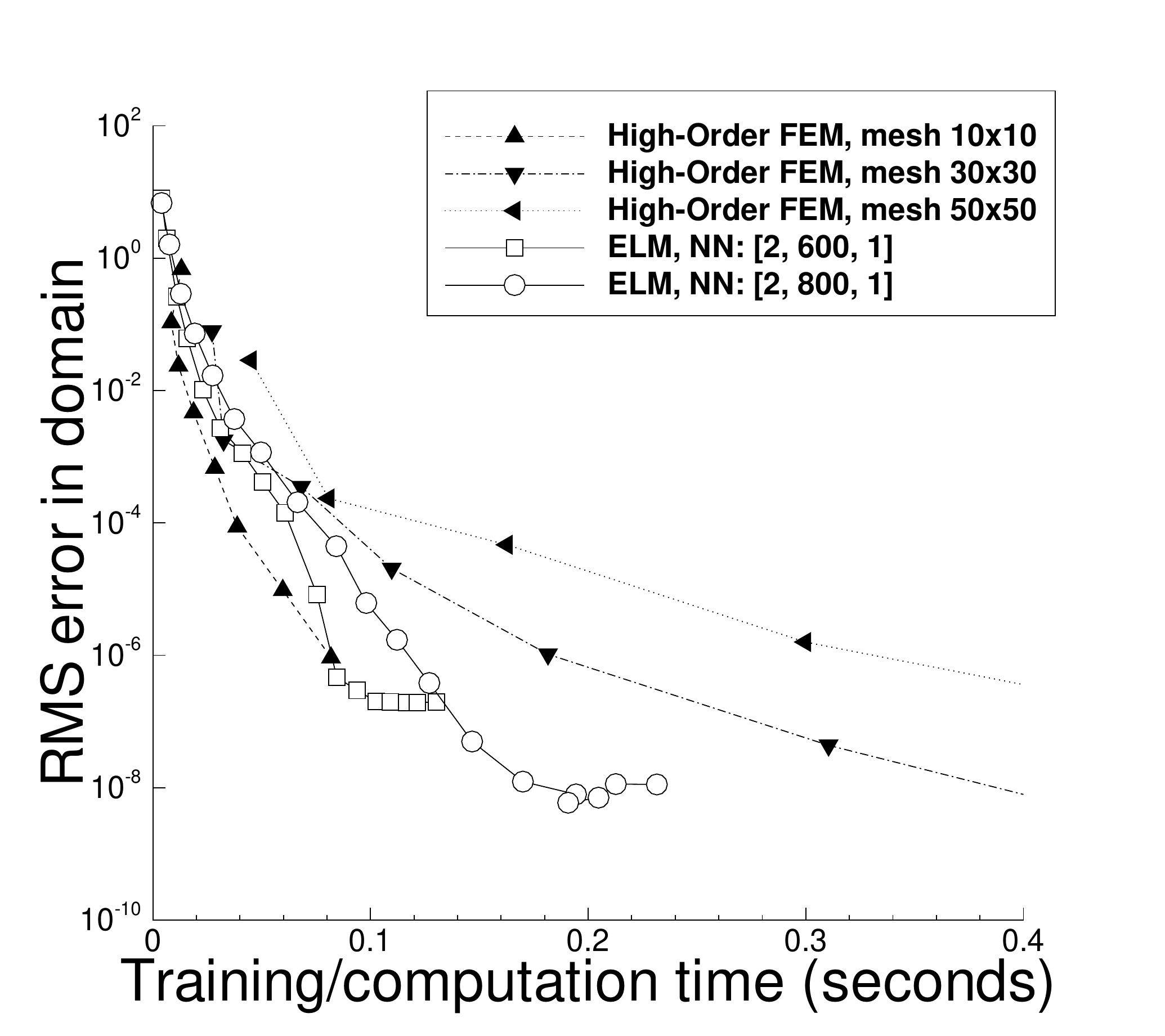}(d)
  }
  \caption{Poisson equation (comparison between ELM and high-order FEM):
    The maximum error (a,c) and the rms error (b,d) in the domain versus
    the computational cost (ELM network training time, FEM computation time)
    of the ELM and the high-order FEM with Lagrange elements of various degrees.
    For FEM,
    in (a,b) the mesh size is varied
    systematically for each given element degree,  and
    in (c,d) the element degree is varied systematically for
    each given mesh size.
    The FEM data in (a,b) with degree=5
    correspond to those of Figures \ref{fg_16}(a,b)
    with degree=5.
    The ELM data in (a,b,c,d) correspond to those of Figures \ref{fg_16}(c,d).
  }
  \label{fg_16_a}
\end{figure}

Figure \ref{fg_16_a} provides a comparison of the computational performance
between the ELM and the high-order FEM with Lagrange elements of higher degrees.
We have conducted two groups of tests with high-order FEM.
In the first group, for a fixed element degree,
we vary the mesh size systematically.
In the second group, for a fixed mesh size, we vary the degree of the
Lagrange elements systematically (between $2$ and $8$).
These two types of tests
approximately correspond to the so-called h-type and p-type
refinements with the high-order hp-finite element
method~\cite{KarniadakisS2005,SzaboB1991,DongK2003}. 

Figures \ref{fg_16_a}(a) and (b) show the maximum and rms errors in the domain
of the high-order FEM, with Lagrange elements of degrees ranging from $3$ to $6$,
versus the FEM computation time in the first group of tests.
The FEM data with degree=5 in these plots correspond
to those of Figures \ref{fg_16}(a,b) with degree=5.
Figures \ref{fg_16_a}(c) and (d) depict the maximum/rms errors
of the high-order FEM, with mesh sizes ranging from $10\times 10$
to $50\times 50$, versus the FEM computation time
in the second group of tests.
In all these plots,
we have included the ELM maximum/rms errors versus the ELM network
training time for comparison, where the ELM data correspond to
those contained in Figures~\ref{fg_16}(c,d) with $M=600$ and $M=800$.

We can make the following observations from Figure~\ref{fg_16_a}.
With the h-type refinement (for a fixed element degree),
there is a cross-over point with respect to the problem size
in the relative performance between ELM and high-order FEM.
For smaller problem sizes (smaller FEM mesh, smaller set of ELM
collocation points), the performance of the ELM and that of the high-order FEM
are largely comparable, with the high-order FEM being a little better.
For larger problem sizes, the ELM outperforms the high-order FEM
markedly (see Figures \ref{fg_16_a}(a,b)). 
With the p-type refinement (for a fixed mesh size),
there is also a cross over in the relative performance between ELM
and high-order FEM with respect to the mesh size.
With a small FEM mesh size, the performance of the high-order FEM
(with varying element degree) and that of the ELM are
comparable, with the high-order FEM being a little better.
With a larger FEM mesh size, the ELM markedly outperforms
the high-order FEM with varying element degree,
especially for larger FEM degrees (see Figures \ref{fg_16_a}(c,d)).
Overall, we observe that the ELM method is 
very competitive compared with the high-order FEM.
For small problem sizes, the ELM performance and the high-order FEM
performance appear largely comparable, with the high-order FEM
oftentimes a little better.
For larger problem sizes, the ELM method
outperforms the high-order FEM. As the problem size becomes
large, ELM can outperform the high-order FEM by a substantial factor.

These observations on the ELM/FEM performance
should be compared with those of~\cite{DongL2020}.
In~\cite{DongL2020}, it is observed that the ELM method can outperform
the classical 2nd-order FEM for larger problem sizes. Compared with
high-order FEM, however, ELM (with the implementation therein)
is observed to be generally not as
competitive~\cite{DongL2020}.
With the improvements in the current work,
especially the use of forward-mode auto-differentiations
for computing the differential operators (see Remark~\ref{rem_12}),
we have significantly increased the ELM computational performance.
As shown above,
the improved ELM herein far outperforms
the classical FEM. Its computational performance
is on par with that of the high-order FEM, and oftentimes
it can outperform the high-order FEM by a substantial margin.


\subsection{Nonlinear Helmholtz Equation}
\label{sec:nonl_helm}

\begin{figure}
  \centerline{
    \includegraphics[width=2in]{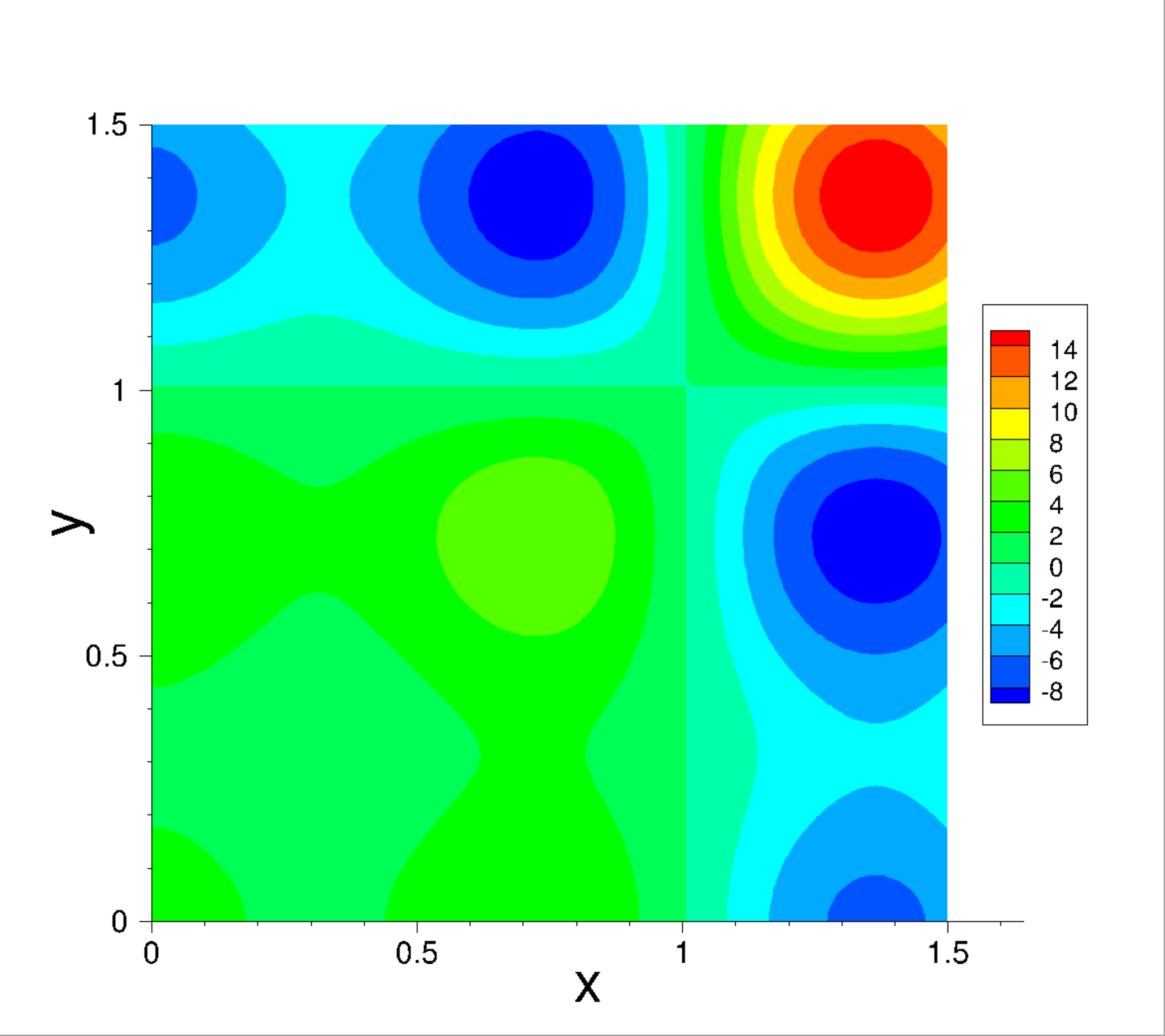}
  }
  \caption{Nonlinear Helmholtz equation: distribution of the exact solution.}
  \label{fg_17}
\end{figure}

We next use a  nonlinear Helmholtz type equation in 2D
to further test the method
for computing the optimal $R_m$ and $\mbs R_m$. We demonstrate
the competitiveness of the ELM method for nonlinear
problems by comparing its performance with
those of the classical and high-order FEMs.
A seed value $25$ has been employed with the random number generators
in Tensorflow and numpy for all the numerical tests in this subsection.

Consider a 2D rectangular domain $\Omega=\{ (x,y)\ |\ 0\leqslant x,y\leqslant 1.5 \}$
and the following boundary value problem on $\Omega$,
\begin{subequations}\label{eq_17}
  \begin{align}
    &
    \frac{\partial^2u}{\partial x^2} + \frac{\partial^2u}{\partial y^2}
    -100 u + 10\cos(2u) = f(x,y), \label{eq_17a} \\
    &
    u(x,0)=g_1(x),\quad u(x,1.5) = g_2(x),\quad
    u(0,y) = h_1(y), \quad u(1.5,y) = h_2(y). \label{eq_17b}
  \end{align}
\end{subequations}
In the above equations, $u(x,y)$ is the field to be solved for,
$f$ is a prescribed source term,
and $g_1$, $g_2$, $h_1$ and $h_2$ are prescribed Dirichlet boundary data.
We choose the source term and the boundary data appropriately
so that the following field satisfies the problem~\eqref{eq_17},
\begin{align}
  &
  u(x,y) = \left[ \frac52\cos\left(\pi x-\frac{2\pi}{5} \right)
    +\frac32\cos\left(2\pi x + \frac{3\pi}{10} \right)
    \right] \left[ \frac52\cos\left(\pi y-\frac{2\pi}{5} \right)
    +\frac32\cos\left(2\pi y + \frac{3\pi}{10} \right)
    \right].
  \label{eq_18}
\end{align}
The distribution of the analytic solution~\eqref{eq_18} in
the $x$-$y$ plane is shown
in Figure \ref{fg_17}.


The settings of the ELM neural network are similar
to those in Section \ref{sec:poisson}.
The input layer contains two nodes, representing $x$ and $y$.
The output layer is linear and contains one node, representing $u$.
The network contains one or more hidden layers, with the Gaussian
activation function for all hidden nodes.
The random hidden-layer coefficients are set based on the Single-Rm-ELM
or Multi-Rm-ELM configurations as described in Section \ref{sec:method}.
The neural network is trained by the NLLSQ-perturb method from~\cite{DongL2020},
as discussed in Section \ref{sec:method}.
The crucial simulation parameters include
the number of training parameters $M$,
the set of $Q=Q_1\times Q_1$ uniform collocation points in the domain,
and the maximum magnitude $R_m$ or $\mbs R_m$ of
the random coefficients.

\begin{figure}
  \centerline{
    \includegraphics[width=2in]{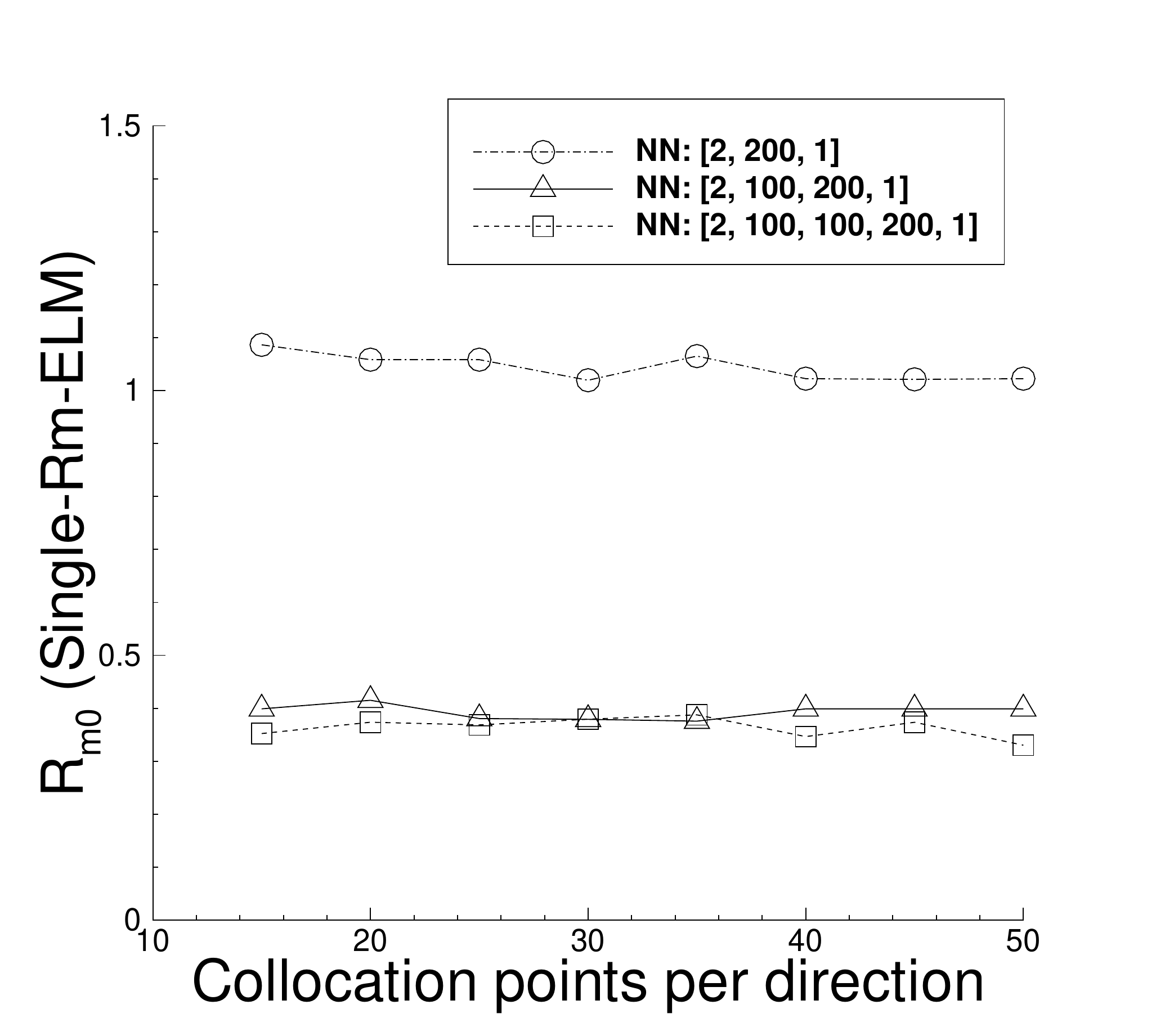}(a)
    \includegraphics[width=2in]{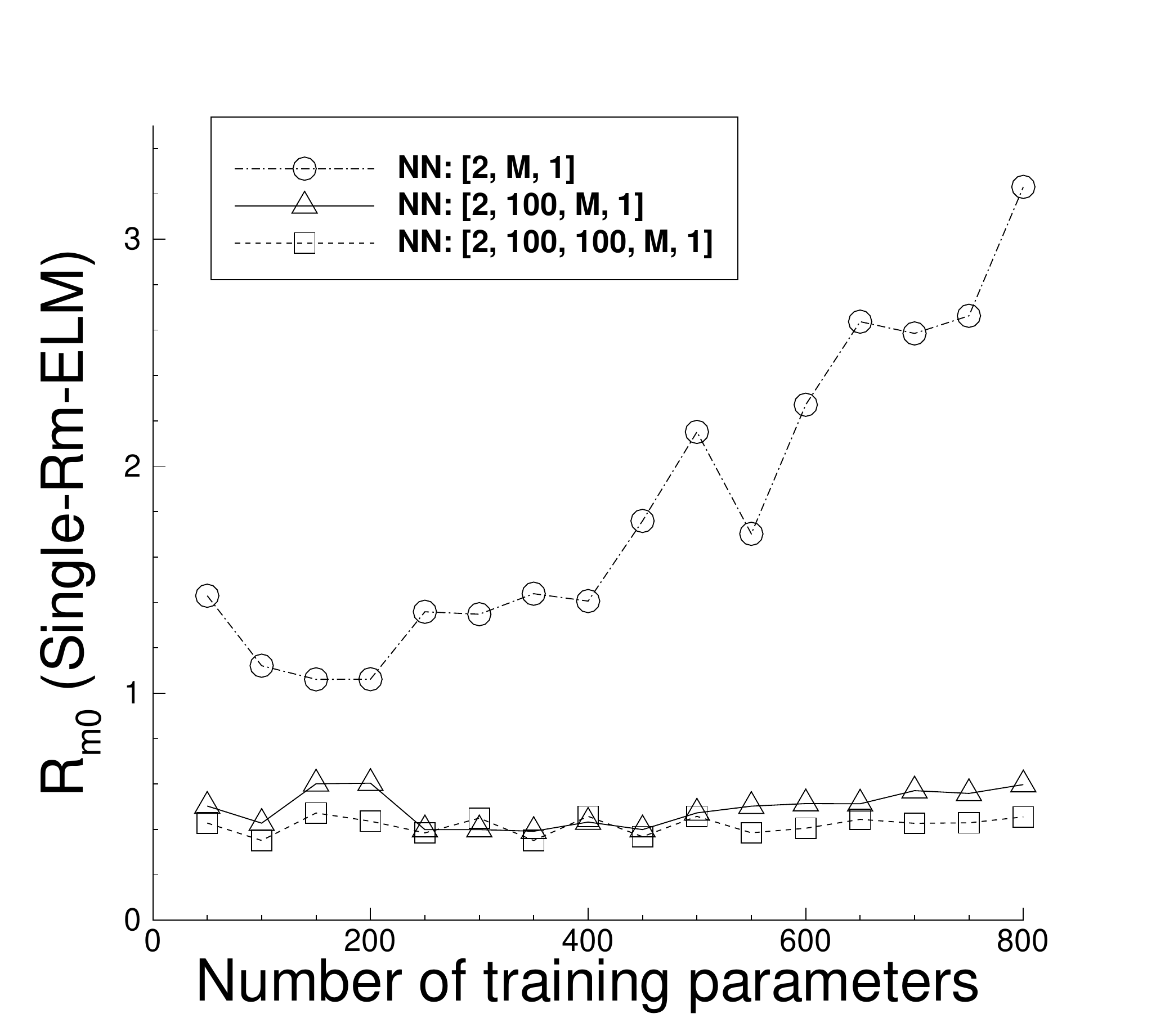}(b)
    \includegraphics[width=2in]{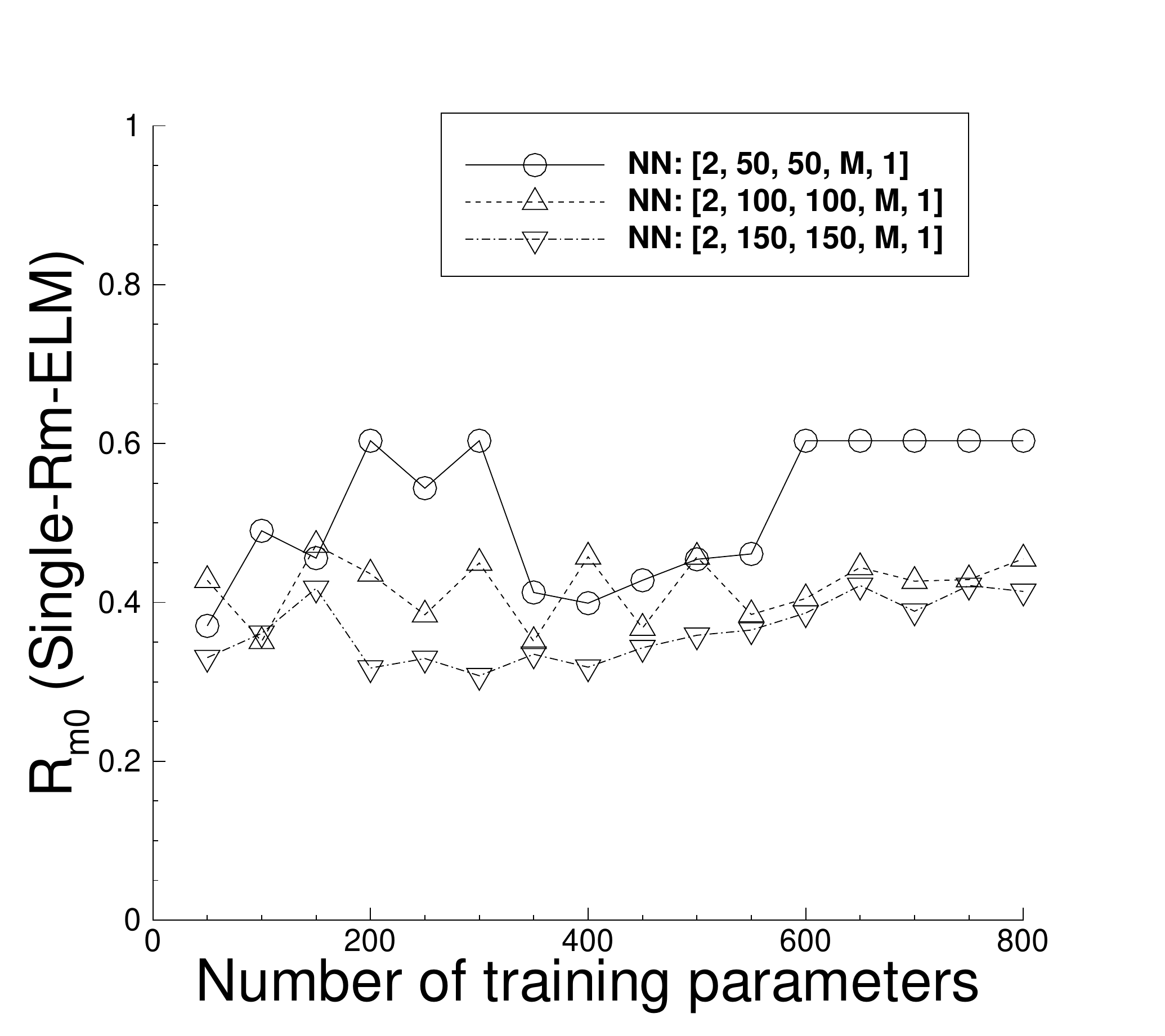}(c)
  }
  \caption{Nonlinear Helmholtz equation (Single-Rm-ELM):
    The optimum $R_{m0}$ versus (a) the number of collocation
    points per direction and (b) the number of training parameters,
    with neural networks of different depth.
    (c) $R_{m0}$ versus the number of training parameters
    with neural networks having the same depth but different width.
    $Q=31\times 31$ in (b,c), varied in (a).
    $M=200$ in (a), varied in (b,c).
    The network architectures are given in the legends.
  }
  \label{fg_18}
\end{figure}

We first look into the Single-Rm-ELM configuration for
assigning the random coefficients in the neural network.
Figure \ref{fg_18} illustrates the characteristics of
the optimum $R_{m0}$ in Single-Rm-ELM obtained by the
differential evolution algorithm.
Note that when computing $R_{m0}$ with differential evolution,
we have turned off the random perturbations and the corresponding
sub-iterations in the NLLSQ-perturb
method, as discussed in Remark~\ref{rem_6a}.
In these tests we consider one to three hidden layers
in the neural network, and vary  the number of
collocation points per direction $Q_1$ or the number of
training parameters $M$  systematically.

Figure \ref{fg_18}(a) depicts the optimum $R_{m0}$ as a function
of $Q_1$, for three neural networks with a fixed number of $200$
training parameters but different depth.
Figures \ref{fg_18}(b) and (c) each depicts the $R_{m0}$ as a function
of  $M$ for three  neural networks.
The three neural networks in plot (b) have different depths,
with one to three hidden layers and with the width fixed at $100$ for
the hidden layers other than the last one.
The three neural networks in plot (c) all contain
three hidden layers, but the width the preceding hidden layers (other than
the last one) varies between $50$ and $150$.
In both (b) and (c), the number of nodes in the last hidden layer (i.e.~$M$)
is varied systematically, and a fixed set of $Q=31\times 31$ uniform
collocation points is employed.
All these results about $R_{m0}$
are obtained with a population size of $4$ and a relative tolerance $0.1$
in the differential evolution algorithm.
The $R_m$ bounds are $[0.1, 3]$ in
the differential evolution algorithm for all the cases  except for
the neural network with the architecture $[2, M, 1]$
in Figure \ref{fg_18}(b), in which the $R_m$ bounds are set to $[0.1, 4]$.

We observe from Figure \ref{fg_18} the same characteristics about $R_{m0}$ for
the nonlinear Helmholtz equation as those for the linear problems
in previous subsections.
For example, $R_{m0}$ is largely independent of the number of collocation
points. It generally decreases with increasing number of layers in
the neural network. There is a large decrease in $R_{m0}$ from one to two
hidden layers in the network, and beyond that the decrease in $R_{m0}$
is almost negligible.

\begin{figure}
  \centerline{
    \includegraphics[width=2in]{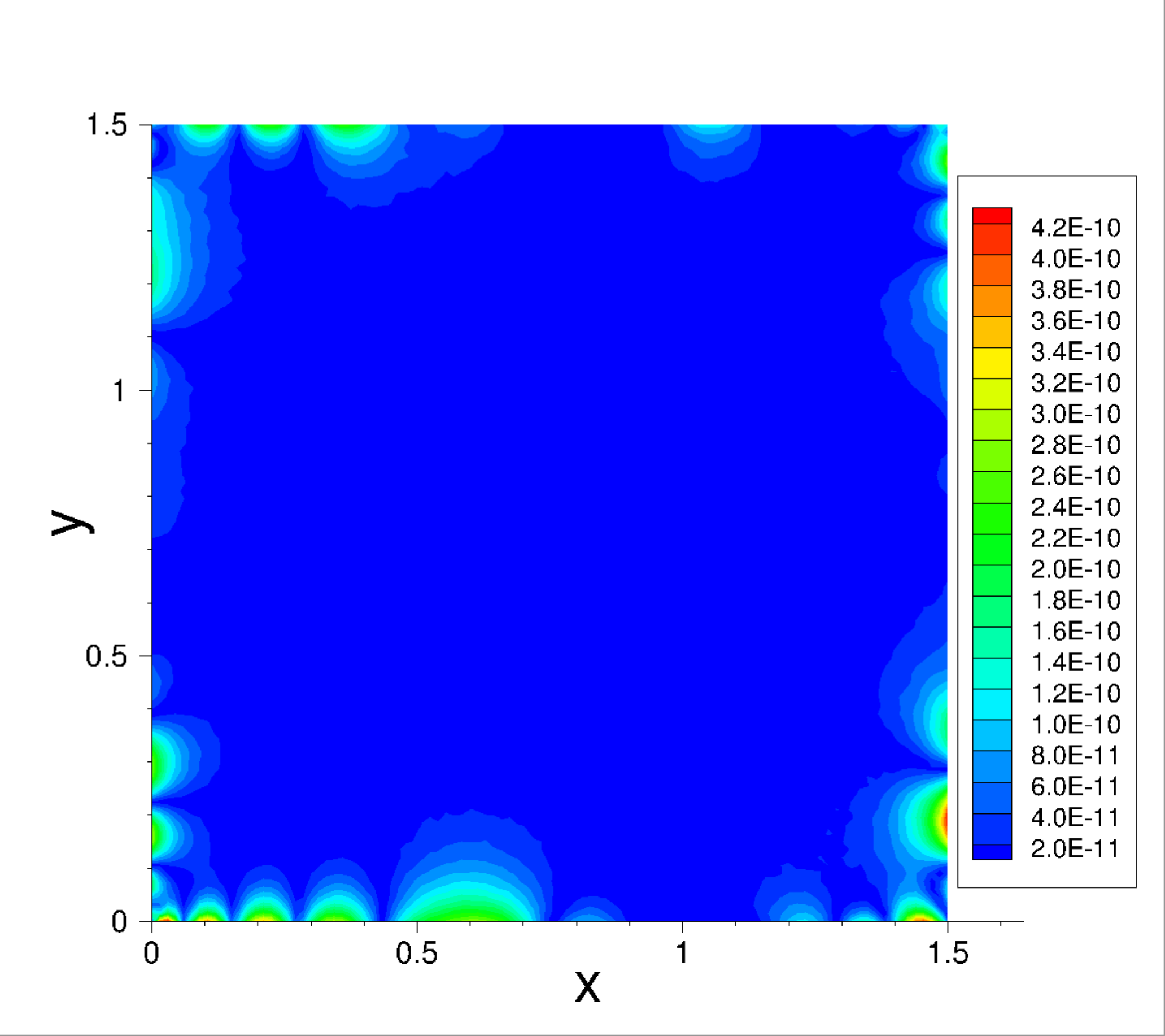}(a)
    \includegraphics[width=2in]{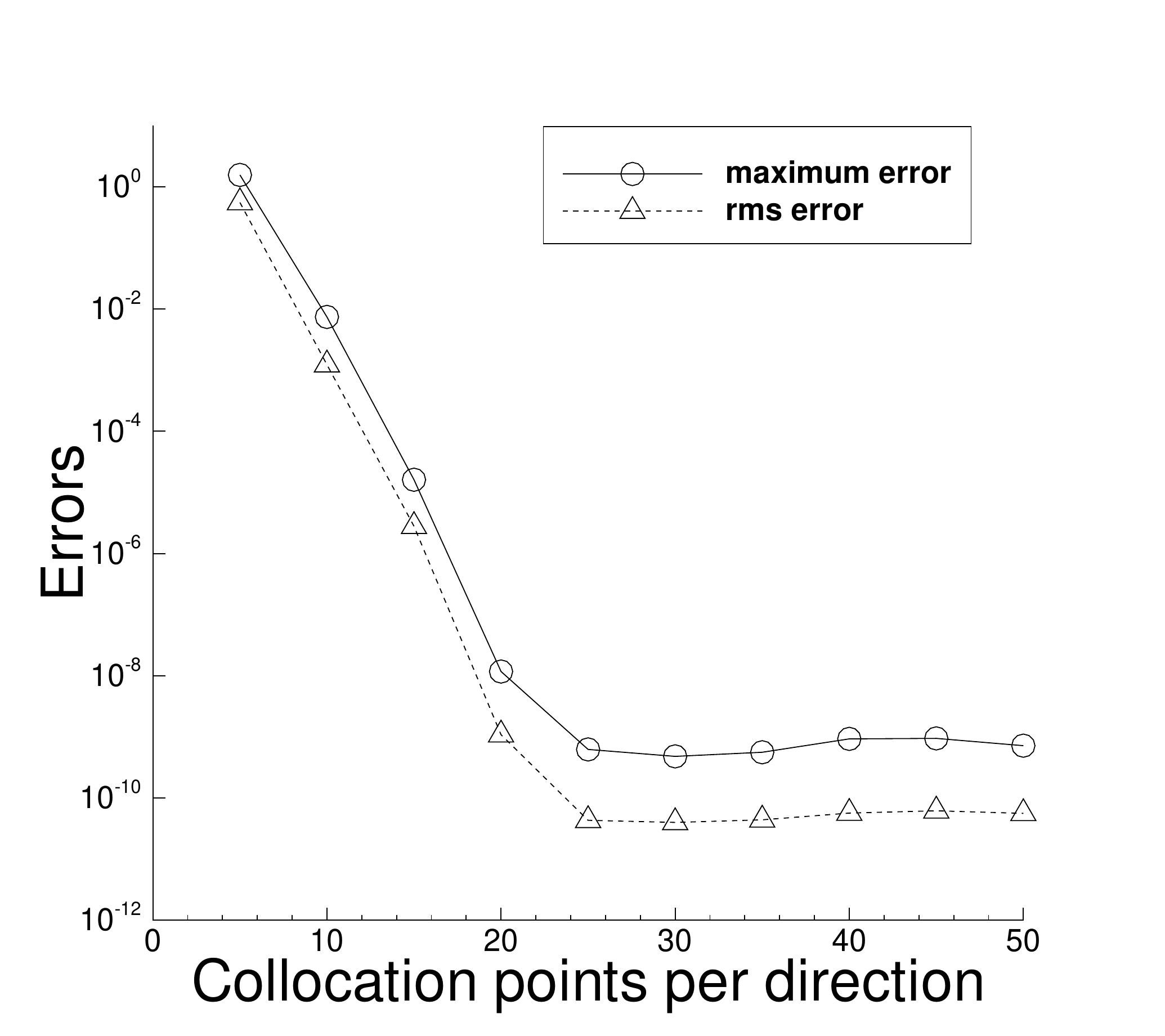}(b)
    \includegraphics[width=2in]{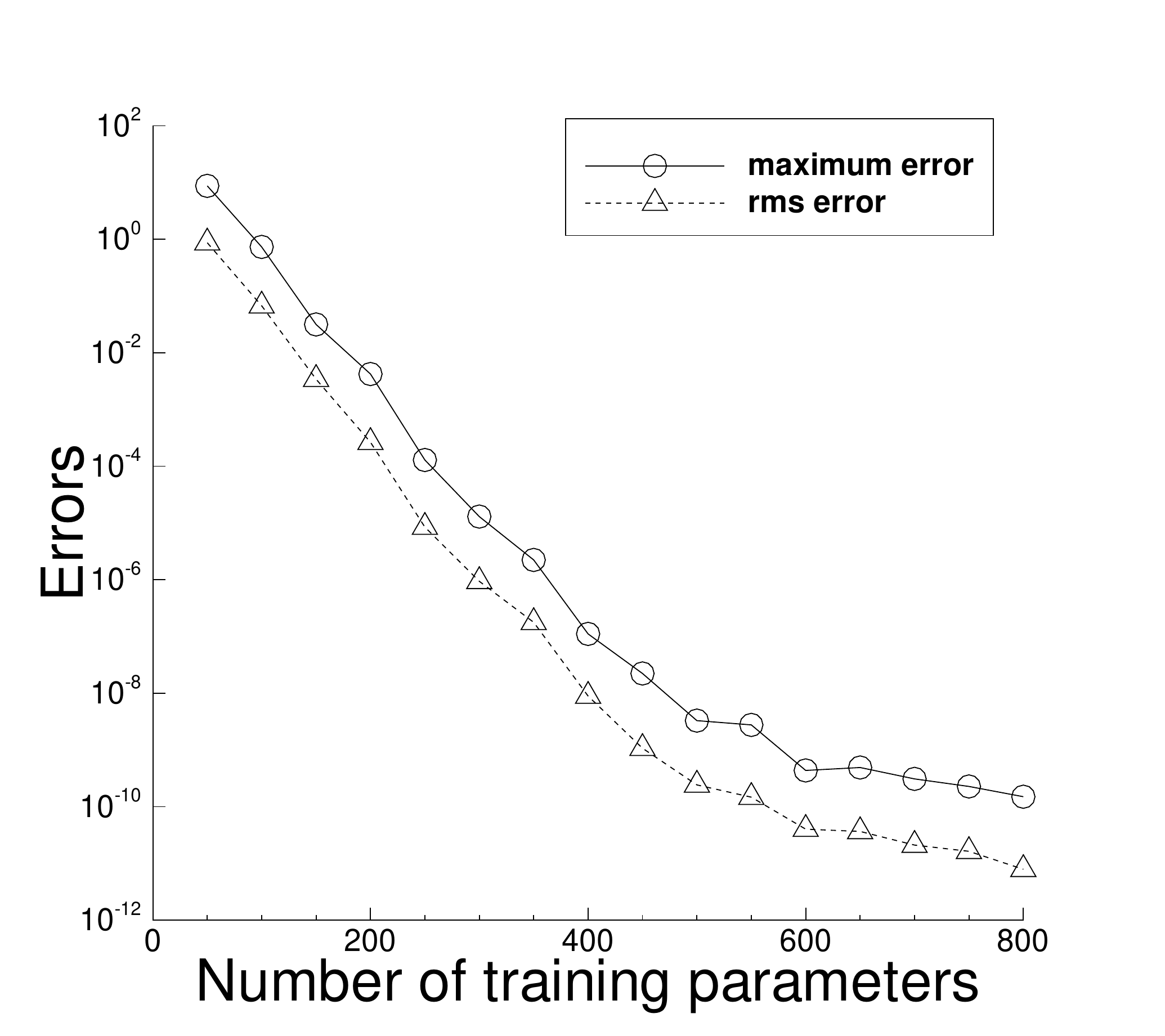}(c)
  }
  \caption{Nonlinear Helmholtz equation (Single-Rm-ELM):
    (a) Absolute error distribution of the Single-Rm-ELM solution.
    The maximum/rms errors in the domain versus
    (b) the number of collocation points per direction, and
    (c) the number of training parameters.
    Network architecture: [2, $M$, 1].
    $Q=31\times 31$ in (a,c), varied in (b).
    $M=600$ in (a,b), varied in (c).
    $R_m=2.1$ in (a,b,c).
  }
  \label{fg_20}
\end{figure}

Figure \ref{fg_20} illustrates the accuracy of the solutions to the nonlinear
Helmholtz equation obtained with Single-Rm-ELM.
In these tests we employ a set of $Q=Q_1\times Q_1$
uniform collocation points, where $Q_1$ is either fixed at $Q_1=31$
or varied systematically,
a neural network with the architecture $[2, M, 1]$, where
$M$ is either fixed at $M=600$
or varied systematically.
We employ a fixed $R_m=2.1$ in these tests, which is close to
the $R_{m0}$  from the differential evolution algorithm
corresponding to $M=600$ and $Q=31\times 31$.
Figure \ref{fg_20}(a) shows the distribution of the absolute
error of the Single-Rm-ELM solution.
It signifies a high solution accuracy,
with the maximum error on the order $10^{-10}$.
Figures \ref{fg_20}(b) and (c) depict the maximum/rms
errors in the domain of the ELM solution as a function
of $Q_1$ and the training parameters $M$, respectively.
The errors decrease exponentially (before saturation) with increasing numbers of
collocation points or training parameters, similar to what has been
observed for the linear problems in previous subsections.

\begin{figure}
  \centerline{
    \includegraphics[width=2in]{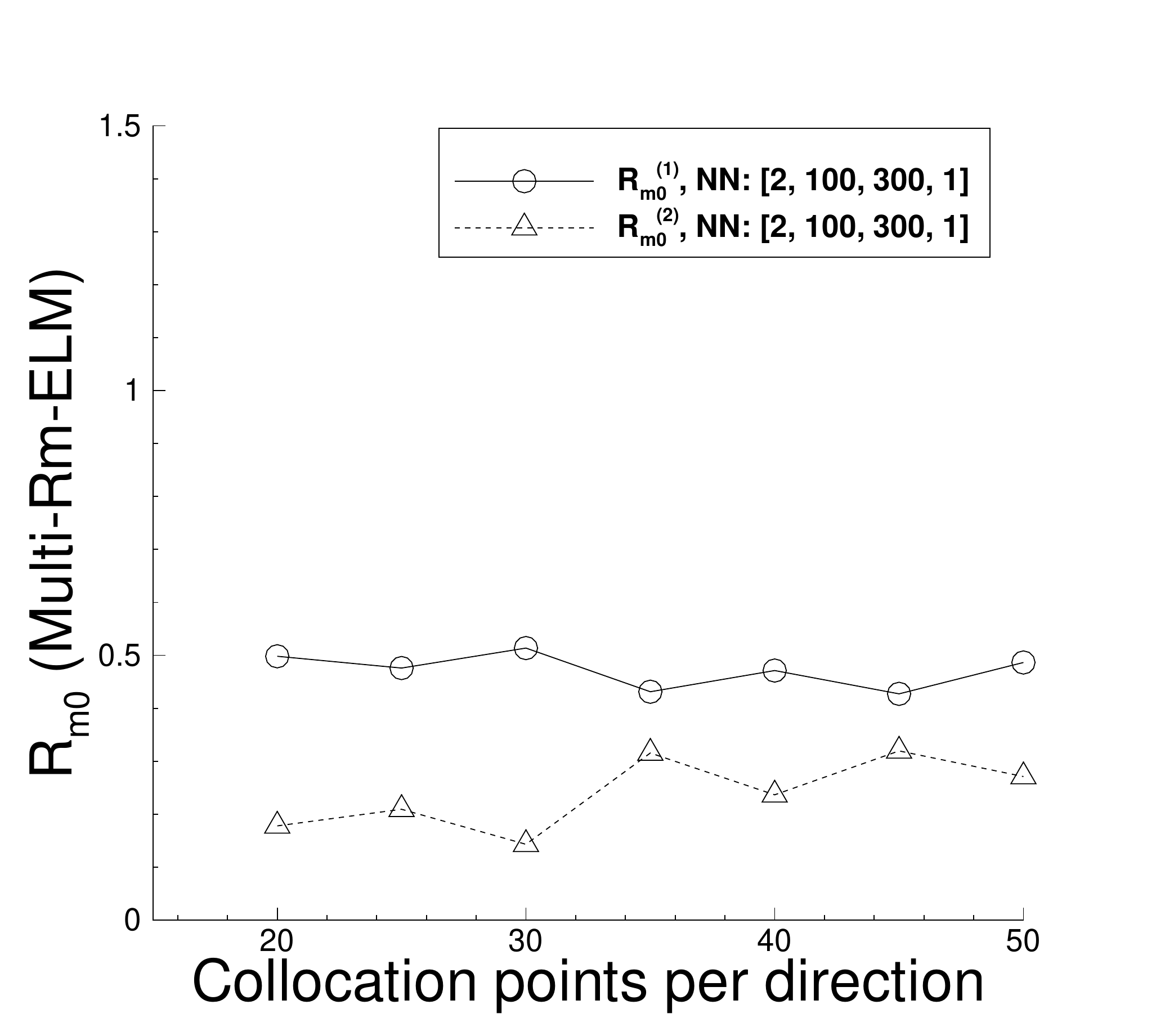}(a)
    \includegraphics[width=2in]{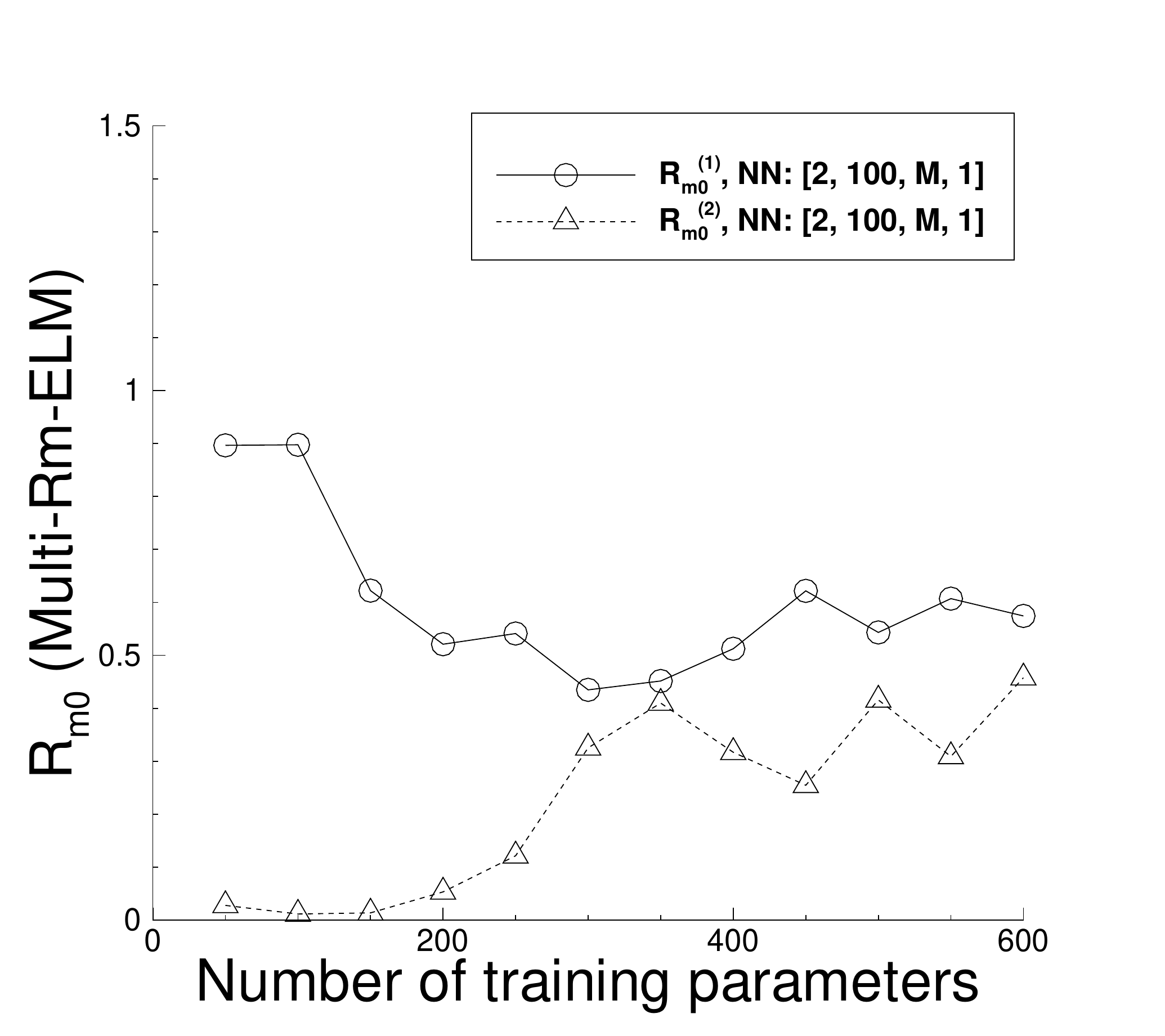}(b)
  }
  \centerline{
    \includegraphics[width=2in]{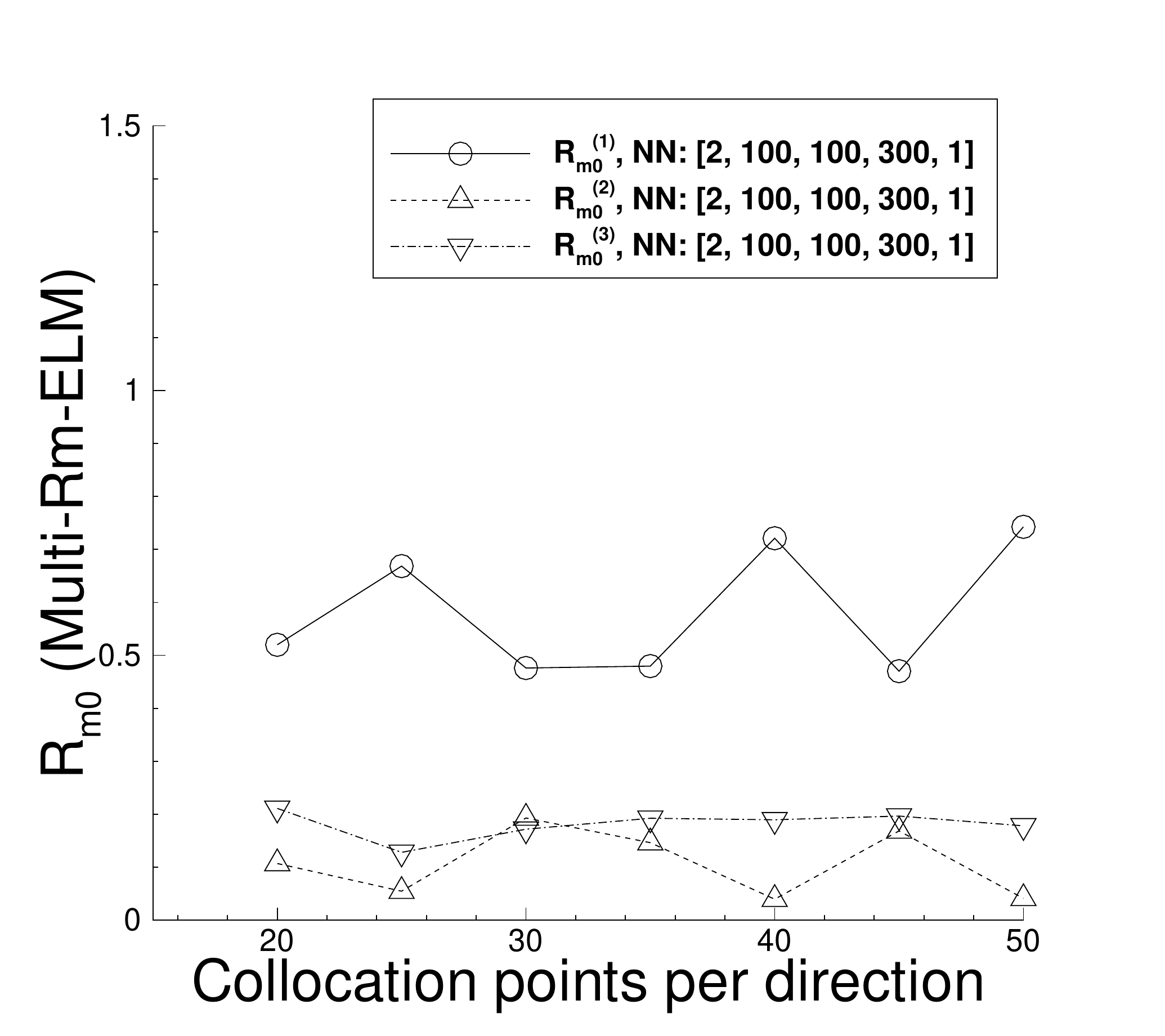}(c)
    \includegraphics[width=2in]{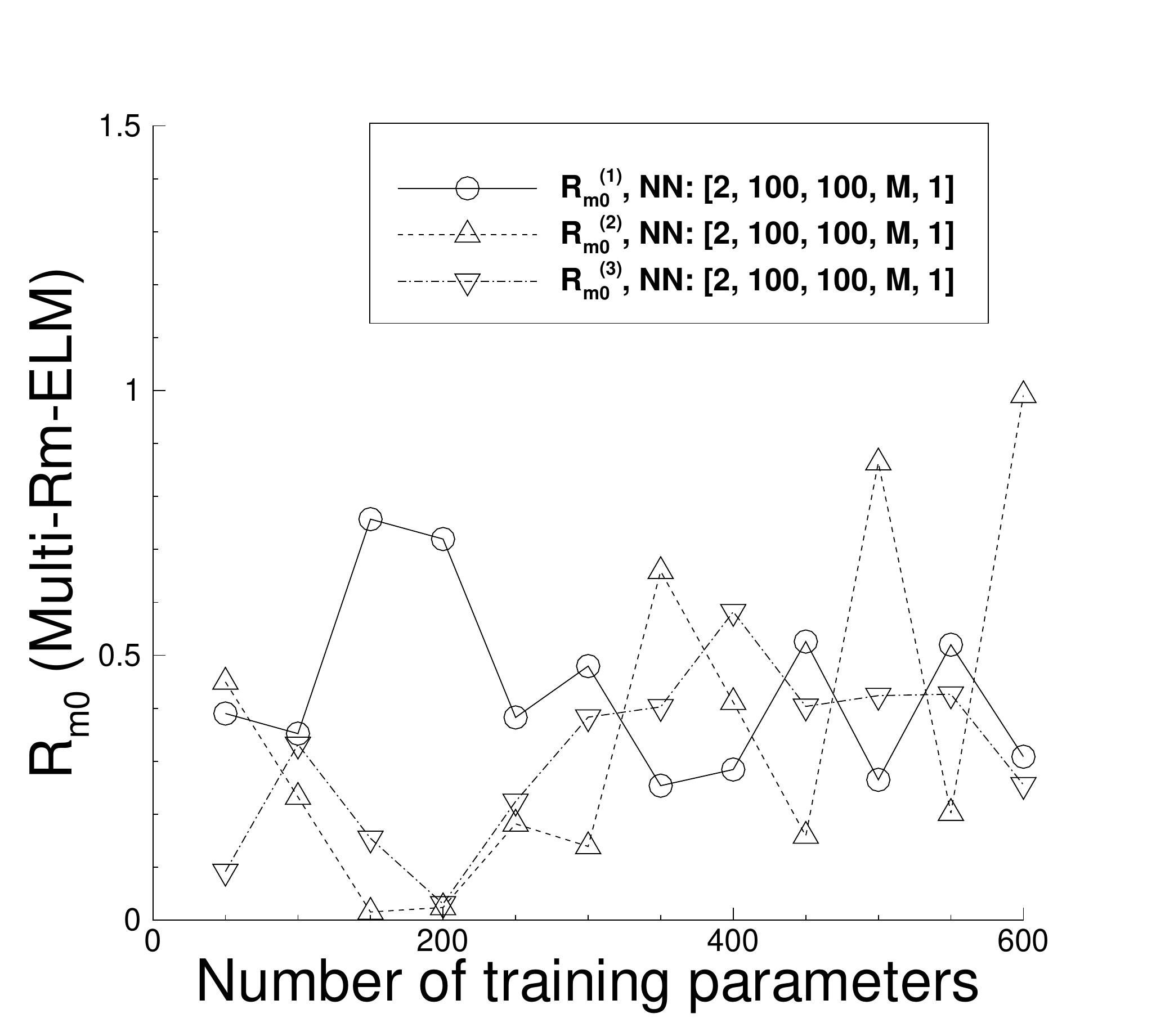}(d)
  }
  \caption{Nonlinear Helmholtz equation (Multi-Rm-ELM):
    The $\mbs R_{m0}$ components versus the number of collocation
    points per direction (a,c) and the number of training parameters (b,d),
    with neural networks having two (a,b) and three (c,d) hidden layers.
    The network architectures are given in the legends.
    $Q=31\times 31$ in (b,d), varied in (a,c).
    $M=300$ in (a,c), varied in (b,d). 
  }
  \label{fg_20_a}
\end{figure}

Let us next consider the Multi-Rm-ELM configuration for setting the random coefficients
in the neural network. Figure \ref{fg_20_a} illustrates the characteristics
of the optimum $\mbs R_{m0}$ obtained with the differential evolution
algorithm, as the number of collocation
points or training parameters is varied.
In these tests we consider two neural networks
with architectures given by $[2, 100, M, 1]$
and $[2, 100, 100, M, 1]$, respectively,
where the number of training parameters
is either fixed at $M=300$ or varied systematically.
A set of $Q=Q_1\times Q_1$ uniform collocation points is employed, where
$Q_1$ is either fixed at $Q_1=31$ or varied systematically.
Figures \ref{fg_20_a}(a) and (b) depict the components of
$\mbs R_{m0}=(R_{m0}^{(1)},R_{m0}^{(2)})$ as a function of $Q_1$ and $M$,
respectively, for the neural network with two hidden layers.
These results are obtained with a population size of $4$, the bounds
$[0.01, 3]$ for both $\mbs R_m$ components, and a
relative tolerance $0.1$ in the differential evolution algorithm.
Figures \ref{fg_20_a}(c) and (d) show the corresponding results
for $\mbs R_{m0}=(R_{m0}^{(1)},R_{m0}^{(2)},R_{m0}^{(3)})$ with the neural network
of three hidden layers,
which are obtained with a population size of $6$, the bounds $[0.01, 2]$
for all $\mbs R_m$ components, and a relative tolerance $0.1$
in the differential evolution.
The $\mbs R_{m0}$ characteristics observed here are quite similar to
those of the linear problems in previous subsections.
The values for the $\mbs R_{m0}$ components fluctuate within a range (generally less
than $1$), and appear more irregular compared with the $R_{m0}$ in Single-Rm-ELM.
The dependence of $\mbs R_{m0}$ on the collocation points
seems generally quite weak. Its relation to the number of training parameters
is quite irregular, especially with more hidden layers in
the neural network.

\begin{figure}
  \centerline{
    \includegraphics[width=2in]{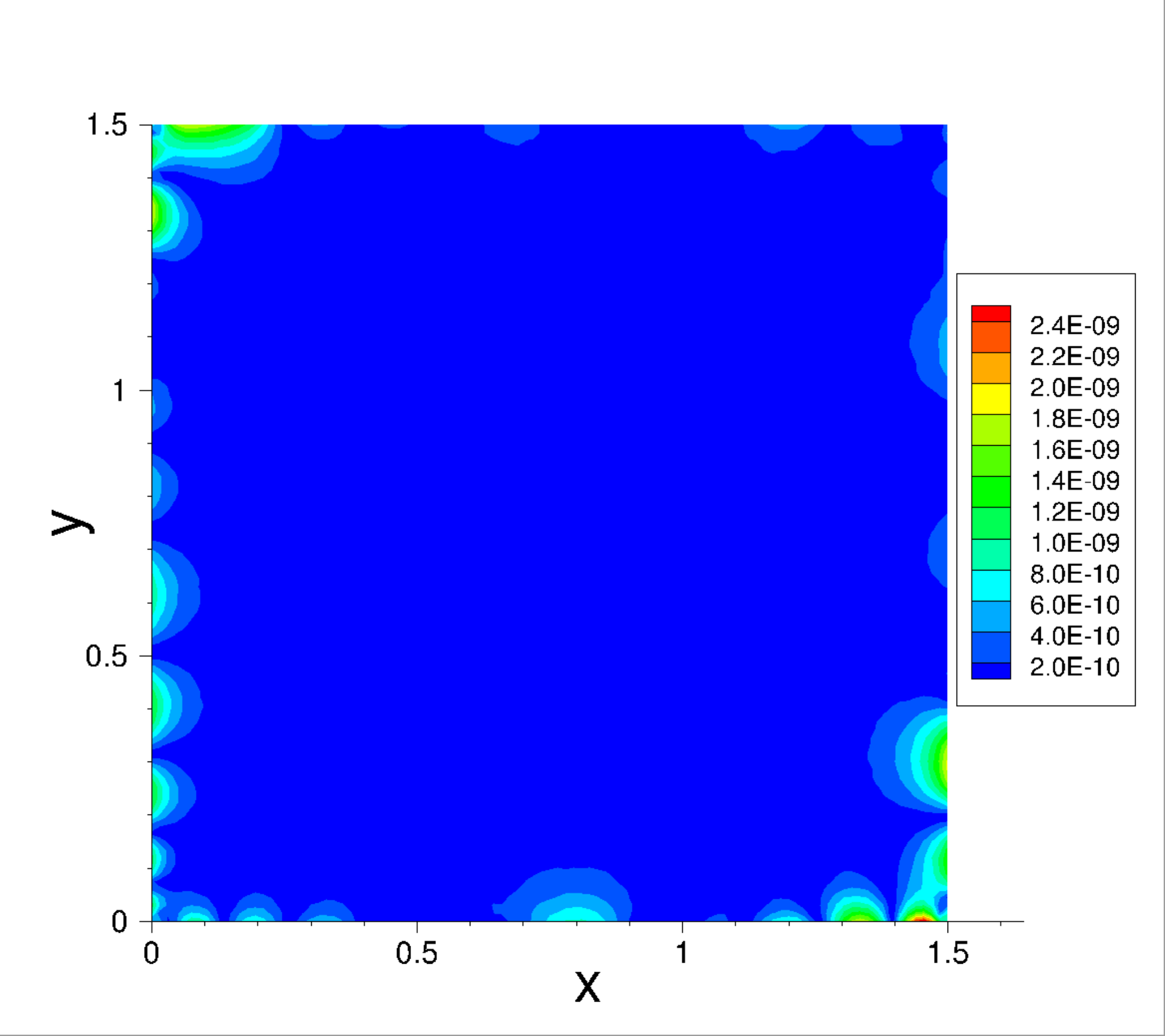}(a)
    \includegraphics[width=2in]{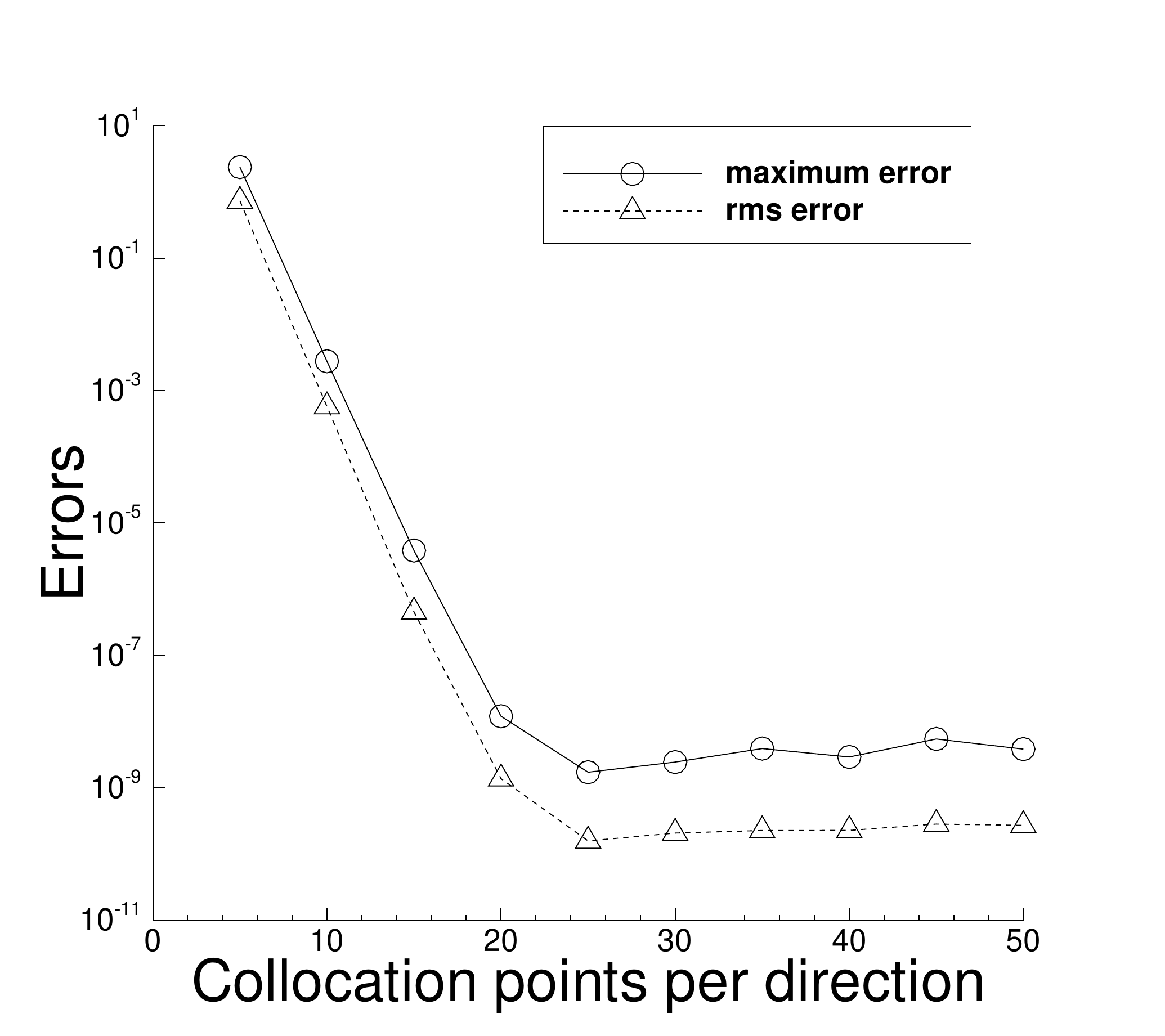}(b)
    \includegraphics[width=2in]{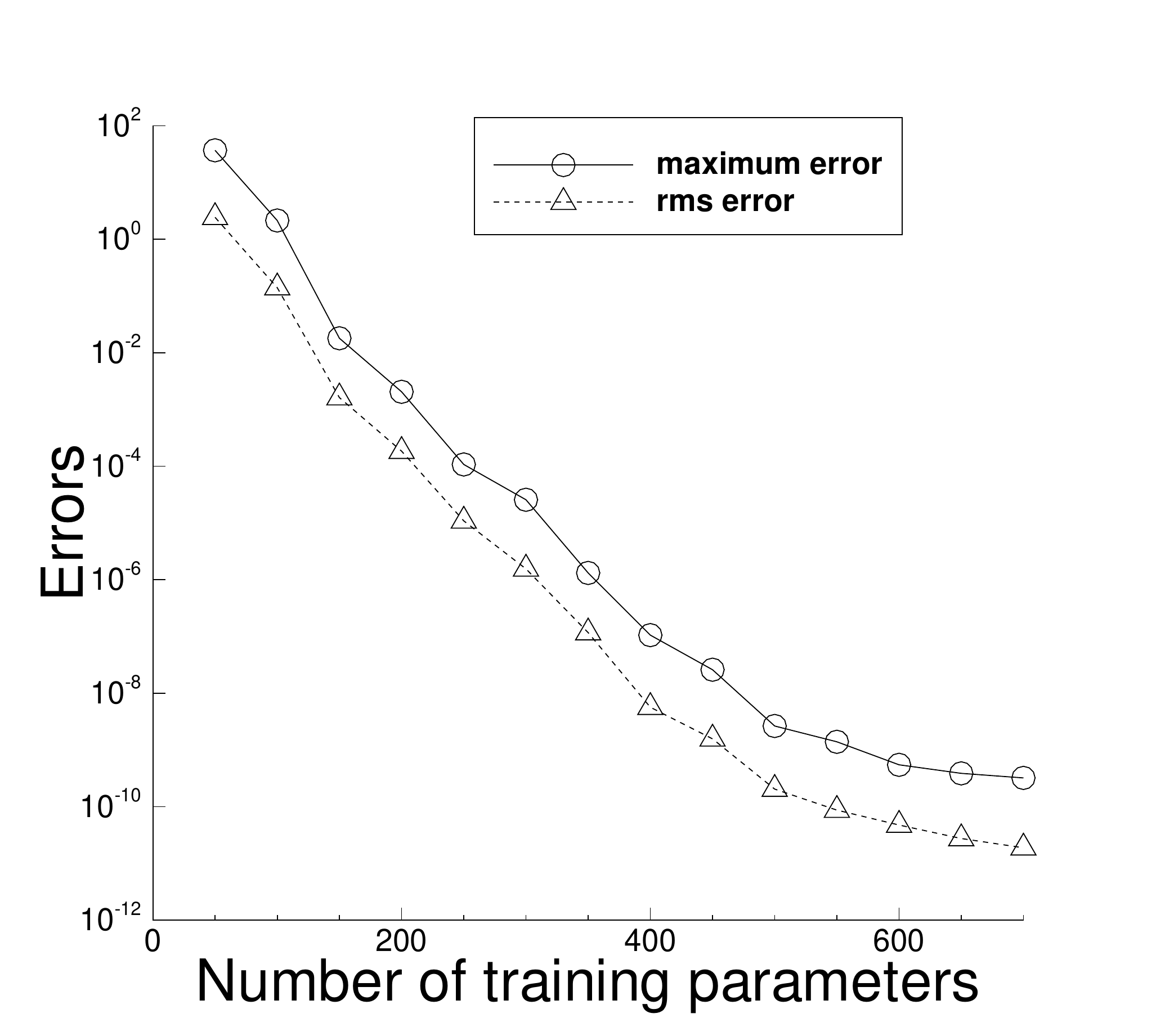}(c)
  }
  \caption{Nonlinear Helmholtz equation (Multi-Rm-ELM):
    (a) Absolute-error distribution of the Multi-Rm-ELM solution.
    The maximum/rms errors in the domain versus (b) the number of
    collocation points per direction, and (c) the number of training parameters.
    Network architecture: [2, 100, $M$, 1].
    $Q=31\times 31$ in (a,c), varied in (b).
    $M=500$ in (a,b), varied in (c).
    $\mbs R_m=(0.55,0.4)$ in (a,b,c).
  }
  \label{fg_21}
\end{figure}

Figure \ref{fg_21} illustrates the accuracy of the Multi-Rm-ELM solutions.
In these tests we employ a neural network
with an architecture $[2, 100, M, 1]$, where the number of training
parameters is either fixed at $M=500$ or varied systematically.
A set of $Q=Q_1\times Q_1$ uniform collocation
points is used, with $Q_1$ fixed at $Q_1=31$ or varied systematically.
We employ a fixed $\mbs R_m=(0.54, 0.4)$ in Multi-Rm-ELM,
which is close to the $\mbs R_{m0}$ obtained with $M=500$ and $Q=31\times 31$.
Figure \ref{fg_21}(a) shows the distribution of the absolute error
of the Multi-Rm-ELM solution corresponding to $Q_1=31$ and $M=500$,
signifying a quite high accuracy with the maximum error on the order $10^{-9}$.
Figures \ref{fg_21}(b) and (c) depict the maximum and rms errors in
the domain of the Multi-Rm-ELM solution versus $Q_1$ and $M$,
respectively, demonstrating the exponential convergence (before saturation)
with respect to these parameters.

\begin{figure}
  \centerline{
    \includegraphics[width=2in]{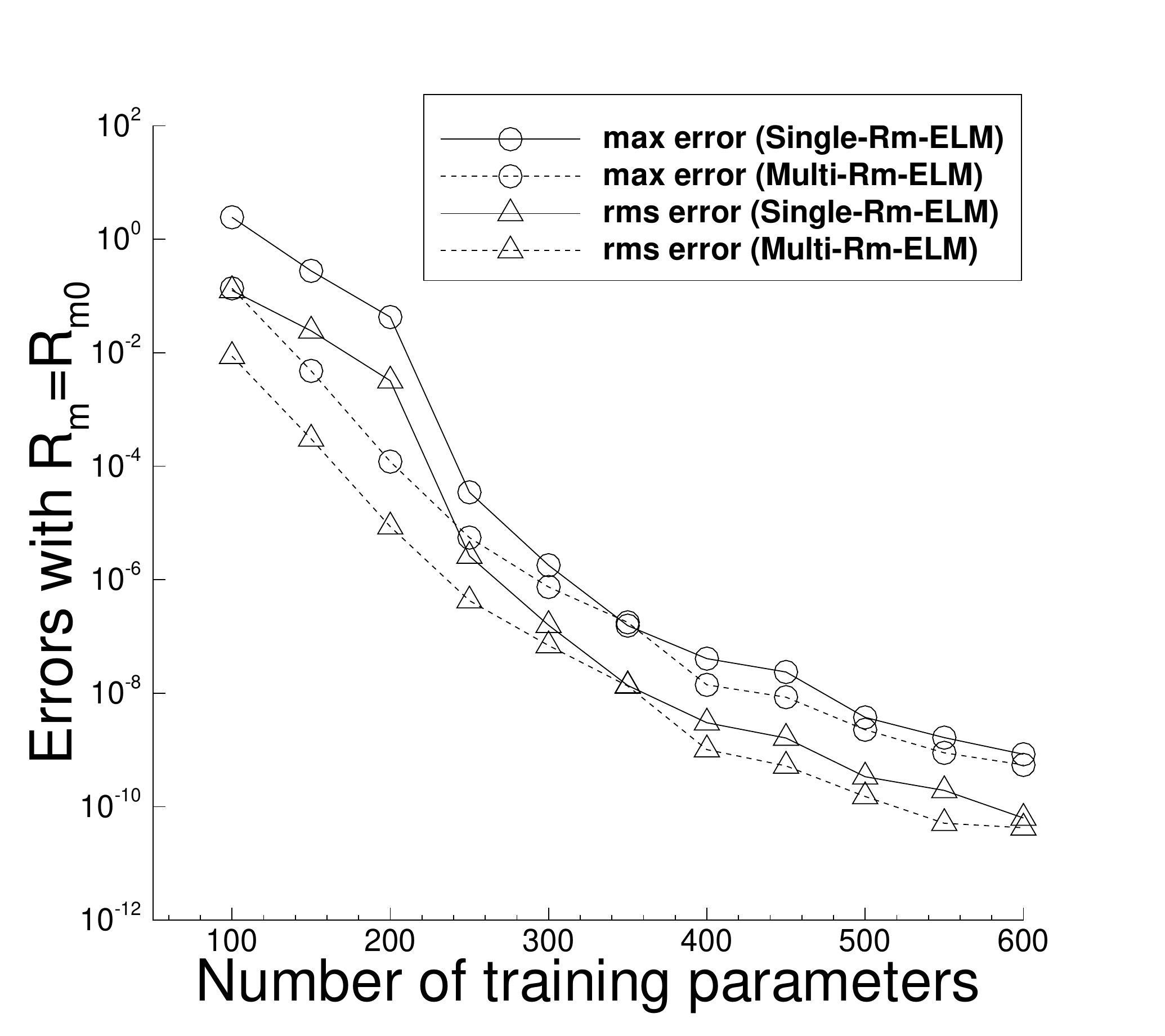}(a)
    \includegraphics[width=2in]{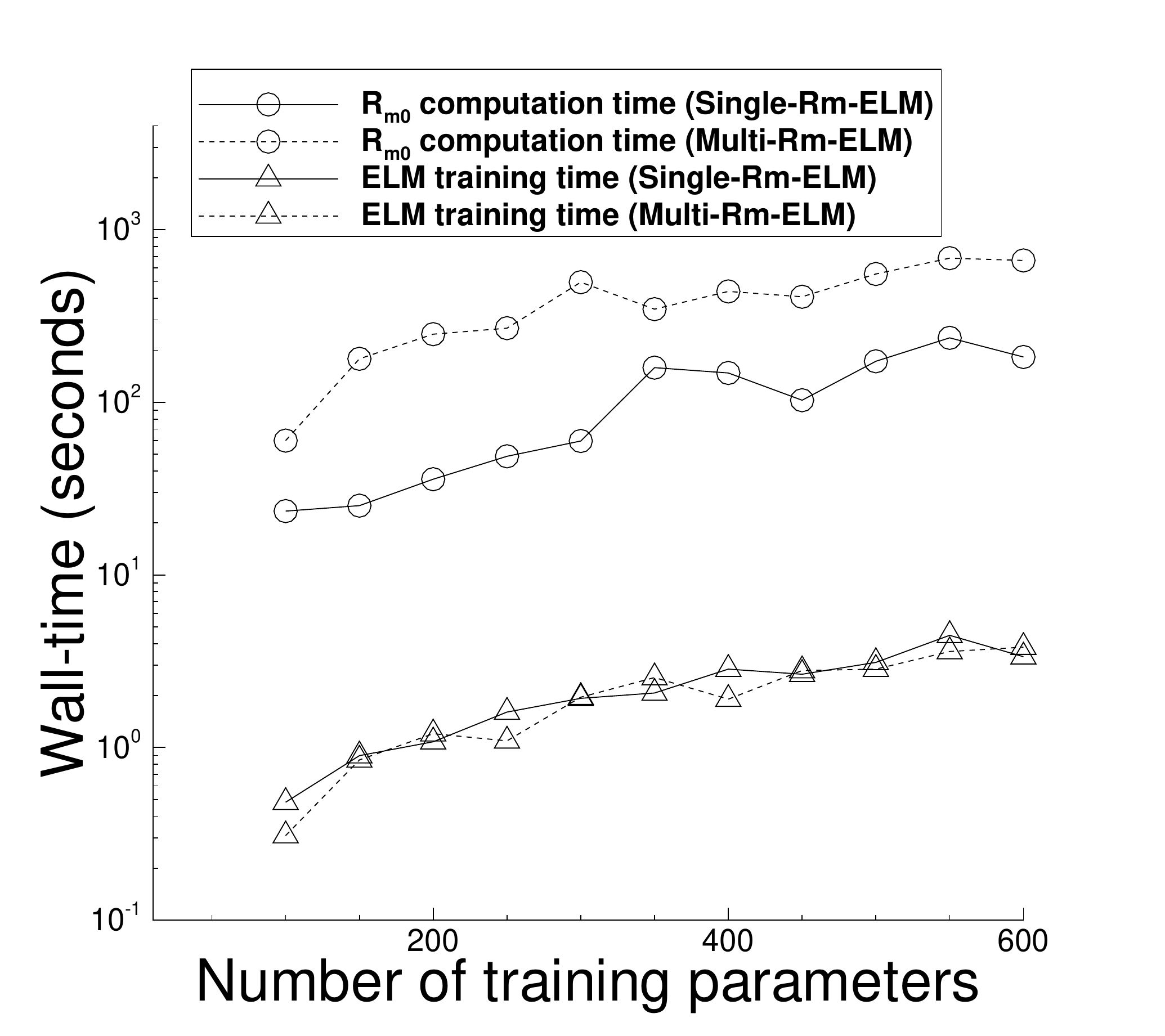}(b)
  }
  \caption{Nonlinear Helmholtz equation:
    (a) The maximum/rms errors in the domain corresponding to $R_m=R_{m0}$ in
    Single-Rm-ELM and $\mbs R_m=\mbs R_{m0}$ in Multi-Rm-ELM, versus
    the number of training parameters ($M$) in the neural network.
    (b) The  $R_{m0}$/$\mbs R_{m0}$ computation time and the ELM
    network training time in Single-Rm-ELM and Multi-Rm-ELM, versus
    the number of training parameters.
    Network architecture: [2, 100, M, 1].
    $Q=31\times 31$ in (a,b).
  }
  \label{fg_22}
\end{figure}

A comparison between Single-Rm-ELM and Multi-Rm-ELM for the nonlinear
Helmholtz equation is provided in Figure \ref{fg_22} with regard to
their accuracies and the $R_{m0}$ ($\mbs R_{m0}$) computation cost.
Here the neural network has an architecture $[2, 100, M, 1]$,
where $M$ is varied between $100$ and $600$ in the tests.
A fixed set of $Q=31\times 31$ uniform collocation points is employed
in the domain. 
For both Single-Rm-ELM and Multi-Rm-ELM, we have employed
a population size of $4$, the $R_m$ ($\mbs R_m$) bounds $[0.01, 3]$,
and a relative tolerance $0.1$ in the differential evolution algorithm.
Figure \ref{fg_22}(a) shows the maximum/rms errors in the domain
of the Single-Rm-ELM and Multi-Rm-ELM solutions obtained with
$R_m=R_{m0}$ ($\mbs R_m=\mbs R_{m0}$), as a function of the
number of training parameters $M$.
The Multi-Rm-ELM produces a consistently better accuracy than Single-Rm-ELM.
Figure \ref{fg_22}(b) shows the $R_{m0}$ ($\mbs R_{m0}$) computation time
in Single-Rm-ELM (Multi-Rm-ELM), as well as the ELM network training time
for a given $R_m$ ($\mbs R_m$), as a function of $M$.
The $\mbs R_{m0}$ computation in Multi-Rm-ELM is notably more costly
than the $R_{m0}$ computation in Single-Rm-ELM,
while the ELM network training cost is essentially the same
for given $R_m$ in Single-Rm-ELM and for given $\mbs R_m$ in
Multi-Rm-ELM.

\begin{figure}
  \centerline{
    \includegraphics[width=2in]{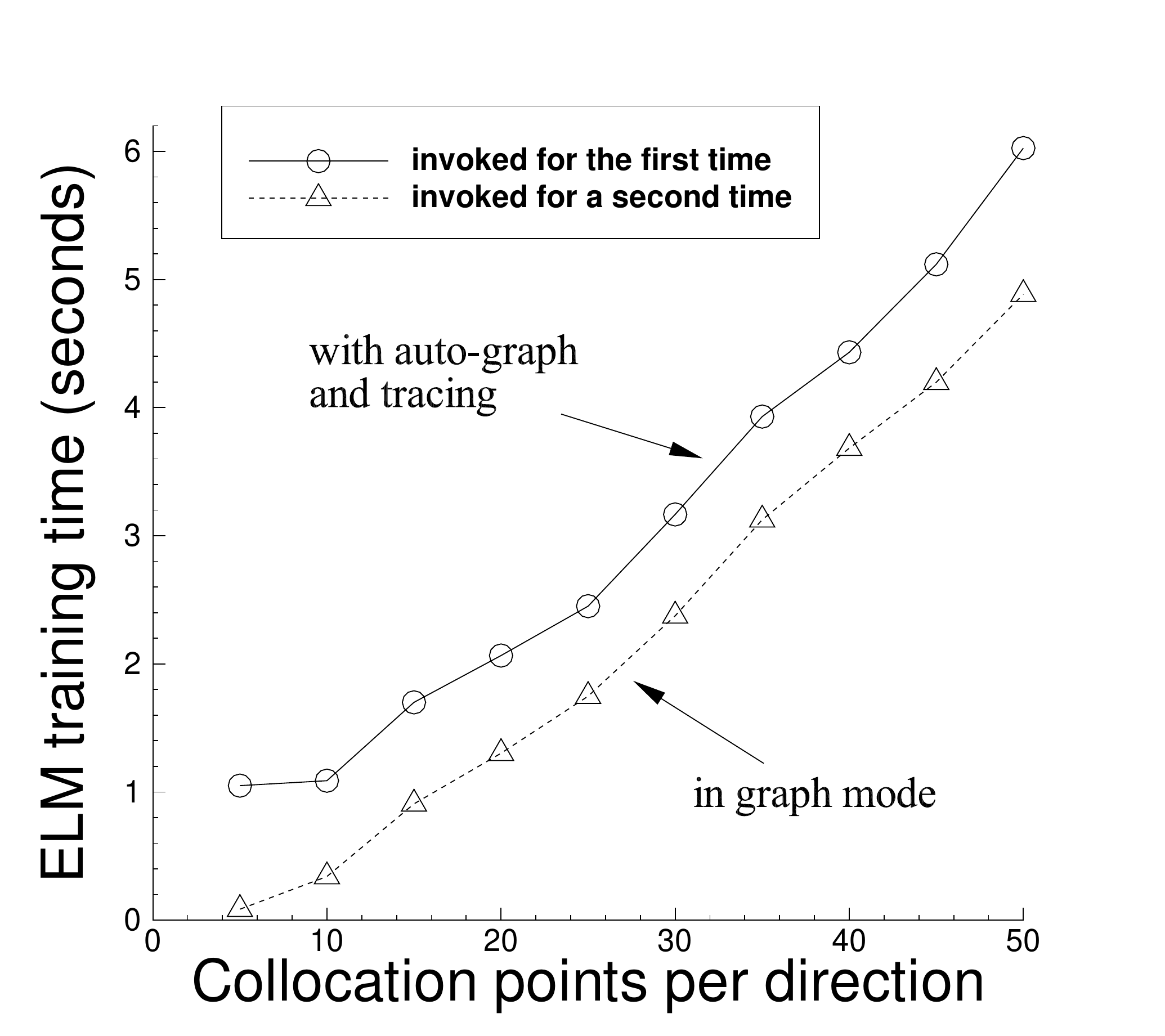}(a)
    \includegraphics[width=2in]{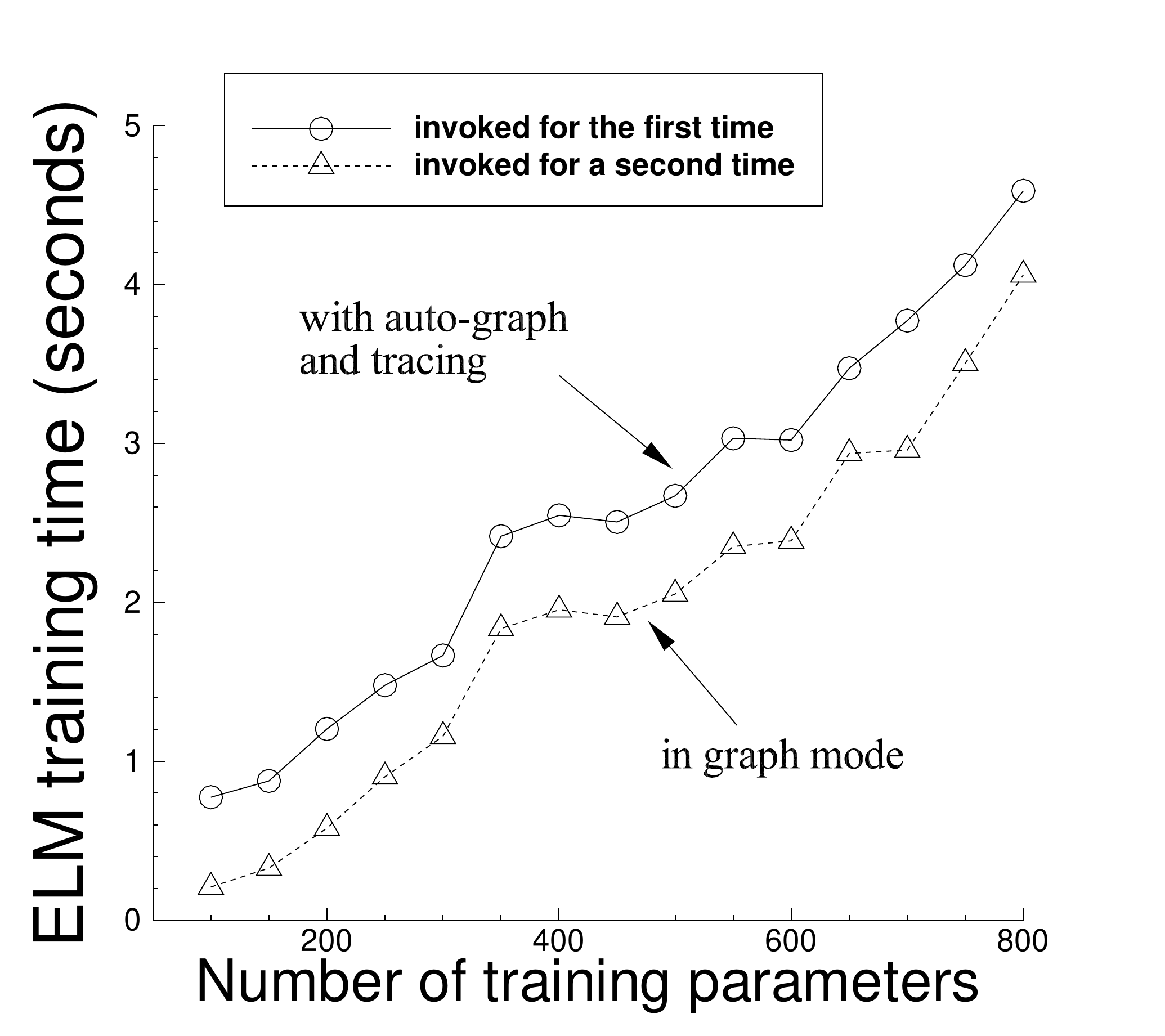}(b)
  }
  \caption{Nonlinear Helmholtz equation (Single-Rm-ELM):
    ELM network training time versus (a) the number of collocation points
    per direction and (b) the number of training parameters,
    obtained with the training routine invoked for the first
    time or subsequently.
    The settings and simulation/network parameters here
    correspond to those of Figures \ref{fg_20}(b,c).
  }
  \label{fg_23}
\end{figure}

Figure \ref{fg_23} illustrates the effect of autograph/tracing and the
computation in the graph mode on the ELM network training time
for the nonlinear Helmholtz equation.
As discussed previously, the Tensorflow Functions for computing
the output of the last hidden layer and the associated differential
operators are slower
when invoked for the first time, because the Tensorflow library
would use the autograph/tracing to
build the computational graphs. But they run much faster when invoked subsequently.
Figures \ref{fg_23}(a) and (b) depict the ELM network training time
with Single-Rm-ELM, obtained with autograph/tracing (invoked for the first time)
or in graph mode (invoked subsequently, no autograph/tracing),
as a function of the number of collocation points per direction and
the number of training parameters, respectively.
The settings and simulation parameters of these two plots correspond
to those of the Figures~\ref{fg_20}(b) and (c), respectively.
In the graph mode the ELM network training time
is markedly reduced.
In the the following comparisons with FEM, the ELM
training time refers to the time obtained in the graph mode
(no autograph/tracing).


\begin{figure}
  \centerline{
    \includegraphics[width=2in]{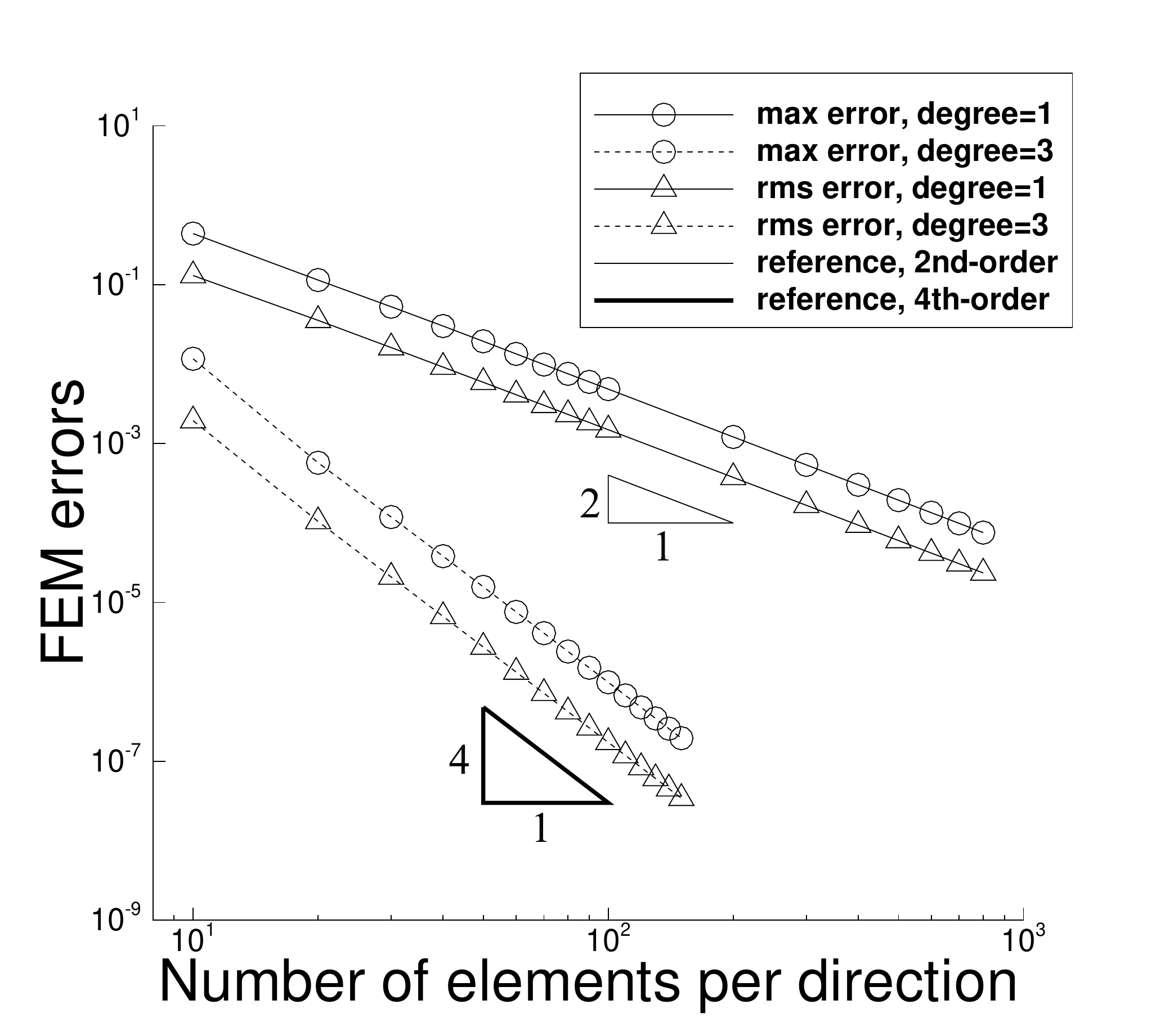}(a)
    \includegraphics[width=2in]{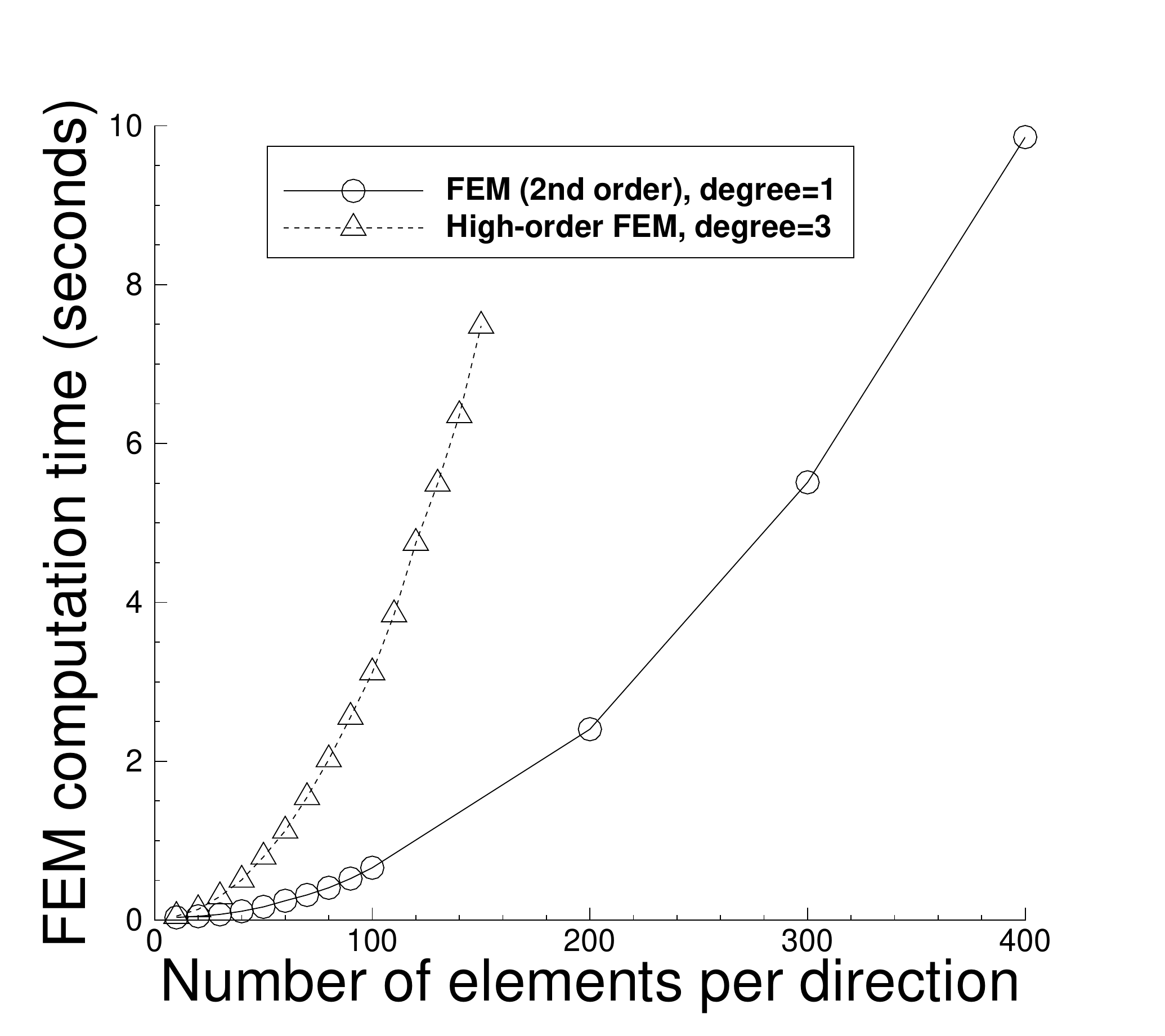}(b)
  }
  \centerline{
    \includegraphics[width=2in]{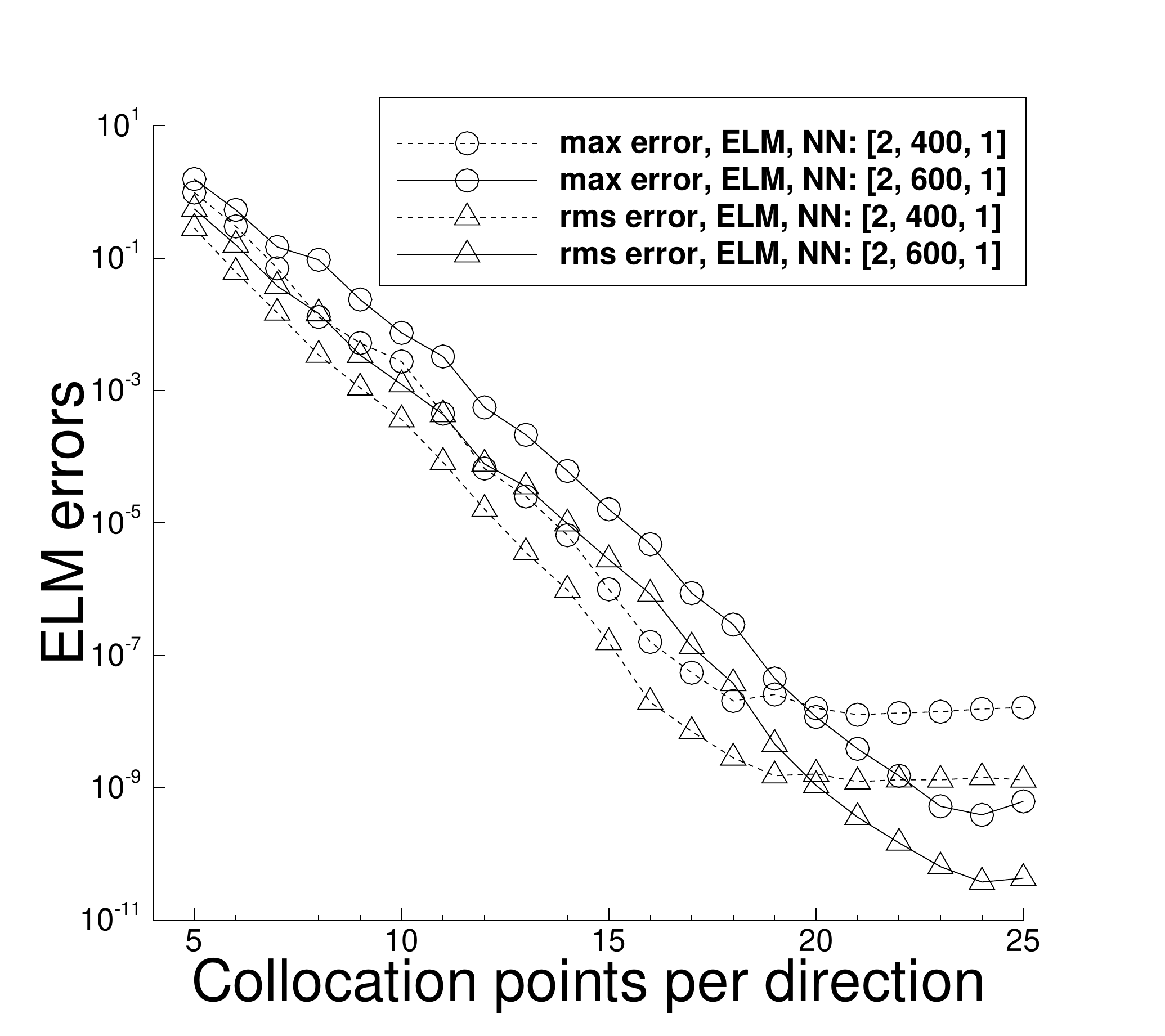}(c)
    \includegraphics[width=2in]{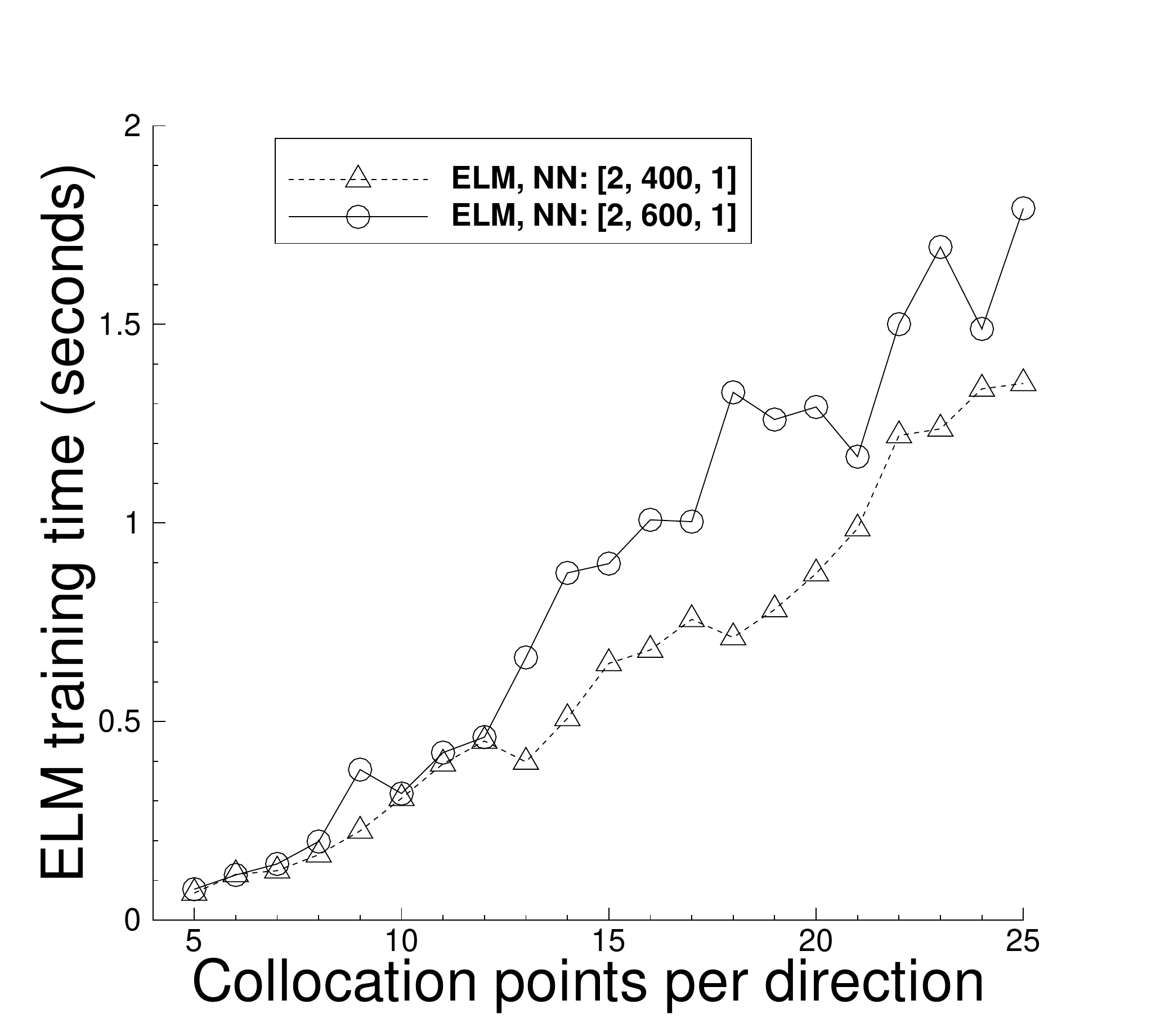}(d)
  }
  \caption{Nonlinear Helmholtz equation (FEM/ELM errors and cost):
    The numerical errors (a) and the computation time (b) of the classical FEM
    (degree=1) and the high-order FEM with Lagrange elements of degree 3,
    versus the number of elements in each direction.
    The numerical errors (c) and the network training time (d) of the ELM method
    versus the number of collocation points in each direction.
    ELM network architecture: $[2, M, 1]$. 
    In (c,d), $R_m=1.5$ for $M=400$, and $R_m=2.1$ for $M=600$.
  }
  \label{fg_26}
\end{figure}

We next compare the computational performance between the ELM method (Single-Rm-ELM
configuration) and the classical and high-order FEM for solving the nonlinear
Helmholtz equation.
With FEM we again use an $N_1\times N_1$ rectangular
mesh (partitioned into $2N_1^2$ triangles), and
the nonlinear Helmholtz equation in weak form is solved by
the Newton's method from the FEniCS library with a relative
tolerance $10^{-12}$.
We would like to mention an implementation detail concerning
the evaluation of the source term and the boundary data
in the system~\eqref{eq_17a}--\eqref{eq_17b}.
When implementing these terms as FEniCS ``Expressions'',
we have employed the element degree plus $4$ as the degree parameter
in these Expressions.
Note that when solving the Poisson equation in Section \ref{sec:poisson}
the degree parameter in the FEniCS Expressions for the source term
and the boundary data is set to be the element degree plus one.
We find that for nonlinear PDEs (nonlinear Helmholtz equation here, and the
Burgers' equation in the next subsection), setting the degree parameter
for these FEniCS Expressions to the element degree plus one is not adequate
with the high-order elements.
When setting it to the element degree plus one, we observe that
one cannot quite obtain
the expected rate of convergence with the high-order FEM, in particular for cases when
the mesh size is not very large.
We have tested various cases by setting the degree parameter in these
FEniCS Expressions to the element degree plus different extra degrees.
We observe that, as the extra degree increases (with the other parameters fixed),
the accuracy of the high-order FEM results
increases significantly initially, and it levels off
as the extra degree increases to $4$ and beyond.
So in this subsection and the next one (Burgers' equation),
we employ the element degree plus $4$ as the degree parameter
when evaluating the FEniCS Expressions for the source term and
the boundary data.

Figure \ref{fg_26} is an overview of the numerical errors of the FEM and ELM
and their computational cost (FEM computation time and ELM network training time)
for solving the nonlinear Helmholtz equation.
Figure \ref{fg_26}(a) shows the maximum/rms errors in the domain
of the classical FEM (linear elements, degree=1) and the high-order FEM
with Lagrange elements of degree=3, as a function of the number of elements
in each direction ($N_1$). The results signify the second-order convergence
rate of the classical FEM and the 4th-order convergence rate of the high-order FEM
with element degree 3.
Figure \ref{fg_26}(b) depicts the corresponding computation time
versus the number of elements per direction for the classical and
high-order FEMs, showing that the cost of the high-order FEM
grows much faster than the classical FEM with increasing mesh sizes.
Figure \ref{fg_26}(c) depicts the maximum/rms errors in the domain
of the ELM method (Single-Rm-ELM configuration) as a function of
the number of collocation points in each direction.
Two neural networks are employed, with architectures
$[2, M, 1]$ with $M=400$ and $M=600$, respectively.
A set of $Q=Q_1\times Q_1$ uniform collocation points is employed,
with $Q_1$ varied systematically between $5$ and $25$.
In ELM the random hidden-layer coefficients are generated 
with $R_m=1.5$ for the neural network with $M=400$ and
$R_m=2.1$ with $M=600$, which are close to the $R_{m0}$ obtained
with the differential evolution algorithm. 
The exponential decrease in the ELM errors are unmistakable.
Figure \ref{fg_26}(d) shows the corresponding ELM network training time
as a function of the number of collocation points.
Here the ELM training time refers to the time obtained in
the graph mode (without autograph/tracing).
One can observe that,
as the number of collocation points increases,
the growth in the ELM training time is nearly linear.

\begin{figure}
  \centerline{
    \includegraphics[width=2in]{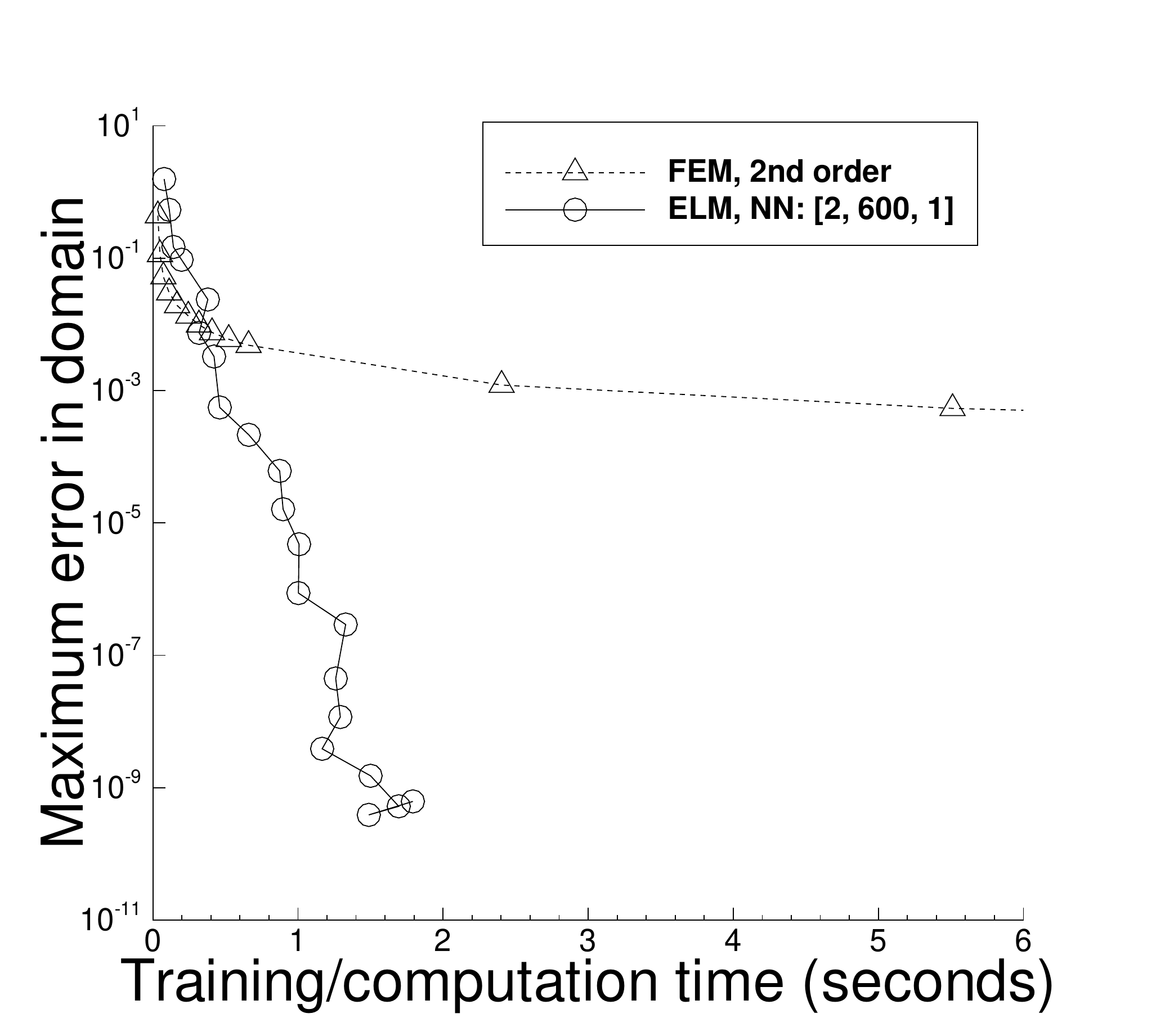}(a)
    \includegraphics[width=2in]{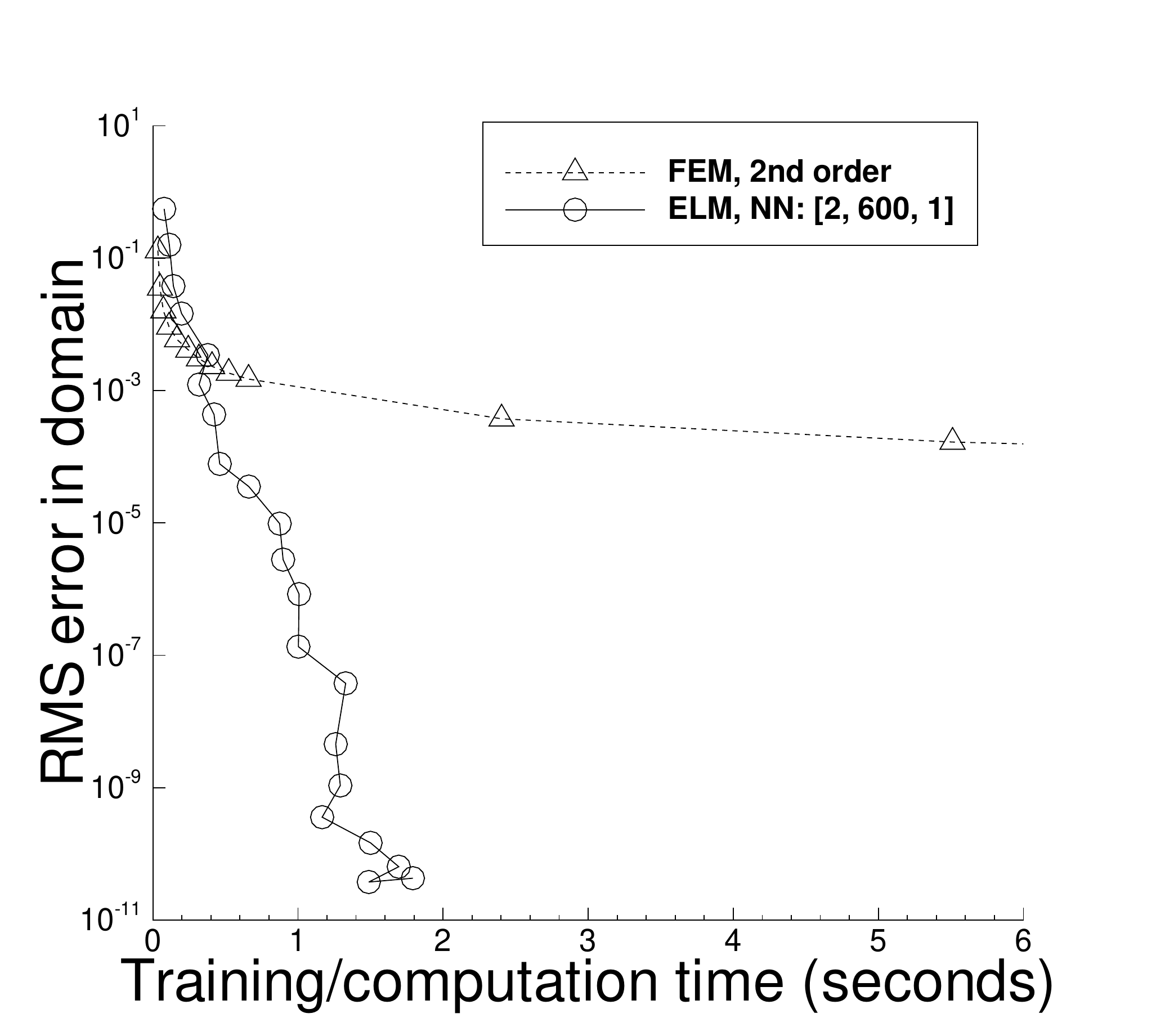}(b)
  }
  \caption{Nonlinear Helmholtz equation (comparison between ELM and classical FEM):
    The maximum error (a) and the rms error (b)  versus the
    computational cost (ELM network training time, FEM computation time)
    of the ELM and the classical FEM.
    The FEM data correspond to those of
    Figures \ref{fg_26}(a,b) with degree$=1$. The ELM data correspond to those
    of Figures \ref{fg_26}(c,d) with $M=600$.
  }
  \label{fg_25}
\end{figure}

Figure \ref{fg_25} compares the computational performance
of the ELM and the classical FEM (2nd-order, linear elements).
It plots the maximum error (plot (a)) and the rms error (plot (b))
in the domain of the ELM (and the classical FEM)
versus the ELM network training time (resp.~the FEM computation time).
The FEM data correspond to those in Figures \ref{fg_26}(a,b)
with degree=1, and the ELM data here correspond to those
in Figures~\ref{fg_26}(c,d) with $M=600$ in the neural network.
The ELM far outperforms the classical FEM in almost all cases,
achieving a considerably better accuracy with the same computational cost/budget
or inducing a considerably smaller cost to achieve the same accuracy.
The exception is in a range of very small problem sizes (FEM mesh size
smaller than around $70\times 70$, ELM collocation points less than
around $10\times 10$, error level above around $5\times 10^{-3}\sim 10^{-2}$,
wall time less than around $0.3$ seconds),
where the ELM performance and the FEM performance
are close, with the FEM a little better.

\begin{figure}
  \centerline{
    \includegraphics[width=2in]{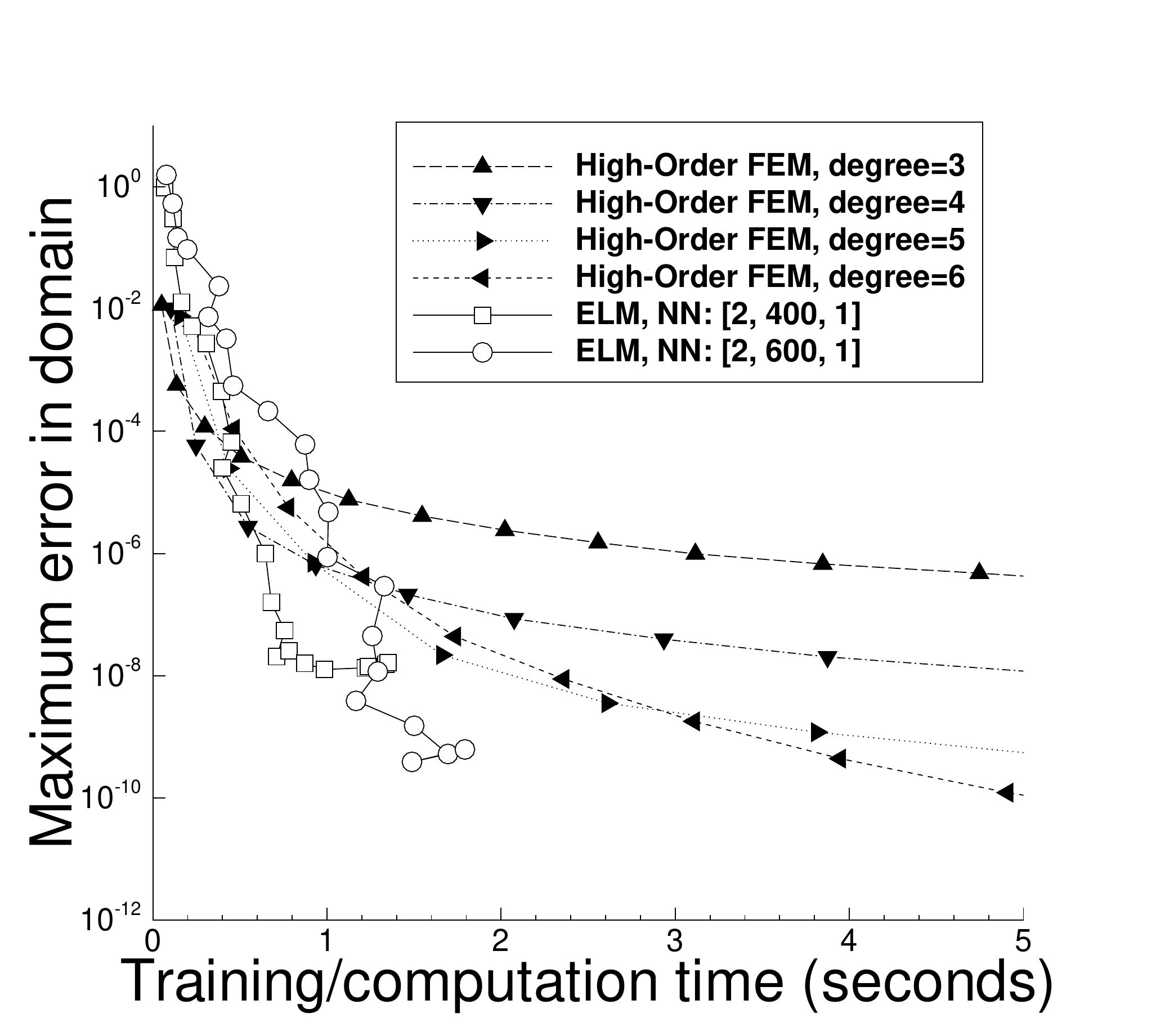}(a)
    \includegraphics[width=2in]{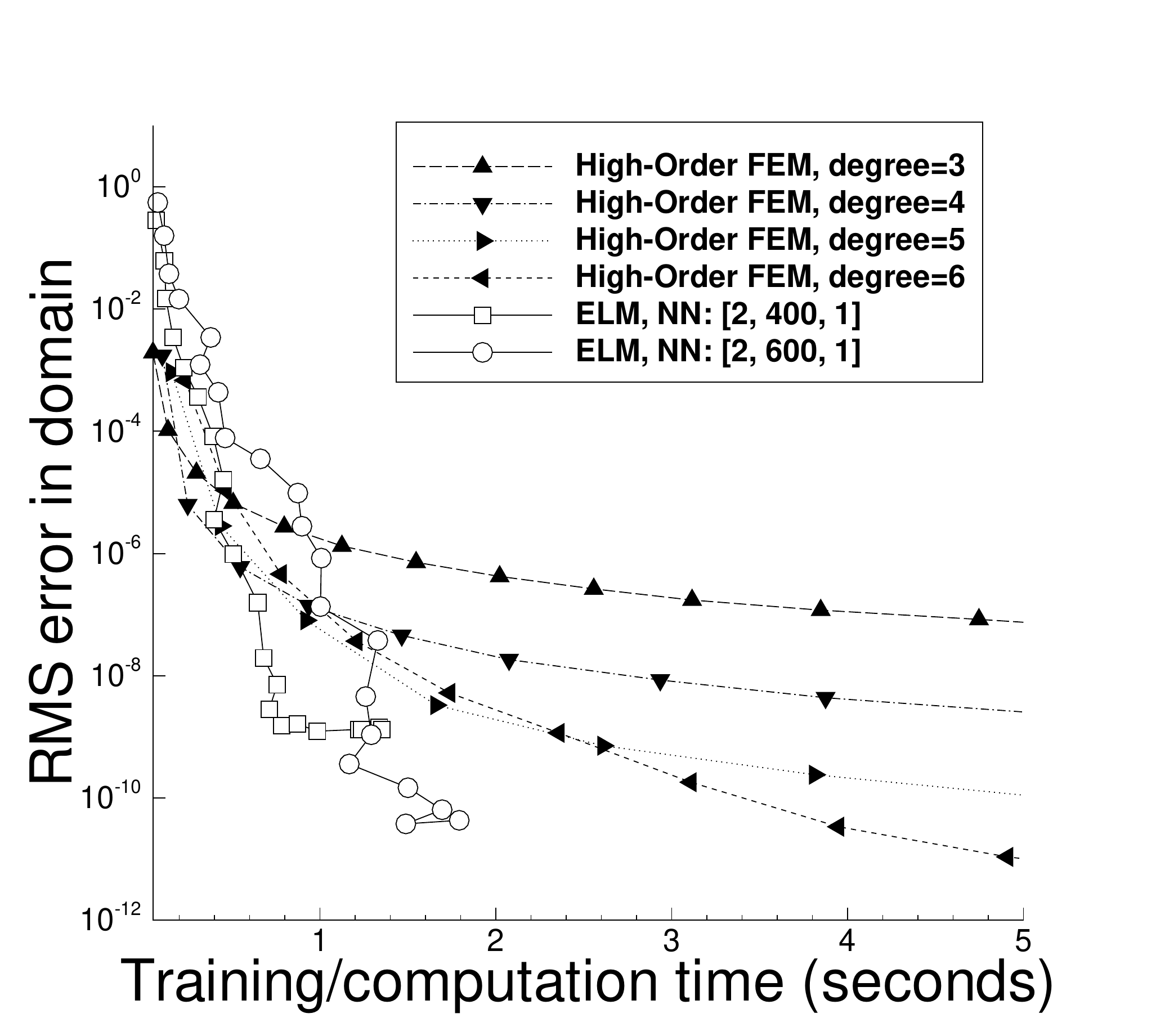}(b)
  }
  \centerline{
    \includegraphics[width=2in]{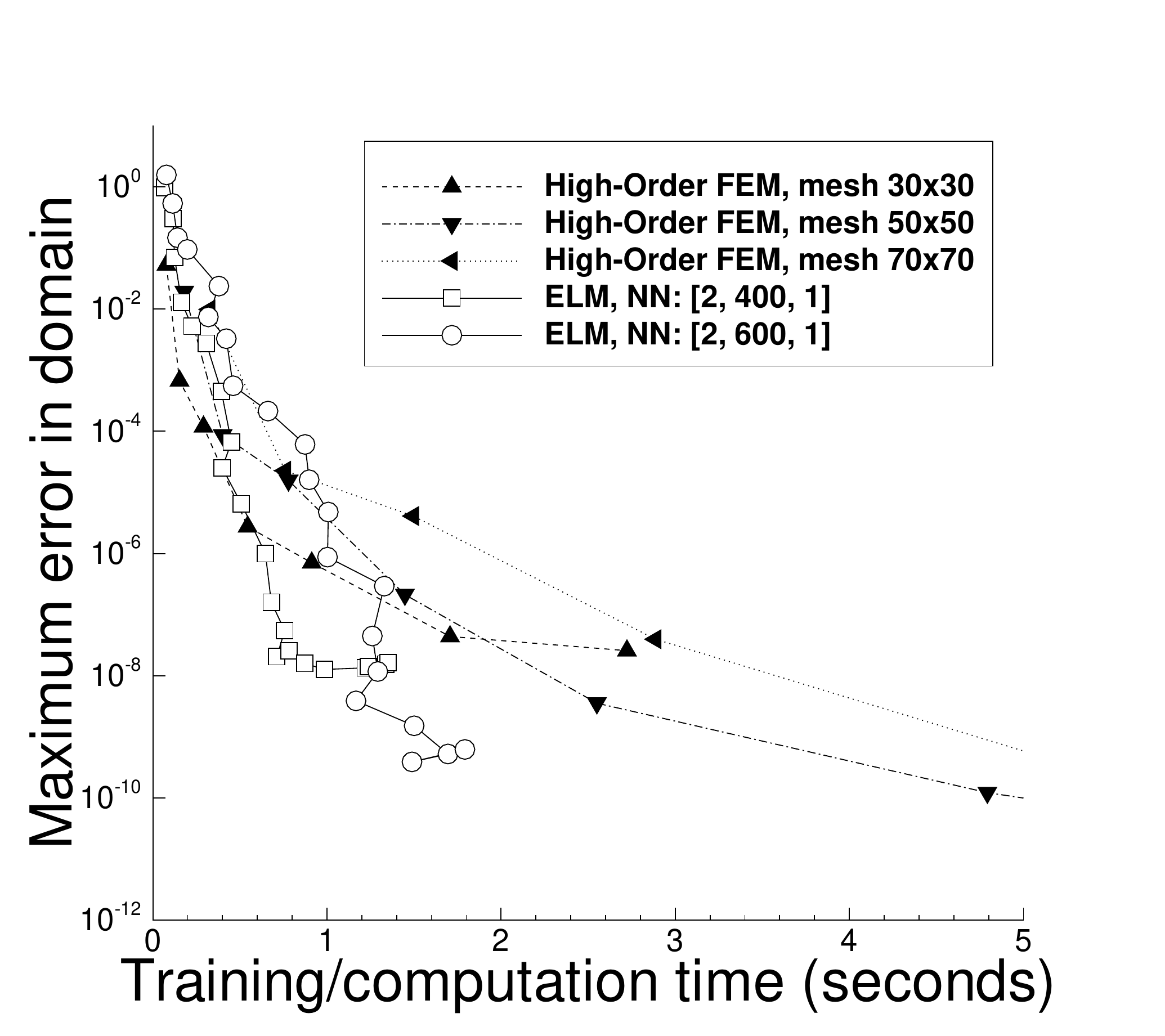}(c)
    \includegraphics[width=2in]{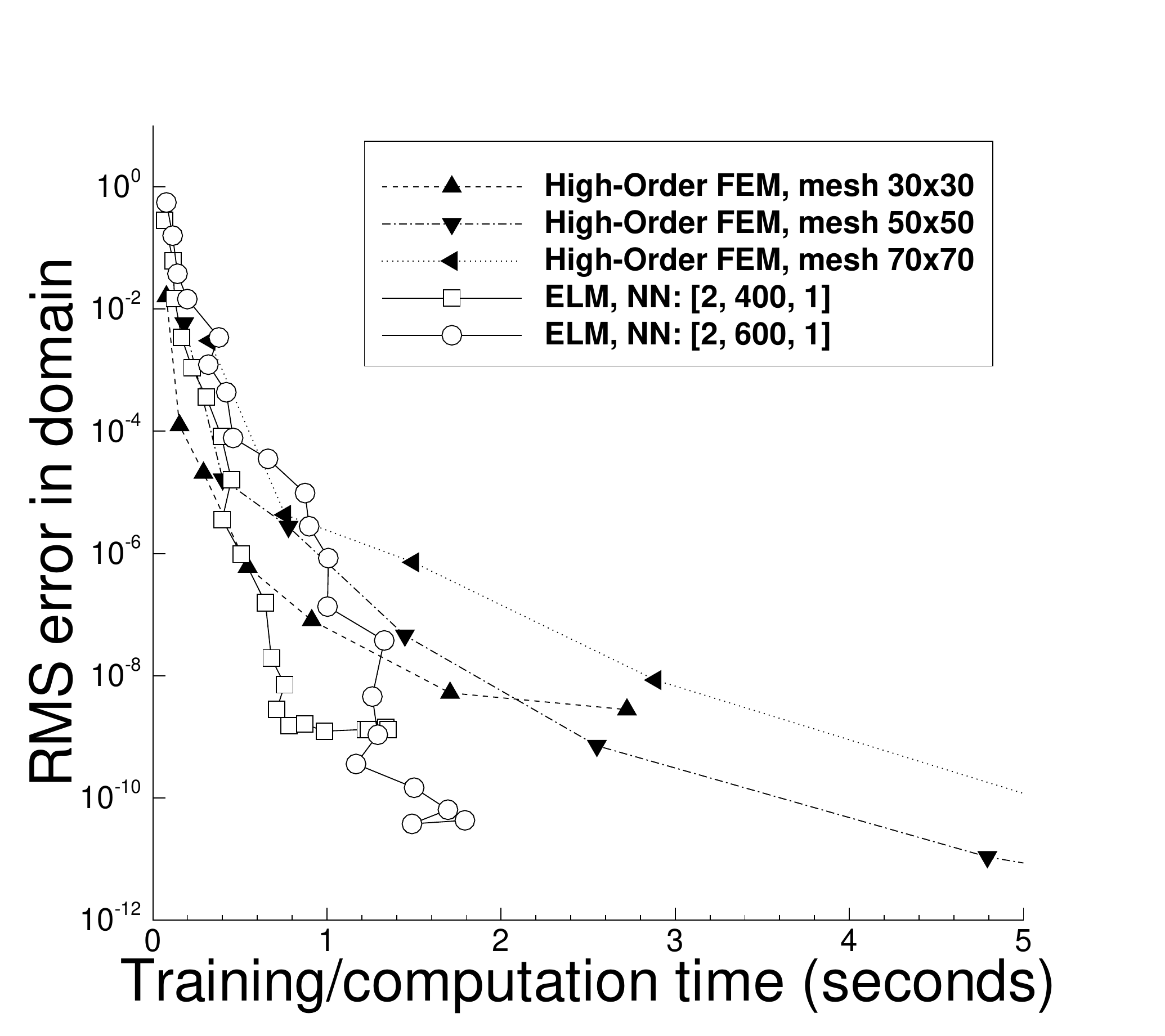}(d)
  }
  \caption{Nonlinear Helmholtz equation (comparison between ELM and high-order FEM):
    The maximum error (a,c) and the rms error (b,d) in the domain versus
    the computational cost (ELM training time, FEM computation time)
    between ELM
    and high-order FEM with  various element degrees. In (a,b)
    the number of elements per direction in the FEM mesh
    is varied systematically for each given element degree. In (c,d)
    the degree of the Lagrange elements is varied systematically
    for each given mesh size.
    The FEM data of degree $3$ in (a,b) correspond to those of
    Figures \ref{fg_26}(a,b)
    with degree $3$. The ELM data in (a,b,c,d) correspond to those
    of Figures \ref{fg_26}(c,d).
  }
  \label{fg_26_a}
\end{figure}

A comparison between ELM and the high-order FEM is provided in
Figure \ref{fg_26_a} for solving the nonlinear Helmholtz equation.
We have again performed the h-type refinement (fix the element degree, vary the
mesh size) and the p-type refinement (fix the mesh size, vary the element degree
between $1$ and $7$) with high-order FEM.
Figures \ref{fg_26_a}(a) and (b) depict the maximum and rms errors, respectively,
of the high-order FEM under the h-type refinements
versus the FEM computation time.
The FEM data for the element degree=3 in these plots correspond to
those in Figures \ref{fg_26}(a,b) with degree=3.
Figures \ref{fg_26_a}(c) and (d) depict the maximum and rms errors of
the high-order FEM under the p-type refinements versus the FEM computation
time. These four plots also include the maximum and rms errors of
the ELM method versus the ELM network training time.
The ELM data here correspond to those in Figures \ref{fg_26}(c,d)
for the two neural networks.

We can make the following observations from Figure \ref{fg_26_a}.
The ELM method outperforms the high-order FEM as the problem size
becomes larger (larger FEM mesh size under fixed element degree,
or larger element degree under fixed mesh size; larger set of ELM collocation points).
With smaller problem sizes (smaller FEM mesh size under fixed element degree,
or smaller element degree under fixed mesh size; smaller set of
ELM collocation points), the computational
performances of ELM and high-order FEM are comparable, with the high-order
FEM being slightly better.
These observations with the nonlinear Helmholtz equation here
are consistent with what has been observed for the Poisson equation (linear)
in the previous subsection.


\subsection{Viscous Burgers' Equation}
\label{sec:burger}

\begin{figure}
  \centerline{
    \includegraphics[height=2in]{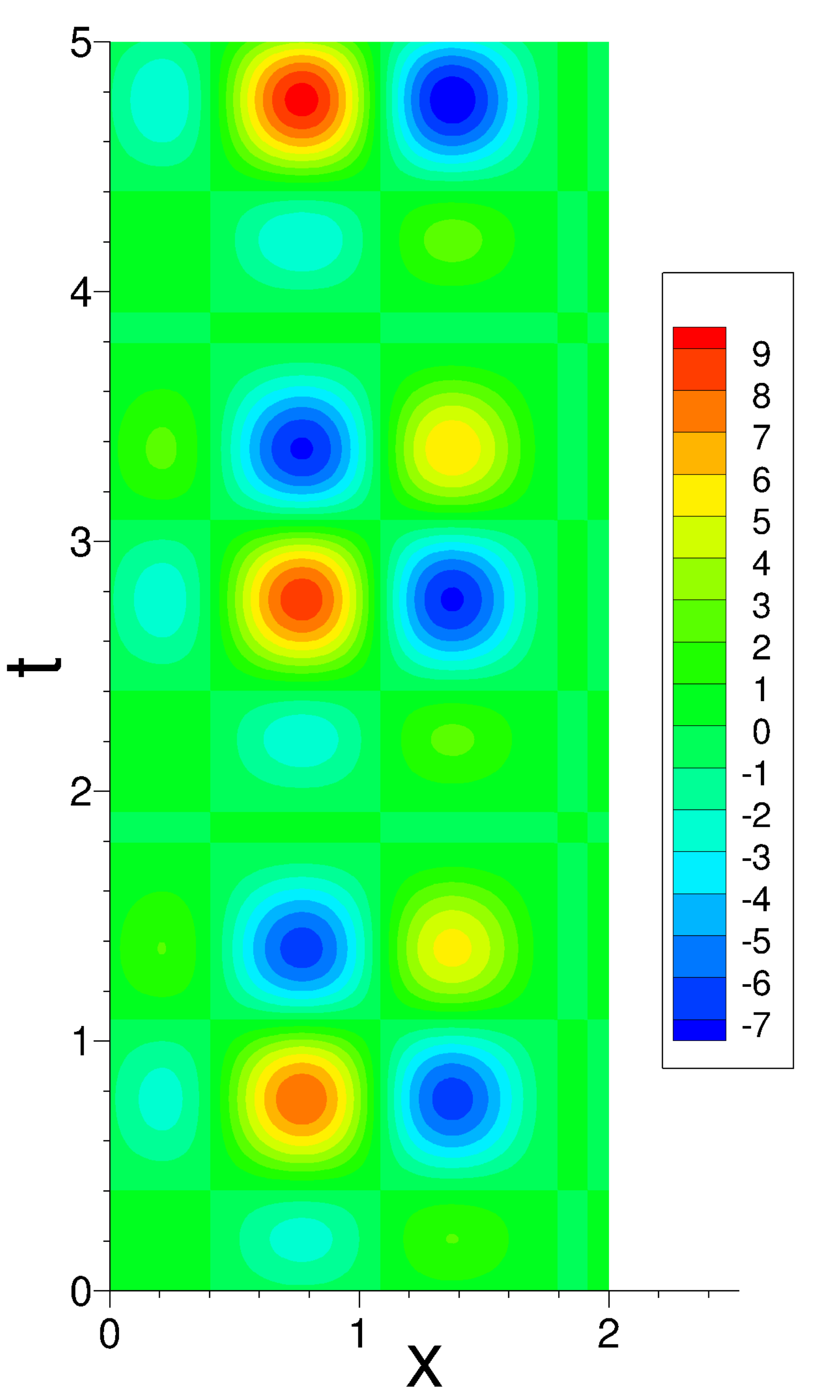}
  }
  \caption{Burgers' equation: distribution of the exact solution.}
  \label{fg_27}
\end{figure}

We next use another nonlinear example, the viscous Burgers' equation,
to test the method for computing the optimal $R_m$ ($\mbs R_m$)
and also compare
the computational performance of the ELM method with the classical and
high-order FEMs. A seed value of $100$ has bee employed for
the random number generators in Tensorflow and numpy for all the numerical
tests in this subsection.

We consider the spatial-temporal domain,
$\Omega=\{(x,t)\ |\ 0\leqslant x\leqslant 2,\ 0\leqslant t\leqslant 5 \}$,
and the following initial-boundary value problem with
the viscous Burgers' equation on $\Omega$,
\begin{subequations}\label{eq_19}
  \begin{align}
    &
    \frac{\partial u}{\partial t} + u\frac{\partial u}{\partial x}
    -\nu\frac{\partial^2u}{\partial x^2} = f(x,t), \label{eq_19a} \\
    &
    u(0,t) = g_1(t), \quad u(2,t) = g_2(t), \label{eq_19b} \\
    &
    u(x,0) = h(x), \label{eq_19c}
  \end{align}
\end{subequations}
where $\nu=0.01$, $u(x,t)$ is the field solution to be sought,
$f(x,t)$ is a prescribed source term, $g_1$ and $g_2$ are
the prescribed Dirichlet boundary condition, and
$h$ denotes the initial distribution.
We choose the source term and the boundary/initial distributions
such that the following manufactured function solves the
system~\eqref{eq_19},
\begin{multline}
  u(x,t) = \left(1+\frac{x}{20} \right)\left(1+\frac{t}{20} \right)
  \left[\frac32\cos\left(\pi x + \frac{7\pi}{20} \right)+ \right. \\
    \left.
     \frac{27}{20}\cos\left(2\pi x - \frac{3\pi}{5} \right)
    \right]
  \left[\frac32\cos\left(\pi t + \frac{7\pi}{20} \right)
    + \frac{27}{20}\cos\left(2\pi t - \frac{3\pi}{5} \right)
    \right].
\end{multline}
Figure \ref{fg_27} illustrates the distribution of this function
in the spatial-temporal plane.


We employ the ELM method (Single-Rm-ELM and Multi-Rm-ELM configurations), together with
the block time marching scheme (see Remark~\ref{rem_7}), to
solve the system~\eqref{eq_19}.
The input layer of the neural network contains two nodes,
representing $x$ and $t$. The linear output layer contains
a single node, representing the solution $u(x,t)$.
The network contains one or multiple hidden layers,
with the Gaussian activation function for all the hidden nodes.
The random hidden-layer coefficients are set according
to the Single-Rm-ELM or Multi-Rm-ELM configurations
from Section \ref{sec:method}.

We partition the spatial-temporal domain into a
number of windows in time (time blocks), and solve the problem on
the time blocks individually and successively~\cite{DongL2020}.
On each time block, the ELM network is trained by the NLLSQ-perturb
method~\cite{DongL2020}. After one time block is computed,
its field solution evaluated at the last time instant is used
as the initial condition for the computation of the next time block.
As discussed in Remark~\ref{rem_7}, we compute the
$R_{m0}$ and $\mbs R_{m0}$ by the differential evolution algorithm
only on the first time block, and we turn off the random perturbations
and the associated subiterations in the nonlinear least squares
 method (NLLSQ-perturb) during the $R_{m0}$ ($\mbs R_{m0}$) computation.

The crucial simulation parameters for this problem include
the time block size (or the number of time blocks),
the number of training parameters $M$ (width of last hidden layer in network),
the set of $Q=Q_1\times Q_1$ uniform collocation points on each time
block, and the maximum magnitude $R_m$ (or $\mbs R_m$) of
the random coefficients.
We employ $20$ uniform time blocks on
the domain $\Omega$, resulting in a time block size $0.25$.
Therefore, the $R_{m0}$ and $\mbs R_{m0}$ are computed on
the first time block, i.e.~by using the spatial-temporal
domain $\Omega_1=\{(x,t)\ |\ x\in[0,2],\ t\in[0,0.25]  \}$.

\begin{figure}
  \centerline{
    \includegraphics[width=2in]{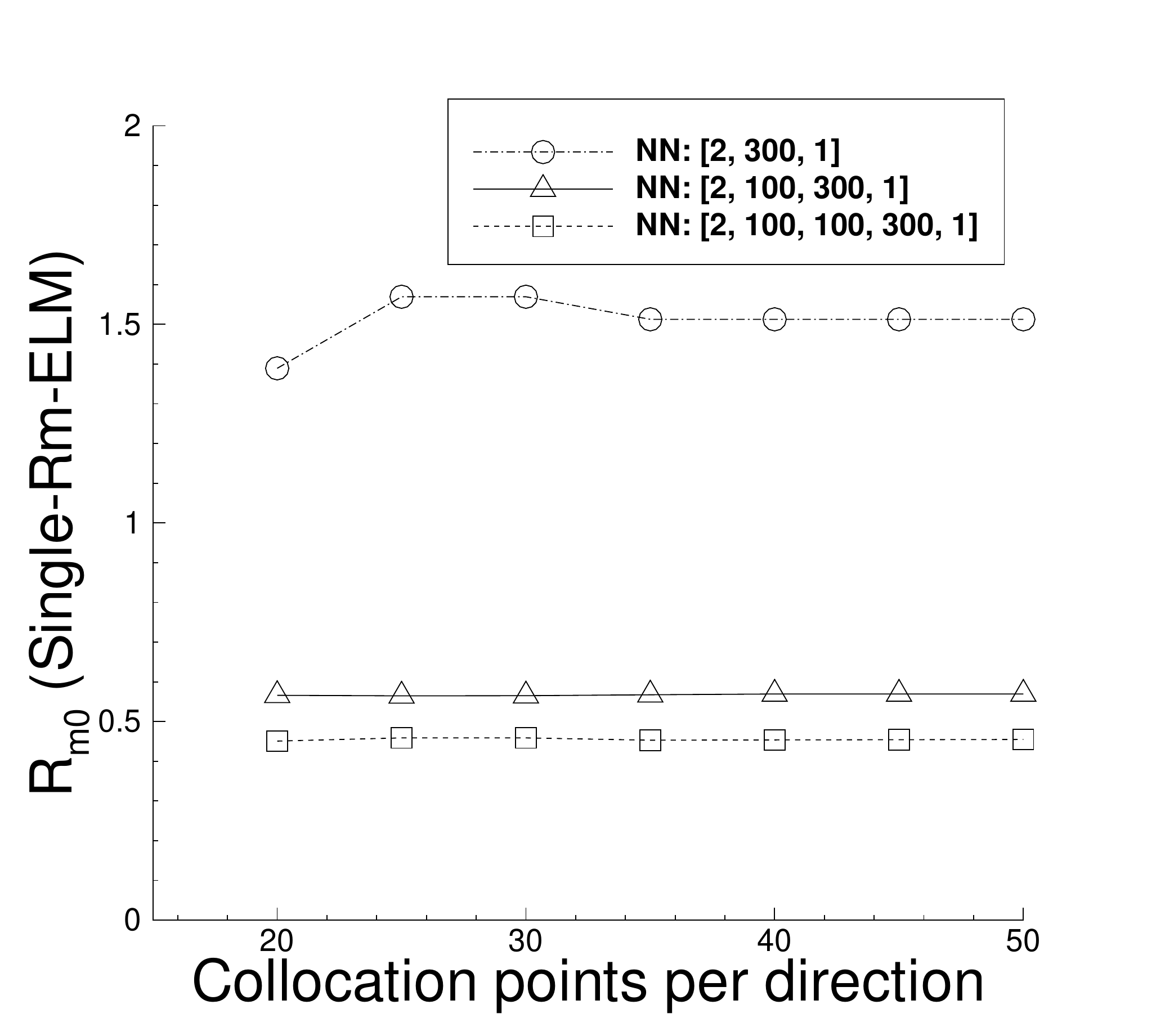}(a)
    \includegraphics[width=2in]{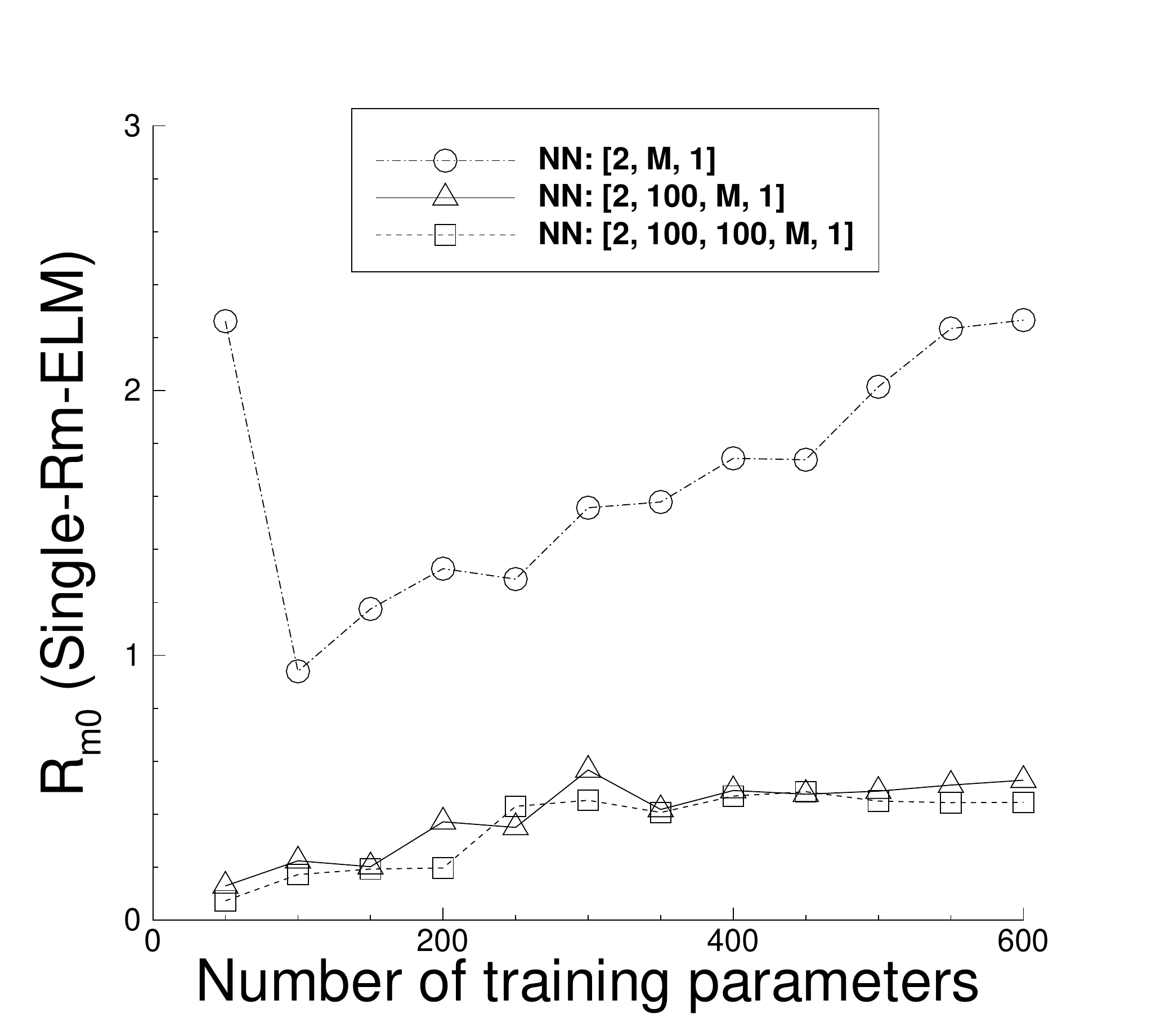}(b)
    \includegraphics[width=2in]{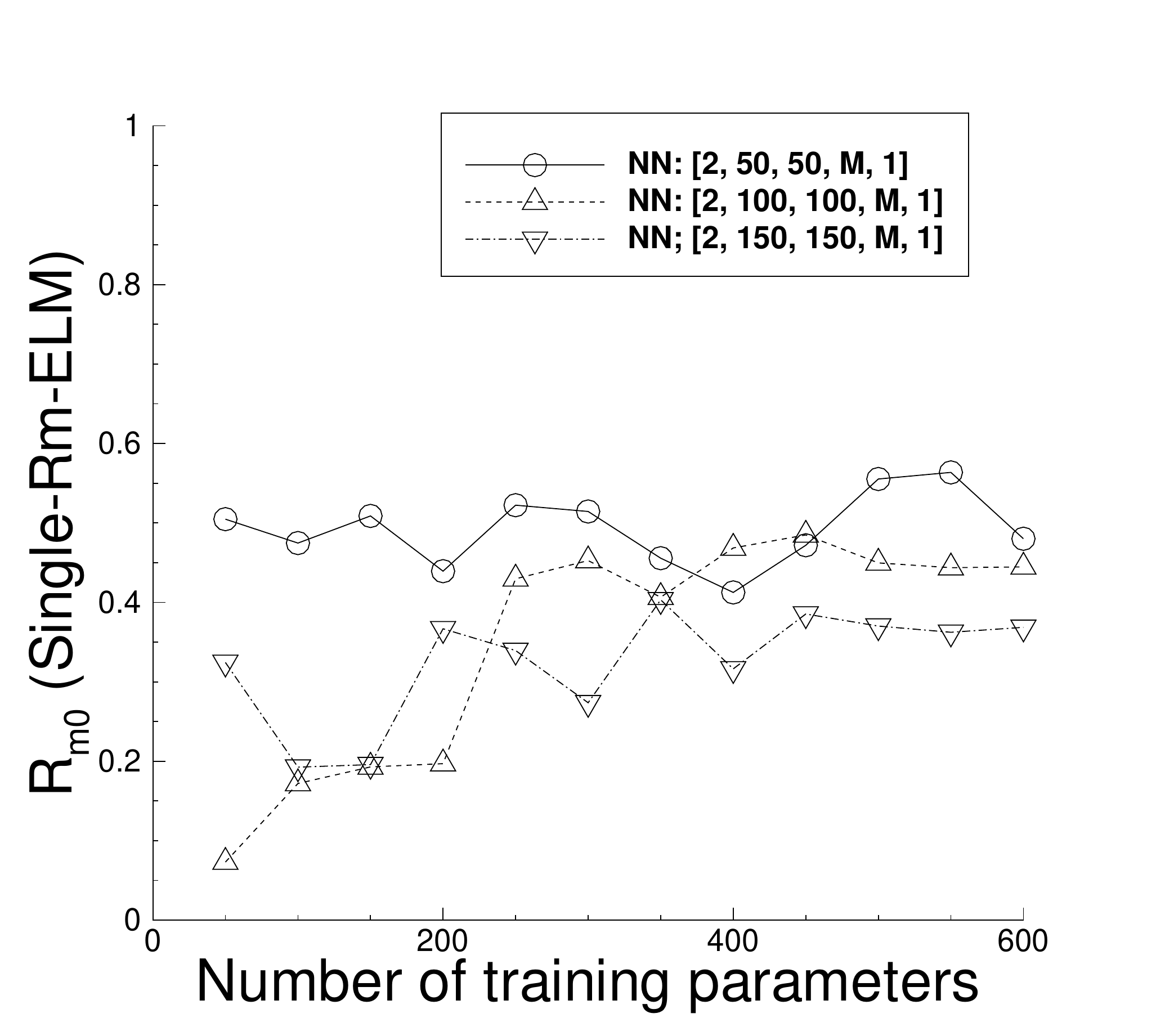}(c)
  }
  \caption{Burgers' equation (Single-Rm-ELM):
    The optimum $R_{m0}$ versus (a) the number of collocation points per
    direction and (b) the number of training parameters,
    with neural networks of different depth. (c) $R_{m0}$ versus the number of
    training parameters with neural networks having the same depth but different
    widths. Domain: $\Omega_1=[0,2]\times [0,0.25]$.
    $Q=31\times 31$ in (b,c), varied in (a). $M=300$ in (a), varied in (b,c).
    The network architectures are given in the legends.
  }
  \label{fg_28}
\end{figure}

We first consider the Single-Rm-ELM configuration,
and Figure \ref{fg_28} illustrates the optimum $R_{m0}$ obtained
with the differential evolution algorithm for the Burgers' equation.
In these tests the computational domain is the spatial-temporal domain of
the first time block $\Omega_1$, and we employ
neural networks with one to three hidden layers.
The number of training parameters is either fixed at $M=300$
or varied systematically.
The set of $Q=Q_1\times Q_1$ uniform collocation points is either fixed at
$Q_1=31$ or varied systematically.

Figure \ref{fg_28}(a) depicts the optimum $R_{m0}$ as a function of
the number of collocation points per direction $Q_1$,
for three neural networks with the same $M=300$ but different depth.
Figures \ref{fg_28}(b) and (c) both depict
the $R_{m0}$ as a function of the number of training parameters $M$,
but for neural networks with different configurations.
The plot (b) is for three neural networks with different depths,
and the plot (c) is for three neural networks  with the same depth
but different widths for the preceding hidden layers.
These results are obtained with a population size of $6$,
the $R_m$ bounds $[0.01, 3]$, and a relative tolerance $0.1$
in the differential evolution algorithm.
The settings and simulation parameters for each plot are
provided in the figure caption.

The $R_{m0}$ characteristics shown by Figure \ref{fg_28} are
consistent with those observed from previous subsections.
For instance, $R_{m0}$ is generally not sensitive to
the number of collocation points
in the domain, especially with more than one hidden layers in
the neural network.
With a single hidden layer in the neural network,
$R_{m0}$ has a notable dependence on the number of training
parameters $M$, and tends to increase with increasing $M$
(Figure \ref{fg_28}(b)).
With two or more hidden layers in the neural network,
$R_{m0}$ only weakly depends on $M$.
$R_{m0}$ tends to decrease with increasing depths in the neural network
or increasing widths of the preceding hidden layers.
When the number of hidden layers increases from one to two,
the reduction in $R_{m0}$ is quite pronounced.
Beyond two hidden layers, on the other hand, there is only a slight
reduction in $R_{m0}$ as the number of hidden layers increases.

\begin{figure}
  \centerline{
    \includegraphics[height=2.0in]{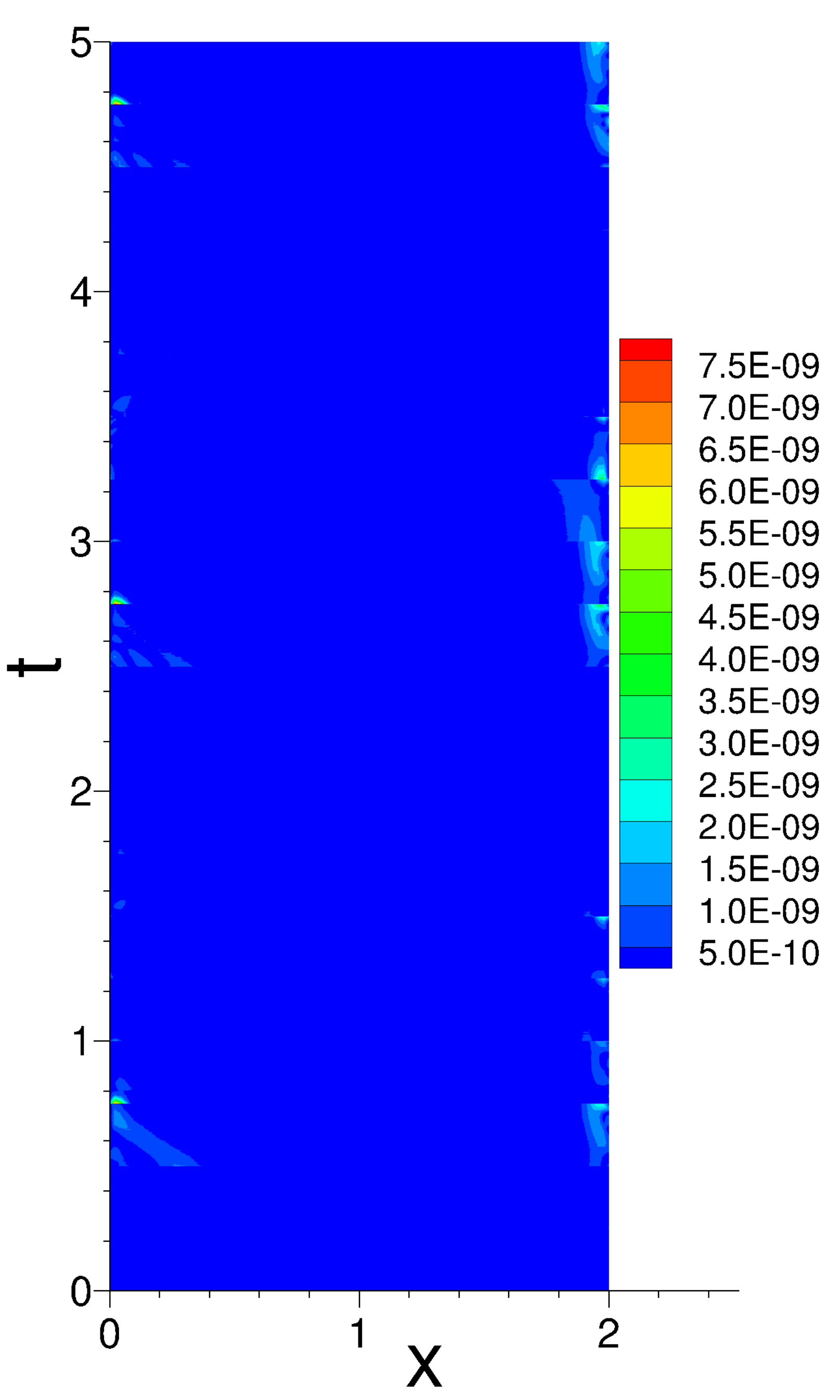}(a)
    \includegraphics[height=2.0in]{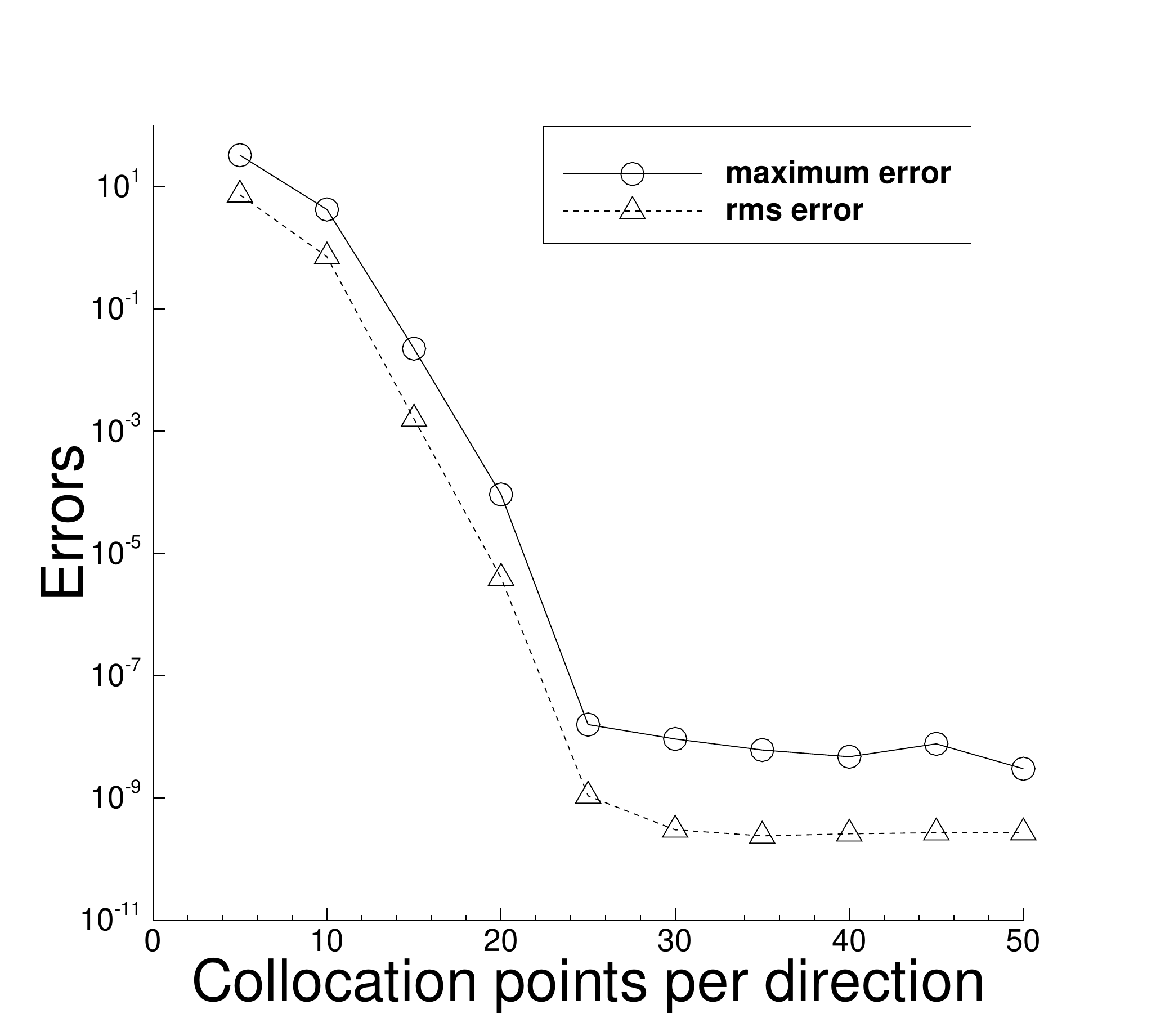}(b)
    \includegraphics[height=2.0in]{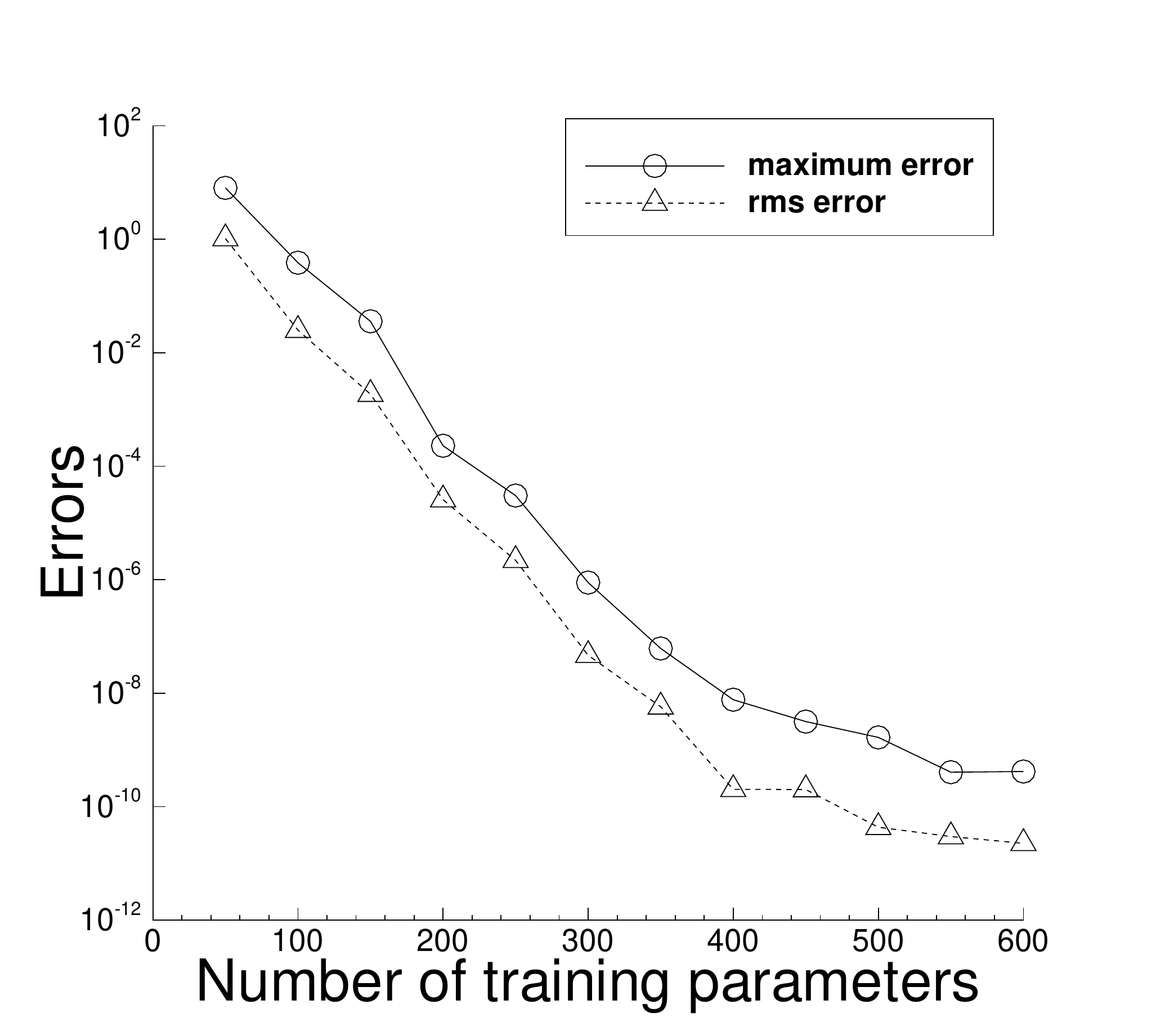}(c)
  }
  \caption{Burgers' equation (Single-Rm-ELM):
    (a) Absolute-error distribution of the Single-Rm-ELM solution on $\Omega$.
    The maximum/rms errors on $\Omega$ versus (b) the number
    of collocation points per direction on each time block,
    and (c) the number of training parameters.
    Domain: $\Omega$, with $20$ time blocks in block time marching.
    Network architecture: [2, $M$, 1].
    $Q=31\times 31$ in (a,c), varied in (b). $M=400$ in (a,b),
    varied in (c). $R_m=2.0$ in (a,b,c).
  }
  \label{fg_29}
\end{figure}

Figure \ref{fg_29}  illustrates the solution accuracy
of the Single-Rm-ELM configuration.
The computational domain here is the entire spatial-temporal domain $\Omega$
($0\leqslant t\leqslant 5$), and we employ $20$ time blocks in
the block time marching scheme, as mentioned before.
The neural network has an
architecture $[2, M, 1]$, where $M$ is either fixed
at $M=400$ or varied systematically.
A set of $Q=Q_1\times Q_1$ uniform collocation points is employed on each time block,
with $Q_1$ either fixed at $31$ or varied systematically.
We employ a fixed $R_m=2.0$ in all these tests,
which is close to the $R_{m0}$ obtained
from the differential evolution algorithm corresponding
to $M=500$ and $Q=31\times 31$.
Figure \ref{fg_29}(a) shows the distribution of the absolute error
of the Single-Rm-ELM solution,
signifying a high accuracy with the maximum error 
on the order $10^{-9}$ in the entire spatial-temporal domain.
Figures \ref{fg_29}(b) and (c) depict the maximum/rms errors in
the overall domain as a function of $Q_1$ and $M$, respectively,
showing the exponential convergence in the numerical errors (before saturation).
The simulation parameters for each plot are provided in
the figure caption.

\begin{figure}
  \centerline{
    \includegraphics[width=2in]{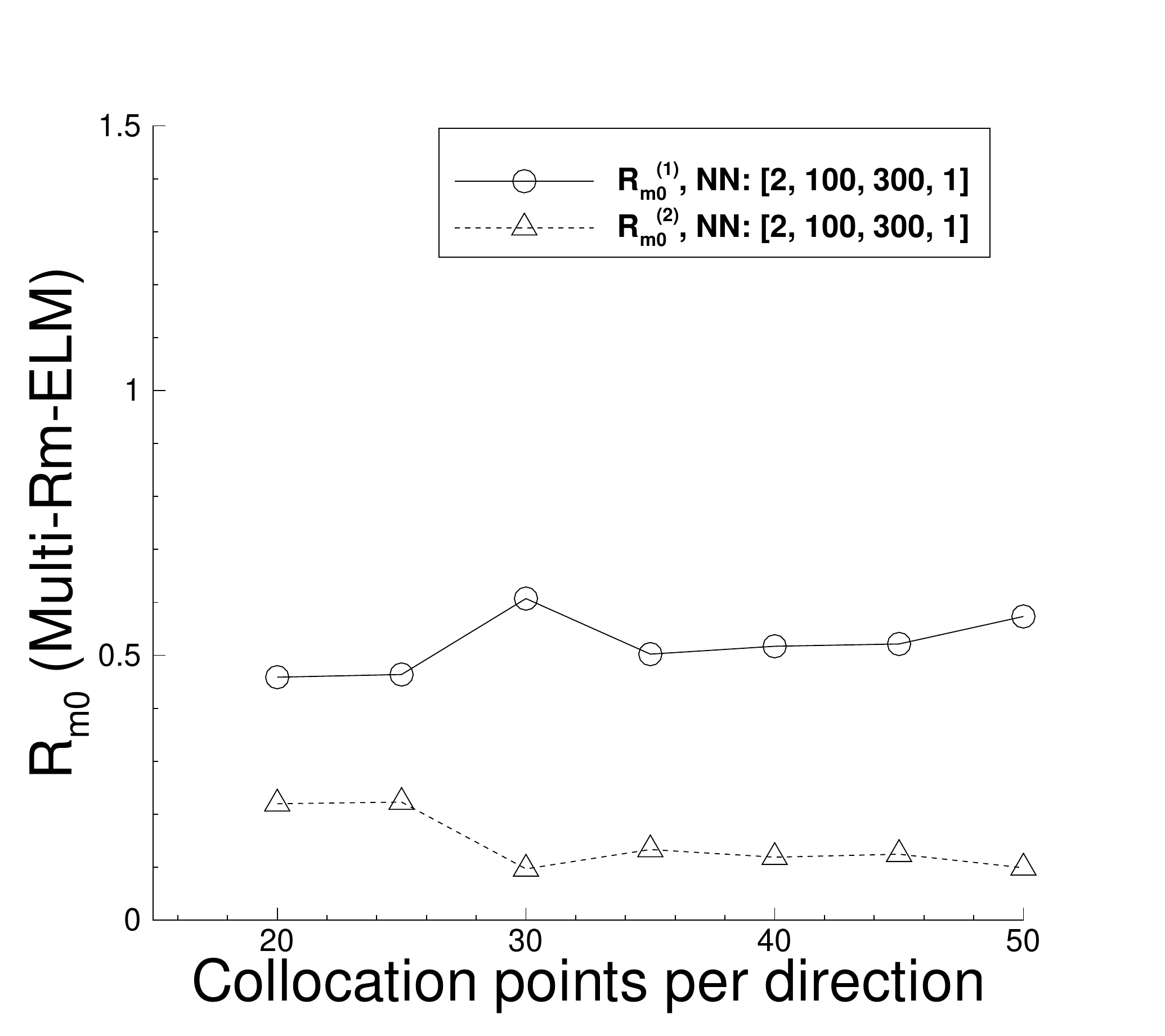}(a)
    \includegraphics[width=2in]{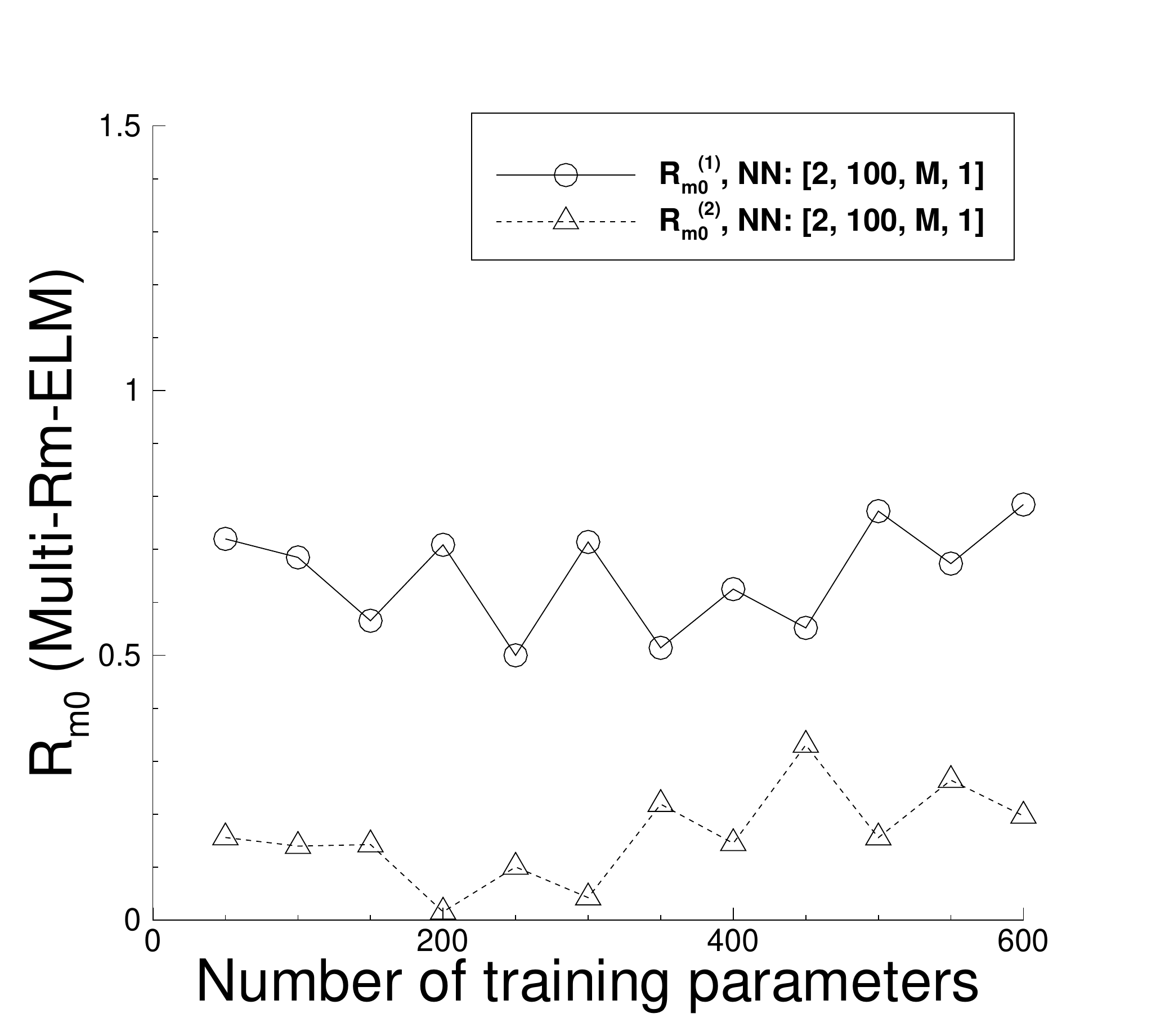}(b)
  }
  \centerline{
    \includegraphics[width=2in]{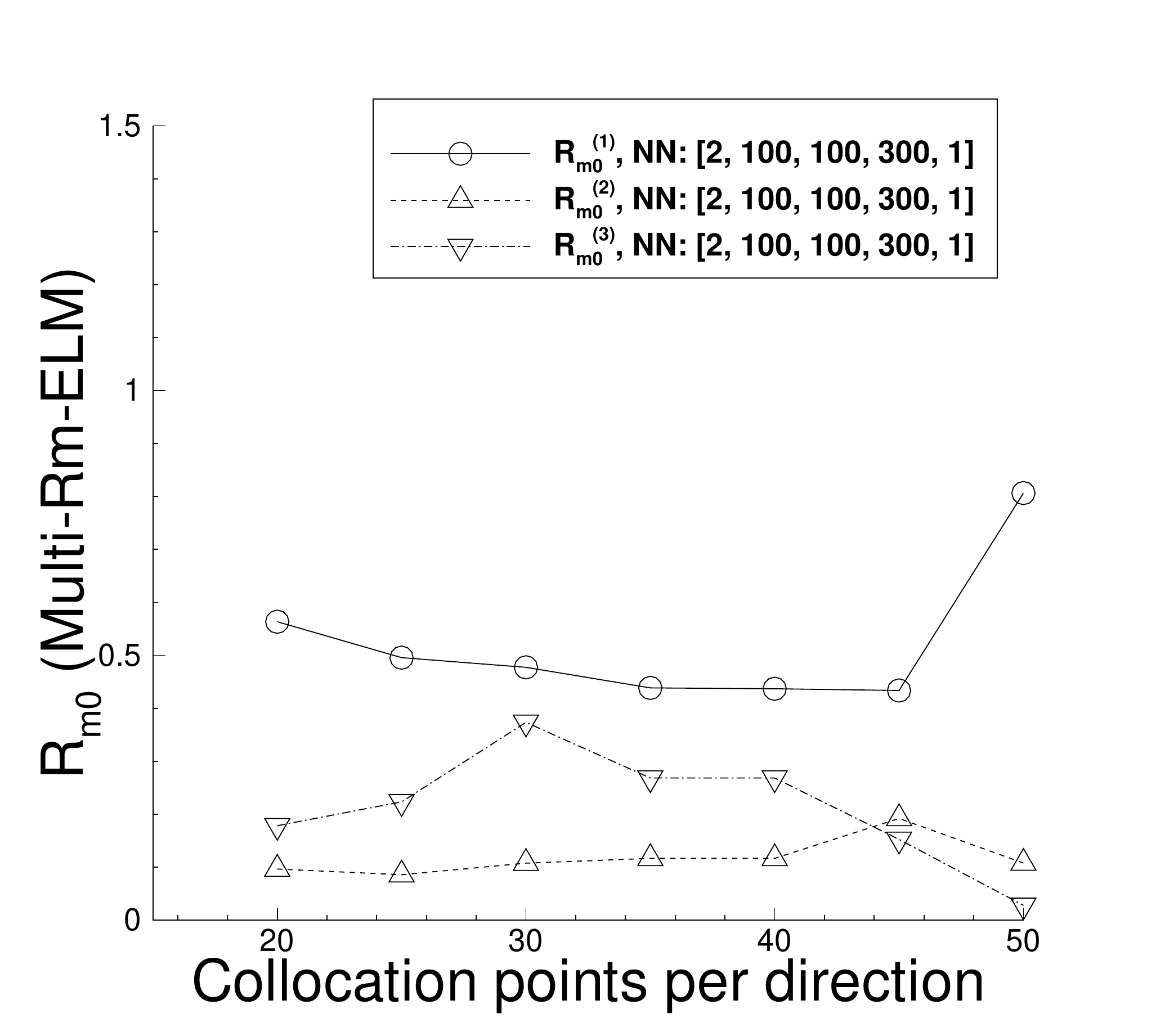}(c)
    \includegraphics[width=2in]{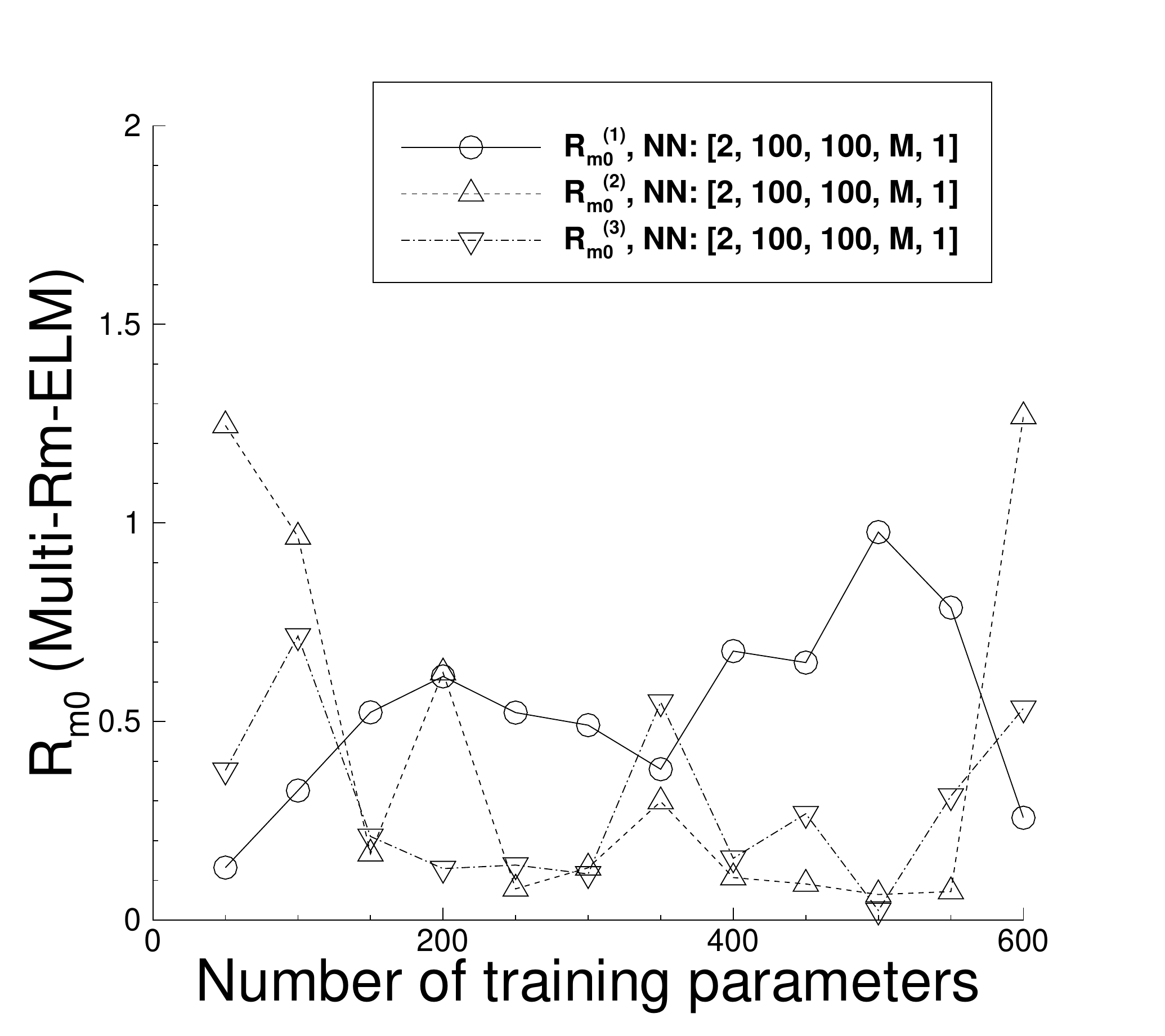}(d)
  }
  \caption{Burgers' equation (Multi-Rm-ELM):
    The optimum $\mbs R_{m0}$ versus the number of
    collocation points per direction (a,c) and the number of training
    parameters (b,d), with neural networks having two (a,b) and
    three (c,d) hidden layers. The network architectures are
    given in the legends.
    Computational domain: $\Omega_1$ ($t\in [0,0.25]$).
    $Q=31\times 31$ in (b,d), varied in (a,c).
    $M=300$ in (a,c), varied in (b,d).
  }
  \label{fg_30}
\end{figure}

Let us next look into the Multi-Rm-ELM configuration  for solving
the Burgers' equation.
The characteristics of the optimum $\mbs R_{m0}$  are illustrated
in Figure \ref{fg_30}.
The computational domain in these tests is the spatial-temporal domain
$\Omega_1$ ($t\in[0, 0.25]$).
We have considered two types of neural networks,
whose architectures are given by
$[2, 100, M, 1]$ and $[2, 100, 100, M, 1]$,
where $M$ is fixed at $300$ or varied systematically.
A uniform set of $Q=Q_1\times Q_1$ collocation points is
employed in the domain, where $Q_1$ is fixed at $31$
or varied systematically.
We employ a population size of $6$, the bounds $[0.01, 3]$
for all the $\mbs R_m$ components, and a relative tolerance
$0.1$ in the differential evolution algorithm for these tests.

Figures \ref{fg_30}(a) and (b) depict the optimum
$\mbs R_{m0}=(R_{m0}^{(1)},R_{m0}^{(2)})$ as a function of $Q_1$ and
$M$, respectively, for the neural networks with two hidden layers.
Figures \ref{fg_30}(c) and (d) depict the corresponding components of the optimum
$\mbs R_{m0}=(R_{m0}^{(1)},R_{m0}^{(2)},R_{m0}^{(3)})$ as a function of
$Q_1$ and $M$ for the neural networks with three
hidden layers.
Overall, the relations of $\mbs R_{m0}$ versus $Q_1$ and $M$
appear to be quite irregular. The relation between $\mbs R_{m0}$ and $Q_1$
seems less irregular, and the dependence appears generally not quite strong. 
On the other hand, the change in
$M$ appears to affect the $\mbs R_{m0}$ components
more strongly, especially with increasing number of hidden layers in
the neural network (see Figure \ref{fg_30}(d)).
These characteristics are similar to those observed in previous subsections
with Multi-Rm-ELM for linear and nonlinear problems.

\begin{figure}
  \centerline{
    \includegraphics[height=2in]{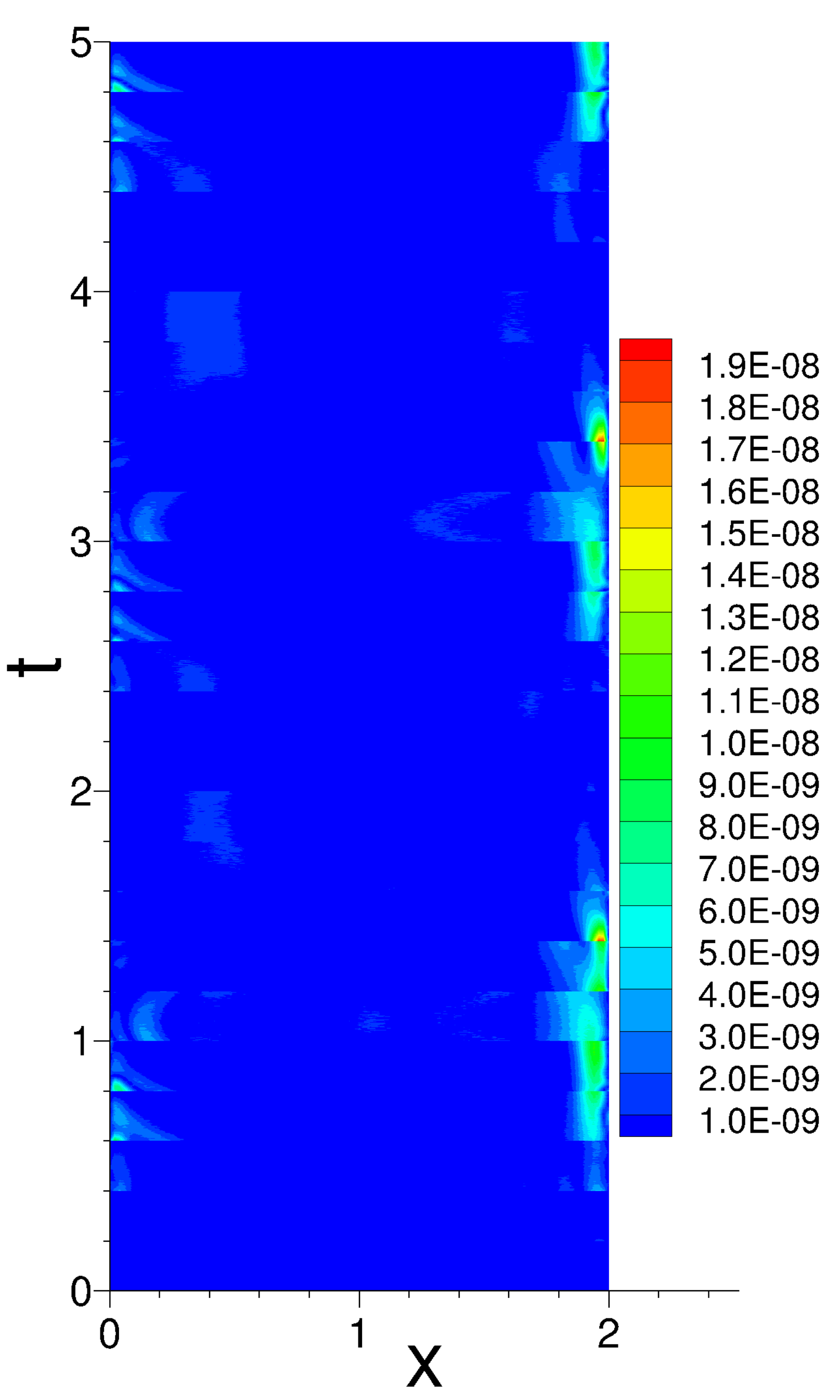}(a)
    \includegraphics[height=2in]{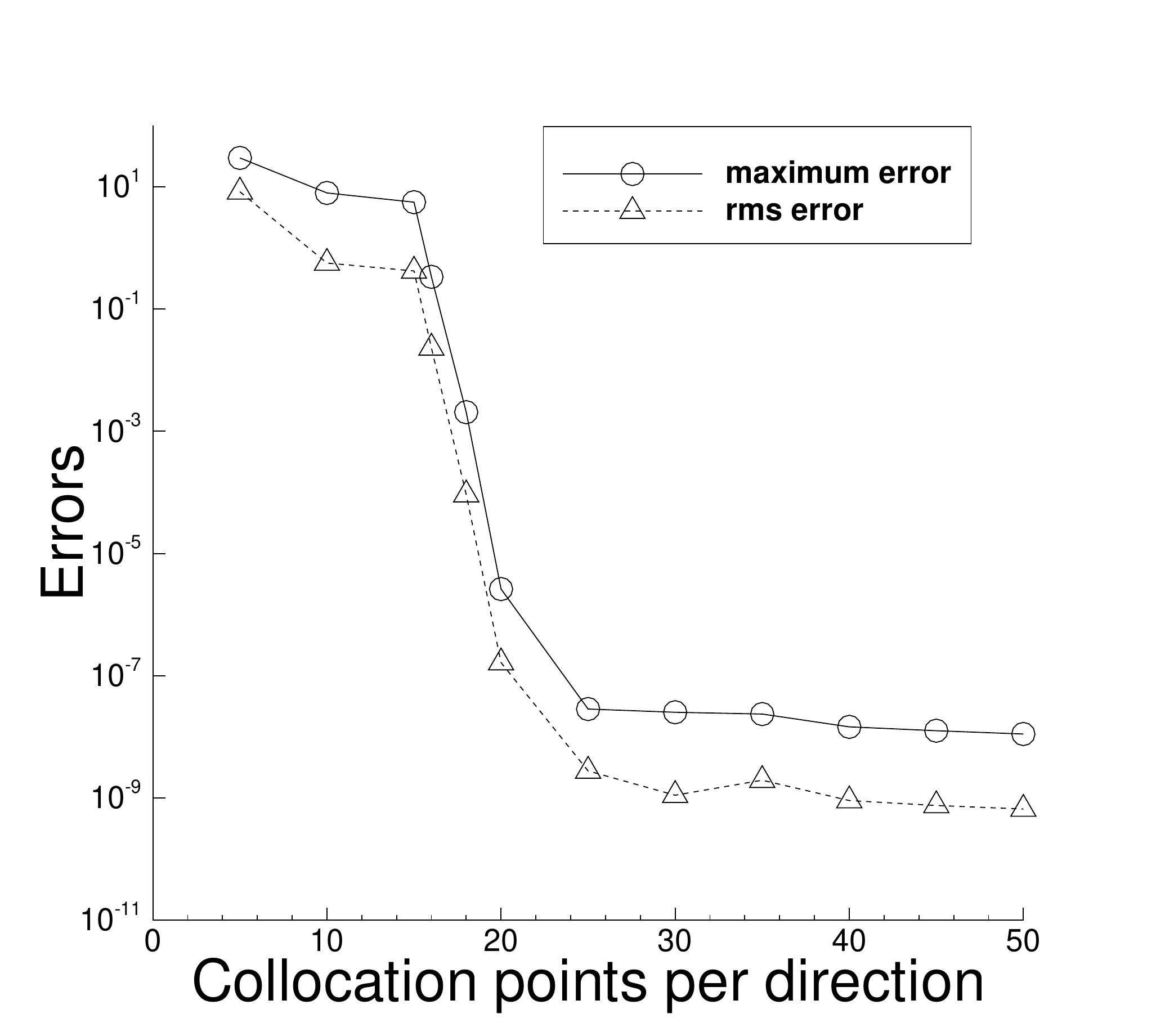}(b)
    \includegraphics[height=2in]{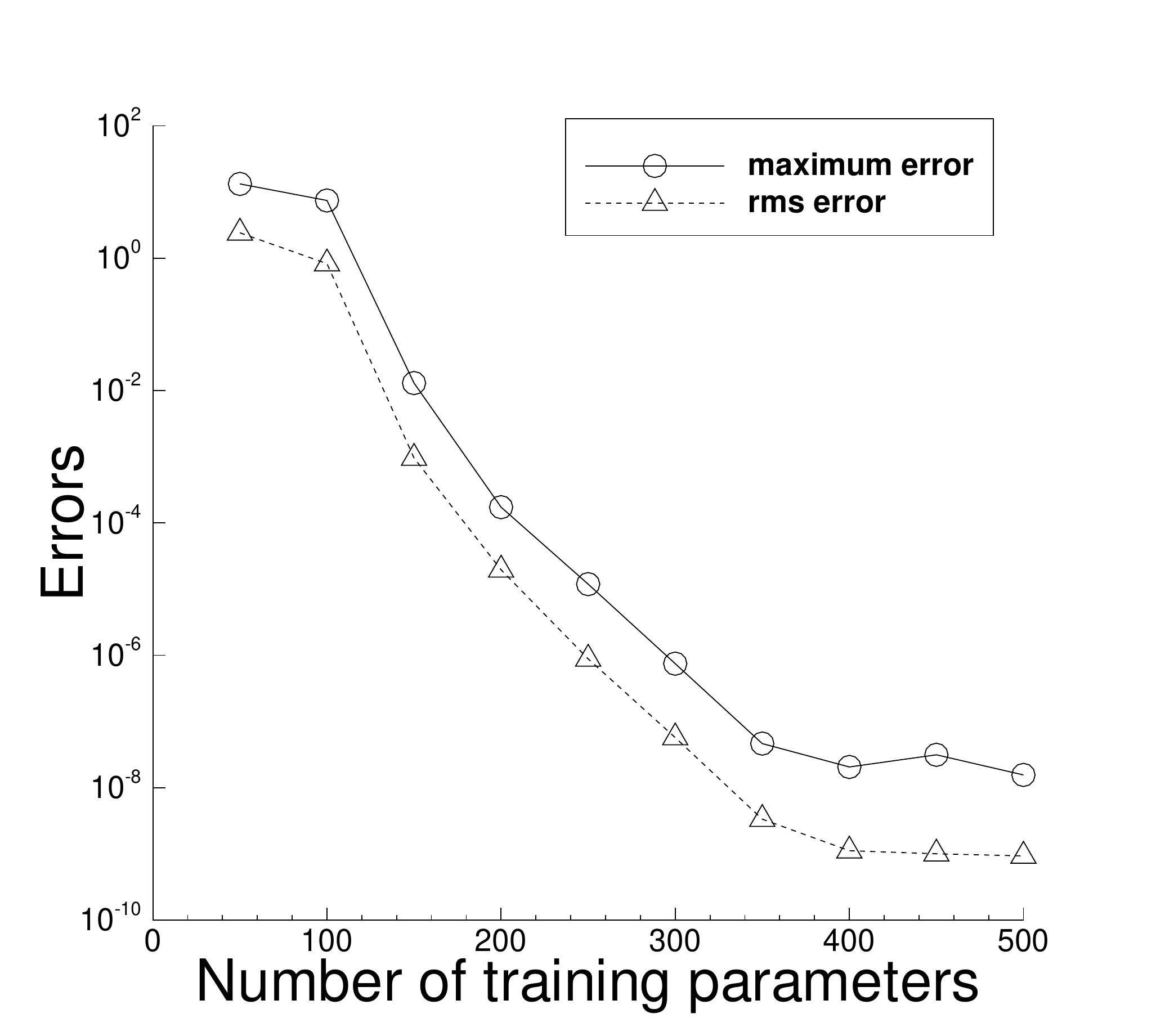}(c)
  }
  \caption{Burgers' equation (Multi-Rm-ELM):
    (a) Absolute-error distribution of the Multi-Rm-ELM solution
    in the spatial-temporal domain.
    The maximum/rms errors on $\Omega$
    versus (b) the number of collocation points per direction
    in each time block, and (c) the number of training parameters.
    Computational domain: $\Omega$, with $25$ time blocks in block time marching.
    Network architecture: [2, 100, $M$, 1].
    $Q=31\times 31$ for each time block in (a,c), varied in (b).
    $M=400$ in (a,b), varied in (c).
    $\mbs R_m=(0.625,0.145)$ in (a,b,c).
  }
  \label{fg_31}
\end{figure}

The solution accuracy of the Multi-Rm-ELM
configuration is illustrated in Figure \ref{fg_31}.
The computational domain in this set of tests is the spatial-temporal domain
$\Omega$ ($t\in[0,5]$), which is partitioned into
$25$ time blocks (block size $0.2$ in time)
in the block time marching scheme with Multi-Rm-ELM. 
The neural network architecture is given by $[2, 100, M, 1]$,
with $M$ fixed at $400$ or varied systematically.
A uniform set of $Q=Q_1\times Q_1$ collocation points is employed
on each time block, where $Q_1$ is fixed at $31$ or varied systematically.
We employ a fixed $\mbs R_m=(0.645, 0.145)$ in these tests,
which is close to the $\mbs R_{m0}$ obtained with
the differential evolution algorithm corresponding to $M=400$
and $Q=31\times 31$.
Figure \ref{fg_31}(a) shows the distribution of the absolute error
of the Multi-Rm-ELM solution on the entire spatial-temporal domain.
The results signify a high accuracy, with a maximum error on
the order $10^{-8}$ in the overall domain.
Figures \ref{fg_31}(b) and (c) depict the maximum error and the rms error
in the overall domain of the Multi-Rm-ELM solution
as a function of $Q_1$ and $M$.
The errors generally appear to decrease exponentially or nearly exponentially
with increasing number of collocation points or training parameters.
But the relation is not that regular. For example,
in Figure \ref{fg_31}(b) as the collocation points
increase from $5\times 5$ to $15\times 15$, there is little decrease
in the errors. However, once beyond that point, there is
a sharp exponential decrease in the errors (before saturation).

\begin{figure}
  \centerline{
    \includegraphics[width=2in]{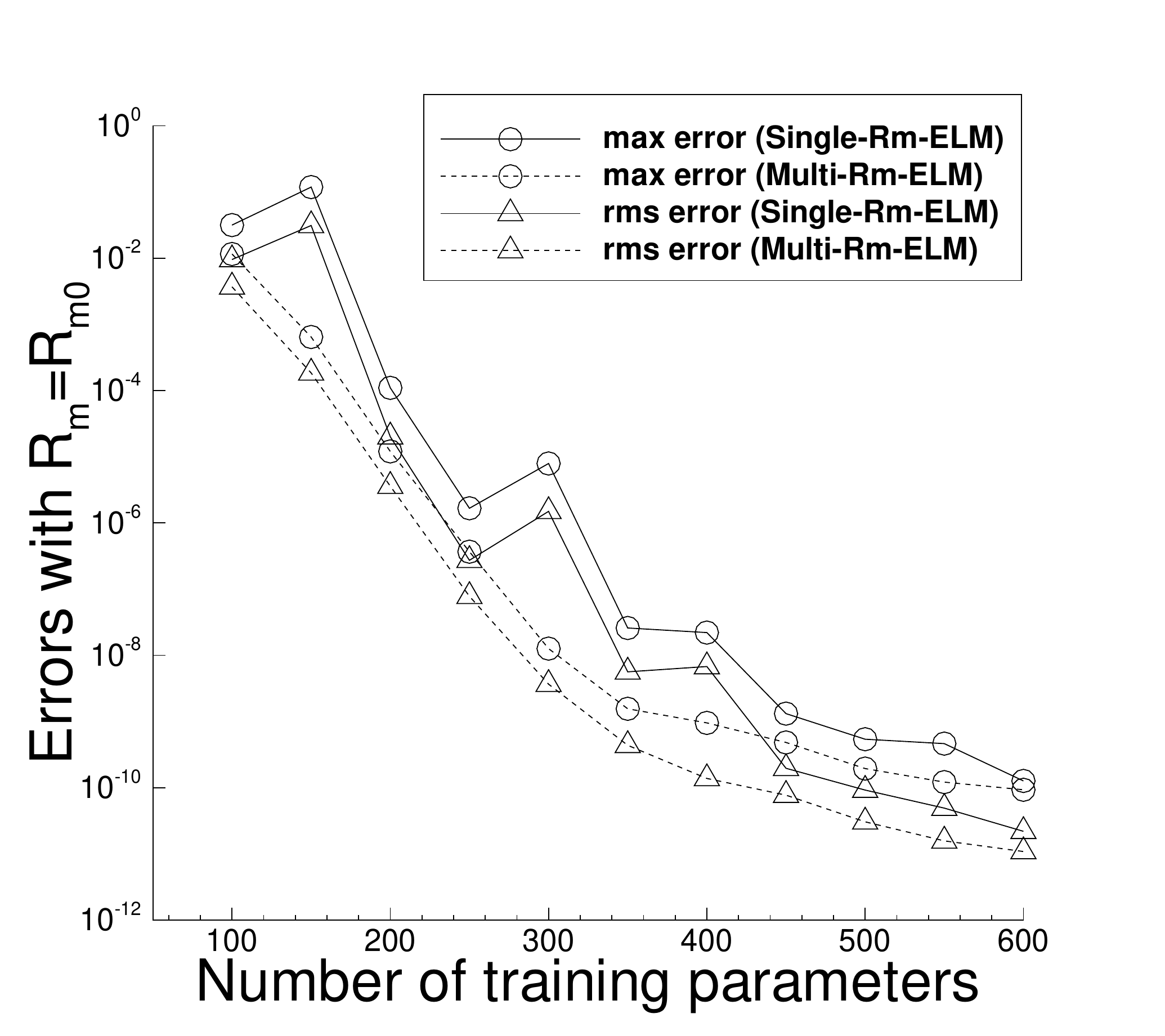}(a)
    \includegraphics[width=2in]{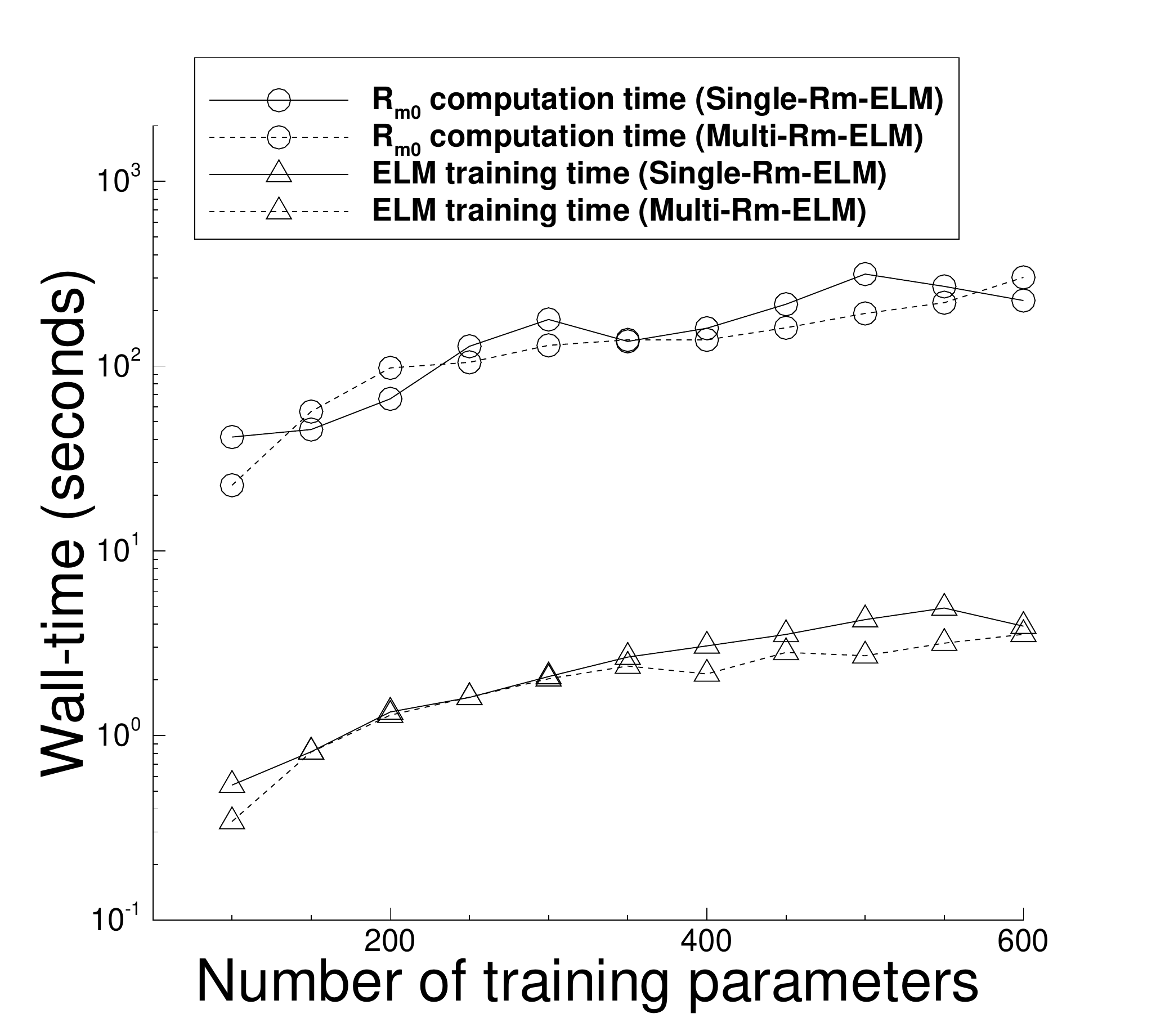}(b)
  }
  \caption{Burgers' equation:
    (a) The maximum/rms errors 
    corresponding to $R_m=R_{m0}$ in Single-Rm-ELM and $\mbs R_m=\mbs R_{m0}$
    in Multi-Rm-ELM, versus the number of training parameters ($M$).
    (b) The  $R_{m0}$ (or $\mbs R_{m0}$) computation time and the ELM network training
    time in Single-Rm-ELM and Multi-Rm-ELM, versus
    the number of training parameters.
    Computational domain: $\Omega_1$ ($t\in[0,0.25]$).
    Network architecture: [2, 100, $M$, 1].
    $Q=31\times 31$ in (a,b).
  }
  \label{fg_32}
\end{figure}

A comparison between Single-Rm-ELM and Multi-Rm-ELM with
regard to their accuracies and the cost for computing $R_{m0}$/$\mbs R_{m0}$
is provided in Figure \ref{fg_32}.
The computational domain is the spatial-temporal domain $\Omega_1$ ($t\in[0,0.25]$).
The neural network has an architecture $[2, 100, M, 1]$ in this group of
tests, where $M$ is varied systematically.
The random hidden-layer coefficients are set based on
the Single-Rm-ELM or Multi-Rm-ELM configurations.
A fixed set of $Q=31\times 31$ uniform collocation points is employed.
Figure \ref{fg_32}(a) shows the maximum/rms errors in the domain
versus the number of training parameters $M$,
corresponding to $R_m=R_{m0}$ in Single-Rm-ELM and $\mbs R_m=\mbs R_{m0}$
in Multi-Rm-ELM.
Here when computing $R_{m0}$ and $\mbs R_{m0}$ we have employed
a population size of $6$, the bounds $[0.01, 3]$, and a
relative tolerance $0.1$ in the differential evolution
algorithm for both Single-Rm-ELM and Multi-Rm-ELM.
One can observe that Multi-Rm-ELM leads to
consistently more accurate results than Single-Rm-ELM.
Figure \ref{fg_32}(b) shows the corresponding cost
for computing the $R_{m0}$/$\mbs R_{m0}$ in Single-Rm-ELM and Multi-Rm-ELM,
as well as the ELM network training time with given $R_m$/$\mbs R_m$,
as a function of $M$.
The $R_{m0}$ computation cost in Single-Rm-ELM and
the $\mbs R_{m0}$ computation cost in Multi-Rm-ELM
appear comparable for the Burgers' equation.
The cost for computing the $R_{m0}$/$\mbs R_{m0}$ is markedly higher than
the ELM network training time for a given $R_m$ or $\mbs R_m$.

\begin{figure}
  \centerline{
    \includegraphics[width=2in]{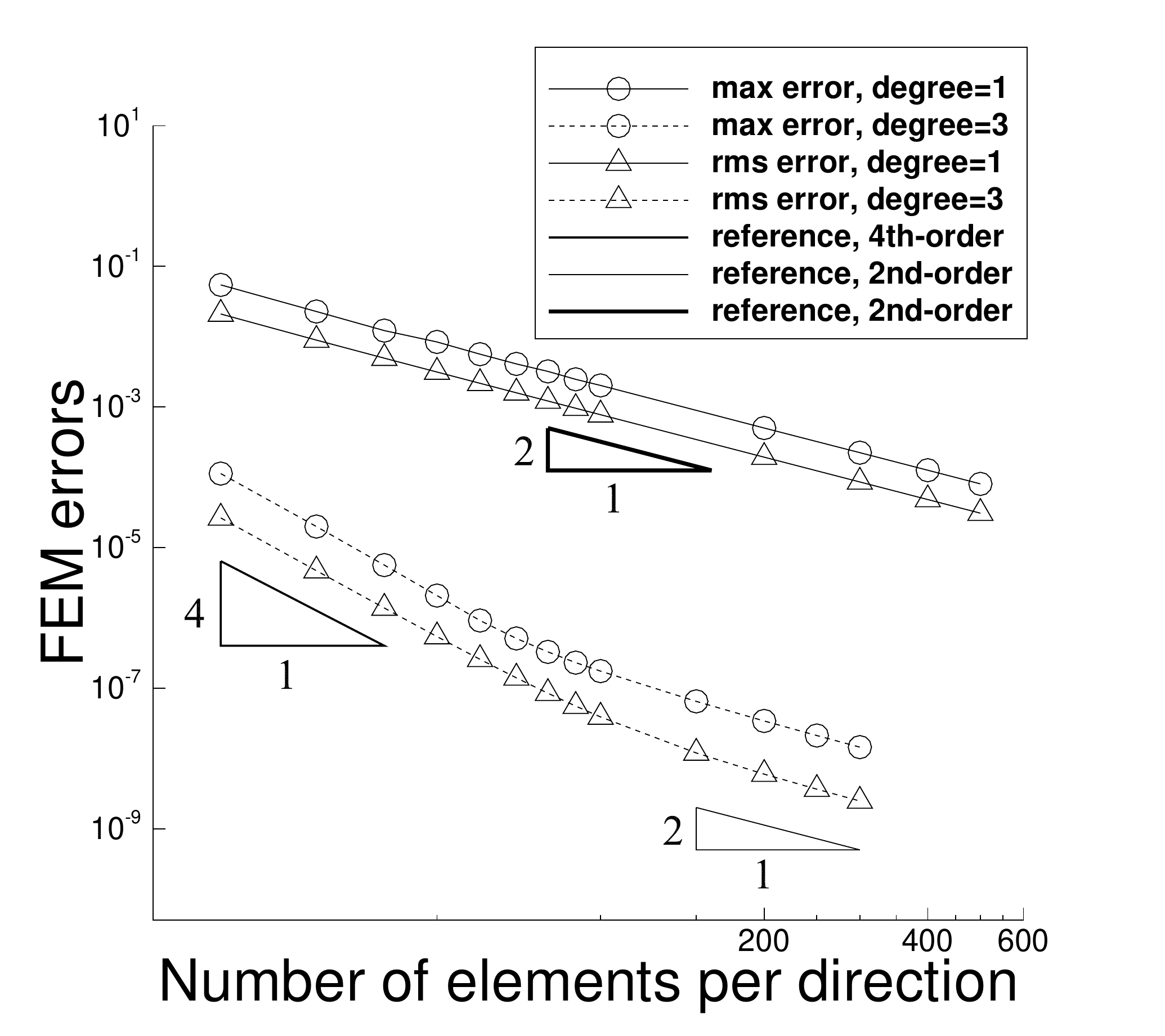}(a)
    \includegraphics[width=2in]{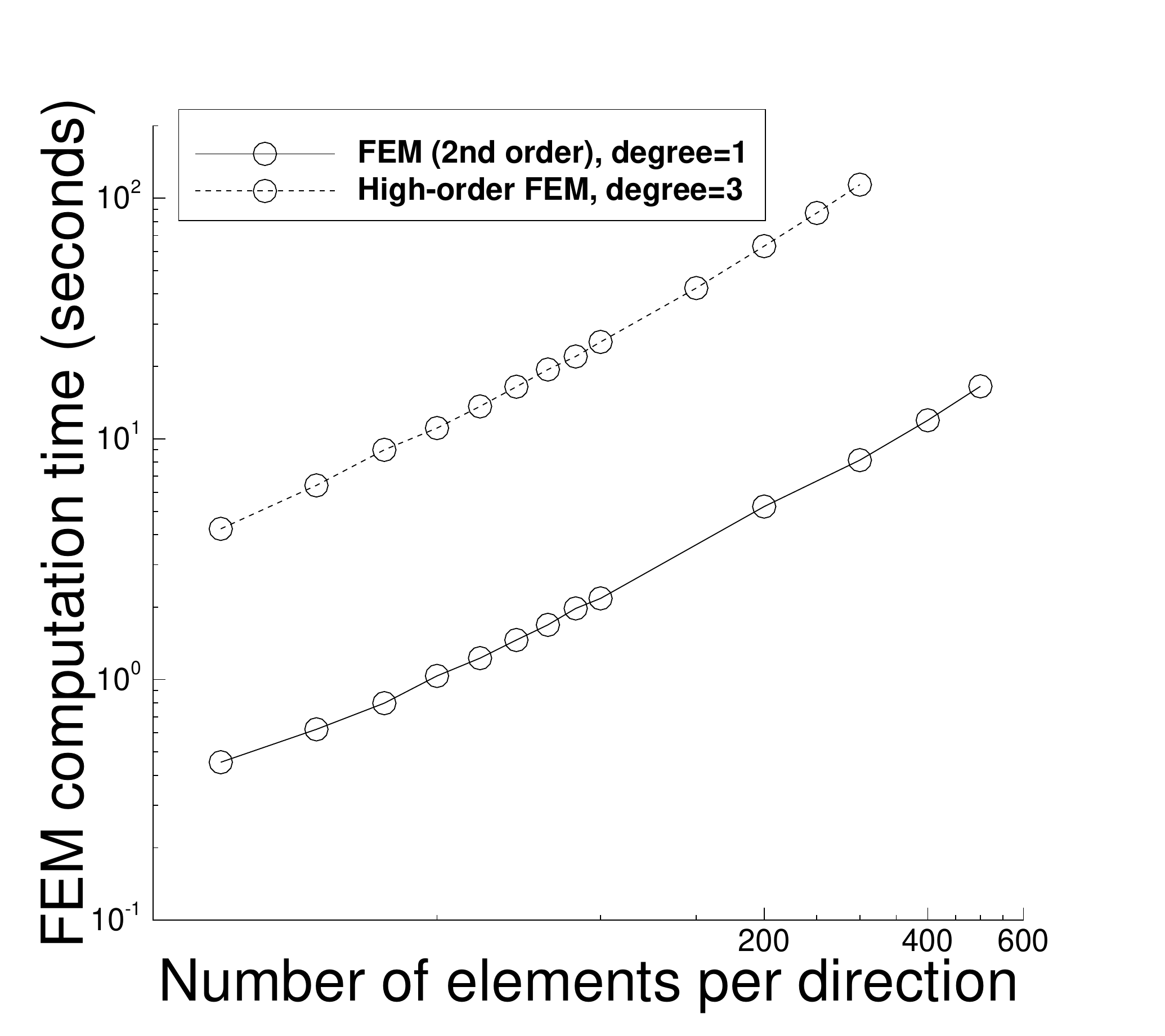}(b)
  }
  \centerline{
    \includegraphics[width=2in]{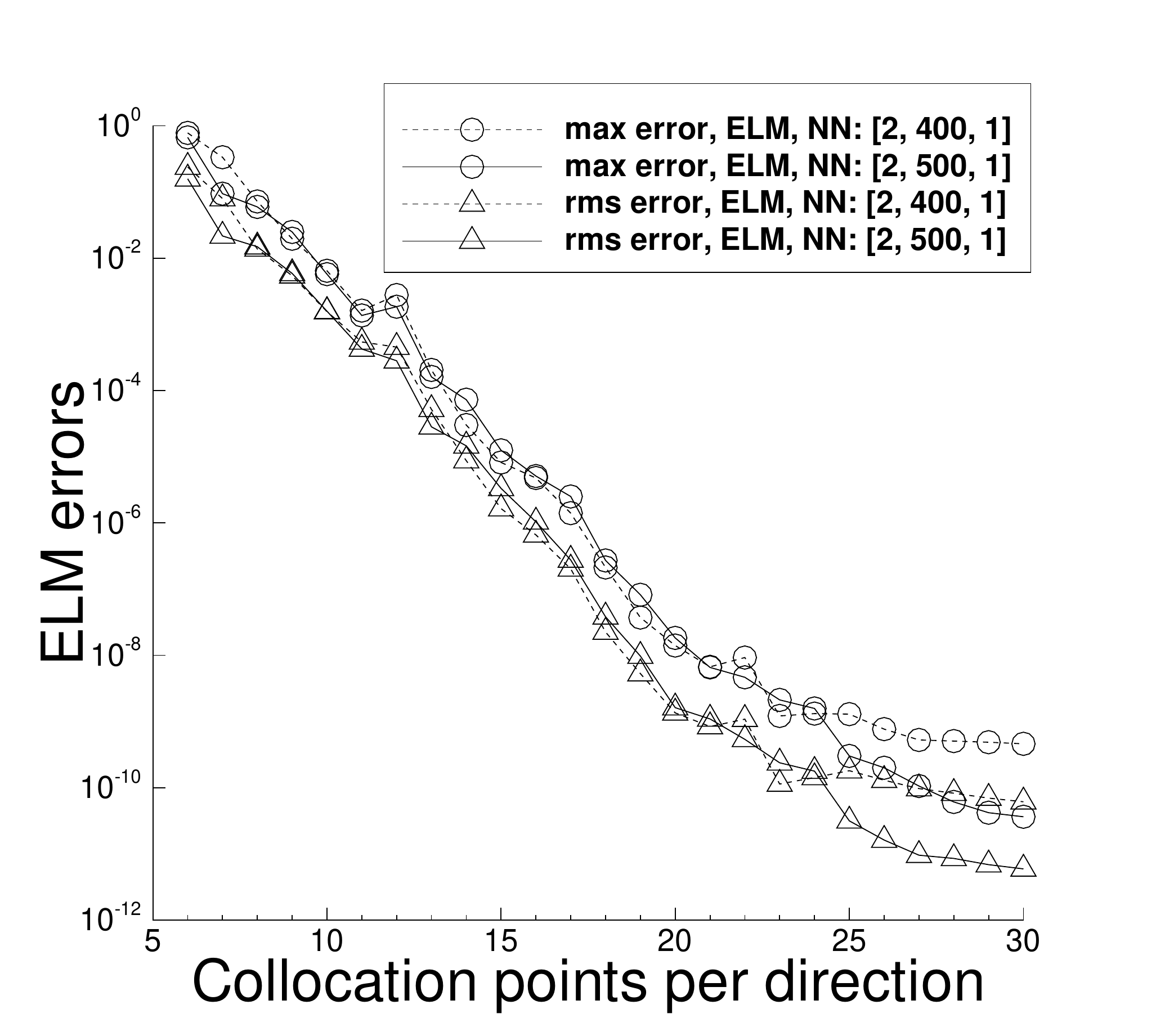}(c)
    \includegraphics[width=2in]{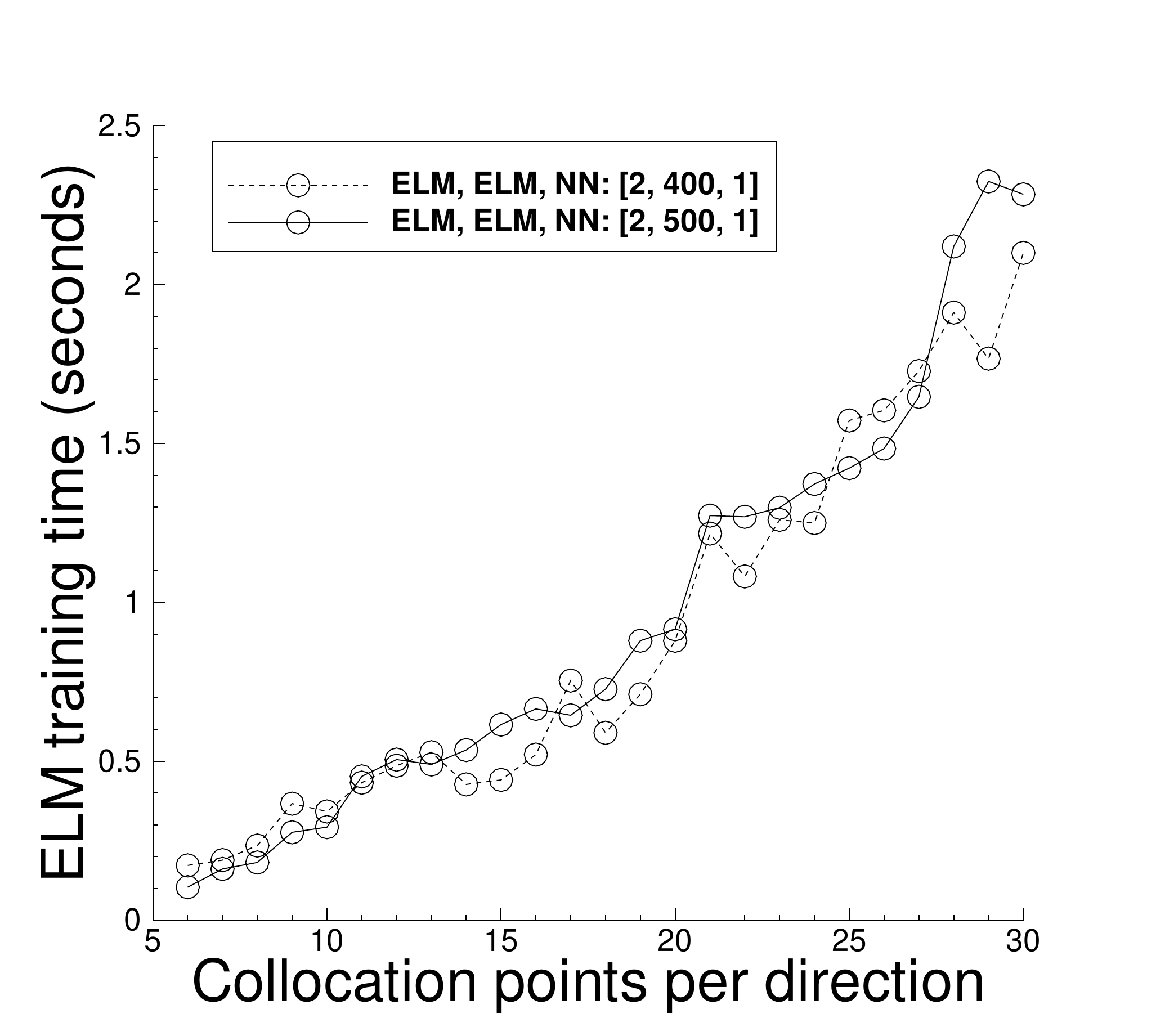}(d)
  }
  \caption{Burgers' equation:
    The numerical errors (a) and the computation time (b) of the classical
    FEM (2nd-order, degree=1) and the high-order FEM with Lagrange elements of
    degree $3$, versus the number of elements in each direction.
    In the FEM tests, as the number of elements increases,
    the time step size ($\Delta t$) is decreased proportionately.
    The numerical errors (c) and the network training time (d) of
    the ELM method versus the number of collocation points
    in each direction of the time block.
    Computational domain: $\Omega_1$ ($t\in[0,0.25]$).
    The network architectures are given in
    the legends.
  }
  \label{fg_33}
\end{figure}

Let us next compare the ELM method (Single-Rm-ELM configuration)
with the FEM (classical and high-order FEMs) for solving
the Burgers' equation.
The computational domain is the spatial-temporal
domain $\Omega_1$ ($t\in[0,0.25]$) in the following tests.

With FEM, we employ a second-order
semi-implicit type time integration scheme
to solve the Burgers' equation (see~\cite{DongL2020}). We discretize the
time derivative term in~\eqref{eq_19a} by the second-order
backward differentiation formula (BDF2), treat the diffusion
term implicitly, and treat the nonlinear term explicitly.
The temporally semi-discretized equation in weak form is then solved in space
by the classical (2nd-order) FEM or high-order FEM with Lagrange
elements, which are implemented using the FEniCS library
with uniform 1D interval mesh. 
The FEM simulation parameters include the time step size $\Delta t$
(or the number of time steps), the number of elements in space,
and the element degree for high-order FEM.
In the FEM tests, we vary the number of elements systematically
and simultaneously vary the number of time steps accordingly,
so that the $\Delta t$ and the element size is
increased or decreased proportionately.
The element degree is also varied with the high-order FEM.
As mentioned in Section~\ref{sec:nonl_helm},
the degree parameter in the FEniCS ``Expressions'' for 
the source term and the boundary/initial data in~\eqref{eq_19}
is set to be the element degree plus $4$ with FEM.

With ELM, we employ one time block in the spatial-temporal domain $\Omega_1$.
We consider two neural networks with the architecture 
$[2, M, 1]$ with $M=400$ and $M=500$, respectively.
A fixed $R_m=2.0$ is used to generate the random hidden-layer
coefficients. We employ a uniform set of $Q=Q_1\times Q_1$ collocation points
on the spatial-temporal domain $\Omega_1$, where
$Q_1$ is varied  systematically in the tests.

Figure \ref{fg_33} provides an overview of the errors and
the computational cost of the FEM and the ELM for solving the Burgers' equation.
Figures \ref{fg_33}(a) and (b) show the maximum/rms errors in $\Omega_1$, and
the computation time,
of the classical FEM (2nd-order, degree=1) and the high-order
FEM with Lagrange elements of degree=3, as a function of
the number of elements in the mesh.
Since the element size and the time step size $\Delta t$ are varied proportionately,
these plots equivalently show the relations of the errors (or computation time)
versus  $\Delta t$.
With the classical FEM, the number of elements varies between $20$
and $500$, and $\Delta t$ is varied proportionately between
$1.25E-3$ and $5.0E-5$ in these data.
With the high-order FEM of degree=3, the number of elements
varies between $20$ and $300$, and $\Delta t$ is varied
proportionately between $1.25E-4$ and $8.33E-6$.
We clearly observe a second-order convergence rate of the classical FEM
with respect to the number of elements, and also with respect to $\Delta t$.
With  Lagrange elements of degree $3$, we observe a 4th-order convergence
rate initially (when the number of elements is not large), which then transitions
to a second-order convergence rate when the number of elements increases
beyond a certain point.
The observed change in the convergence rate with high-order FEM is due to the interplay
and the dominance of the spatial truncation error
or the temporal truncation error in different
regimes. When the spatial error dominates, what one observes is
the actual spatial convergence rate (4th-order with element degree $3$).
When the temporal error dominates, on the other hand,
what one observes is the second-order convergence rate with respect to $\Delta t$,
because the spatial error becomes insignificant compared with the temporal error in
this case.
Figure \ref{fg_33}(b) signifies that computational cost of
the high-order FEM is markedly larger than that of the classical FEM.

Figures \ref{fg_33}(c) and (d) depict the maximum/rms errors and the network training time
of the ELM as a function of the number of collocation points in each direction ($Q_1$)
with the two neural networks.
Here the ELM training time is the time obtained
in the graph mode (no autograph/tracing).
We observe the familiar exponential decrease in the ELM errors.
The plot (d) indicates that the ELM network training time grows nearly
linearly with increasing number of collocation points.

\begin{figure}
  \centerline{
    \includegraphics[width=2in]{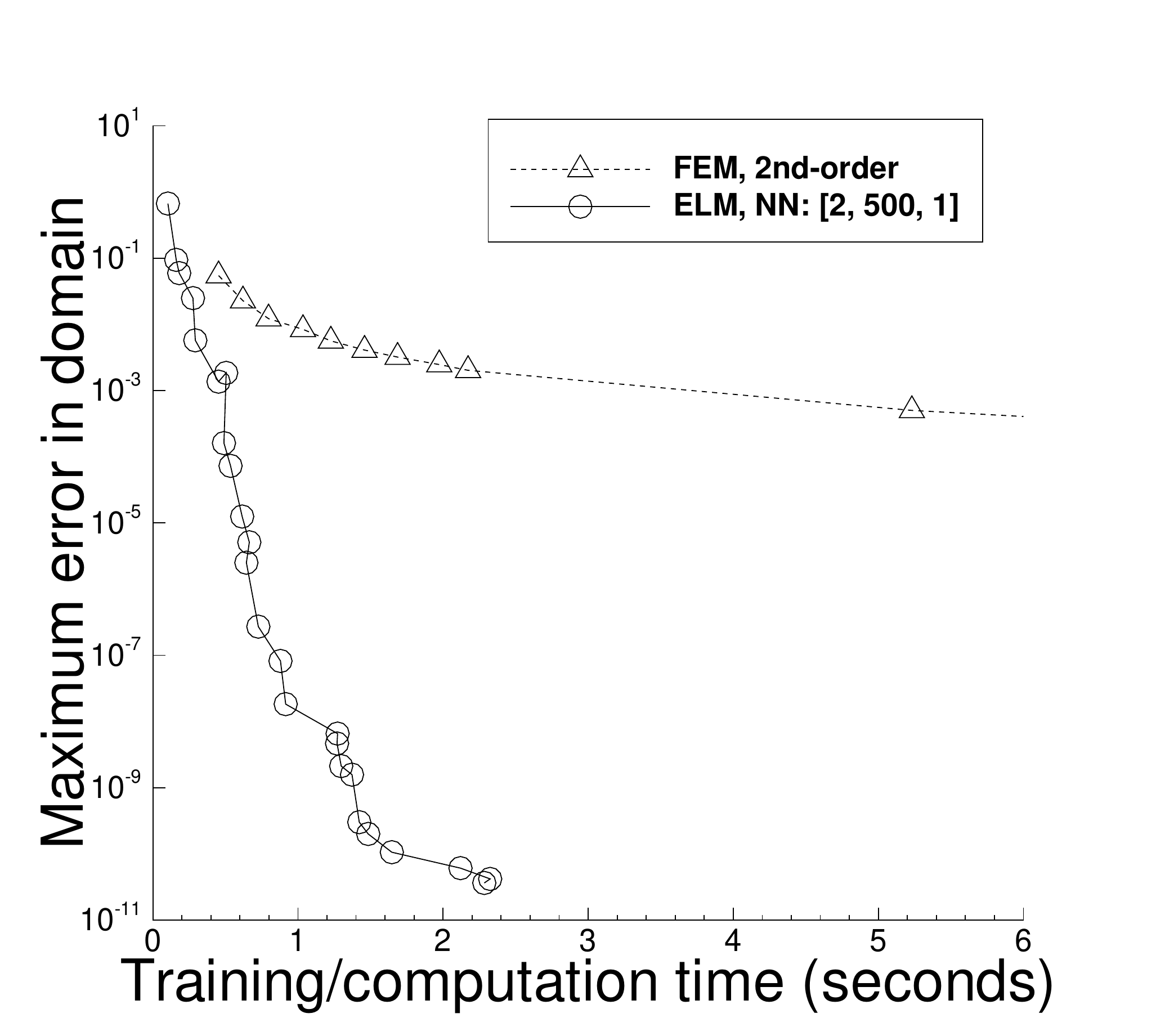}(a)
    \includegraphics[width=2in]{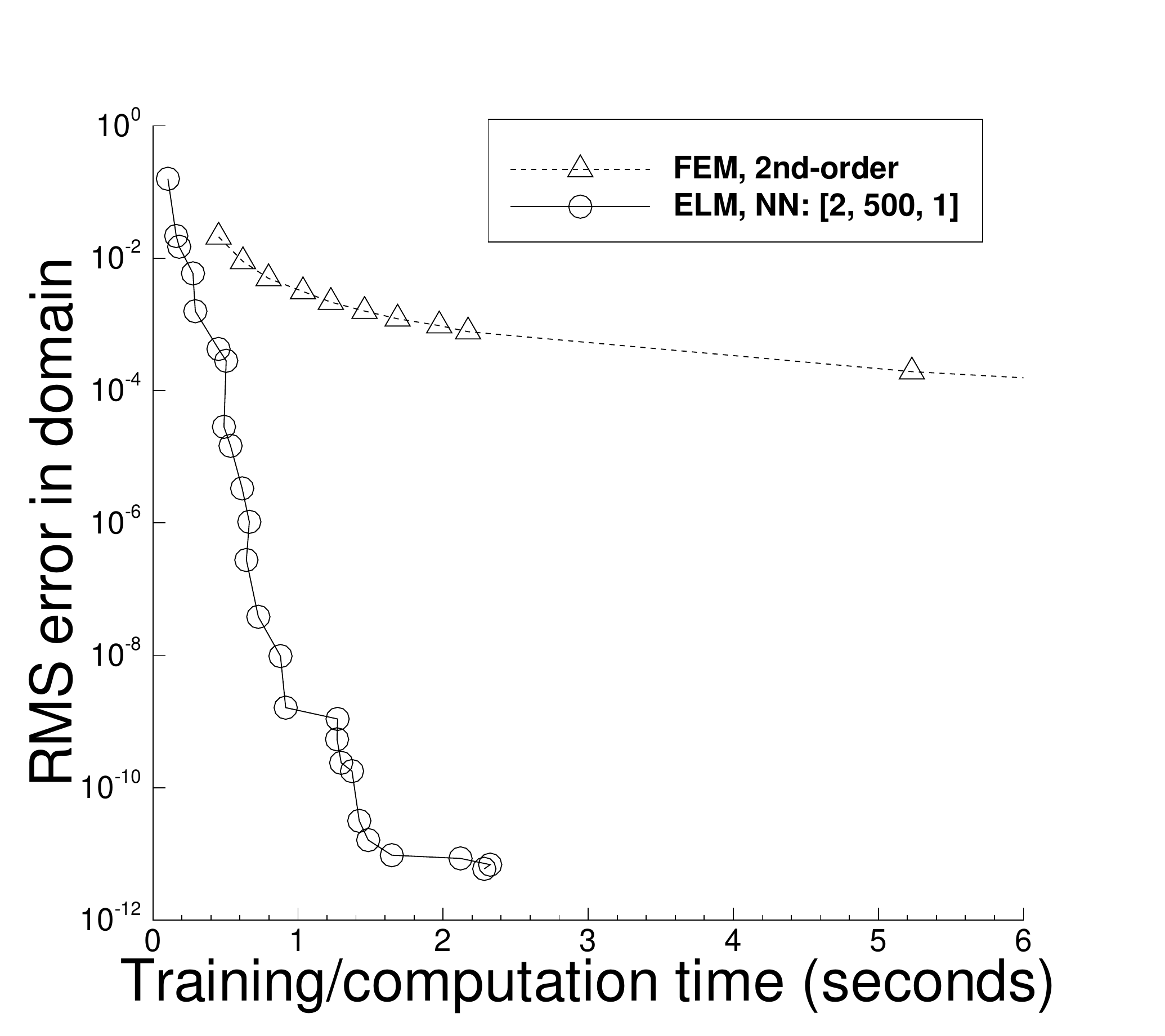}(b)
  }
  \caption{Burgers' equation (comparison between ELM and classical FEM):
    The maximum errors (a) and the rms errors (b) in the spatial-temporal
    domain versus the computational cost (ELM training time, FEM computation time)
    between ELM and the classical FEM. The FEM data
    correspond to those in Figures \ref{fg_33}(a,b) with degree=1. The ELM data
    correspond to those in Figures \ref{fg_33}(c,d) with $M=500$.
  }
  \label{fg_34}
\end{figure}

Figure \ref{fg_34} compares the computational performance between
the ELM method and the classical FEM for the Burgers' equation.
The two plots show the maximum errors and the rms errors in
the overall spatial-temporal domain of the ELM and classical FEM solutions
versus their computational cost (ELM network training time, FEM
computation time).
The FEM data here correspond to those in Figures \ref{fg_33}(a,b)
with degree=1, and the ELM data correspond to those in
Figures \ref{fg_33}(c,d) with the neural network
architecture $[2, 500, 1]$.
One can observe that the ELM method consistently and
far outperforms the classical FEM.

\begin{figure}
  \centerline{
    \includegraphics[width=2in]{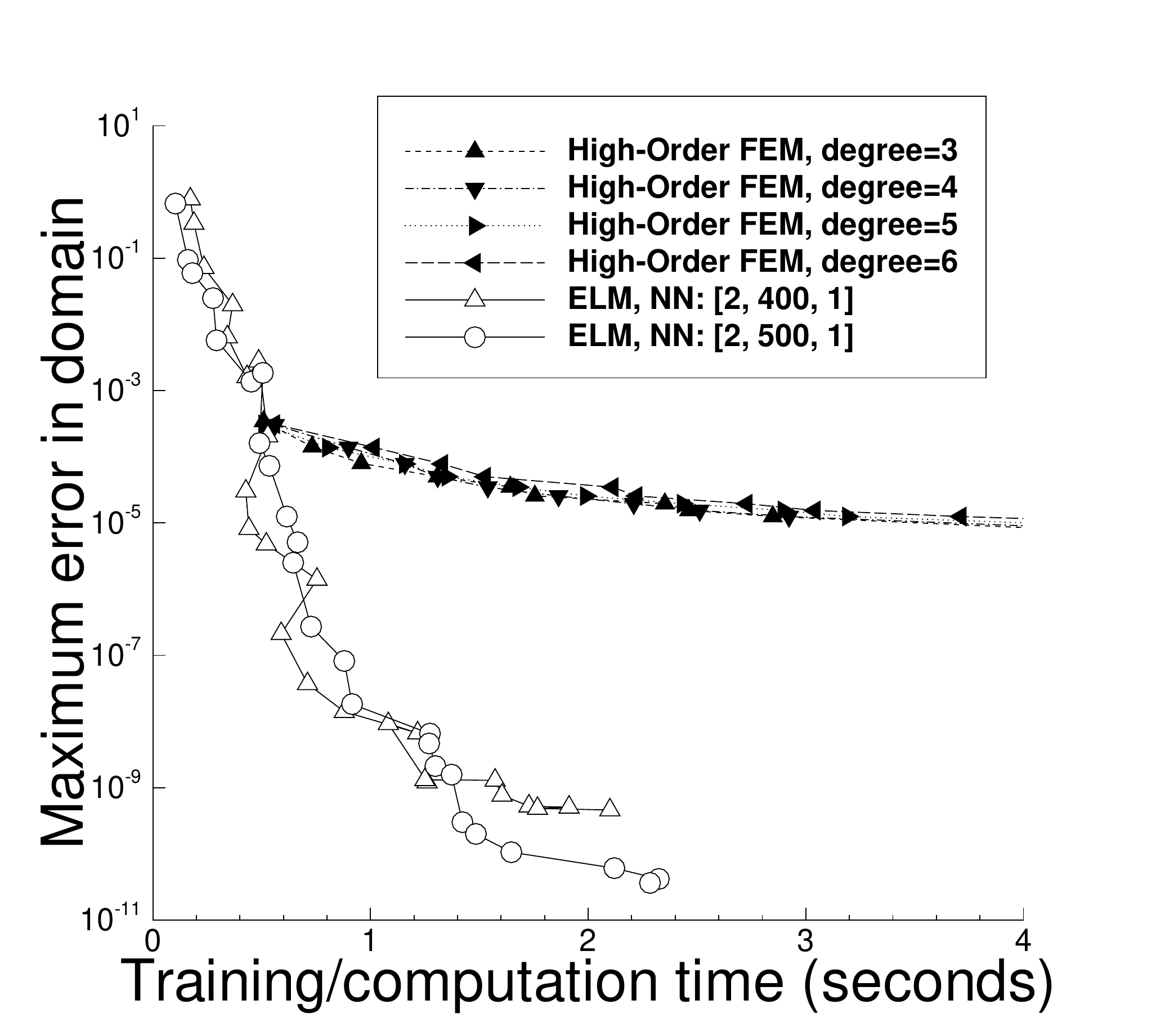}(a)
    \includegraphics[width=2in]{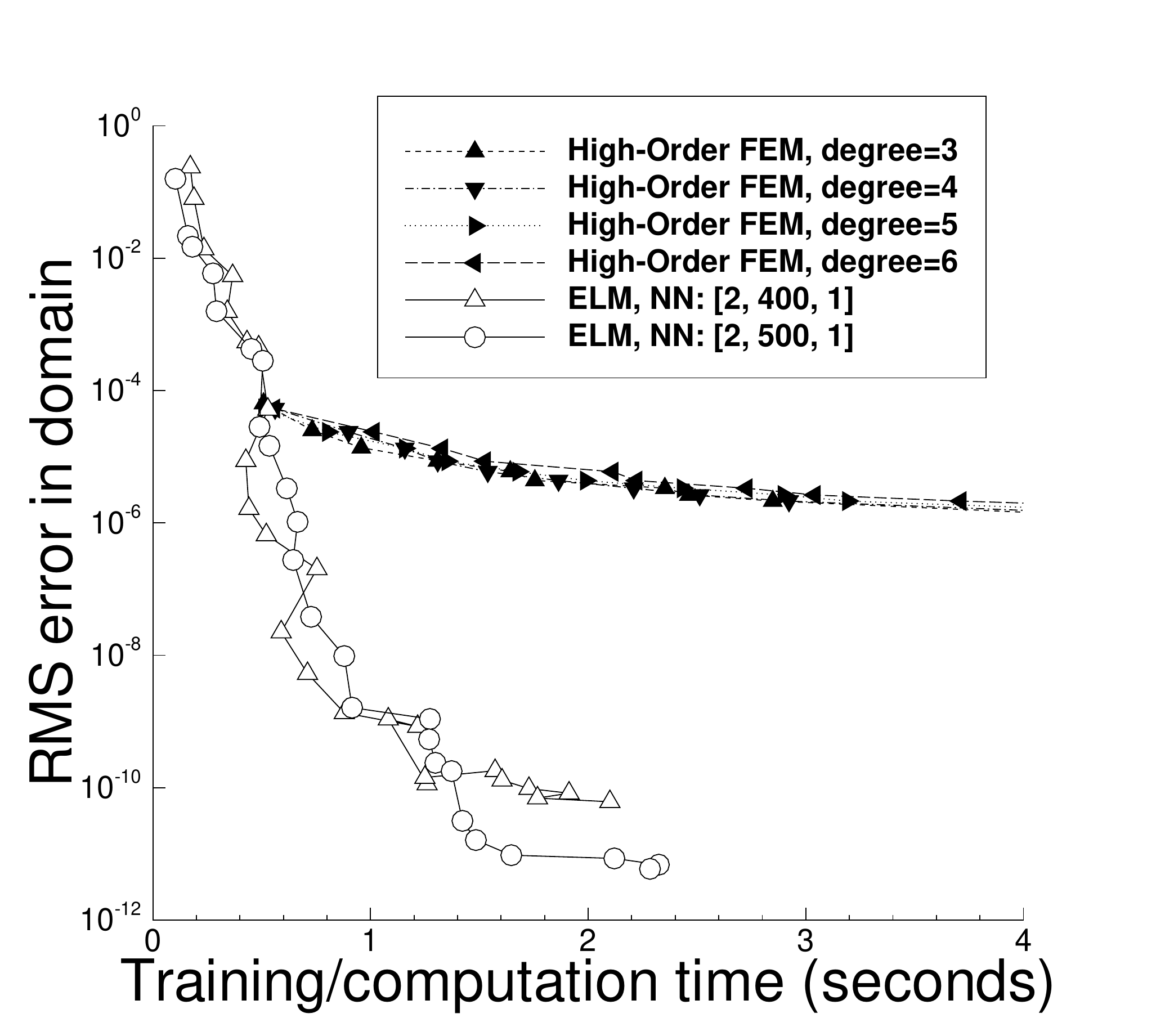}(b)
  }
  \centerline{
    \includegraphics[width=2in]{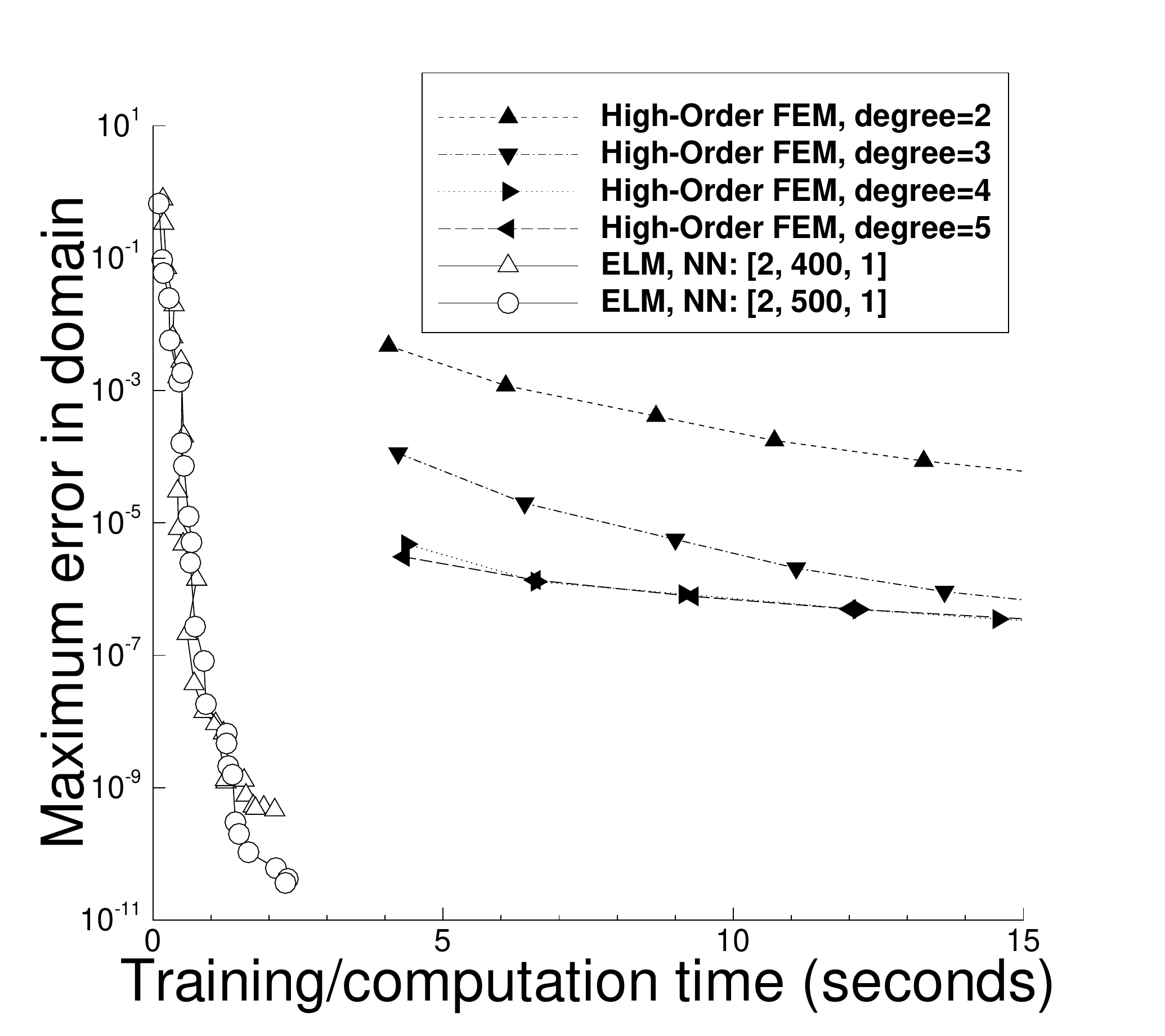}(c)
    \includegraphics[width=2in]{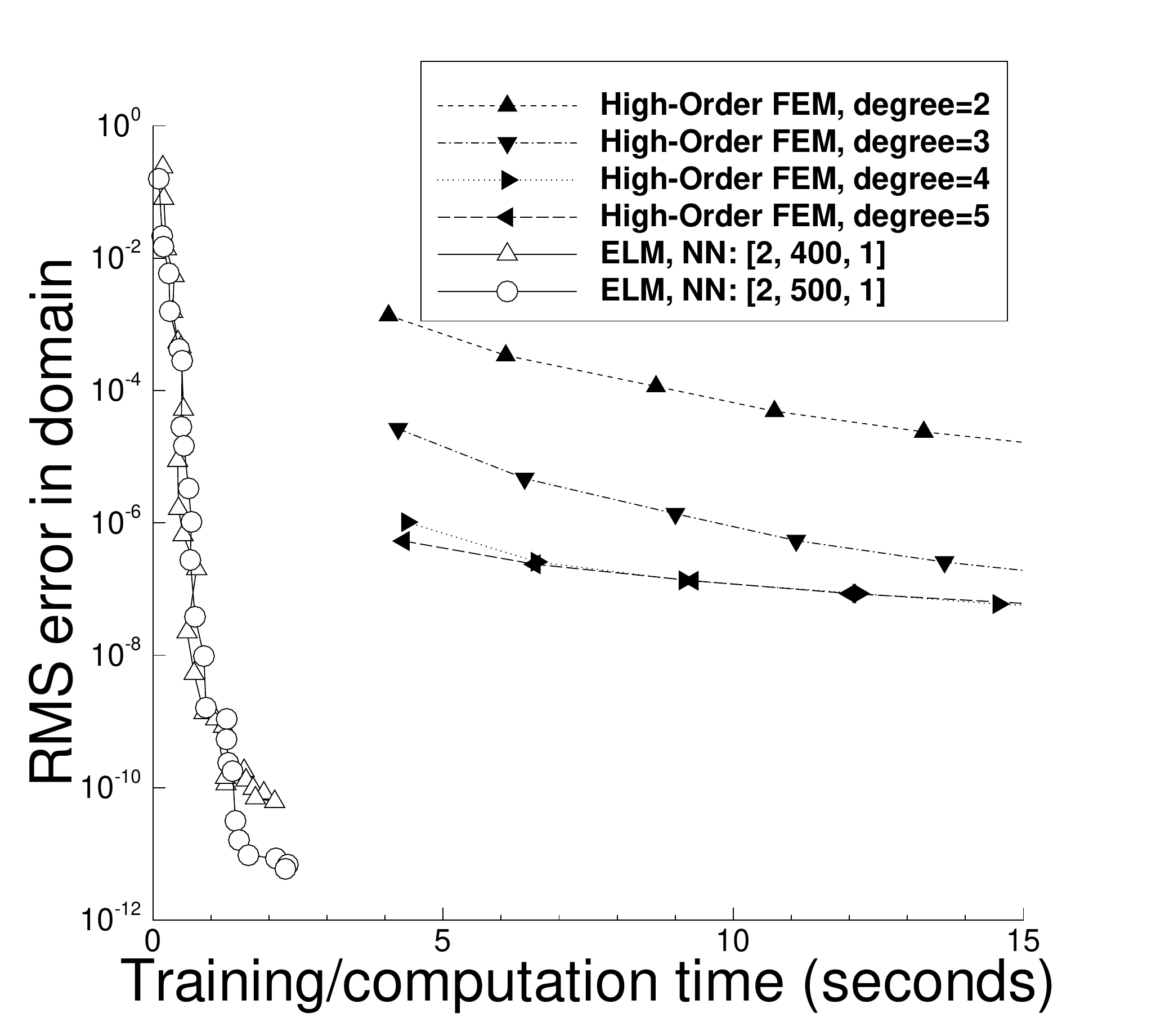}(d)
  }
  \caption{Burgers' equation (comparison between ELM and high-order FEM):
    The maximum error (a,c) and the rms error (b,d) in the
    spatial-temporal domain versus the computational cost (ELM training time,
    FEM computation time) between ELM and high-order FEM.
    Computational domain: $\Omega_1$ ($t\in[0,0.25]$).
    The ELM data correspond to those in Figures \ref{fg_33}(c,d).
    In the FEM tests, the element size and the time step size ($\Delta t$)
    are decreased simultaneously and proportionately for each given element degree.
    In (a,b), the number of FEM elements varies between
    $20$ and $500$, and $\Delta t$ varies between $1.25E-3$ and $5E-5$.
    In (c,d), the number of FEM elements varies between $20$ and $100$, and
    $\Delta t$ varies between $1.25E-4$ and $2.5E-5$.
    The FEM data with degree=3 in (c,d) correspond to a portion of the data in
    Figures \ref{fg_33}(a,b) with degree=3.
  }
  \label{fg_35}
\end{figure}

Figure \ref{fg_35} is a comparison of the computational
performance between the ELM and the high-order FEM
for solving the Burgers' equation.
The computational domain here is the spatial-temporal domain $\Omega_1$
($t\in[0,0.25]$).
Two sets of FEM tests are conducted here, in both of which the element size
and the time step size ($\Delta t$) are reduced simultaneously
and proportionately for a given element degree.
In the first set of tests, the number of elements vary between $20$ and $500$,
and the time step size varies between $\Delta t=1.25E-3$ and $\Delta t=5.0E-5$
accordingly. In the second set, the number of elements vary between $20$ and $100$,
and the time step size varies between $\Delta t=1.25E-4$ and $\Delta t=2.5E-5$
accordingly.
The ELM data correspond to those given in Figures \ref{fg_33}(c,d).

Figures \ref{fg_35}(a) and (b) show the maximum errors and the rms errors,
respectively,
of the high-order FEM and the ELM solutions in the spatial-temporal domain
versus the FEM computation time and the ELM network training time
for the first set of FEM tests.
We observe that the FEM curves corresponding to different element degrees
essentially overlap with one another. This is because in this set of
tests the $\Delta t$ is relatively large and the temporal truncation error
dominates. So in this case increasing the element degree barely
affects the total FEM error.
The data in Figures \ref{fg_35}(a,b) demonstrate that the ELM consistently
outperforms the high-order FEM, and by a considerable margin as
the problem size increases. 

Figures \ref{fg_35}(c) and (d) show the maximum and rms errors
of the high-order FEM versus the computation time for the second set of
FEM tests, together with the ELM errors versus the ELM training time.
We can observe that the FEM error generally decreases as the
element degree increases (e.g.~from $2$ to $3$ and $4$), and that
the FEM error remains essentially the same as the element degree
increases to $4$ and beyond.
This is because the $\Delta t$ is smaller here than in the first set of FEM tests,
and so the spatial truncation error dominates the FEM total error, at least
with the smaller element degrees.
As the element degree increases, the spatial truncation error
is reduced rapidly and the temporal truncation error gradually becomes dominant.
At this point, further increase in the FEM element degree will not notably
affect the total FEM error.
Because of the smaller $\Delta t$ values in the second set of FEM tests,
a significantly larger number of time steps need to be
computed in the FEM simulations, resulting in an overall increased FEM
computation time. 
We can observe from Figures \ref{fg_35}(c,d) that
the ELM way outperforms the high-order FEM for this set of tests.


The above comparisons show that the ELM method combined with block time
marching is effective and efficient for solving time dependent PDEs.
It is considerably more efficient than the FEM (classical and high-order FEMs)
combined with the commonly-used second-order time stepping scheme,
in terms of the accuracy and the incurred computational cost.
These comparisons are conducted on a relatively small
temporal domain ($t\in [0,0.25]$).
When the temporal dimension of the spatial-temporal domain increases
(i.e.~for longer-time simulations), the advantage of the ELM combined
with block time marching becomes even more prominent.

\section{Concluding Remarks}
\label{sec:summary}

In extreme learning machines (ELM) the hidden-layer coefficients
of the neural network
are pre-set to uniform random values generated on the interval $[-R_m,R_m]$,
where the maximum magnitude $R_m$ of the random coefficients
is a user-provided constant (hyperparameter),
and the output-layer coefficients are trained by a linear or nonlinear
least squares computation~\cite{DongL2020}.
More accurate ELM results have been observed to be
associated with a range of moderate
values for $R_m$ (see~\cite{DongL2020}).
In the current paper, we have presented a method for computing
the optimal or near-optimal value for the $R_m$ constant
for solving partial differential equations (PDE).
The presented method is based on the differential evolution algorithm,
and seeks the optimal $R_m$ by minimizing the norm of the residual
vector of the linear or nonlinear algebraic system that results
from the ELM representation of the PDE solution and that corresponds to the 
ELM least squares solution to the system.
This  method  amounts to
a pre-processing procedure. It determines a near optimal value
for $R_m$, which can be used in ELM for
solving linear or nonlinear PDEs.
In practice, we observe that any value in a neighborhood of the returned
$R_{m0}$ from the method can be used in the ELM
simulation and leads to comparable accuracy.
This is because, as shown in~\cite{DongL2020},
there is usually a range of $R_m$ values that lead to
good accuracy with ELM.

We have investigated two configurations in ELM for setting the random hidden-layer
coefficients, Single-Rm-ELM and Multi-Rm-ELM.
The Single-Rm-ELM configuration corresponds to the conventional ELM,
in which the weight/bias coefficients for all the hidden layers
of the neural network are assigned to random values generated on
$[-R_m,R_m]$, with a single $R_m$ constant.
In the Multi-Rm-ELM configuration, the weight/bias coefficients
in the $l$-th hidden layer, $1\leqslant l\leqslant L-1$ with $(L-1)$ denoting
the total number of hidden layers,
are set to random values generated on $[-R_m^{(l)},R_m^{(l)}]$.
Therefore, the maximum magnitudes of the random coefficients in different
hidden layers may be different in Multi-Rm-ELM, and they
are characterized by the vector
$\mbs R_m=(R_m^{(1)},R_m^{(2)},\dots,R_m^{(L-1)})$.

We have computed the optimal $R_m$ in Single-Rm-ELM and the optimal
$\mbs R_m$ in  Multi-Rm-ELM
using the method developed here for a number of linear and nonlinear PDEs.
We have the following observations
about the Single-Rm-ELM and Multi-Rm-ELM and
their respective optimum $R_{m0}$ and $\mbs R_{m0}$:
\begin{itemize}
\item
  The optimum $R_{m0}$ of Single-Rm-ELM is largely independent of
  the number of collocation points. $R_{m0}$  only weakly depends
  on the number of training parameters for neural networks with two or
  more hidden layers. For neural networks
  with a single hidden layer, the dependence of $R_{m0}$ on the number of
  training parameters ($M$) is stronger, and $R_{m0}$ tends to increase
  with increasing $M$ (when $M$ is not very small).

\item
  $R_{m0}$ generally decreases with increasing number of hidden layers
  in the neural network. There is a fairly large drop in
  $R_{m0}$ from a single hidden layer to two hidden layers.
  Beyond two hidden layers,
  the decrease in $R_{m0}$ is only slight and can oftentimes be negligible
  as the number of hidden layers further increases.

\item
  $R_{m0}$ has only a very weak (oftentimes negligible)
  dependence on the number of nodes
  in the hidden layers preceding the last hidden layer.
  $R_{m0}$ tends to decrease slightly with increasing width of
  the preceding hidden layers. 

\item
  The optimum $\mbs R_{m0}$ of Multi-Rm-ELM tends to exhibit a 
  relationship that is not quite regular
  with respect to the ELM simulation parameters.
  However, the trend exhibited by $\mbs R_{m0}$ appears reminiscent of
  what has been observed about the optimum $R_{m0}$ of Single-Rm-ELM.
  For example, $\mbs R_{m0}$ depends only weakly on (or nearly independent of)
  the number of collocation points, and appears to generally increase
  with increasing number of training parameters.

\item
  The Multi-Rm-ELM configuration with $\mbs R_m=\mbs R_{m0}$
  leads to consistently more accurate simulation
  results than the Single-Rm-ELM configuration with $R_m=R_{m0}$,
  under otherwise identical conditions.
  On the other hand, the $\mbs R_{m0}$ computation cost in Multi-Rm-ELM
  is generally higher
  than the $R_{m0}$ computation cost in Single-Rm-ELM.
  
\end{itemize}


We have made several improvements to the implementation of the ELM method
in the current work. The most crucial change lies in the adoption of
a forward-mode auto-differentiation for computing the differential operators
associated with the output fields of the last hidden layer of the
neural network. This is implemented using the ``ForwardAccumulator'' in
Tensorflow. In contrast, these differential operators were computed
by the default reverse-mode auto-differentiation (``GradientTape'') of
Tensorflow in the previous work~\cite{DongL2020}. 
These improvements result in a significant boost to the computational
performance of ELM.

Another aspect of the current contribution is a systematic comparison
of the computational performance between the current ELM method and the classical
and high-order finite element methods (FEM) for solving linear and nonlinear
PDEs. The ELM method employs the improved implementation
and the near optimal $R_m$ obtained from the differential
evolution algorithm.
The classical FEM (second-order, linear elements) and the high-order FEM
are implemented based on the FEniCS library by employing the
Lagrange elements of degree one or higher degrees. 
By looking into the ELM/FEM accuracy and their computational cost
(FEM computation time, ELM network training time) for a number of linear/nonlinear
PDEs, we have the following observations:
\begin{itemize}
\item
  For stationary (i.e.~time-independent) PDEs, the ELM far outperforms
  the classical FEM if the problem size is not very small.
  For very small problem sizes (small FEM mesh, small number of ELM training
  collocation points), the computational performance of the ELM and the classical FEM is
  close, with the classical FEM a little better.

\item
  For stationary PDEs, there is a crossover point in the relative performance
  between ELM and the high-order FEM
  with respect to the problem size (FEM mesh size or element degree,
  ELM collocation points).
  For smaller problem sizes (smaller FEM mesh in h-type refinements or smaller
  element degree in p-type refinement; smaller number of ELM collocation points),
  the ELM and high-order FEM are close in computational performance,
  with the high-order FEM appearing slightly better.
  As the problem size becomes larger, the ELM markedly outperforms the high-order FEM.

\item
  For time-dependent PDEs, the ELM method combined with the block time
  marching scheme consistently and significantly outperforms both the classical
  and the high-order FEMs
  (combined with a time-stepping scheme).
  
\end{itemize}

These performance comparisons demonstrate that
the neural network-based ELM method is computationally competitive
compared with not only the classical second-order FEM
but also the high-order FEM based on high-order polynomials.
The ELM exceeds  the classical FEM
by a considerable margin in terms of computational performance.
The ELM method delivers a comparable performance to high-order FEM for
smaller problem sizes.
For larger problem sizes, the ELM performance exceeds the performance of
high-order FEM.
The ELM method is more efficient than or as efficient as  the high-order FEM.

Can artificial neural networks provide a competitive method
for scientific computing and in particular for numerical PDEs?
Can one devise a neural network-based method for approximating PDEs
that can outcompete the traditional
numerical techniques?
These questions have motivated the current effort and also
our recent work in~\cite{DongL2020}.
The ELM type methods developed in~\cite{DongL2020} and the current work
for solving PDEs seek a different approach from
the existing DNN-based PDE solvers, in order to achieve high
accuracy and competitive computational performance.
Our methods attempt to exploit the randomization
of a subset of the network weights
in order to simplify the the optimization task of the network training, and
more importantly we train the neural network by a linear or nonlinear
least squares computation (rather than the gradient descent type algorithms).
The exponential convergence behavior (for smooth solutions)
with respect to the number of
training data points and training parameters and the high accuracy
exhibited by these methods are reminiscent of the traditional
high-order methods such as the spectral, spectral element
or hp-finite element type
techniques. 

The current work and our recent work in~\cite{DongL2020} provide
strong evidence that
the answer to the above questions seems indeed to be positive. 
Our previous work~\cite{DongL2020}
demonstrates that the ELM type method can be more competitive than
or as competitive as the classical second-order FEM.
The importance of the current work lies in that it further shows that
the ELM type method can be more competitive than or as competitive as 
the high-order FEM in terms of the accuracy and computational cost.
These studies collectively instigate a neural network-based
accurate, efficient and competitive technique for numerical approximation
of PDEs in computational science and engineering applications.


\section*{Acknowledgement}
This work was partially supported by
NSF (DMS-2012415, DMS-1522537).

\bibliographystyle{plain}
\bibliography{elm1,elm,mypub,dnn1,dnn,sem,obc}

\begin{thebibliography}{10}

\bibitem{Alabaetal2019}
P.A. Alaba, S.I. Popoola, L.~Olatomiwa, M.B. Akanle, O.S. Ohunakin, E.~Adetiba,
  O.D. Alex, A.A.A. Atayero, and W.M.A.W. Daud.
\newblock Towards a more efficient and cost-sensitive extreme learning machine:
  a state-of-the-art review of recent trend.
\newblock {\em Neurocomputing}, 350:70--90, 2019.

\bibitem{BaydinPRS2018}
A.G. Baydin, B.A. Pearlmutter, A.A. Radul, and J.M. Siskind.
\newblock Automatic differentiation in machine learning: a survey.
\newblock {\em J. Mach. Learn. Res.}, 18:1--43, 2018.

\bibitem{CaiCLL2020}
Z.~Cai, J.~Chen, M.~Liu, and X.~Liu.
\newblock Deep least-squares methods: an unsupervised learning-based numerical
  method for solving elliptic {PDE}s.
\newblock {\em Journal of Computational Physics}, 420:109707, 2020.

\bibitem{CalabroFS2021}
F.~Calabro, G.~Fabiani, and C.~Siettos.
\newblock Extreme learning machine collocation for the numerical solution of
  elliptic {PDEs} with sharp gradients.
\newblock {\em Computer Methods in Applied Mechanics and Engineering},
  387:114188, 2021.

\bibitem{Cotter1990}
N.E. Cotter.
\newblock The stone-weierstrass theorem and its application to neural networks.
\newblock {\em IEEE Transactions on Neural Networks}, 4:290--295, 1990.

\bibitem{Courant1943}
R.L. Courant.
\newblock Variational methods for the solution of problems of equilibrium and
  vibration.
\newblock {\em Bulletin of the American Mathematical Society}, 49:1--23, 1943.

\bibitem{Dong2015clesobc}
S.~Dong.
\newblock A convective-like energy-stable open boundary condition for
  simulations of incompressible flows.
\newblock {\em Journal of Computational Physics}, 302:300--328, 2015.

\bibitem{Dong2018}
S.~Dong.
\newblock Multiphase flows of {N} immiscible incompressible fluids: a
  reduction-consistent and thermodynamically-consistent formulation and
  associated algorithm.
\newblock {\em Journal of Computational Physics}, 361:1--49, 2018.

\bibitem{DongK2003}
S.~Dong and G.E. Karniadakis.
\newblock P-refinement and p-rethreads.
\newblock {\em Computer Methods in Applied Mechanics and Engineering},
  192(19):2191--2201, 2003.

\bibitem{DongL2020}
S.~Dong and Z.~Li.
\newblock Local extreme learning machines and domain decomposition for solving
  linear and nonlinear partial differential equations.
\newblock {\em Computer Methods in Applied Mechanics and Engineering},
  387:114129, 2021.
\newblock (also arXiv:2012.02895).

\bibitem{DongL2021}
S.~Dong and Z.~Li.
\newblock A modified batch intrinsic plascity method for pre-training the
  random coefficients of extreme learning machines.
\newblock {\em Journal of Computational Physics}, 445:110585, 2021.
\newblock (also arXiv:2103.08042).

\bibitem{DongN2020}
S.~Dong and N.~Ni.
\newblock A method for representing periodic functions and enforcing exactly
  periodic boundary conditions with deep neural networks.
\newblock {\em Journal of Computational Physics}, 435:110242, 2021.

\bibitem{DongS2012}
S.~Dong and J.~Shen.
\newblock A time-stepping scheme involving constant coefficient matrices for
  phase field simulations of two-phase incompressible flows with large density
  ratios.
\newblock {\em Journal of Computational Physics}, 231:5788--5804, 2012.

\bibitem{DwivediS2020}
V.~Dwivedi and B.~Srinivasan.
\newblock Physics informed extreme learning machine (pielm) $-$ a rapid method
  for the numerical solution of partial differential equations.
\newblock {\em Neurocomputing}, 391:96--118, 2020.

\bibitem{EY2018}
W.~E and B.~Yu.
\newblock The deep {R}itz method: a deep learning-based numerical algorithm for
  solving variational problems.
\newblock {\em Communications in Mathematics and Statistics}, 6:1--12, 2018.

\bibitem{EndresSF2018}
S.C. Endres, C.~Sandrock, and W.W. Focke.
\newblock A simplicial homology algorithm for {L}ipschitz optimization.
\newblock {\em Journal of Global Optimization}, 72:181--217, 2018.

\bibitem{FabianiCRS2021}
G.~Fabiani, F.~Calabro, L.~Russo, and C.~Siettos.
\newblock Numerical solution and bifurcation analysis of nonlinear partial
  differential equations with extreme learning machines.
\newblock {\em Journal of Scientific Computing}, 89:44, 2021.

\bibitem{FreireRB2020}
A.L. Freire, A.R. Rocha-Neto, and G.A. Barreto.
\newblock On robust randomized neural networks for regression: a comprehensive
  review and evaluation.
\newblock {\em Neural Computing and Applications}, 32:16931--16950, 2020.

\bibitem{GillMW2021}
P.E. Gill, W.~Murray, and M.H. Wright.
\newblock {\em Numerical Linear Algebra and Optimization}.
\newblock SIAM, 2021.

\bibitem{GoodfellowBC2016}
I.~Goodfellow, Y.~Bengio, and A.~Courville.
\newblock {\em Deep Learning}.
\newblock The MIT Press, 2016.

\bibitem{Haykin1999}
S.~Haykin.
\newblock {\em Neural Networks: A Comprehensive Foundation}.
\newblock Prentice Hall, 1999.

\bibitem{HeX2019}
J.~He and J.~Xu.
\newblock {MgNet}: A unified framework for multigrid and convolutional neural
  network.
\newblock {\em Science China Mathematics}, 62:1331--1354, 2019.

\bibitem{HornikSW1989}
K.~Hornik, M.~Stinchcombe, and H.~White.
\newblock Multilayer feedforward networks are universal approximators.
\newblock {\em Neural Networks}, 2:359--366, 1989.

\bibitem{HornikSW1990}
K.~Hornik, M.~Stinchcombe, and H.~White.
\newblock Universal approximation of an unknown mapping and its derivatives
  using multilayer feedforward networks.
\newblock {\em Neural Networks}, 3:551--560, 1990.

\bibitem{HuangHSY2015}
G.~Huang, G.B. Huang, S.~Song, and K.~You.
\newblock Trends in extreme learning machines: a review.
\newblock {\em Neural Networks}, 61:32--48, 2015.

\bibitem{HuangZS2006}
G.-B. Huang, Q.-Y. Zhu, and C.-K. Siew.
\newblock Extreme learning machine: theory and applications.
\newblock {\em Neurocomputing}, 70:489--501, 2006.

\bibitem{HuangCS2006}
G.B. Huang, L.~Chen, and C.-K. Siew.
\newblock Universal approximation using incremental constructive feedforward
  networks with random hidden nodes.
\newblock {\em IEEE Transactions on Neural Networks}, 17:879--892, 2006.

\bibitem{JagtapKK2020}
A.D. Jagtap, E.~Kharazmi, and G.E. Karniadakis.
\newblock Conservative physics-informed neural networks on discrete domains for
  conservation laws: applications to forward and inverse problems.
\newblock {\em Computer Methods in Applied Mechanics and Engineering},
  365:113028, 2020.

\bibitem{Karniadakisetal2021}
G.E. Karniadakis, G.~Kevrekidis, L.~Lu, P.~Perdikaris, S.~Wang, and L.~Yang.
\newblock Physics-informed machine learning.
\newblock {\em Nature Reviews Physics}, 3:422--440, 2021.

\bibitem{KarniadakisS2005}
G.E. Karniadakis and S.J. Sherwin.
\newblock {\em Spectral/hp element methods for computational fluid dynamics,
  2nd edn.}
\newblock Oxford University Press, 2005.

\bibitem{KharazmiZK2019}
E.~Kharazmi, Z.~Zhang, and G.E. Karniadakis.
\newblock Variational physics-informed neural networks for solving partial
  differential equations.
\newblock {\em arXiv:1912.00873}, 2019.

\bibitem{KrishnapriyanGZKM2021}
A.S. Krishnapriyan, A.~Gholami, S.~Zhe, R.M. Kirby, and M.W. Mahoney.
\newblock Characterizing possible failure modes in physics-informed neural
  networks.
\newblock {\em arXiv:2109.01050}, 2021.

\bibitem{LagarisLF1998}
I.E. Lagaris, A.C. Likas, and D.I. Fotiadis.
\newblock Artificial neural networks for solving ordinary and partial
  differential equations.
\newblock {\em IEEE Transactions on Neural Networks}, 9:987--1000, 1998.

\bibitem{LagarisLP2000}
I.E. Lagaris, A.C. Likas, and D.G. Papageorgiou.
\newblock Neural-network methods for boundary value problems with irregular
  boundaries.
\newblock {\em IEEE Transactions on Neural Networks}, 11:1041--1049, 2000.

\bibitem{LeeK1990}
H.~Lee and I.~Kang.
\newblock Neural algorithms for solving differential equations.
\newblock {\em Journal of Computational Physics}, 91:110--117, 1990.

\bibitem{LiTWL2020}
K.~Li, K.~Tang, T.~Wu, and Q.~Liao.
\newblock {D3M}: A deep domain decomposition method for partial differential
  equations.
\newblock {\em IEEE Access}, 8:5283--5294, 2020.

\bibitem{Li1996}
X.~Li.
\newblock Simultaneous approximations of mulvariate functions and their
  derivatives by neural networks with one hidden layer.
\newblock {\em Neurocomputiing}, 12:327--343, 1996.

\bibitem{LinYD2019}
L.~Lin, Z.~Yang, and S.~Dong.
\newblock Numerical approximation of incompressible {N}avier-{S}tokes equations
  based on an auxiliary energy variable.
\newblock {\em Journal of Computational Physics}, 388:1--22, 2019.

\bibitem{MeadeF1994}
A.J. Meade and A.A. Fernandez.
\newblock The numerical solution of linear ordinary differential equations by
  feedforward neural networks.
\newblock {\em Math. Comput. Modeling}, 19(12):1--25, 1994.

\bibitem{MeadeF1994b}
A.J. Meade and A.A. Fernandez.
\newblock Solution of nonlinear ordinary differential equations by feedforward
  neural networks.
\newblock {\em Math. Comput. Modeling}, 20(9):19--44, 1994.

\bibitem{PanghalK2020}
S.~Panghal and M.~Kumar.
\newblock Optimization free neural network approach for solving ordinary and
  partial differential equations.
\newblock {\em Engineering with Computers}, Early Access, February 2020.

\bibitem{RaissiPK2019}
M.~Raissi, P.~Perdikaris, and G.E. Karniadakis.
\newblock Physics-informed neural networks: a deep learning framework for
  solving forward and inverse problems involving nonlinear partial differential
  equations.
\newblock {\em Journal of Computational Physics}, 378:686--707, 2019.

\bibitem{RuddF2015}
K.~Rudd and S.~Ferrari.
\newblock A constrained integration ({CINT}) approach to solving partial
  differential equations using artificial neural networks.
\newblock {\em Neurocomputing}, 155:277--285, 2015.

\bibitem{Samaniegoetal2020}
E.~Samanaiego, C.~Anitescu, S.~Goswami, V.M. Nguyen-Thanh, H.~Guo, K.~Hamdia,
  X.~Zhuang, and T.~Rabczuk.
\newblock An energy approach to the solution of partial differential equations
  in computational mechanics via machine learning: concepts, implementation and
  applications.
\newblock {\em Computer Methods in Applied Mechanics and Engineering},
  362:112790, 2020.

\bibitem{ScardapaneW2017}
S.~Scardapane and D.~Wang.
\newblock Randomness in neural networks: an overview.
\newblock {\em WIREs Data Mining Knowl. Discov.}, 7:e1200, 2017.

\bibitem{SirignanoS2018}
J.~Sirignano and K.~Spoliopoulos.
\newblock {DGM}: A deep learning algorithm for solving partial differential
  equations.
\newblock {\em Journal of Computational Physics}, 375:1339--1364, 2018.

\bibitem{StornP1997}
R.~Storn and K.~Price.
\newblock Differential evolution -- {A} simple and efficient heuristic for
  global optimization over continuous spaces.
\newblock {\em Journal of Global Optimization}, 11:341--359, 1997.

\bibitem{SzaboB1991}
B.~Szabo and I.~Babushka.
\newblock {\em Finite Element Analysis}.
\newblock John Wiley \& Sons, Inc., 1991.

\bibitem{TangDH2015}
J.~Tang, C.~Deng, and G.B. Huang.
\newblock Extreme learning machine for multilayer perceptron.
\newblock {\em IEEE Transactions on neural networks and learning systems},
  32(2):392--404, 2015.

\bibitem{TangWL2021}
K.~Tang, X.~Wan, and Q.~Liao.
\newblock Adaptive deep density estimation for fokker-planck equations.
\newblock {\em arXiv:2103.11181}, 2021.

\bibitem{TisseraM2016}
M.D. Tissera and M.D. McDonnell.
\newblock Deep extreme learning machines: supervised autoencoding architecture
  for classification.
\newblock {\em Neurocomputing}, 174:42--49, 2016.

\bibitem{WanW2020}
X.~Wan and S.~Wei.
\newblock {VAE-KRnet} and its applications to variational {B}ayes.
\newblock {\em arXiv:2006.16431}, 2020.

\bibitem{WangYP2020}
S.~Wang, X.~Yu, and P.~Perdikaris.
\newblock When and why {PINN}s fail to train: a neural tangent kernel
  perspective.
\newblock {\em arXiv:2007.14527}, 2020.

\bibitem{WangL2020}
Y.~Wang and G.~Lin.
\newblock Efficient deep learning techniques for multiphase flow simulation in
  heterogeneous porous media.
\newblock {\em Journal of Computational Physics}, 401:108968, 2020.

\bibitem{Werbos1974}
P.J. Werbos.
\newblock Beyond regression: new tools for prediction and alaysis in the
  behavioral sciences.
\newblock {\em PhD Thesis, Harvard Univeristy, Cambridge, MA}, 1974.

\bibitem{WinovichRL2019}
N.~Winovich, K.~Ramani, and G.~Lin.
\newblock Conv{PDE-UQ}: Convolutional neural networks with quantified
  uncertainty for heterogeneous elliptic partial differential equations on
  varied domains.
\newblock {\em Journal of Computational Physics}, 394:263--279, 2019.

\bibitem{YangD2020}
Z.~Yang and S.~Dong.
\newblock A roadmap for discretely energy-stable schemes for dissipative
  systems based on a generalized auxiliary variable with guaranteed positivity.
\newblock {\em Journal of Computational Physics}, 404:109121, 2020.
\newblock (also arXiv:1904.00141).

\bibitem{YentisZ1996}
R.~Yentis and M.E. Zaghoul.
\newblock {VLSI} implementation of locally connected neural network for solving
  partial differential equations.
\newblock {\em IEEE Trans. Circuits Syst. I}, 43:687--690, 1996.

\bibitem{YuKK2017}
Y.~Yu, R.M. Kirby, and G.E. Karniadakis.
\newblock Spectral element and hp methods.
\newblock {\em Encyclopedia of Computational Mechanics, John Wiley and Sons,
  NY}, 1:1--43, 2017.

\bibitem{ZangBYZ2020}
Y.~Zang, G.~Bao, X.~Ye, and H.~Zhou.
\newblock Weak adversarial networks for high-dimensional partial differential
  equations.
\newblock {\em Journal of Computational Physics}, 411:109409, 2020.

\bibitem{ZhengD2011}
X.~Zheng and S.~Dong.
\newblock An eigen-based high-order expansion basis for structured spectral
  elements.
\newblock {\em Journal of Computational Physics}, 230:8573--8602, 2011.

\end{thebibliography}

\end{document}